\documentclass{stefano}

\usepackage{thesis_commands}

\begin{document}

\begin{titlepage}

\begin{center}

\begin{minipage}{0.15\textwidth}
\includegraphics[width=\textwidth]{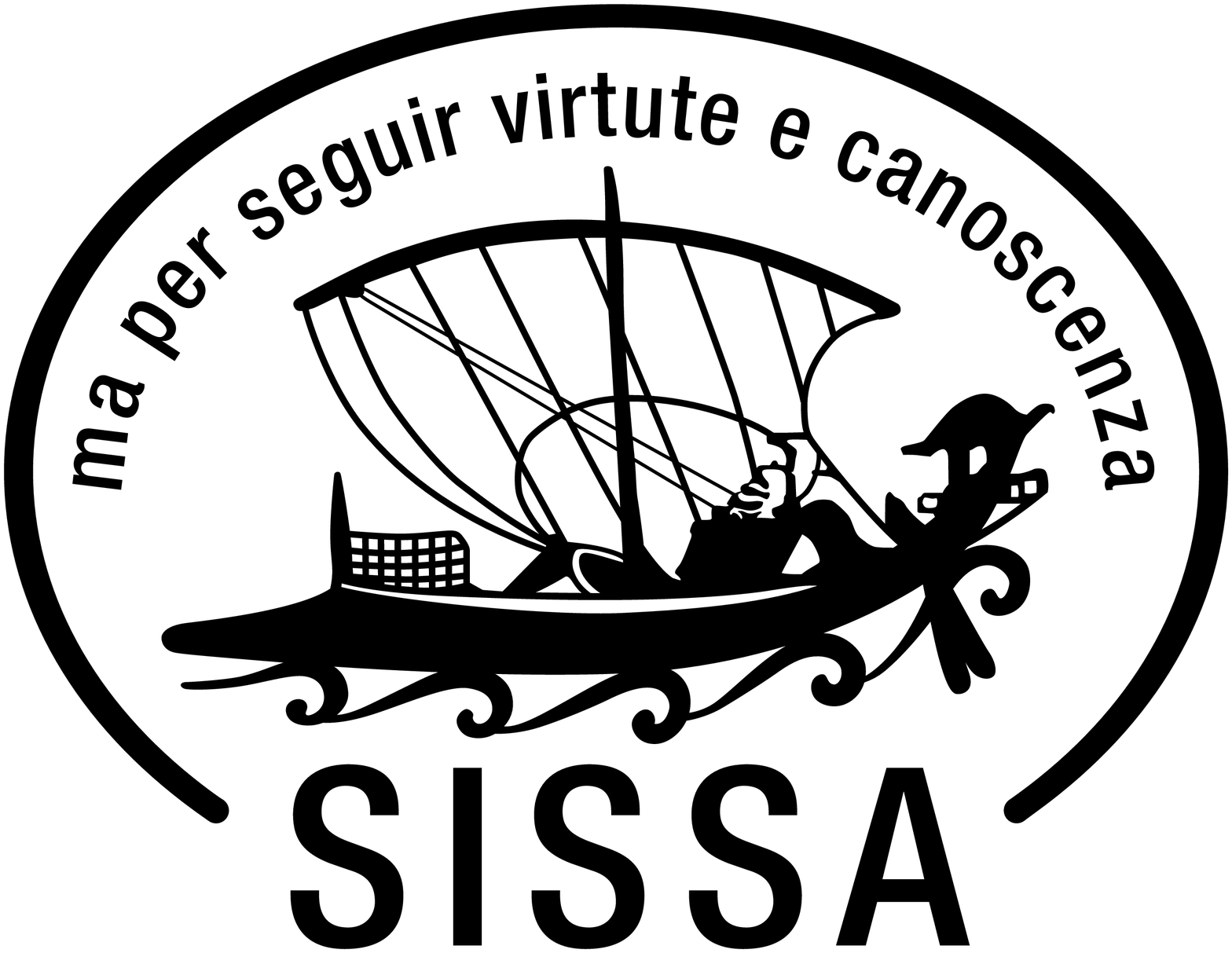} 
\end{minipage}\hspace{0.02\textwidth}
\begin{minipage}{0.82\textwidth}
 \textsc{\Large Scuola Internazionale Superiore di Studi Avanzati}\\[1ex]
\textsc{\Large International School for Advanced Studies}
\end{minipage}
\\[5cm]

\textsc{\LARGE thesis submitted for the degree of\\[1ex]Doctor Philosophi\ae}\\[2.5cm]

\setlength{\fboxrule}{0.4pt}
\framebox{
\begin{minipage}{\textwidth}
\centering
\huge \bfseries
\vspace{2ex}
Analogue~gravitational~phenomena\\[1ex]in Bose--Einstein~condensates
\vspace{2ex}
\end{minipage}
}\\[5cm]

\LARGE
\begin{minipage}{0.58\textwidth}
\textsc{Candidate}\\[1ex]
{\bfseries Stefano Finazzi}
\end{minipage}
\begin{minipage}{0.4\textwidth}
\textsc{Supervisor}\\[1ex]
{\bfseries Prof.~Stefano Liberati}
\end{minipage}

\end{center}

\end{titlepage}

\blankpage

\thispagestyle{empty}
\vspace*{5cm}
\begin{flushright}
\begin{minipage}{0.35\textwidth}
{\large\it A Donata e ai miei genitori}
\end{minipage}
\end{flushright}

\clearpage

\blankpage

\begin{abstract}
\phantomsection
\addcontentsline{toc}{chapter}{Abstract}
\setcounter{page}{5}

Analogue gravity is based on a simple and striking observation: In some physical systems there exist perturbations that, under certain conditions, are governed by the same equation describing the propagation of fields in curved spacetimes.
However, while quantum phenomena like the emission of thermal radiation from astrophysical black holes are hardly detectable, analogue systems can provide the unique possibility to experimentally test these effects.
Originally driven by this motivation, the study of analogies between quantum field theory in curved space and condensed matter has now produced a fruitful interchange leading to new and deep insights in both fields.
For instance, effects due to trans-Planckian physics can be naturally addressed in analogue systems because their microphysics is well known and under control.

In particular, Bose--Einstein condensates (BECs) are very rich and powerful systems in which this analogy can be implemented also at a quantum level, since they have a high degree of quantum coherence, they are very cold, and their speed of sound is very low, making them easy to handle in a laboratory.
In this Thesis, we analyze the phenomenology of spontaneous particle creation in BEC flows with various velocity profiles. In the presence of a single horizon, we investigate how the temperature of Hawking-like radiation is fixed by geometrical properties of the flow. We also perform a detailed analysis of flows that couple black and white acoustic horizons. In these settings, new interesting phenomena appear, such as an instability known in general relativity as black hole laser effect.

Analogue systems can also serve as toy models for an emergent gravity scenario. From the analogue gravity perspective, we address one of the most challenging problems in modern physics, regarding the nature and the value of the cosmological constant. The analogy between gravity and BECs provides some hints about how to properly attack the computation of this constant. Furthermore a new analogue system based on a relativistic BEC is proposed. This system allows to map into an acoustic metric a wider class of geometries and suggests a new possible way to emerge gravity from a flat Lorentz invariant background.

Finally, this field of research has been producing a wide set of tools that can be profitably applied to solve general relativity problems apparently unrelated to the analogue model framework. As an example, we study the stability of a special class of geometries allowing superluminal travel, by using the language borrowed from analogue gravity.\\[2em]

\begin{center}
{\bf Date of PhD defense: 5 July 2011} 
\end{center}

\end{abstract}

\thispagestyle{empty}

\vspace*{\fill}

\begin{center}
\bf Based on\\
\end{center}

\begin{itemize}
\item SF, S. Liberati, and C. Barcel\'o. Semiclassical instability of dynamical warp drives. {\it Phys.~Rev.~D}~{\bf 79}, 124017 (2009).
\item SF, S. Liberati, and C. Barcel\'o. Superluminal warp drives are semiclassically unstable. {\it J. Phys.: Conf. Ser.} {\bf 222}, 012046 (2010).
\item C. Barcel\'o, SF, and S. Liberati. Semiclassical instability of warp drives. {\it J. Phys.: Conf. Ser.} {\bf 229}, 012018 (2010).
\item  S. Fagnocchi, SF, S. Liberati, M. Kormos, and A. Trombettoni. Relativistic Bose--Einstein condensates: a new system for analogue models of gravity. {\it New~J.~Phys.}~{\bf 12}, 095012 (2010).
\item SF and R. Parentani. Black hole lasers in Bose--Einstein condensates. {\it New~J.~Phys.}~{\bf 12}, 095015 (2010).
\item SF and R. Parentani. Spectral properties of acoustic black hole radiation: Broadening the horizon. {\it Phys.~Rev.~D}~{\bf 83}, 084010 (2011).
\item SF and R. Parentani. On the robustness of acoustic black hole spectra. {\it J.~Phys.:~Conf.~Ser.} {\bf 314} 012030 (2011, published after PhD defense).
\item SF, S. Liberati, and C. Barcel\'o. Semiclassical Warp-Drive Instability. In {\it Twelfth Marcel Grossmann Meeting on General Relativity}, edited by T. Damour, R. T. Jantzen, and R. Ruffini, p. 2369 (World Scientific, Singapore, 2012, published after PhD defense).
\item SF, S. Liberati, and L. Sindoni. Cosmological Constant: A Lesson from Bose-Einstein Condensates. {\it Phys. Rev. Lett.} {\bf 108} 071101 (2012, published after PhD defense).
\end{itemize}

\vspace{1em}

\begin{center}
 \bf Other publications extracted from this Thesis after PhD defense
\end{center}

\begin{itemize}
\item A. Coutant, SF, S. Liberati, and R. Parentani. Impossibility of superluminal travel in Lorentz violating theories. {Phys. Rev. D} {\bf 85}, 064020 (2012).
\item SF, S. Liberati, and L. Sindoni. The analogue cosmological constant in Bose-Einstein condensates: a lesson for quantum gravity. arXiv:1204.3039 [gr-qc], to appear in {\it Proceedings of the II Amazonian Symposium on Physics}.
\end{itemize}

\vspace{1em}

\begin{center}
 \bf Awards
\end{center}

\begin{itemize}
\item Second prize of the 2009 FQXi (Foundational Questions Institute) essay contest {\it What is Ultimately Possible in Physics?}\\
 C. Barcel\'o, SF, and S. Liberati. On the impossibility of superluminal travel: the warp drive lesson. arXiv:1001.4960 [gr-qc].
\item Member of the FQXi since January 2010.
\end{itemize}

\vspace*{\fill}

\pagestyle{headernonumber}

\setcounter{tocdepth}{3}

\phantomsection
\addcontentsline{toc}{chapter}{Contents}
\tableofcontents

\cleardoublepage

\phantomsection					
\addcontentsline{toc}{chapter}{Introduction}	
\chapter*{Introduction}				
\label{chap:intro}				
\chaptermark{Introduction}			

Analogies in physics have often provided deep insight and inspiration to deal with fundamental problems. In the last 30 years, {\it analogue models of gravity}~\cite{livrev} had this role with respect to pressing issues in gravitation theory, such as the mechanism of Hawking radiation, the fate of Lorentz invariance at ultra-short distances and the nature of spacetime and gravity as possibly emergent phenomena.

The general idea behind analogue models of gravity is that in many physical systems it is possible to identify suitable excitations that propagate as fields on a curved spacetime (CS). In particular, there is a wide class of condensed matter systems that admit a hydrodynamical description within which it is possible to show that acoustic disturbances propagate on an effective geometry, the so called ``acoustic metric''~\cite{unruh}. In fact, the propagation of excitations in the hydrodynamical regime can be described through a relativistic equation of motion in a CS.  So, even starting from non-relativistic equations, one can show that excitations are endowed with a Lorentz invariant dynamics where the Lorentz group is generally associated with an invariant speed coinciding with the speed of sound. Furthermore, these systems generically show a breakdown of such acoustic regime leading to Lorentz violations at high energies (as expected from the Newtonian nature of the fundamental equations).

The main motivation initially stimulating the research activity on this topic was the possibility of experimentally studying gravitational phenomena otherwise impossible to detect, in particular black hole evaporation via Hawking radiation (Unruh, 1981,~\cite{unruh}). Unfortunately, this possibility was more theoretical than practical because, as Unruh noticed, the expected temperature for an acoustic black hole of 1~mm radius and a speed of sound of 300~m/s was extremely low, $3\times 10^{-7}~\mbox{K}$.
This was probably one of the reasons why this important seminal paper remained almost unnoticed until the nineties, when the analogy between acoustic propagation in fluids and propagation of fields in a CS was rediscovered by Visser~\cite{Visser:1993ub}.
Furthermore, the interest in analogue gravity (AG) was boosted by a second motivation, namely the possibility of using the analogy between acoustics and quantum field theory (QFT) in CS as a theoretical tool to understand the behavior of spacetimes at very small scales and to check whether unknown modified physics may or may not modify the phenomenology of processes such as particle production in CS. Even if Unruh had immediately realized that the acoustic geometry had to break down at the atomic or molecular scale ($10^{-10}~\mbox{m}$), just as the concept of a smooth spacetime was expected to breakdown at Planck's scale ($10^{-35}~\mbox{m}$), Jacobson~\cite{TJ91} was the first to seriously take this analogy as a guidance to investigate how black hole evaporation is affected by ultrashort-distance effects.

Unfortunately, contacts with the condensed matter community remained quite weak for some years, despite the proposal of $^3$He by Volovik and Jacobson~\cite{volovikvortex,JacobsonVolovik} as a first condensed matter system to implement the analogy. They estimated a Hawking temperature of about $5~\mu\mbox{K}$, which should have been large enough to allow an experimental detection of the Hawking flux.
Then, around year 2000, the number of physical systems proposed as analogue models of gravity started to grow. In particular Garay and collaborators~\cite{BEC,Garay} proposed to use Bose--Einstein condensates (BECs), which seem so far the most promising systems for the experimental detection of phenomena such as Hawking radiation. Many other systems have been proposed to simulate both spacetimes showing black hole horizons and cosmological metrics.
Among the former, in addition to the cited superfluid Helium and BECs, we recall Fermi gases~\cite{fermi}, slow light~\cite{slow_leonhardt,slow_reznik,slow_US}, non-linear electromagnetic waveguides~\cite{waveguide_SU,waveguide_leonhardt} and ion rings~\cite{ions_prl,ions_long}.
Concerning the simulation of cosmological metrics, BECs have definitely been the favorite physical systems~\cite{livrev,expandinguniverse_blv,expandinguniverse_ff,expandinguniverse_silke,expandinguniverse_piyush}.

Finally, in the very last years, the first experiments in BECs~\cite{technion}, water~\cite{germain_exp,Silke_exp}, and fused silica~\cite{fiber_exp} have been realized with the objective of reproducing curved geometries and to study Hawking radiation. New possibilities of experimentally detecting Hawking radiation have also been opened by~\cite{carusotto1}, where it was proposed to measure correlations between Hawking particles and their partners. As confirmed by numerical simulations~\cite{carusotto2}, the correlation signal is not masked by the uncorrelated thermal noise. This technique potentially permits to detect Hawking radiation even when the Hawking temperature is smaller than the temperature of the condensate.

While the above subjects have represented what we might call the main stream of research on analogue models, in more recent years there has been growing attention to their application as toy models for emergent gravity scenarios~\cite{two-componentBEC,two-componentBEC2,gravdynam,scalargravity,birefringence}. The latter are driven by the idea that gravity could be an intrinsically classical/large-scale phenomenon similar to a condensed matter state made of many fundamental constituents~\cite{BLH}. In this sense, gravity would not be a fundamental interaction but rather a large-scale effect, which emerges from a quite different dynamics of some elementary quantum objects. In this sense, many examples can be brought up, starting from the causal set proposal~\cite{causalset,causalset_emergence}, passing to group field theory~\cite{oriti,oriti2} or the recent quantum graphity models~\cite{konopka1,konopka2,konopka3}, and other approaches (see, \eg,~\cite{dreyer_fall,dreyer_emergent,Padma}).
Analogue models, in particular BECs, have represented a very interesting test field as they are in many ways ideal systems where the mechanism of emergence of symmetries, spacetime, and possibly gravitation-like dynamics can be
comprehensively studied and understood. In particular, much attention has been devoted in the recent past to Lorentz symmetry breaking and its possible role in leading to viable mechanisms of emergent dynamics (see, \eg, the related discussion in~\cite{SIGRAV}).
\\

This is the very active and stimulating context in which the work of this Thesis has been realized. For the first time since Unruh's paper in 1981~\cite{unruh}, the community of AG has involved theoretical physicists with gravitational, QFT and quantum gravity (QG) background, theoretical condensed matter people, and experimentalists in BECs, fluid dynamics, and optics. Even if initially proposed as tools to confirm theoretical predictions in gravity such as black hole evaporation, analogue models are now stimulating the investigation of new phenomena in condensed matter with a bidirectional knowledge exchange between different field of physics. Furthermore, the techniques developed in the last 30 years in the context of AG, turn out to be helpful to address problems in gravity and analogue systems are very interesting toy models of QG.
In this Thesis we present our contribution in all these directions.

\phantomsection										%
\addcontentsline{toc}{section}{Open issues in quantum field theory in curved spacetime}	%
\sectionmark{Open issues in QFT in CS}							%
\section*{Open issues in quantum field theory in curved spacetime}			%
\sectionmark{Open issues in QFT in CS}							%
\label{sec:open_QFT_CS}									%

One of the main motivations at the basis of the analogue model program is the possibility to address problems of QFT in CS, using well known physical systems to check predictions of QFT in CS and as toy models for the construction of a QG theory.
In this section we summarize some of the most challenging open issues in QFT in CS, that are directly addressed in this Thesis or for which we propose a viable approach through AG models.

\phantomsection								%
\addcontentsline{toc}{subsubsection}{The trans-Planckian problem}	%
\subsubsection*{The trans-Planckian problem}				%

There are situations where very small scales (high frequencies) can be probed by low frequency measures. As a first example, Hawking radiation comes from regions arbitrarily close to the horizon of black holes, where Hawking particles must have originated at sub-Planckian wavelengths~\cite{thooft,TJ91}, because of the exponential blueshift suffered close to the horizon in the standard analysis.
Secondly, the physical wavelengths of comoving scales which correspond to the present large-scale structure of the Universe were smaller than the Planck length before inflation~\cite{martinbrandenberger}.
Therefore, the usual computation of the spectrum of the emitted Hawking particles, in the first case, and of the cosmological fluctuations, for the second one, might be affected by unknown physics at sub-Planckian scales.

This is a very serious issue because black-hole thermodynamics~\cite{BCH} resides on the existence of thermal Hawking radiation and, more importantly also from an observational point of view, the present structure of the Universe was originated by quantum fluctuations in the inflationary epoch.
A possible way to address this problem is to look for the robustness of such phenomena, when various modifications to the ordinary low energy physics are implemented at Planckian energies.

Analogue gravity provides a rich set of powerful physical systems where these issues can be directly investigated and tested. In fact, every real system looking like a continuum at large scales shows discrete structure at atomic or molecular scales. Thus, these systems can be used to provide deep insight in the investigation of modifications of physical laws at trans-Planckian scales.

\phantomsection								%
\addcontentsline{toc}{subsubsection}{Casual properties of spacetime}	%
\subsubsection*{Casual properties of spacetime}				%

General relativity (GR) allows the existence of spacetimes with very exotic causal structures, for instance with close timelike curves~\cite{Hawking-Ellis}. Said with more popular and evocative words, GR allows the existence of time machines, together with the well-known related causal paradoxes.

Various solutions to these paradoxes have been proposed by addressing those issues from very different approaches~\cite{visser-chronology}:
\begin{itemize}
 \item Chronology protection conjecture~\cite{hawking-chronology}: time machines are forbidden by some physical mechanism.
  This conjecture has not been proved yet, because (1) we are not able to perform a self-consistent calculation taking into account the back-reaction of the renormalized stress-energy tensor on a given spacetime, (2) the Kay--Radzikowski--Wald theorem \cite{krw}, which is probably the most important results in this direction, implies the breakdown of the renormalization procedure of the stress-energy tensor on chronological horizons (which are just a special sort of Cauchy horizons).
 \item Novikov's consistency proposal~\cite{novikov,novikovself}: closed timelike curves are permitted, and a principle of self-consistency must consequently be added to physical laws. Thus, a local solution to the equations of physics can occur only if it can be extended to a global solution defined in the whole universe. This imply that, given a certain Cauchy ipersurface, initial data are not completely arbitrary but must be chosen accordingly to the self-consistency principle.
 \item Radical alterations of our description of the spacetime structure: spacetime should be described by a non-Hausdorff manifold to allow for chronology violation and to permit that the present has multiple futures or multiple pasts~\cite{deutsch}.
\end{itemize}

Furthermore, the possibility of building time machines is closely related to the feasibility of superluminal travel~\cite{rolnik,everett}. Even assuming the light postulate, stating that nothing can locally travel faster than light, in the context of GR one could in principle create spacetime shortcuts such as wormholes or warp drives. This would allow to travel from one point to another in a time much smaller than the time needed by a light ray to connect the same two points without passing through the shortcut (see~\cite{visser-book} for a complete review on this topic). Opportunely combining two superluminal travels (or either moving  or desynchronizing the mouths of a wormhole), it is actually possible to create a time machine.
Unfortunately, it seems a general fact that a large amount of energy-condition violating matter is needed to sustain spacetime shortcuts. Even if forms of energy that violate energy conditions are well known, such as in Casimir effect, it would be very unlikely to obtain the required amount to keep open a wormhole or to speed up a warp drive for any practical purpose. However,
there is no theoretical reason preventing the realization of such shortcuts, thus one may wonder whether they would be forbidden by some physical mechanism. Given the connection between spacetime shortcuts and time machines, such a mechanism would possibly represent a low level implementation of the chronology protection conjecture.

Finally, spacetimes allowing superluminal travel through the presence of warp-drive bubbles possess Cauchy horizons~\cite{wd}. Namely, there is a region of spacetime (beyond the Cauchy horizon) where the values of fields are not completely determined by initial data at early times and the possibility of making prediction breaks down beyond the horizon. Cauchy horizons are also found in the interior of Kerr--Newman black holes, where they are known to be unstable~\cite{Simpson:1973ua,poissonisrael,markovicpoisson}. This is not in contradiction with the Fulling--Sweeny--Wald theorem~\cite{fsw}, saying that the Hadamard singularity structure of the two-point function of a scalar field is preserved by Cauchy evolution, which precisely fails at the Cauchy horizon.
In Chap.~\ref{chap:warpdrive} we show how the relation among the instability of Cauchy horizons, superluminal travel, and time machines can be investigated using techniques developed for analogue models of gravity.

\phantomsection										%
\addcontentsline{toc}{subsubsection}{Nature and value of the cosmological constant}	%
\subsubsection*{Nature and value of the cosmological constant}				%

The cosmological constant~\cite{carroll} $\Lambda$ appears in Einstein's equation [adopting the signature $(-,+,+,+)$]
\begin{equation}
 R_{\mu\nu}-\frac{1}{2}g_{\mu\nu}R+\Lambda\,g_{\mu\nu}=\frac{8\pi G}{c^4}T_{\mu\nu}
\end{equation}
as a source term present even in the absence of matter and with the symmetries of the vacuum ($T_{\mu\nu}^\Lambda\propto g_{\mu\nu}$).
Because of this property it is usually interpreted as a ``vacuum energy''.
Unfortunately, if one calculates the value of this constant as the zero-point energy of quantum fields, the result differs from the observed one by 120 orders of magnitude. As we shall discuss in Chap.~\ref{chap:cosmoconstant}, supersymmetric theories~\cite{susy} only partially correct this huge discrepancy, that still remains very large.
The only option we have to explain observations is an extremely fine tuning~\cite{finetuning,rovelli} in the renormalization of this constant.

This discrepancy between the theoretically computed and the measured value of $\Lambda$ is a strong signal that something is missing in our comprehension of the origin and the nature of this constant. This suggests that the problem is related to the failure of effective field theory (EFT) at Planck scales. Indeed, a proper computation would require a full knowledge of a quantum theory of gravity. Since we do not have such a theory, we can use analogies to learn how the theoretical computation of such a constant should be performed once the microscopic theory is fully under control.
In particular, in BECs with $U(1)$-symmetry breaking~\cite{gravdynam}, not only phonons propagate as fields in a curved geometry, but also a dynamics for the analogue metric can be defined. Namely, the analogue of the non-relativistic limit of Einstein's equations can be derived, together with a cosmological constant term, therefore allowing the direct computation of this constant from microphysics and its comparison with the standard zero-point energy~\cite{cosmobec}.

\phantomsection											%
\addcontentsline{toc}{subsubsection}{Naturalness problem for Lorentz violating theories}	%
\subsubsection*{Naturalness problem for Lorentz violating theories}				%

The continuous spacetime structure is expected to break down at a scale of the order of the Planck length and such a breakdown is expected to be related to the breaking of Lorentz invariance at the same scale. Let us then assume that the field propagators are proportional to
\begin{equation}
 \frac{\ii}{\om^2+c^2k^2+O(k^3/K^3)},
\end{equation}
where $1/K$ is of the order of the Planck length. If such a field is coupled to some other field, it can been shown~\cite{Collins}, that one-loop self-interaction diagrams cause the appearance of modifications proportional to the coupling constant $g$ down to the $k^2$ term. This would not be a problem if there were only two fields, since one might always absorb this term by renormalizing the speed of light $c$. However, if there were more then two fields, coupled with different constants $g$, the additional $k^2$ term would be different for each field. As a consequence, different fields would perceive different speed of lights. This implies that Lorentz violation would appear at relatively low energies, even if it was originally introduced at very high momentum $K$ on operators with higher-than-two order. This is the so called naturalness problem.

Thus, it is interesting to look for an alternative way to emerge gravity from a more fundamental structure, in order to eliminate or at least tame the naturalness problem. Apparently, this problem is related to the fact that the theory is not Lorentz invariant at high energy. Analogue gravity can provide some hints on how to address this issues. In Chap.~\ref{chap:relbec} we propose a new analogue system~\cite{relbec}, where phonons live in a curved geometry with local Lorentz invariance (the limit velocity is the speed of sound) at low energy, while this symmetry is broken at higher energy. At very high energy gravity eventually disappears and global Lorentz invariance is recovered (Minkowski spacetime where the limit velocity is the speed of light). We think that such a system is a proficient toy model where the naturalness problem may be somehow tamed.

\phantomsection										%
\addcontentsline{toc}{section}{Analogue models as a test field and operating framework}	%
\sectionmark{A test field and operating framework}					%
\section*{Analogue models as a test field and operating framework}			%
\addcontentsline{toc}{section}{A test field and operating framework}			%
\label{sec:AG}										%

All the issues described in the above section can be profitably addressed in the context of AG, mainly using real physical systems as toy models for the investigations of spacetime properties, related to possible unknown modifications of physical laws at trans-Planckian scales and various implementations of high energy Lorentz violation. In particular, we sketched how the trans-Planckian problem is naturally described in the context of AG, in physical systems where the microphysics is fully under control, such as BECs. Furthermore, since the procedure in which the continuous geometry emerges is well known in these models, we can use them also to investigate how the corresponding emergence phenomenon takes place in spacetime, leading to Einstein's equations with an already encoded cosmological constant.

However, AG is not only toy models for the study of QG, even if this is one of the most important theoretical motivations to do research in this field. {\it In primis}, AG was born of the need for a real physical system to test QFT in CS phenomena.
As a natural fact, one of the main objective of this research is to extend the set of spacetimes that one can reproduce with these systems. Actually, even if it is relatively simple to find analogue models reproducing general features of interesting spacetimes, such as horizons or expanding universes, it is still rather difficult to exactly reproduce (even theoretically) fundamental metrics such as those describing Schwarzschild's or Kerr's black holes (see Sec.~\ref{sec:acousticmetrics}).

Furthermore, Unruh's original motivation of realizing physical systems to experimentally detect Hawking radiation is no more an abstract possibility. Also thanks to Unruh himself~\cite{Silke_exp}, a few experiments in water~\cite{Silke_exp}, BEC~\cite{technion}, and fused silica~\cite{fiber_exp} have been realized. The first one led to the detection of the classical stimulated version of Hawking radiation. In the second experiment, a stable sonic horizon was realized for the first time in a BEC, but no measures of Hawking effect have been performed so far. In the third one, a couple of horizons were created using a moving refractive index perturbations and Hawking radiation was claimed to have been detected, but this result is still controversial~\cite{fiber_comment,fiber_reply}.

Finally, the analogue model program has created new stimuli for the investigation of new condensed matter systems, also providing tools for their description. Nowadays, thanks to the possibility of experimental realizations, interest in analogue systems has increased not only in relation to GR and QFT, but also as very interesting and rich physical systems {\it per se}.
The first example in this sense is the quite successful description, although not exact, through the analogy with particle production in expanding/contracting universes~\cite{Calzetta:2002zz}, of the emission of jets of atoms observed in a BEC where the sign of interaction were suddenly inverted~\cite{Donley}.
On the opposite, one can even describe phenomena as Hawking radiation in BEC, without using any analogy with gravity~\cite{MacherBEC,carusotto2}. This approach have led also to the description of the exponential amplification of spontaneous emitted radiation in BEC flows with two sonic horizons~\cite{bhlasers}, see Chap.~\ref{chap:bhlaser}, inspired by the black hole laser effect which was firstly described for generic scalar fields with modified dispersion relation in geometries with a black and a white horizon~\cite{cj}.
In hydrodynamics, this exchange of knowledge with gravity has produced new interesting results about wave-blocking and the description of new kinds of ``horizons'', such as the blue horizon in water flows, where ordinary waves are converted in capillary waves~\cite{germain_exp,germainulf}.

\phantomsection						%
\addcontentsline{toc}{section}{Plan of the Thesis}	%
\sectionmark{Plan of the Thesis}			%
\section*{Plan of the Thesis}				%
\sectionmark{Plan of the Thesis}			%
\label{sec:plan}					%

Part~\ref{part:QFTandAM} is a quite broad introduction about AG.
In Chap.~\ref{chap:general} we introduce the analogy between gravity and other physical system, focusing on fluid dynamics and the related phenomenology. In Chap.~\ref{chap:warpdrive} we discuss an application of AG techniques to the study of a class of exotic spacetimes, warp drives, allowing superluminal travel. In Chap.~\ref{chap:physics} we present the three physical systems for which experiments have been realized so far. In that chapter, we show how the analogy between a continuous spacetime and the actual physical models breaks down at scales where the microscopic properties of the systems show up. We discuss also the problems related to the robustness of phenomena such as Hawking radiation, once these short scale behaviors are taken into account.

In Part~\ref{part:BECphenomenology}, we investigate the properties of Hawking radiation and, more generally, of particle production in spacetimes with one or two horizons in a specific physical system, namely BEC, whose complete theoretical description is presented in Appendix~\ref{app:BEC}. Differently from Part~\ref{part:QFTandAM}, we do not reside on the analogy with gravity, but all the results are obtained using the BEC fundamental equations. We find that Hawking radiation is robust under certain conditions (Chap.~\ref{chap:hawking}). Furthermore, we discuss systems with two horizons in Chaps.~\ref{chap:warpdriveBEC} and~\ref{chap:bhlaser}, motivated by both theoretical and experimental reasons.
In particular, in Chap.~\ref{chap:bhlaser}, we study the configuration corresponding to the experimental set up of~\cite{technion}, where a white hole and a black hole horizon are coupled. As first discussed in~\cite{cj}, we confirm the presence of dynamical instabilities, commonly called black hole laser effect, in such a system.
The structure of the numerical code used for those investigations is described in Appendix~\ref{app:code}.

Finally, in Part~\ref{part:BECemergent}, we discuss analogue models from the standpoint of emergent gravity. In Chap.~\ref{chap:cosmoconstant} we use a particular BEC model where it is possible to define an analogue gravitational dynamics to investigate the mechanism in which an analogue cosmological constant emerges from the microphysical properties.
In Chap.~\ref{chap:relbec} we propose a new analogue system based on a BEC whose atoms are described by a relativistic Lagrangian. The main feature of this model is the absence of a preferred direction of time as in non-relativistic systems. Because of that, relativistic BECs are interesting toy models to study the possible emergence of gravity from a more fundamental Minkowskian structure, without absolute time.

\cleardoublepage

\pagestyle{header}

\part{Quantum Field Theory in~Curved~Spacetime and~Analogue~Gravity}
\label{part:QFTandAM}

\chapter{General properties}	
\label{chap:general}		

In the \nameref{chap:intro} of this Thesis, we broadly discussed the motivations and the objectives at the basis of analogue gravity (AG), showing how this field of research can provide deeper understanding both of gravity and of the systems in which the analogy is implemented.
In this chapter, we introduce the basic concepts of AG, beginning with a very simple example. In Sec.~\ref{sec:generalmetric}, we show how the propagation of sound in fluids is described, under certain conditions, by the same equation describing the propagation of a scalar field in a curved spacetime (CS).
Then, we present some gravitational phenomena that have a corresponding phenomenon in other physical system, such as Hawking radiation or particle production in an expanding universe.

However, in this chapter, we just focus on the similarities between quantum field theory (QFT) in CS and analogue systems, in a very general framework. Then, in Chap.~\ref{chap:warpdrive} we present an application of some tools developed in the context of AG to the study of the semiclassical instability of dynamical warp drives.
The discussion about the differences between analogue systems and gravity are instead the subject of Chap.~\ref{chap:physics}, where we consider three specific systems: surface waves in water, phonons in Bose--Einstein condensates (BECs) and photons in media with moving refractive index perturbations.

\sectionmark{Effective geometries}			%
\section{From fluid dynamics to effective geometries}	%
\label{sec:generalmetric}				%
\sectionmark{Effective geometries}			%

The simplest example of analogue spacetime is acoustics in fluids~\cite{unruh, Visser:1993ub,P-G,Visser:1998qn,livrev}.  Sound waves propagating in a fluid are indeed dragged by the flow, which acts as a CS bending light rays. This analogy appears clearer when considering the extreme situation of a supersonic flow, where sound cannot propagate upstream. If a fluid is sped up from subsonic to supersonic, sound cannot escape from the supersonic region back to the subsonic one and it is trapped in the same way as light is trapped in a black hole.
To translate this physical analogy in mathematical language one must show that propagation of sound in a moving flows is described by the same equation governing light propagation in a CS. This is true under certain conditions that we discuss below.

In the geometrical acoustic approximation, a fluid is characterized by its velocity $\bv(t,\bx)$ and the local speed of sound $c(t,\bx)$. In the reference frame of the laboratory, the equation of motion of sound waves is
\begin{equation}
 \frac{\dd\bx}{\dd t}=c\,\bn+\bv,
\end{equation}
where $\bn$ is the unit vector pointing in the direction of propagation of the wave. Its normalization reads
\begin{equation}
 \frac{1}{c^2}\left(\frac{\dd\bx}{\dd t}-\bv\right)\cdot\left(\frac{\dd\bx}{\dd t}-\bv\right)=1,
\end{equation}
that is
\begin{equation}
 -(c^2-v^2)\,\dd t^2-2\bv\cdot\dd\bx\,\dd t+\dd\bx\cdot\dd\bx=0.
\end{equation}
This equation describes the null line element of a light ray propagating in a curved geometry. From this condition it is possible to determine the general line element
\begin{equation}\label{eq:lineelement}
 \dd s^2=\Omega^2\left[-(c^2-v^2)\,\dd t^2-2\bv\cdot\dd\bx\,\dd t+\dd\bx\cdot\dd\bx\right],
\end{equation}
up to a conformal factor $\Omega(t,\bx)$.

The above derivation is valid for every system, where perturbations can be described in eikonal approximation. However, it allows to determine only the sound cones, that is the conformal (and causal) structure of the spacetime. In order to fix also the conformal factor $\Omega$ and to obtain the full acoustic metric, one has to pick a particular physical system and demonstrate that, in that model, there exist some field $\phi$ and a metric $g_{\mu\nu}$, such that the dynamics of $\phi$ is described by a Klein--Gordon equation in a curved background (see, for example,~\cite{Hawking-Ellis,MisnerThorneWheeler,Wald}):
\begin{equation}
 \label{eq:KGfluid}
 \frac{1}{\sqrt{|g|}}\partial_\mu \left(\sqrt{|g|}g^{\mu\nu}(t,\bx)\partial_\nu\phi\right)=0,
\end{equation}
where $g$ is the determinant of the metric tensor $g_{\mu\nu}$,
\begin{align}
 \partial_0&=\partial/\partial t,\\
 \partial_i&=\partial/\partial {x_i},\quad i=1,\dots,d,
\end{align}
and $d$ is the number of spatial dimensions.

Defining the metric density
\begin{equation}\label{eq:deff}
 f^{\mu\nu}\equiv\sqrt{|g|}g^{\mu\nu},
\end{equation}
Eq.~\eqref{eq:KGfluid} becomes
\begin{equation}\label{eq:fieldeq}
 \partial_\mu \left(f^{\mu\nu}(t,\bx)\partial_\nu\phi\right)=0.
\end{equation}

To demonstrate that sound waves are actually described by the above equation with a suitable choice of $f^{\mu\nu}$, we start from the continuity equation (mass conservation)
\begin{equation}\label{eq:continuity}
 \pdt\rho+\nx\cdot(\rho\bv)=0
\end{equation}
and Euler's equation
\begin{equation}\label{eq:euler}
 \rho(\pdt+\bv\cdot\nx)\bv=-\nx p,
\end{equation}
which are the fundamental equations of fluid dynamics for a fluid with  \emph{no viscosity} and \emph{no external forces}~\cite{Landau-Lifshitz}.
Since
\begin{equation}
 \bv\times(\nx\times\bv)=\frac{1}{2}\nx v^2-(\bv\cdot\nx)\bv,
\end{equation}
Euler's equation can be written as
\begin{equation}\label{eq:euler2}
 \pdt\bv=-\frac{\nx p}{\rho}-\frac{1}{2}\nx v^2+\bv\times(\nx\times\bv).
\end{equation}
To proceed with the derivation of the acoustic metric it is necessary to assume that the fluid is \emph{locally irrotational}, that is the vorticity $\nx\times\bv$ vanishes. This also implies that, locally, $\bv$ is the gradient of a scalar function, called the velocity potential\footnote{Flow configurations with global vortexes can still be described in this formalism by allowing for a multivalued velocity potential $\phi$.}
\begin{equation}\label{eq:vpotential}
 \bv=-\nx\phi.
\end{equation}
Note that $\phi$ is defined only up to a function of time $t$.
The last needed assumption is \emph{barotropicity}, \ie, $p=p(\rho)$ is a function only of the density $\rho$. Under this condition the differential of the enthalpy density $h$ reads
\begin{equation}\label{eq:enthalpy}
 \dd h\equiv\frac{\dd p(\rho)}{\rho}=\frac{\dd p}{\dd\rho}\frac{\dd \rho}{\rho}=c^2\frac{\dd \rho}{\rho},
\end{equation}
where $c$ is the velocity of sound
\begin{equation}\label{eq:csound}
 c\equiv\sqrt{\frac{\dd p}{\dd\rho}}.
\end{equation}
Integrating Eq.~\ref{eq:euler2} with respect to $\bx$ and using the gauge freedom of $\phi$ to absorb the arbitrary constant of integration that is a function of time, we obtain Bernoulli equation
\begin{equation}\label{eq:bernoulli}
 -\pdt\phi+h+\frac{1}{2}(\nx\phi)^2=0.
\end{equation}

We now linearize Eqs.~\eqref{eq:continuity} and~\eqref{eq:bernoulli} around some background values $\rho_0$, $p_0$, and $\phi_0$, of the dynamical variables $\rho$, $p$, and $\phi$, and indicate with subscript $1$ the first order deviations with respect to the background. With this parametrization we separate a long range, slowly varying component (the background) from small-scale, high-frequency, and small-amplitude component, that is sound. Thus, to perform this expansion we require
\begin{equation}
p_1\ll p_0,\qquad
\rho_1\ll \rho_0,\qquad
\phi_1\ll \phi_0.
\end{equation}
With these assumptions we obtain
\begin{gather}
\pdt\rho_1+\nx\cdot(\rho_1\bv_0-\rn\nx\phi_1)=0,\\
-\pdt\phi_1+h_1-\bv_0\cdot\nx\phi_1=0.
\end{gather}
From Eq.~\eqref{eq:enthalpy}, one obtains $h_1=c^2\rho_1/\rn$, so that the second equation yields
\begin{equation}
 \rho_1=\frac{\rn}{c^2}(\pdt+\bv_0\cdot\nx)\phi_1.
\end{equation}
Eliminating $\rho_1$, the first equation becomes
\begin{equation}
 \left[(\pdt+\nx\cdot\bv_0)\frac{\rho}{c^2}(\pdt+\bv_0\cdot\nx)-\nx\cdot\rn\nx\right]\phi_1=0,
\end{equation}
where the differential operators $\pdt$ and $\nx$ act on everything on their right. This equation is precisely in the form of Eq.~\eqref{eq:fieldeq} with
\begin{equation}
f^{\mu\nu}=\frac{\rn}{c^2}
 \begin{pmatrix}
 -1	&	-{\bv_0}^T	\\
 -\bv_0	&	c^2{\mathbbm 1}_{d\times d}-\bv_0\otimes{\bv_0}^T
 \end{pmatrix}.
\end{equation}
After some tedious manipulations [see Eq.~\eqref{eq:deff}], one obtains the contravariant metric tensor
\begin{equation}
g^{\mu\nu}=\frac{1}{c^2}\left(\frac{\rn}{c}\right)^{-2/(d-1)}
 \begin{pmatrix}
 -1	&	-{\bv_0}^T	\\
 -\bv_0	&	c^2{\mathbbm 1}_{d\times d}-\bv_0\otimes{\bv_0}^T
 \end{pmatrix}
\end{equation}
and the acoustic metric~\cite{P-G}
\begin{equation}\label{eq:fluidmetric}
 g_{\mu\nu}=\left(\frac{\rn}{c}\right)^{2/(d-1)}
  \begin{pmatrix}
 -(c^2-v_0^2)	&	-{\bv_0}^T\\
 -\bv_0	&	{\mathbbm 1}_{d\times d}
 \end{pmatrix},
\end{equation}
which gives the line element~\eqref{eq:lineelement}, with conformal factor
\begin{equation}\label{eq:conformalfactor}
 \Omega(t,\bx)=\left(\frac{\rn(t,\bx)}{c(t,\bx)}\right)^{1/(d-1)}.
\end{equation}

We conclude with some brief comments about this result. First, we summarize the assumptions on the fluid that we made to obtain the acoustic metric:
\begin{itemize}
 \item no viscosity,
 \item no local vorticity,
 \item barotropicity,
 \item small-amplitude perturbations.
\end{itemize}
The first three are conditions on the fluid, while the fourth one is a requirement on the amplitude of perturbations needed for the linear-sound approximation to be valid. If some disturbance does not satisfy this hypothesis, it must be treated as a solution of the full hydrodynamics equations and, mathematically speaking, it must be incorporated in the background component.

Second, let us comment on the structure of the acoustic metric itself~\eqref{eq:fluidmetric}. Given the above relations between the quantities describing the fluid (pressure, density, velocity), it is possible to show~\cite{livrev} that the flow is completely characterized by two quantities, that one can choose to be the velocity potential $\phi(t,\bx)$ and the speed of sound $c(t,\bx)$.
On the general relativity (GR) side instead, the most general metric is a rank 2 symmetric tensor in $d+1$ dimensions, having $(d+1)(d+2)/2$ independent components, from which one must subtract $d+1$ changes of coordinates. In total one remains with $d(d+1)/2$ degrees of freedom.
This means that the analogue metric can reproduce all the possible GR geometries only in $1+1$ dimensions, while for $d\geq2$ only a subset of the whole set of GR metrics has an acoustic counterpart~\cite{livrev}.

However, in $1+1$ dimensions, the metric tensor $g_{\mu\nu}$ itself is not well defined, because the conformal factor diverges for $d\to1$. Nevertheless, when describing the propagation of a field in a CS, the fundamental object characterizing the geometry is not the covariant metric $g_{\mu\nu}$, but the contravariant metric density $f^{\mu\nu}$, which directly enters the field equation~\eqref{eq:fieldeq}~\cite{livrev}. This object is well defined in an arbitrary number of dimensions as its derivation does not resides on the specific value of $d$. This implies, for instance, that phenomena related only to the kinematic of the quantum field, like Hawking radiation (see Sec.~\ref{subsec:hawking}), are present and well defined for every value of $d$. Finally, note that this problem concerns only intrinsically one-dimensional systems. For a three-dimensional system with axial symmetry, which one may experimentally use to simulate a $1+1$ geometry, the conformal factor is of course well defined, as well as the metric tensor.

Third, the effective metric $g_{\mu\nu}$ emerges from a Newtonian spacetime endowed with Galilean invariance. While the molecules of the fluid live in a flat space with absolute time, the perturbations on top of the fluid do not perceive this Newtonian world but rather live in a locally Minkowskian spacetime, \ie\/ satisfying local Lorentz invariance.
However, the presence of an underlying spacetime structure has some relevant implications on the structure of the emerging geometry which can be reproduced with a fluid model. In fact, since one can use (and actually, this is the most natural choice) the time of the laboratory as the time coordinate in the description of the acoustic geometry, it is evident that some causal properties (for instance, stable causality~\cite{Hawking-Ellis,Wald}) are inherited from the Newtonian spacetime of the lab. This also implies that some interesting spacetimes affected by causal pathologies (\eg, geometries with closed time-like curves~\cite{Hawking-Ellis,visser-book,hawking-chronology, Hawking:1991pk, Cassidy:1997kv, Hawking:2002yr, visser-chronology} cannot be reproduced with such an analogue system.

Finally, we remark that a given flow does reproduce a geometry only in a fixed set of coordinates. Changes of coordinates, which would make perfectly sense for a hypothetical phononic observer (that is an observer that can do measurements only through phonons and phononic interactions), do not have a direct interpretation from the point of view of the laboratory, because the lab ``world'' feels the Newtonian spacetime and not the emergent one.
However, even within the above limitations, one can still reproduce very relevant spacetimes with nontrivial causal structures, such as horizons, ergoregions, or superluminal bubbles, as we shall see in what follows.

We conclude this introductory section noting that a stationary spacetime is reproduced by a steady flow. In that case, $K=\pdt$ is a Killing vector for the acoustic metric. The condition identifying flows that correspond to static metrics is instead less trivial. A metric is static when it is stationary and the time-space cross-terms vanish. To this aim, starting from Eq.~\eqref{eq:lineelement}, one must perform a change of coordinates
\begin{equation}
 \dd t'=\dd t+\frac{\bv\cdot\dd\bx}{c^2-v^2},
\end{equation}
such that the line element becomes
\begin{equation}
 \dd s^2=\Omega^2\left[-(c^2-v^2)\,\dd t'^2+\left(\delta_{ij}+\frac{v^i v^j}{c^2-v^2}\right)\dd x^i\dd x^j\right].
\end{equation}
However, this change of coordinates is properly defined only if $\dd t'$ is integrable, namely if $\dd t'$ is an exact form. In practice, this means that one must find a function $t'(t,x)$, such that $\nx t'=\bv/(c^2-v^2)$. Given that the curl of the gradient of any function vanishes,
\begin{equation}
 \nx\times\left[\frac{\bv}{c^2-v^2}\right]=0.
\end{equation}
Since we have already assumed that the flow is irrotational ($\nx\times\bv=0$), this condition reduces to
\begin{equation}\label{eq:static}
 \bv\times\nx|c^2-v^2|=0,
\end{equation}
\ie, $\bv$ is parallel to $\nx(c^2-v^2)$.

\section{Acoustic metrics}	%
\label{sec:acousticmetrics}	%

\subsection{Horizons and ergoregions}	%
\label{subsec:horizons}			%

As discussed in the previous section, with fluid models one can reproduce a wide zoology of geometries with nontrivial causal structures. In this section, we therefore present some specific examples corresponding to static spherically symmetric black holes, rotating black holes, and warp drives. However, before analyzing these systems in details, it is worth understanding how to cast the concepts of ergoregion, trapped surface and horizon in the formalism of the acoustic analogy.

Differently from general spacetimes in GR, for acoustic analogues there is a preferred reference frame given by the laboratory spatial and time coordinates. In particular there is a natural definition of ``at rest'', with respect to the lab, and there is a preferred time foliation fixed by the lab time coordinate $t$. Moreover, when the flow is steady, the vector $K=\pdt$ is a Killing vector for the emergent geometry. As a consequence, an ergoregion is naturally defined as a region where the vector $K$ becomes spacelike and an ergosurface is the surface where $K$ is null. From Eq.~\eqref{eq:fluidmetric}, the norm of $K$ is
\begin{equation}
 g_{\mu\nu}K^{\mu}K^{\nu}\propto-(c^2-v^2),
\end{equation}
indicating that $\pdt$ is spacelike when the flow is supersonic.\footnote{For the sake of simplicity, we have suppressed the subscript 0 of $v$. We will adopt this notation throughout this Thesis, when it is not source of confusion.}
In this region, the flow is so strong that sound cannot propagate upstream, but it is dragged by the fluid.
Correspondingly, in GR, in an ergoregion nothing can stay ``at rest'' with respect to fixed stars, even if it is subject to infinitely strong forces~\cite{MisnerThorneWheeler,Hawking-Ellis,Wald}. 

Furthermore, the presence of a preferred time foliation allows to define the concept of trapped surface in a straightforward way, without the technicalities usually required for the standard and general definition~\cite{Hawking-Ellis}. Then, one can bypass the definition of expansion of null geodesic congruence and directly define a trapped surface as a closed two-surface that sound can cross only in one direction. If sound is dragged inward or outward by the flow, the surface is outer-trapped or inner-trapped, respectively. More formally, an outer-trapped (respectively, inner-trapped) surface is a surface where the normal component of the fluid velocity is greater than the local speed of sound and inward (respectively, outward) pointing.

A trapped region is then simply defined as the region containing all the trapped surfaces and the apparent horizon is its boundary. As a direct consequence of the above definitions, the apparent horizon is the surface where the normal component of the fluid velocity is equal to the local speed of sound. When the direction of the velocity is inward, the horizon is the boundary of a ``dumb hole'', \ie, it is the analogue of a~\emph{black} horizon, when it is outward, it is the analogue of a~\emph{white} horizon.

Future and past event horizons are defined in the same way as in GR, as the boundaries of the regions from which null geodesics cannot escape or cannot enter, respectively. However, this definition implies the knowledge of the behavior of null geodesics in the whole spacetime (till \scrip\/ or \scrim, respectively). In the following chapters, we will deal only with apparent horizons, that however coincide with event horizons when the geometry is stationary.

\subsection{The surface gravity}	%
\label{subsec:surfacegracity}		%

The presence of a preferred direction of time allows also to unambiguously define the concept of surface gravity. To this aim, in GR one needs to identify a timelike Killing vector which becomes null on the horizon. Instead, in analogue models one can always define the surface gravity with respect to the vector field $\pdt$, both in the case of steady flow (when $\pdt$ is a Killing vector) and in the more general case when the flow is not steady.

We compute here the surface gravity $g_{\rm H}$ in the simple case (see~\cite{P-G} for the general case) when the acoustic metric describes a static spacetime [see Eq.~\eqref{eq:static}]. This represents a special case of stationary spacetime, so that $K=\pdt$ is a Killing vector and the flow is steady.
Roughly speaking, the surface gravity is the acceleration, measured from infinity, felt by an observer at rest (with respect to the foliation defined by $K$) close to the horizon. The normalized vector tangent to the world line of this observer is
\begin{equation}
 V=\frac{K}{||K||},
\end{equation}
where, since $K$ is timelike
\begin{equation}
||K||=\sqrt{-K^\mu K_\mu}=\Omega\sqrt{c^2-v^2}.
\end{equation}
The four-acceleration of the observer is
\begin{equation}
 A^\mu=V^\nu\nabla_\nu V^\mu=\frac{1}{2}\frac{\nabla^\mu||K||^2}{||K||^2},
\end{equation}
where we used $\nabla_\mu K_\nu-\nabla_\nu K_\mu=0$. The proper acceleration is $||A||$, whereas the acceleration measured using the Killing time $t$ is
\begin{equation}
 \frac{\dd\tau}{\dd t}||A||=||K||\,||A||=||K||\sqrt{g^{\mu\nu}A^\mu A^\nu}=\frac{1}{2}\frac{\bv\cdot\nx(c^2-v^2)}{v}+O(c^2-v^2),
\end{equation}
since $\bv$ is parallel to $\nabla(c^2-v^2)$ as a consequence of Eq.~\eqref{eq:static}.
Taking the limit on the horizon, one obtains the surface gravity
\begin{equation}
 g_{\rm H}=\frac{1}{2}
 \left.
 \frac{\partial(c^2-v^2)}{\partial n}
 \right|_{\rm H},
\end{equation}
where $\bn$ is the unit vector normal to the horizon, that is parallel to $\bv$. For this reason, also $v=c$ on the horizon and
\begin{equation}
 g_{\rm H}=c_{\rm H}\,\left.\frac{\partial|c-v|}{\partial n}\right|_{\rm H}.
\end{equation}
With a bit of abuse of terminology we will call the surface gravity $\kappa$ the following quantity with dimension of a frequency
\begin{equation}\label{eq:kappa}
 \kappa\equiv\frac{g_{\rm H}}{c_{\rm H}}=\left.\frac{\partial|c-v|}{\partial n}\right|_{\rm H}.
\end{equation}

This quantity is fundamental because it fixes the temperature of the spectrum of the particles emitted in the Hawking process. In principle, one can apply to a phonon field propagating in an effective geometry with acoustic horizon the same argument used by Hawking~\cite{Hawking74, Hawking75} for a scalar field in a geometry with a black horizon. The Hawking temperature of the created phonons is computed below and given in Eq.~\eqref{eq:hawking_fluid},
from which it is clear the advantage of having introduced the frequency $\kappa$. In what follows we will work in units where the temperature has dimension of energy, that is $k_{\rm B}=1$.

The computation of the surface gravity in the general case of a stationary acoustic spacetime requires a bit more involved analysis. Eventually, the result is completely similar to the static one~\cite{P-G}
\begin{equation}
 g_{\rm H}=c_{\rm H}\,\left.\frac{\partial|c-v_\perp|}{\partial n}\right|_{\rm H},
\end{equation}
where $v_\perp=\bn\cdot\bv$ is the component of the velocity orthogonal to the horizon.

\subsection{Static black holes}		%
\label{subsec:schwarzschild}		%

The simplest black hole metric one can think about in GR is Schwarzschild's metric, which is the most general spherically symmetric  vacuum solution of Einstein's equations. The problem of finding an acoustic analogue of such a simple geometry is not as trivial as it might appear.

The standard way to tackle this problem is to look for a flow matching the Painlev\'e--Gullstrand~\cite{Painleve,Gullstrand,Lemaitre} representation of the Schwarzschild geometry
\begin{equation}\label{eq:PG}
\dd s^2 = 
- \left(1-\frac{2GM}{r}\right) \dd t^2 
\pm \sqrt{\frac{2GM}{r}}\, \dd r \, \dd t 
+ \dd r^2 + r^2\left( \dd\theta^2 + \sin^2\theta \; \dd\phi^2 \right).
\end{equation}
Unfortunately, the trivial guess
\begin{align}
  c&=\mbox{const.},\\
  \rho&=\mbox{const.},\\
  v&=\sqrt{\frac{2GM}{r}},
\end{align}
which would perfectly map the metric of Eq.~\eqref{eq:PG} onto the metric~\eqref{eq:fluidmetric}, does not satisfy the continuity equation~\eqref{eq:continuity} for steady flows [$\nx\!\cdot(\rho\bv)=0$]. The best one can do~\cite{P-G} is to take $c$ to be position independent and to fix $\rho\propto r^{-3/2}$, by imposing the continuity equation in spherical symmetry, such that the acoustic conformal factor is
\begin{equation}
\Omega^2\propto r^{-3/2}.
\end{equation}
In this way, one can reproduce a geometry which is conformal to Schwarzschild's one. This is fine when one is interested in conformal invariant quantities as the surface gravity or the Hawking temperature. However, while the Hawking mechanism is not affected by the conformal factor, the propagation of non-conformally coupled fields to the geometry depends on the value of $\Omega$.

Nevertheless, in the appendix of~\cite{silke}, it is shown that the isotropic version of Schwarzschild's geometry
\begin{equation}
 \dd s^2 = 
- \frac{\left(1-GM/2r\right)^2}{\left(1+GM/2r\right)^4} \dd t^2 
+ \left(1+\frac{GM}{2r}\right)^2\left[\dd r^2 + r^2\left( \dd\theta^2 + \sin^2\theta \; \dd\phi^2 \right)\right]
\end{equation}
can be casted into acoustic form, with the choice
\begin{align}
 v&=0,\\
 \rho&\propto1-\left(\frac{GM}{2r}\right)^2,\\
 c^2&\propto\frac{\left(1-GM/2r\right)^2}{\left(1+GM/2r\right)^6},
\end{align}
and adjusting the pressure using the definition of $c$ [see Eq:~\eqref{eq:csound}]. Although it is theoretically possible to set up a fluid with $\rho$ and $c$ specified by the above equations, we remark that a quite involved potential must be introduced in Euler's equation in order to obtain such profiles, making the practical realization of this system rather complicated.

Even if, in the end, it is possible to \emph{exactly} map the Schwarzschild geometry on an acoustic metric, this example shows how finding the acoustic counterpart of an acoustic geometry is generally not a trivial task. For this reason, it is important to explore new possible systems to extend the set of reproducible metrics (see Chap.~\ref{chap:relbec}).

\subsection{Acoustic causal structure}  %
\label{subsec:causalstructure}		%

\begin{figure}
 \centering
 \includegraphics[width=0.7\textwidth]{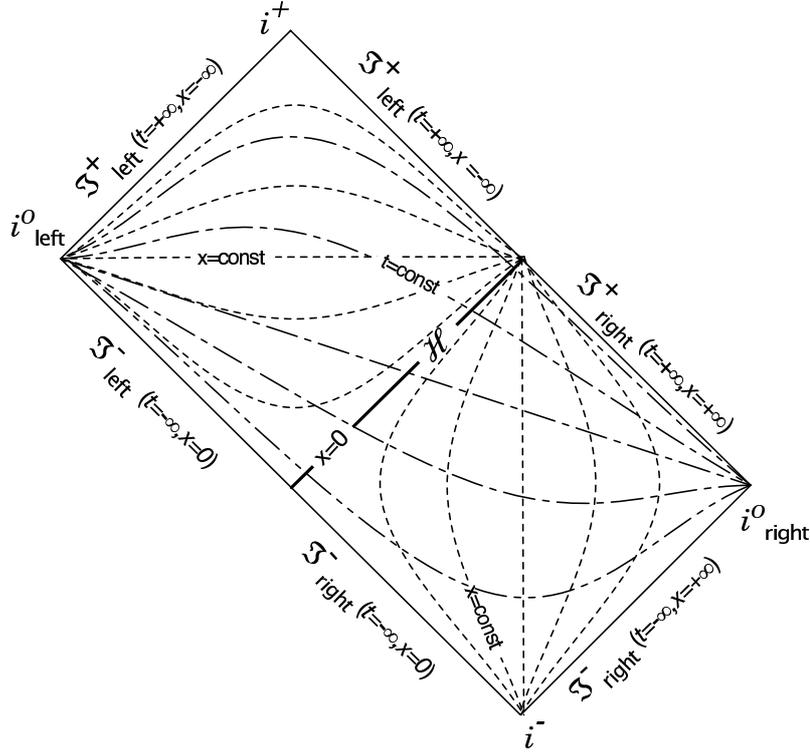}
 \caption{Penrose diagram of an acoustic black hole (From~\cite{causalstructure}). Small-dashed (respectively, dotdashed) lines are constant $x$ (respectively, $t$) curves.}
 \label{fig:bhpen}
\end{figure}

In general, one is often interested in phenomena that depend only on the spacetime causal structure, even when one cannot exactly cast a GR metric in acoustic form. In this case, it is enough to map the metric only up to a conformal factor. In fact, this task is usually much easier as we have just shown for the Painlev\'e--Gullstrand representation of the spherically symmetric black-hole.
A systematic study of acoustic causal structures was performed in~\cite{causalstructure}. As a propaedeutic exercise to the nontrivial construction of the causal structure of a dynamical warp drive (see Sec.~\ref{subsec:warpdrive}), we briefly summarize the steps to obtain the causal structure of an eternal acoustic black hole (see Fig.~\ref{fig:bhpen}).

To build the Penrose diagram for a given geometry we first need to reduce to $1+1$ dimensions. This is of course possible for a Schwarzschild black hole because of the spherical symmetry.\footnote{In rotating black holes one usually restricts the analysis to the equatorial plane and further reduces the number of dimensions to $1+1$, thanks to axial symmetry.} In $1+1$ dimensions the line element of Eq.~\eqref{eq:lineelement} simply reduces to
\begin{equation}\label{eq:1plus1element}
 \dd s^2=\Omega^2\left[-(c^2-v^2)\,\dd t^2-2v\,\dd x\,\dd t+\dd x^2\right].
\end{equation}
Then, defining retarded and advanced null coordinates
\begin{equation}\label{eq:dudv}
\begin{aligned}
 du&\equiv F(t,x)\left(\dd t-\frac{\dd x}{c(t,x)+v(t,x)}\right),\\
 dw&\equiv G(t,x)\left(\dd t+\frac{\dd x}{c(t,x)-v(t,x)}\right),
\end{aligned}
\end{equation}
Eq.~\eqref{eq:1plus1element} becomes
\begin{equation}
 \dd s^2=-\frac{\Omega^2}{F\,G}\left(c^2-v^2\right)\,\dd u\,\dd w
\end{equation}
and, in the case of a stationary geometry, one can choose $F=G=1$, because $c$ and $v$ do not depend on time. Assuming a left-going flow, that is $v(x)<0$, an acoustic horizon is present where $c+v=0$. We then expand this small quantity in a neighborhood of the horizon $x_{\rm H}$, using the definition of surface gravity of Eq.~\eqref{eq:kappa}:
\begin{equation}
 c(x)+v(x)=\kappa(x-x_{\rm H})+ O\left((x-x_{\rm H})^2\right),
\end{equation}
so that, from the point of view of an observer to the right of $x_{\rm H}$, the horizon corresponds to a black hole configuration.\footnote{For an observer to the right of $x_{\rm H}$, a white hole configuration would correspond to a right moving fluid with $v(x)-c(x)=-\kappa(x-x_{\rm H})+ O\left((x-x_{\rm H})^2\right)$.}

Integrating the differentials of Eq.~\eqref{eq:dudv} in a neighborhood of the horizons, we obtain
\begin{equation}
\begin{aligned}
u&\sim t-\frac{1}{\kappa}\,\ln|x-x_{\rm H}|,\\
w&\sim t+\frac{x}{2\,c}.%
\end{aligned}
\end{equation}
Since $u$ diverges when approaching $x_{\rm H}$, on both sides, it is not possible to cover the whole spacetime with $w$ and $u$ coordinates. Moreover, the $u$ coordinate does not uniquely identify a point in spacetime because of the symmetry $x-x_{\rm H}\to-(x-x_{\rm H})$. However, there exist null retarded coordinates that are regular on the horizons and uniquely identify the spacetime. For instance, the new coordinate
\begin{equation}\label{eq:U}
 U\propto\sign(x)\,\ee^{-\kappa u},
\end{equation}
satisfies both conditions.
Indeed, close to the horizon,
\begin{equation}
 U\sim (x-x_{\rm H})\,\ee^{-\kappa t}
\end{equation}
and $U$ is monotonic both in $r$ and $t$.
As a final step, with the usual compactification (see, for example,~\cite{Hawking-Ellis}), one defines two new null coordinates $\mathcal{U}$ and $\mathcal{V}$, both ranging in a compact set:
\begin{equation}
\begin{aligned}
  \mathcal{U}=\arctan(U),\\
  \mathcal{W}=\arctan(w).
 \end{aligned}
\end{equation}
\begin{figure}
 \centering
 \includegraphics[width=0.62\textwidth]{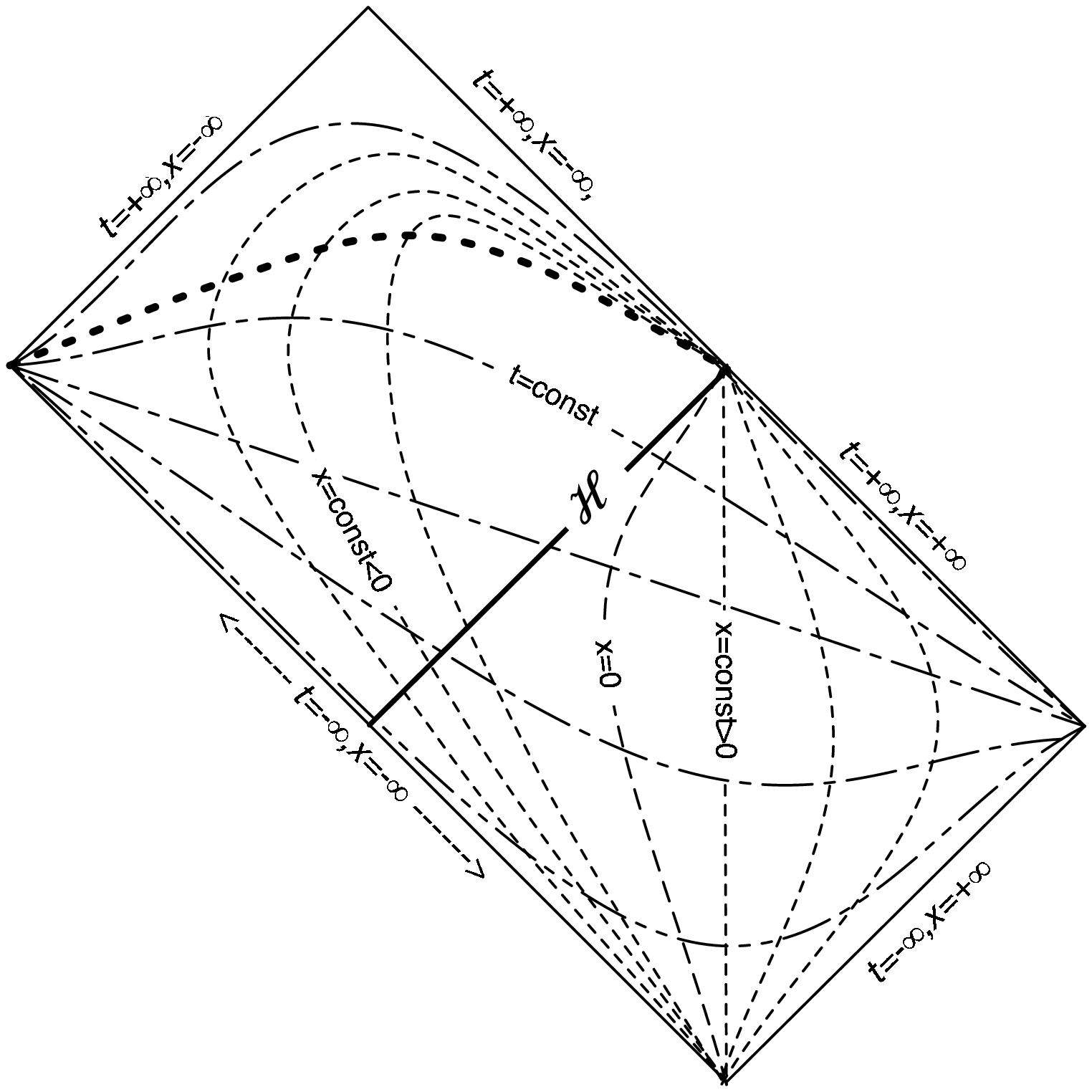}
 \caption{Conformal diagram of the switching on of an acoustic black hole (From~\cite{causalstructure}).
Note that the lines $x=\mbox{const}$ become null at the apparent
horizon (dashed line).}
 \label{fig:bhpen_dyn}
\end{figure}

Such a construction yields the Penrose diagram of Fig.~\ref{fig:bhpen} for a realistic flow with a ``dumb'' horizon without any singularity in the velocity profile (for example without the singularity at $r=0$ of the Schwarzschild geometry) and with two asymptotically flat regions ($v$ becomes constant for $x\to\pm\infty$). The horizon $\hor$ is placed at $x_{\rm H}=0$ and, by construction, no singularity is present beyond the horizon. Since the spacetime is stationary (steady flow), the apparent horizon $\hor$ coincides with the event horizon.

The construction of the Penrose diagram of the analogue of a collapsing black hole (Fig.~\ref{fig:bhpen_dyn}), that is an acoustic black hole whose horizon is created at a finite time, is straightforward. One has just to ``glue'' together a Minkowskian causal structure at $t\to-\infty$ with the acoustic black hole causal structure of Fig.~\ref{fig:bhpen}. In this case, one can show that the coordinate $U$ defined in Eq.~\eqref{eq:U} is an affine null parameter at $t\to-\infty$, while $u$ is affine at late times:
\begin{align}
 U&\sim t-\frac{x}{c},\qquad t\to-\infty,\quad x\to-\infty,\\
 u&\sim t-\frac{x}{c},\qquad t\to+\infty,\quad x\to+\infty.
\end{align}

\begin{figure}
 \centering
 \includegraphics[width=0.7\textwidth]{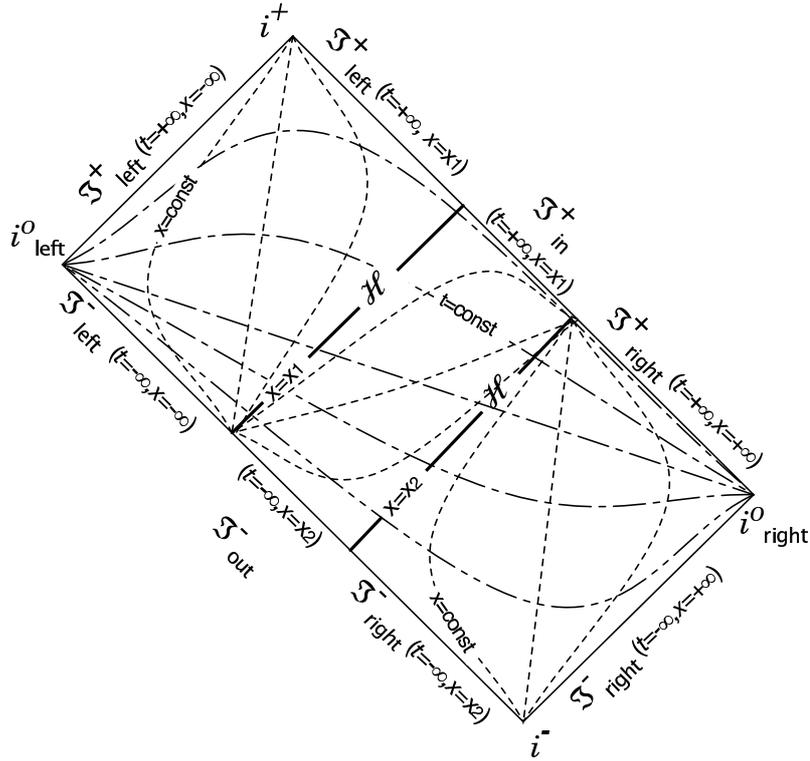}
 \caption{Penrose diagram of an acoustic white-hole--black-hole pair (From~\cite{causalstructure}). Small-dashed (dotdashed) lines are constant $x$ ($t$) curves.}
 \label{fig:laser-pen}
\end{figure}

Finally, in Fig.~\ref{fig:laser-pen} we show the causal structure of a white-hole--black-hole pair, in preparation of Chap.~\ref{chap:bhlaser}, where this system in studied in a BEC. A white-hole--black-hole pair is realized by a one-dimensional flow (we take $v<0$) with a central supersonic region and two asymptotically flat subsonic regions, thus presenting two horizons at $x_1$ and $x_2$. As we will see in details in Chap.~\ref{chap:bhlaser}, the left horizon $x_1$ behaves like a white hole for an observer in the left asymptotic region, while the right horizon $x_2$ behaves as a black hole for an observer in the right asymptotic region. Also in the acoustic geometry emerging from this system there are no singularities.

Remarkably, right-going rays propagating in the internal supersonic region are swept leftward towards $x_1$, but cannot cross it because the flow is subsonic for $x<x_1$ (of course, when the flow is subsonic right-going rays $u$ propagate rightward).

\subsection{Rotating black holes}	%
\label{subsec:kerr}			%

The attempts to find some analogue of rotating black hole geometries are mainly motivated by the hope of measuring the analogue of super-radiance~\cite{superrad}, an amplification of the modes due to the presence of an ergoregion~\cite{zeldovich}.
Unfortunately, the Kerr metric, describing a rotating black hole (vacuum solution of Einstein's equations with axial symmetry) cannot be casted in acoustic form, because of a very fundamental reason. While there is no foliation of Kerr's geometry with conformally flat spatial sections~\cite{garatprice,kroonprl,kroon}, the spatial sections of any acoustic geometry described by the metric of Eq.~\eqref{eq:fluidmetric} are indeed explicitly conformal invariant.

However, with a quite complicated change of coordinates, it is possible to cast the equatorial plane of Kerr's black hole in acoustic form~\cite{silke}. Using a simple vortex flow~\cite{P-G}, one can nevertheless reproduce the most relevant features of rotating black hole geometries, \ie, the presence of an ergosurface distinguished from an event horizon.

For an incompressible fluid, the continuity equation~\eqref{eq:continuity} and the no-vorticity condition imply that both the radial and tangential components of the fluid velocity are proportional to $1/r$. The line element is therefore
\begin{equation}
 \dd s^2=-c^2 \dd t^2+\left(\dd r-\frac{A}{r}\dd t\right)^2+\left(r\,\dd \theta-\frac{B}{r}\dd t\right)^2+\dd z^2,
\end{equation}
where we dropped the conformal factor that is position-independent because of incompressibility. Equivalently
\begin{equation}
 \dd s^2=-\left(c^2- \frac{A^2+B^2}{r^2}\right)\dd t^2-2\frac{A}{r}\dd t\,\dd r-2B\,\dd t\,\dd \theta+\dd r^2+r^2\,\dd \theta^2+\dd z^2.
\end{equation}
The ergosphere is the region where the Killing vector $K=\pdt$ becomes timelike. Its radius is fixed by putting to $0$ the $tt$-component of the metric:
\begin{equation}
 r_{\rm e}=\frac{\sqrt{A^2+B^2}}{c},
\end{equation}
while the acoustic horizon is placed where the radial velocity equals the speed of sound:
\begin{equation}
 t_{\rm h}=\frac{|A|}{c}\leq r_{\rm e}.
\end{equation}
When $A>0$ the vortex reproduces an acoustic white hole, while for $A<0$ it reproduces an acoustic black hole.

\subsection{Warp drives}	%
\label{subsec:warpdrive}	%

In this section, we show how the techniques developed for the analysis of the causal structure of acoustic spacetimes, presented in Sec.~\ref{subsec:causalstructure}, can be used to investigate the properties of warp-drive geometries~\cite{wd}.
Since Alcubierre introduced them \cite{alcubierre}, warp drives have been certainly one of the most studied spacetime geometries among those requiring exotic matter for its existence (see \cite{lobo2007} for a recent review about ``exotic spacetimes'').
They were immediately recognized as interesting configurations for two main reasons.

First, they provide, at least theoretically, a way to travel at superluminal speeds:
A warp drive can be described as a spheroidal bubble separating an almost flat internal region from an external
asymptotically flat spacetime with respect to which the bubble moves at an arbitrary speed. The idea is that, although
on top of a spacetime nothing can move with a speed greater than that of light, spacetime itself has no \emph{a priori} restriction on the speed with which it can be stretched.
In principle, if a huge amount of exotic matter (\ie, violating energy conditsion) were available (see beginning of Chap.~\ref{chap:warpdrive}), one might build a sort of railway which would stretch the spacetime in a properly synchronized manner, generating a warp-drive bubble that travels faster than light. However, the first construction and synchronization of the railway should proceed at subluminal speed, since nothing can move outside the lightcone.

Second, they provide an exciting ground for testing our comprehension of GR and QFT in CS (for instance when investigating warp-drive implications for causality \cite{everett}). In fact, a time machine~\cite{everett} could be built through a couple of superluminal warp drives traveling in opposite directions. Thus, the investigation of such geometries and the study of their stability properties can shed some light on the feasibility of spacetimes violating chronology and on the possible existence of chronology protection mechanisms~\cite{hawking-chronology}.

However, by a simple argument, it is easy to see that the geometry of a couple of warp drives used to travel in time cannot be put in acoustic form, even if the warp drive metric can. In fact, time travel is a coordinate independent concept (it resides on the existence of close timelike curves). In particular, if time travel existed in a spacetime describable by an acoustic metric, there would be closed timelike curves with respect to the laboratory time. This would imply that time travel would be possible in a lab! Of course, given that the laboratory lives in a Galilean invariant world, spacetimes with closed timelike curves cannot be described by acoustic metrics. In a more formal language, causal-related pathologies cannot arise in analogue spacetimes, since they inherit the property of stable causality from the underlying Newtonian spacetime, which has an absolute time.
Nevertheless the properties of acoustic metrics provide us with a lot of information about single warp-drive geometries (see Chap.~\ref{chap:warpdrive}).

\subsubsection{Eternal superluminal warp drives}	%
\label{subsec:etWD}					%

In their original form \cite{alcubierre}, the warp-drive geometries are described by the simple expression
\begin{equation}\label{eq:3Dalcubierre}
 ds^2=-c^2 dt^2+\left[dx-v(r)dt\right]^2+dy^2+dz^2,
\end{equation}
where $r\equiv \sqrt{(x-v_0 t)^2+y^2+z^2}$ is the distance from the center of the bubble {and} $v_0$ the warp-drive velocity. Here and thereafter $v=v_0 f(r)$ {with} $f$ a suitable smooth function satisfying $f(0)=1$ and $f(r)\to0$ for $r\to\infty$. 

The warp-drive metric~\eqref{eq:3Dalcubierre} is explicitly in fluid-like form [see Eq.~\eqref{eq:fluidmetric}].
We can therefore investigate the causal structure of a warp drive, following the method presented in \cite{causalstructure} and reviewed in Sec.~\ref{subsec:causalstructure}, for acoustic spacetimes.
We begin from an eternal warp drive, moving at constant velocity, and then we study a dynamic situation in which a warp-drive bubble is accelerated.

In 1+1 dimensions Alcubierre's metric~\eqref{eq:3Dalcubierre} reduces to
\begin{equation}\label{eq:alcubierre}
 ds^2=-c^2 dt^2+\left[dx-v(r)dt\right]^2.
\end{equation}
Here $r$ is defined as the signed distance from the center of the bubble, $r\equiv x-v_0 t$. Again, we define $v(r)=v_0f(r)$ but $f$ is now taken to be defined also for negative values of $r$. Its boundary conditions will 
be $f(0)=1$ and $f(r) \to 0$ for $r \to \pm \infty$.
For illustrative purposes, let us choose the following simple bell-shaped function:
\begin{equation}\label{eq:f_wardprive}
 f(r)=\frac{1}{\cosh\left(r/a\right)}
\end{equation}
and $v_0>c$, that is, our warp drive is superluminal. We stress that all the results of this section and Chap.~\ref{chap:warpdrive} do not depend on the particular choice of the function $f$. They are still valid providing that $f$ satisfies the above conditions. We have chosen a particular form for it just for simplicity.
Now, by using $(t,r)$ coordinates the metric Eq.~\eqref{eq:alcubierre} reads
\begin{equation}\label{eq:warpdrivemetric}
 ds^2=-c^2 dt^2+\left[dr-\bar{v}(r)dt\right]^2~,\qquad \bar{v}(r)=v(r)-v_0.
\end{equation}
Note that $\bar{v}<0$ because the warp drive is right going ($v>0$) but $v(r)\leq v_0$.

By definition, at any time the center of the bubble is located at $r=0$.
The causal structure of this metric can be analyzed following \cite{causalstructure}. Let us define the warp-drive Mach number $\alpha\equiv v_0/c$. The shift velocity becomes
\begin{equation}\label{eq:fluidvelocity}
 \bar{v}(r)=\alpha c\left[\frac{1}{\cosh\left(r/a\right)}-1\right].
\end{equation}
Two horizons appear when $\alpha>1$. Their locations are found by putting $\bar{v}$ equal to $-c$,
\begin{equation}\label{eq:sonicpoints}
r_{1,2}=\mp a \ln\left(\beta+\sqrt{\beta^2-1}\right),\qquad\beta\equiv\frac{\alpha}{\alpha-1}>1.
\end{equation}
Because of our simple profile choice, the two horizons are symmetrically located with respect to $r=0$. In more general situations $r_1$ and $r_2$ will be completely arbitrary satisfying only $r_1<r_2$.

\begin{figure}[t]
\centering
 \includegraphics[width=0.7\textwidth]{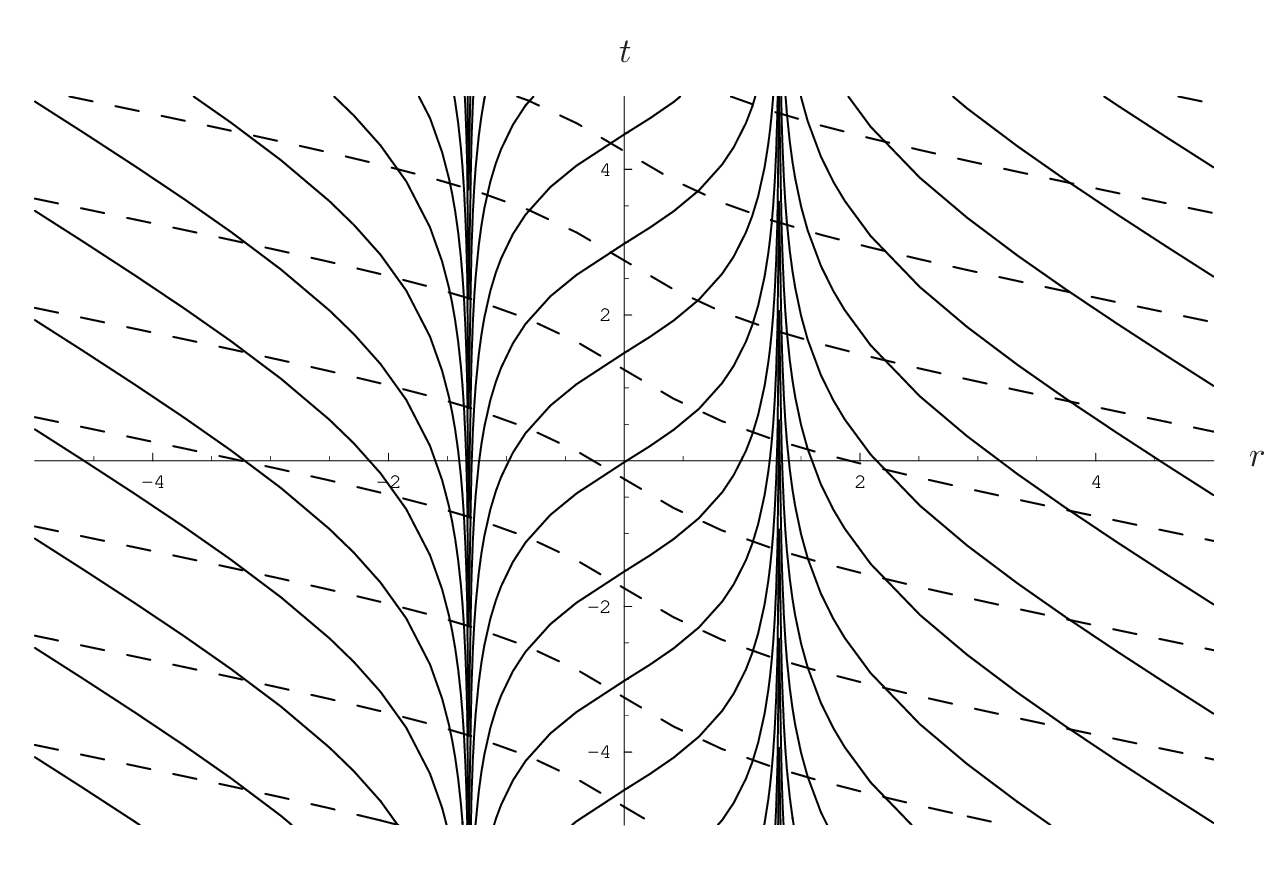}
 \caption{Lines of constant $u$ (solid lines) and $w$ (dashed lines) for the Alcubierre warp drive, $a=c=1$, $\alpha=2$.}
 \label{fig:wd-uw}
\end{figure}

Right- and left-going null coordinates $u$ and $v$ can be defined as
\begin{align}
 du &\equiv dt - \frac{dr}{c+\bar{v}(r)}, \label{eq:du}\\
 dw &\equiv dt + \frac{dr}{c-\bar{v}(r)}. \label{eq:dw}
\end{align}
We note that the spacetime is divided into three distinct regions: $\rm I$ ($r<r_1$), $\rm II$ ($r_1<r<r_2$), and $\rm III$ ($r>r_2$). There are no $u$ rays connecting these regions, while $w$ rays cross all the regions. When $r$ approaches the horizons, $u$ diverges logarithmically. Integrating these equations we find
\begin{align}
 u_{\rm I} &= t+\frac{r}{c\left(\alpha-1\right)}
	-\frac{a\beta}{c\left(\alpha-1\right)\sqrt{\beta^2-1}}\ln\left[\frac{e^{-(r-r_1)/a}-1}{e^{-(r-r_2)/a}-1}\right],
\\
 u_{\rm II} &= t+\frac{r}{c\left(\alpha-1\right)}
	-\frac{a\beta}{c\left(\alpha-1\right)\sqrt{\beta^2-1}}\ln\left[\frac{1-e^{-(r-r_1)/a}}{e^{-(r-r_2)/a}-1}\right],
\\
 u_{\rm III} &= t+\frac{r}{c\left(\alpha-1\right)}
	-\frac{a\beta}{c\left(\alpha-1\right)\sqrt{\beta^2-1}}\ln\left[\frac{1-e^{-(r-r_1)/a}}{1-e^{-(r-r_2)/a}}\right],
\\
 w &= t+\frac{r}{c\left(\alpha+1\right)}
	+\frac{2a\gamma}{c\left(\alpha+1\right)\sqrt{1-\gamma^2}}\arctan\left(\frac{e^{r/a}-\gamma}{\sqrt{1-\gamma^2}}\right),
\end{align}
where $\beta$ is defined in Eq.~\eqref{eq:sonicpoints} and $\gamma\equiv\alpha/\left(\alpha+1\right)<1$.
In Fig.~\ref{fig:wd-uw} we plot the  lines of constant $u$ and $w$.
The horizon at $r=r_1$ corresponds to $u_{\rm I}\to +\infty$ and to $u_{\rm II}\to +\infty$, while the horizon
at $r=r_2$ corresponds to $u_{\rm II}\to -\infty$ and to $u_{\rm III}\to -\infty$.

We define a signed surface gravity on our two horizons:
\begin{equation}
 \kappa_{1,2}
  \equiv {\left.\frac{dv(r)}{dr}\right|}_{r=r_{1,2}
 }=\pm\frac{c\left(\alpha-1\right)\sqrt{\beta^2-1}}{a\beta}\equiv\pm \kappa.
\end{equation}
So doing, the surface gravity associated with the first horizon is positive $\kappa_1=\kappa>0$, while
the one associated with the second horizon is negative $\kappa_2=-\kappa<0$.\footnote{In the definition of 
surface gravity, more correctly one should multiply the velocity derivative by $c$ to get an acceleration. However we use this slightly modified definition [see Eq.~\eqref{eq:kappa} in Sec.~\ref{subsec:surfacegracity}] to avoid the appearance of too many constant factors $c$ in our formulas.}
As we noticed for the positions of $r_1$ and $r_2$, in a general situation in which $f$ is not symmetric, the two surface gravities may have different absolute values. However, the one associated with the first horizon (respectively, second horizon) will be always positive (respectively, negative). We will soon show that these two horizons represent a black and a white horizon, respectively.
Here on, we will consider the two surface gravities to have the same absolute value. Retaining different absolute values will not lead to more general results, but will just make the notation heavier.

As in Sec. 6 of \cite{causalstructure} (see also Sec.~\ref{subsec:causalstructure}), one can find some transformation $U_i=U_i(u_i)$ that close to the horizons behaves as
\begin{eqnarray}
&&U_{\rm I}(u_{\rm I} \to +\infty) \simeq  U_{\rm I} + A_{\rm I} e^{-\kappa u_{\rm I}},
\\
&&U_{\rm II}(u_{\rm II} \to \pm \infty) \simeq U_{\rm II\pm} \mp A_{\rm II\pm} e^{\mp \kappa u_{\rm II}},
\\
&&U_{\rm III}(u_{\rm III} \to -\infty) \simeq U_{\rm III} - A_{\rm III} e^{\kappa u_{\rm III}},
\end{eqnarray}
where $U_{\rm I},U_{\rm II+},U_{\rm II-},\mbox{ and }U_{\rm III}$ are arbitrary constants and $A_{\rm I},A_{\rm II+},A_{\rm II-},\mbox{ and }A_{\rm III}$ are positive constants. It is possible to choose these transformations such that they match on the horizons, obtaining a global $U$ varying regularly from $+\infty$ to $-\infty$ as $r$ varies from $-\infty$ to $+\infty$.  {This matching will become natural when dealing with dynamical configurations in which the warp drive is created by accelerating the bubble from an initial zero velocity}. 
In order to do that we can choose $U_{\rm I}=U_{\rm II+}\equiv U_{\rm BH}$ and $U_{\rm II-}=U_{\rm III} \equiv U_{\rm WH}<U_{\rm BH}$.
In this way the three transformations have the following domains:
\begin{align}
 u_{\rm I} \in (-\infty, +\infty)\quad &\Rightarrow\quad U_{\rm I} \in (+\infty, U_{\rm BH}),\\
 u_{\rm II} \in (+\infty, -\infty)\quad &\Rightarrow\quad U_{\rm II} \in (U_{\rm BH}, U_{\rm WH}),\\
 u_{\rm III} \in (-\infty, +\infty)\quad &\Rightarrow\quad U_{\rm III} \in (U_{\rm WH}, -\infty).
\end{align}

\begin{figure}[t]
\centering
 \includegraphics[width=0.75\textwidth]{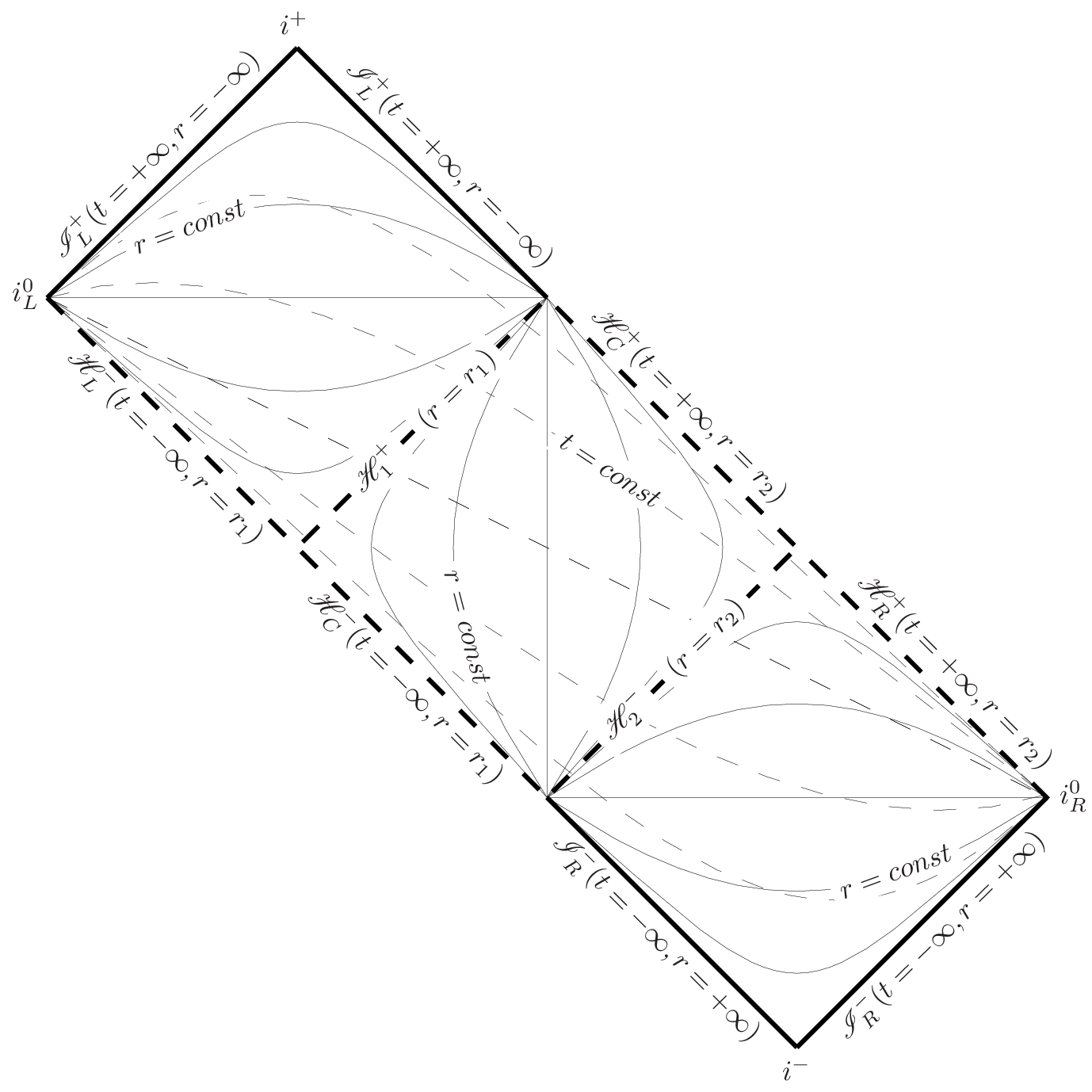}
 \caption{Penrose diagram of an eternal warp drive. Lines of constant $r$ (solid lines) and of constant $t$ (dashed lines). Future and past horizons at $r=r_1,r_2$ (heavy dashed lines). The geometry can be extended to the future of $\hor_C^+$ and $\hor_R^+$ and to the past of $\hor_C^-$ and $\hor_L^-$.}
 \label{fig:wd-pen}
\end{figure}

For instance, specific transformations having the required properties are the following:
\begin{align}
 &U_I=\frac{1}{2} + e^{-\kappa u_I},\\
 &U_{\rm II}=\frac{1}{2}\tanh\left(\frac{\kappa u_{\rm II}}{2}\right),\\
 &U_{\rm III}=-\frac{1}{2} - e^{\kappa u_{\rm III}}.
\end{align}
Now we can bring the right and left infinities to a finite distance by using a compactifying transformation like
\begin{align}
 \mathcal{U}_{\rm I} &\equiv \arctan(U_{\rm I}),\\
 \mathcal{U}_{\rm III} &\equiv \arctan(U_{\rm III}),\\
 \mathcal{W} &\equiv \arctan(w).
\end{align}
The Penrose diagram for this spacetime is plotted in Fig.~\ref{fig:wd-pen}. Notice that the diagram does not correspond
to a maximal analytical extension but to a particular patch of the total spacetime. The dashed lines signal the locations at which the geometry can be extended.
The two external regions $r>r_2$ and $r<r_1$ appear to an observer living inside the bubble as an eternal white hole and an eternal black hole, respectively. For this reason, we call \hplus~and \hminus, respectively, the \emph{black horizon} and the \emph{white horizon} of the warp drive. Notice however that, for the inner observers in an eternal configuration, at $r_1$ there is also a white horizon [$\hor_C^-(t=-\infty, r=r_1)$ in the diagram] and at $r_2$ a black horizon [$\hor_C^+(t=+\infty, r=r_2)$ in the diagram]. The geometry can be extended through these two null lines. We do not picture the extended regions as there are ambiguities in the prescription of the matter distribution in those other ``universes.''\footnote{In the analysis of the maximal extention of the Schwarzschild or Kerr spacetimes this problem is not present as one considers only vacuum solutions of the Einstein equations.}

\subsubsection{Dynamic superluminal warp drives}	%
\label{subsec:dynWD}					%

What happens to the causal structure when we consider the creation of a superluminal warp drive starting from initially flat spacetime? For concreteness, we study a simple case in which we reach the final velocity $v=v_0$ at a finite time which we take to be $t=0$. We modify the metric of Eq.~\eqref{eq:alcubierre} introducing a switching factor $\delta(t)$:
\begin{equation}\label{eq:dynamicalcubierre}
 ds^2=-c^2 dt^2+\left[dx-v(r,t)dt\right]^2,
\end{equation}
where
\begin{equation}
v(r,t)=v_0\delta(t)f(r),
\end{equation}
with  $f(r)$ defined in Eq.~\eqref{eq:f_wardprive} and
\begin{equation}
 \delta(t)\equiv
 \left\{
  \begin{aligned}
    &e^{t/\tau}\qquad &\text{if} \quad t<0,  \\
    &1\qquad &\text{if}\quad t \geq 0.
  \end{aligned}
 \right.
\end{equation}
Again, we can change coordinates, keeping the center of the bubble at rest ($r=0$). This can be obtained by defining
\begin{equation}
 dr\equiv dx-v_0\delta(t) dt.
\end{equation}
This is an exact differential form and can be integrated to get
\begin{equation}
 r=
 \left\{
  \begin{aligned}
    &x-v_0\tau\left[e^{t/\tau}-1\right]\qquad &\text{if} \quad t<0,  \\
    &x-v_0 t\qquad &\text{if}\quad t\geq 0.
  \end{aligned}
 \right.
\end{equation}
The metric becomes
\begin{equation}\label{eq:dynamicfluidmetric}
\begin{split}
 &ds^2 =-c^2 dt^2+\left[dr-\hat{v}(r,t)dt\right]^2,\\
 &\hat{v}(r,t) =v_0\delta(t)\left[f(r)-1\right],
\end{split}
\end{equation}
and the apparent horizons associated with the $t$~slicing are found by setting $\hat{v}=-c$. In this case a solution does not exist for any value of $t$, so that the apparent horizons are created at infinity at some finite $t_{\rm H}$. We show this below. Let us write the equation for the apparent horizons in the following form:
\begin{equation}\label{eq:eqsonic}
 f(r)=1-\frac{c}{v_0\delta(t)}.
\end{equation}
The function $f$ takes all the values between $0$ and $1$. In particular, $f(r)\to 0$ for $r\to\pm\infty$ and $f(0)=1$. Besides, the right-hand side of Eq.~\eqref{eq:eqsonic} is a monotonic function of $t$, such that, for $t\to-\infty$, $1-c/\left(v_0\delta(t)\right)\to-\infty$ and reaches the value $1-c/v_0>0$ for $t \geq 0$. As a consequence, there exists a time $t_{\rm H}<0$ 
so that, for $t>t_{\rm H}$, there are always two solutions of Eq.~\eqref{eq:eqsonic}, corresponding to a black and a white horizon. These horizons are born at $t=t_{\rm H}$ at spatial infinity and at $t=0$ they settle at two fixed positions $r_1$ and $r_2$.

\begin{figure}
\centering
\includegraphics[width=.75\textwidth]{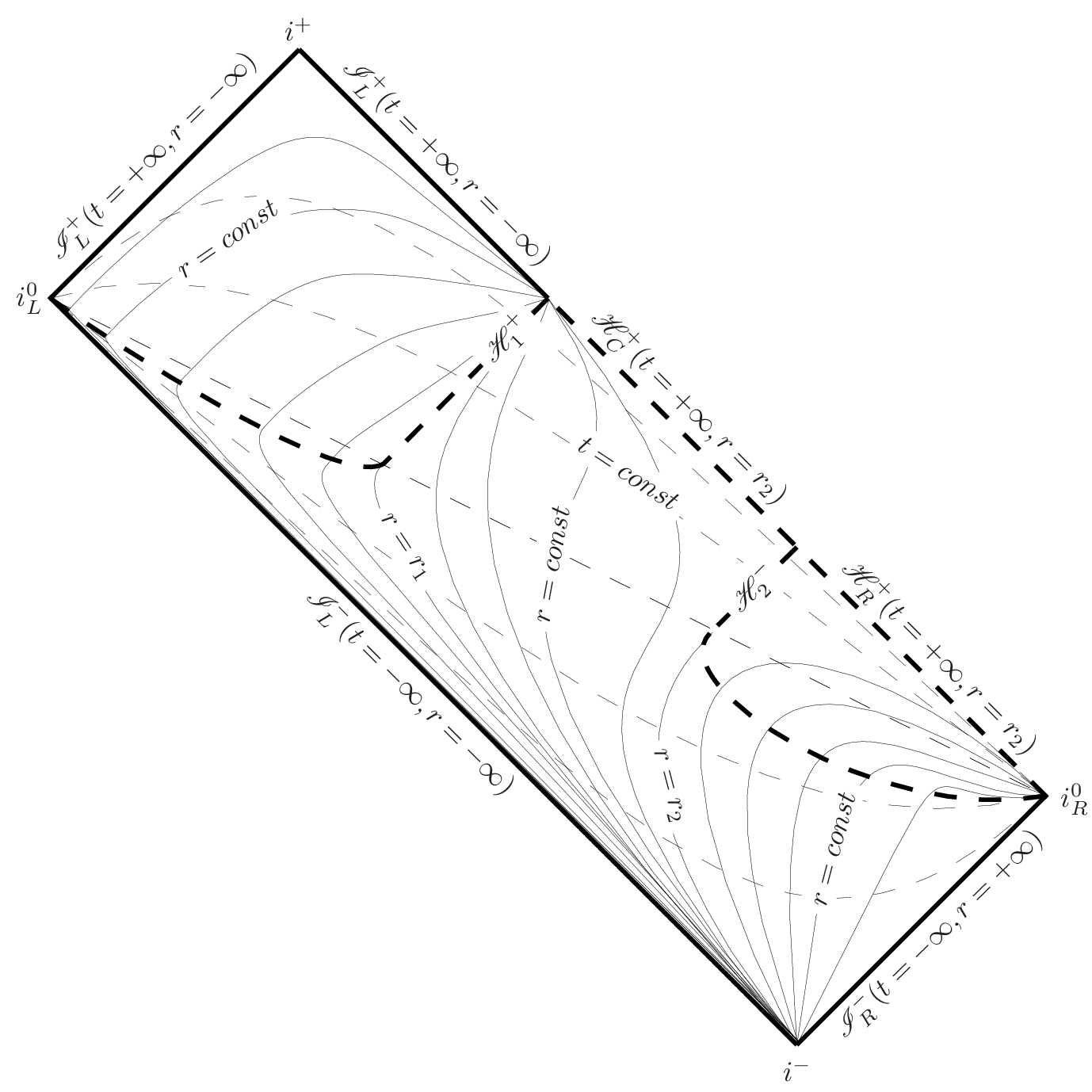}
 \caption{Penrose diagram of a dynamic warp drive. Lines of constant $r$ (solid lines) and of constant $t$ (dashed lines). The lines of constant $r$ become null at the apparent horizons (heavy dashed lines).}
 \label{fig:dynwd-pen}
\end{figure}
Keeping these points in mind, we are able to build the Penrose diagram for the dynamic warp drive~(Fig.~\ref{fig:dynwd-pen}). At early times the metric is approximately Minkowskian, because $\delta(t)\to0$ for $t\to-\infty$. 
Therefore, the causal structure is initially Minkowskian. Then, it progressively changes till $t=0$. At this time one has built a stationary warp drive, just as in the previous section.
After this time, the Penrose diagram looks exactly equal to that in Fig.~\ref{fig:wd-pen}. The final Penrose diagram is just obtained by gluing together the two behaviors. Again, we are not drawing an analytically extended diagram but only the relevant patch for the analysis that follows in Chap.~\ref{chap:warpdrive}. Given that a timelike observer can reach   \scribub~and $\hor_R^+$ in a finite proper time, the geometry can be extended in the future, beyond these lines. This is actually a subtle point. In fact,  \scribub~and $\hor_R^+$ (which are linked to the formation of a white horizon) are on the boundary of the Cauchy development of \scrim. In this sense, they are Cauchy horizons given that initial data are assigned only on \scrim. Hence, as noticed in Sec.~\ref{subsec:etWD}, a possible extension would not be unique. In any case such an extension will not be relevant for what follows, given that we shall limit ourselves to investigate the asymptotic behavior of the RSET associated with the onset of the superluminal warp drive.

Let us stress that there are other possible ways to create a dynamical warp drive. In Sec.~\ref{sec:kink}, we shall choose a different interpolation between the Minkowski and the warp-drive metrics, in which the horizons appear both at finite time and at finite space positions.

\subsection{Cosmological metrics}	%
\label{subsec:cosmo}			%

In addition to geometries with horizons, analogue models can reproduce cosmological metrics. The interest in the study of this kind of acoustic metric is mainly oriented to the analysis of particle creation in expanding or contracting universes (see Sec.~\ref{subsec:universe})~\cite{expandinguniverse_blv, Barcelo:2003wu,  expandinguniverse_ff, Fedichev:2003bv, Fedichev:2003dj, expandinguniverse_piyush,Lidsey:2003ze, Weinfurtner:2004mu,silke_thesis,expandinguniverse_silke,silke_guide}. Moreover, for the first time, the analogy revealed its power not only in the investigation of theoretical issues related to gravity, but, conversely, to explain an experimental result in condensed matter (the Bosenova phenomenon~\cite{Donley}, see Sec.~\ref{subsec:universe}) using techniques of QFT in CS~\cite{Calzetta:2002zz}.

Starting from the acoustic metric of Eq.~\eqref{eq:fluidmetric}, there are two simple ways to reproduce a Friedmann--Robertson--Walker (FRW) metric either by changing the velocity $\bv$ of the fluid (for instance, making it expanding) or by keeping the fluid at rest and changing only the speed of sound.
In the first case, one can take a purely radial fluid velocity $\bv=(\dot b/b){\bf r}\equiv H(t){\bf r}$, with $b(t)$ a scale factor depending only on time. Defining a new radial coordinate $r_b\equiv r/b$, the line element of Eq.~\eqref{eq:lineelement} is
\begin{equation}\label{eq:lineelement_calcfrw}
 \dd s^2=\frac{\rho}{c}\left[-c^2\dd t^2+b^2\left(\dd r_b^2+r_b^2\dd\Omega^2\right)\right].
\end{equation}
From the continuity equation~\eqref{eq:cont}
\begin{equation}
 \rho(t)=\frac{\rn}{b^3(t)},
\end{equation}
where $\rn$ is a constant. Introducing the new time variable
\begin{equation}\label{eq:frwtime}
 \tau\equiv\int \sqrt{\rho c}\,\dd t
\end{equation}
and defining
\begin{equation}
 a(\tau)\equiv b\sqrt\frac{\rho}{c},
\end{equation}
the line element~\eqref{eq:lineelement_calcfrw} reads\
\begin{equation}\label{eq:lineelementflatfrw}
 \dd s^2=-\dd\tau^2+a(\tau)^2\left(\dd r_b^2+r_b^2\dd\Omega^2\right),
\end{equation}
corresponding to the metric of a FRW universe.

In the second case, keeping the fluid at rest, the line element of the acoustic metric is simply
\begin{equation}
 \dd s^2=-\rho\,c\,\dd t^2+\frac{\rho}{c}\dd\bx^2.
\end{equation}
Defining the proper time $\tau$ as in Eq.~\eqref{eq:frwtime} and the scale factor as $a(\tau)\equiv\sqrt{\rho/c}$, one obtains again the metric of Eq.~\eqref{eq:lineelementflatfrw}.

In both cases, one can simulate only FRW universes with flat spatial sections ($k=0$)~(see, for instance,~\cite{Hawking-Ellis}). We show in Chap.~\ref{chap:relbec}~\cite{relbec} that, using the emergent metric of a relativistic fluid (that is satisfying Lorentz invariance instead of Galilean invariance), the set of reproducible metrics extends to FRW universes with hyperbolic spatial sections ($k=-1$).

\section{Phenomenology}		%
\label{sec:phenomenon}		%

As discussed in the~\nameref{chap:intro}, the main motivation for introducing analogue models of gravity is the necessity to look for systems where various gravitational phenomena can occur, with the twofold objective of improving our theoretical understanding and of experimentally detecting phenomena that cannot be tested otherwise.
In this section we very briefly review the phenomenology of analogue models, stressing similarities and differences with respect to proper gravitational phenomena. We will also present how the study of these models has been further stimulated by the possibility to use them in the investigation of the robustness and the possible modifications of these phenomena when allowing for high-energy/small-scale modifications of the continuous spacetime structure at Planck scale. In fact, all analogue systems break the emergent local Lorentz invariance at the atomic or molecular scale. Differently from gravity and QFT, in these systems the microscopic physics is completely under control, making them a good test field for the analysis of these issues.

Before passing to the analysis of the specific phenomenology, we just briefly recall how it is possible to produce particles from vacuum when a quantum field lives in a curved geometry~\cite{birreldavies}. Let us take for simplicity a scalar field $\hat\phi$ minimally coupled to gravity. Its dynamics is described by
\begin{equation}\label{eq:generalKG}
 \left(\Box_g-m^2\right)\hat\phi=\left(\frac{1}{\sqrt{-g}}\partial_\mu\sqrt{-g}\,g^{\mu\nu}\partial_\nu-m^2\right)\hat\phi=0.
\end{equation}
To define particles, the geometry must have some timelike Killing vector. Different Killing observes expand the field using their own frequency eigenmodes, corresponding to different Killing vectors. Let us take two such observers and expand the field in their normalized frequency eigenmodes ${u_i}$ and $\bar{u}_i$, respectively
\begin{align}
 \hat\phi(x) &= \sum_i \hat a_i u_i(x)+\hat a_i^\dag u_i^*(x),\\
 \hat\phi(x) &= \sum_i \hat{\bar{a}}_i \bar{u}_i(x)+\hat{\bar{a}}_i^\dag \bar{u}_i^*(x),
\end{align}
where $x=(t,\bx)$ and, if $\partial_t$ and $\partial_{\bar t}$ are the timelike Killing vectors of the two observers, respectively, ${u_i}$ and $\bar{u_i}$ satisfy
\begin{align}
 \pdt u_i(x)&= -\ii \omega_i u_i(x),\\
 \partial_{\bar t}\bar u_i(x) &= -\ii\bar\omega_i\bar{u}_i(x).
\end{align}
In general ${u_i}$ and $\bar{u_i}$ are related by a transformation of the form
\begin{equation}
 \bar{u}_j=\sum_i\left(\alpha_{ji}u_i+\beta_{ji}u^*_i\right),
\end{equation}
where the terms $\beta_{ji}u^*_i$ allow for mixing of positive and negative energy modes. If they do not vanish, a mode containing only positive frequencies with respect to the time $t$ will also contain negative frequencies with respect to the time $\bar{t}$. This is the crucial point at the origin of spontaneous particle creation. The transformation on the modes induces a transformation on the operators $\hat a$ and $\hat a^\dag$:
\begin{equation}
 \hat a_i=\sum_j\left(\alpha_{ji}\hat{\bar{a}}_j+\beta_{ji}^*\hat{\bar{a}}_j^\dag\right).
\end{equation}
Defining the vacuum state $\ket{\bar0}$ for modes $\bar{u}_j$ as usual
\begin{equation}
 \hat{\bar{a}}_j\ket{\bar{0}}=0~,\quad \ket{\bar{1}_j}=\hat{\bar a}_j^\dag\ket{\bar0},\quad\dots,
\end{equation}
the operators $\hat a_i$ do not annihilate $\ket{\bar0}$, provided $\beta_{ji}\neq0$,
\begin{equation}
 \hat a_i\ket{\bar{0}}=\sum_j\beta_{ji}^*\ket{\bar1_j}\neq0,
\end{equation}
and the expectation value of the particle number operator $\hat N_i=\hat a^\dag_i \hat a_i$ is
\begin{equation}\label{eq:particlenumber}
 \bra{\bar0}\hat N_i\ket{\bar0}=\sum_j{\left|\beta_{ji}\right|}^2.
\end{equation}
Thus, for the unbarred observer, the state $\ket{\bar0}$, which is vacuum for the barred observer, is full of particles when the $\beta$ coefficients do not vanish, that is when there is a mixing between negative and positive frequency modes.

As anticipated, this is the crucial feature underlying particle pair creation. The values of the coefficients $\beta$ is fixed by the properties of the metric $g_{\mu\nu}$. This derivation is completely kinematic, in the sense that no information about the dynamics of the spacetime is needed, nor on how the field backreacts on the geometry. This means that it can be repeated as it is, every time there is a quantum field whose behavior is governed by an equation of the form~\eqref{eq:generalKG}. Furthermore, if two completely different systems are nevertheless described by the same metric $g_{\mu\nu}$, the values of the $\beta$ and $\alpha$ coefficients are exactly the same, leading to the same emission of particles. This argument is at the basis of the investigation of the phenomenology of 
analogue models.

\subsection{Particle production in expanding universes}	%
\label{subsec:universe}					%

Even if in the rest of the Thesis we will deal with particle creation in black/white-hole-like geometries, we start with a brief section on particle production in expanding universes for two reasons. First, they are easier to deal with. Second, they are the only case in which AG has been used backward to explain a condensed matter experiment~\cite{Donley,Calzetta:2002zz} with standard tools of QFT in CS, although this interpretation is still controversial.

\begin{figure}
 \psfrag{c}[c][c]{$C(\eta)$}
 \psfrag{e}[c][c]{$\eta$}
 \psfrag{m}[c][c]{$A-B$}
 \psfrag{p}[c][c]{$A+B$}
 \psfrag{in}[c][c]{In region}
 \psfrag{out}[c][c]{Out region}
 \includegraphics[width=\textwidth]{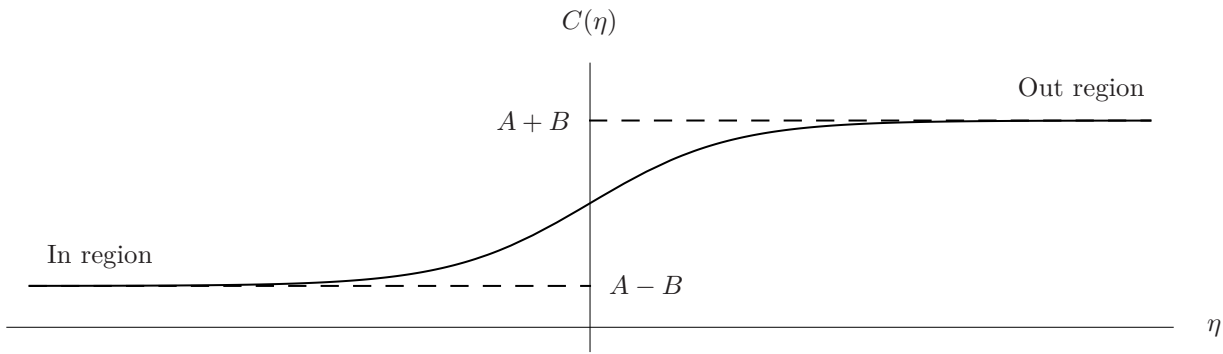}
 \caption{Scale factor of the metric of Eq.~\eqref{eq:simpleFRW}. $C$ is constant for both $\eta\to-\infty$ and $\eta\to+\infty$.}
 \label{fig:simpleFRW}
\end{figure}

Just to focus the problem, let us take a two-dimensional universe whose scale factor is constant both at early and late times, such that there are early- and late-time Killing vectors (see Fig.~\ref{fig:simpleFRW}):
\begin{equation}\label{eq:simpleFRW}
 ds^2=C(\eta)\left(d\eta^2-dx^2\right),\quad C(\eta)=A+B\tanh(\rho\eta).
\end{equation}
One can find two complete sets of modes, solutions of Eq.~\eqref{eq:generalKG}, which behave as plane waves at $\eta\to\pm\infty$, respectively
\begin{gather}
  \lim_{t\to-\infty}u_k^{\rm in}(\eta) = (8\pi^2\omega_{\rm in})^{-1/2}e^{-i\omega_{\rm in}\eta+ikx}~,\quad\omega_{\rm in}=\sqrt{k^2+m^2(A-B)},\\
  \lim_{t\to+\infty}u_k^{\rm out}(\eta) = (8\pi^2\omega_{\rm out})^{-1/2}e^{-i\omega_{\rm out}\eta+ikx}~,\quad\omega_{\rm out}=\sqrt{k^2+m^2(A+B)},
\end{gather}
and compute the transformation connecting the two sets
\begin{equation}
 u_k^{\rm in}(\eta,x)=\alpha_{k}u_k^{\rm out}(\eta,x)+\beta_{k}{u}^{{\rm out}*}_{-k}(\eta,x).
\end{equation}
Using Eq.~\eqref{eq:particlenumber}, one obtains the occupation number of out-particles, when there are no particles in the initial state
\begin{equation}\label{eq:createdparticles}
 \bra{0_{\rm in}}\hat N_k^{\rm out}\ket{0_{\rm in}}=|\beta_k|^2=\frac{\sinh^2(\pi\omega_+/\rho)}{\sinh(\pi\omega_{\rm in}/\rho)\sinh(\pi\omega_{\rm out}/\rho)},
\end{equation}
where
\begin{equation}
 \omega_\pm=\frac{1}{2}\left(\omega_{\rm out}\pm\omega_{\rm in}\right).
\end{equation}
Equation~\eqref{eq:createdparticles} illustrates well how particles are actually created in an expanding universe.

Going back to AG, a lot of models have been considered so far~\cite{expandinguniverse_blv, Barcelo:2003wu,  expandinguniverse_ff, Fedichev:2003bv, Fedichev:2003dj, expandinguniverse_piyush,Lidsey:2003ze, Weinfurtner:2004mu,silke_thesis,expandinguniverse_silke,silke_guide}. Both analytical results and simulations confirm that all these systems must show particle creation.
More interestingly, the analogy with gravitational phenomena was used in~\cite{Calzetta:2002zz} to explain the experimental results of~\cite{Donley}. In that experiment a condensate of a few milion $^{85}\mbox{Rb}$ atoms at a temperature of $3\,\mbox{nK}$ is made unstable by tuning the interactions between atoms. Reverting the sing of interaction from repulsive to attractive, after a short waiting time, the condensate collapses. At the same time, it loses a fraction of atoms in a ``nova''-like burst. It is possible to show that the dynamics of quantum excitations in such a collapsing BEC is closely similar to the behavior of a quantum field in a contracting universe (that is the time reversal of the above example).

Moreover, the analogy becomes even deeper. If after some time, but before the condensates has time to stabilize, the sign of the interaction is again suddenly reversed and made repulsive, jets of atoms with very low energy are emitted. Using again the analogy with universe expansion, in~\cite{Calzetta:2002zz} it was shown that large wavelength modes are amplified, being outside the analogue cosmological horizon. When the condensate start expanding after the last inversion of the interaction sign, those amplified modes go back into the horizon. They stop growing and start oscillating, that is particle are created. Since their wavelength is large, they have a very low frequency.

However, this simple model is still controversial and cannot explain all the details of the Bosenova phenomenon. Indeed, altough this analogy describes well the overall qualitative features of the collapse and also provides some quantitative fits with experimental data, the authors themselves recognize that their model is not satisfactory in predicting the bursts and that the backreaction of the excitations on the condensate should be considered in a more accurate analysis.
However, this result is an important indication that tools borrowed from QFT, GR, and cosmology can be profitably used to understand in an elegant way phenomena in other fields of physics, reinforcing the solidity of the AG program.

To conclude this section, it is fair to say that excitations in condensates by a time-modulation of the scattering length or of the effective atomic mass are well-known by the cold atom community~\cite{ETH1,ETH2,patterns,faraday} and have been independently explained without any analogy with quantum field theory in curved spacetime~\cite{interpretation}. More information can be found in~\cite{carusottocasimir}.

\subsection{Hawking radiation}	%
\label{subsec:hawking}		%

Hawking radiation was discovered in 1974~\cite{Hawking74,Hawking75} as a thermal black body radiation emitted from black holes, with temperature
\begin{equation}\label{hawking_blackhole}
 T_{\rm H}=\frac{\hbar\,c^3}{8\pi\,GM\,k_B}\approx 6.17\times10^{-8}\frac{M_\odot}{M}\,\mbox{K}.
\end{equation}
For the Hawking flux to overcome the cosmic microwave background, whose temperature is $3\,\mbox{K}$, the emitting black hole should have a mass of $10^{22}\,\mbox{kg}$ and a radius of about $7\times10^{-6}\,
\mbox{m}$. Unfortunately, we have observational evidence only of stellar mass black holes or supermassive black holes (millions of solar masses).

However, the importance of this phenomenon goes further beyond the fascinating result that black holes are not really black (even if their emission is very faint), once quantum effects are taken into account. Indeed, they are required to thermally emit by black-hole thermodynamics laws~\cite{BCH,bekenstein} as a consistency condition. This explains the efforts made in the last decades to find some system where this effect could be revealed.

Although the original Hawking derivation was performed in a spacetime with a collapsing star that eventually ends up in a black hole by forming a horizon, it is easy to see that the only fundamental assumption is that the spacetime has an apparent horizon, no matter what are the laws governing the geometry dynamics~\cite{Visser:2001kq}. This means that one can in principle reproduce Hawking radiation in a laboratory, provided that one has a system where there is some kind of perturbation that can be described by a relativistic quantum field (for example phonons in a BEC, see Chap.~\ref{chap:hawking}), propagating in an effective geometry with some sort of horizon. Moreover, the very fundamental process at the basis of Hawking radiation is a general form of vacuum instability through particle pair creation. This phenomenon takes place whenever there is some region of space where negative norm modes can propagate~\cite{Primer}. In particular it was found~\cite{MGR,stephens,BPS,parentanibrout} a close analogy between the Hawking mechanism and pair production in static electric fields~\cite{schwinger}.

Even if the mechanism at the basis of Hawking radiation is much more general than what may initially appear, the application of the analogy to fluid mechanics or condensed matter analogues is not completely straightforward because of some important differences between a relativistic field propagating in a CS and phonons on top of a fluid. After the discovery of the Hawking effect, it was soon realized~\cite{Unruh:1976db} that the original calculation resided on the validity of QFT up to arbitrary high energies. Indeed, if a quantum of finite energy is traced back from future infinity backward in time, it gets exponentially blueshifted while approaching the horizon. This implies that possible modifications related to unknown physics at Planck scale could strongly influence the Hawking process. One might object that the infinite blueshift suffered by quanta at the horizon is just an effect related to the system of reference and that it can be eliminated by choosing a different observer. However, this eliminates the trans-Planckian problem only in one direction, nothing changes in the orthogonal direction and the problem is even worse in the opposite one, because it is not possible to boost away an $s$-wave.

In this sense, analogue systems provide a clear example where the geometry breaks down at scales as small as the atomic or molecular scale.
To reproduce the Hawking derivation for the simple acoustic black hole, whose geometry was studied in Sec.~\ref{subsec:causalstructure}, one can use the results given at the begininning of this section for the computation of the spectrum of the emitted particles.
We define two observers, one at early times, before the creation of the horizon, and the other one at late times, far from the horizon ($x\to+\infty$). The late-time observer expands the field $\hat\phi$ in eigenmodes
\begin{equation}
 u_\omega^{\rm out}\propto e^{-i\omega u},\\
\end{equation}
whereas the early-time observer will use his own frequency eigenmodes
\begin{equation}
  u_{\omega'}^{\rm in}\propto e^{-i\omega' U}=e^{-i\omega' p(u)},
\end{equation}
where $U=p(u)$ is a function which fixes the coordinate transformation and behaves as in Eq.~\eqref{eq:U} for $u\to+\infty$.

From this relation, one can compute the transformation rules of the modes $u_{\omega'}^{\rm in}$ and $u_{\omega}^{\rm out}$ and, finally, obtain the occupation number of out-particles in the in-vacuum state (Eq.~\eqref{eq:particlenumber})
\begin{equation}\label{eq:fluidboseeinstein}
 N_\omega=\sum_{\omega'}|\beta_{\omega'\omega}|^2\propto\frac{1}{e^{\hbar\omega/k_B \Th}-1},
\end{equation}
where $\Th$ is the acoustic Hawking temperature
\begin{equation}\label{eq:hawking_fluid}
 k_{\rm B}T_{\rm H}=\frac{\hbar g_{\rm H}}{2\pi c_{\rm H}}=\frac{\hbar \kappa}{2\pi}.
\end{equation}
Eq.~\eqref{eq:fluidboseeinstein} represents a Bose--Einstein distribution of particles, corresponding to a Planck spectrum with temperature $\Th$.

Apparently nothing peculiar appears in the derivation of Eq.~\eqref{eq:fluidboseeinstein}. However, to obtain this result, we started from the relation $U=p(u)$ whose asymptotic behavior is given by Eq.~\eqref{eq:U}. The form of $p(U)$ actually resides only on the particular choice of the fluid metric of Eq.~\eqref{eq:fluidmetric}, which was directly derived from fluid dynamics equations~\eqref{eq:continuity} and~\eqref{eq:euler}. This means that this result is valid as long as one can describe the system using fluid dynamics equations [plus the assumptions summarized after Eq.~\eqref{eq:conformalfactor}], which describe the fluid only at scales much longer than the molecular one, that is in the hydrodynamical approximation. If Hawking effect really depended on the spacetime structure at any arbitrary small scale (arbitrary high frequency), analogue models should be a good place to check this possibility. For this reason, a lot of efforts have been devoted both to directly study real physical systems, where perturbations are described by a non-relativistic dispersion relation at very high frequencies, and to investigate more general situations, where deviations from exact local Lorentz symmetry are parametrized in a way that mimics the high energy behavior of some analogue models.

Finally, other issues must be taken into account when planning possible experimental measurements of the Hawking effect. In fact, because of the faintness of this radiation, it is very difficult to find a system where this effect is not masked by noise. In particular, since the expected Hawking spectrum is thermal, it is even more difficult to detect this signal, because it lacks any peculiar feature.

In Chap.~\ref{chap:physics} we shall discuss how these aspects show up in real physical systems. We shall discuss both the expected robustness of analogue Hawking radiation, taking into account the breaking of the effective geometry at molecular scales, and
the experimental limits in possible measurements. Furthermore, Part~\ref{part:BECphenomenology} is devoted to the analysis of Hawking radiation and related phenomena in the specific case of BECs.

\chapter{An~application~of~analogue~gravity: The~warp-drive~semiclassical~instability}	
\label{chap:warpdrive}									
\chaptermark{Warp-drive instability}							

In Sec.~\ref{subsec:warpdrive} we introduced the warp-drive geometry and studied the causal structure of both an eternal and a dynamic warp drive. We stressed there the importace of such systems as a test field for the investigation of fundamental issues in gravity. In particular, they can improve our understanding of causal properties of spacetime, since, in principle, one might use a couple of superluminal warp drives to build a time machine~\cite{everett}.

From the point of view of analogue gravity this system is a nice example where techniques developed for analogue models can be used to investigate properties of a spacetime in general relativity (GR). In this chapter we study the stability properties of warp drives in presence of quantum effects that we treat at a semiclassical level. In Chap.~\ref{chap:warpdriveBEC} we shall extend the present analysis to a case in which a modified dispersion relation is present. In that case we shall work with a real analogue system, using a Bose--Einstein condensate to create the analogue metric of a warp drive.
\\

After the warp-drive spacetimes~\cite{alcubierre} were proposed, their most investigated aspect has been the amount of exotic matter (\ie~energy-conditions-violating matter) that would be required to support them.\footnote{While initially it was supposed that exotic matter was needed only for superluminal warp drives ($v_0>c$), it was later recognized \cite{lobovisser2004a,lobovisser2004b} that energy-conditions-violating matter is needed also for subluminal speeds. This points out that the need of exotic matter is peculiar of the warp-drive geometry itself, not appearing only in the superluminal regime.} It was soon realized that this was not only related to the size of the warp-drive bubble but also determined by the thickness of the bubble walls~\cite{pfenningford}. It was found that, if the exoticity was provided by the quantum nature of a field, satisfying therefore the so-called quantum inequalities (QI),\footnote{See~\cite{roman} for a review about quantum inequalities applied to some exotic spacetimes.} then the violations of the energy conditions would have to be confined to Planck-size regions, making the bubble-wall thickness $\Delta$ to be, accordingly, of Planck size [$\Delta \leq 10^2\,(v_0/c)\,\,L_{\rm P}$, where $v_0$ is the bubble velocity and $L_{\rm P}$ is the Planck length]. However, it can be shown that very thin walls require very large amounts of exotic matter: \eg, in order to support a warp-drive bubble with a size of about $100$ m and propagating at $v_0\approx c$, one would need a total negative energy 
$|E| \gg \,10^{11} M_\odot$.\footnote{If one could somehow avoid the QI, then it would be possible to built warp drives with much larger wall thickness. For example, for $\Delta\simeq 1$~m,  {one would only need} $|E|\gtrsim1/4\,M_\odot$.}
Perspectives for warp-drive engineering can be improved by resorting to a modified warp-drive configuration with a reduced surface area but the same bubble volume~\cite{broeck}.
The total amount of negative energy required to support these warp drives becomes quite small (for example, $|E_-|\approx 0.3 M_\odot$ for a $100$ m-radius bubble, although one has to add as well some positive energy outside the bubble $E_+\approx 2.5 M_\odot$), bringing the warp drive closer to a realistic solution albeit still far from foreseeable realizations.

{Regarding the feasibility of warp-drive configurations, a parallel line of research has focused on the study of their robustness
against the introduction of quantum corrections to GR. In particular, in \cite{hiscock} it was studied what would be the effect of having semiclassical corrections in the case of an eternal superluminal warp drive.}
There, it was noticed that to an observer within the warp-drive bubble, the backward and forward walls (along the direction of motion) {look, respectively, like the horizon of a black hole and of a white hole}.
By imposing over the spacetime a quantum state which is vacuum at the null infinities (\ie, what one may call the analogue of the Boulware state for an eternal black hole), it was found that the renormalized stress-energy tensor (RSET).
{diverges} at the horizons.\footnote{We shall work in the Heisenberg representation so that only operators, not the states, evolve in time.}
{Independently of the availability of exotic matter to build the warp drive in the first place, the existence of a divergence of the RSET at the horizons would be telling us that it is not possible to create a warp-drive geometry within the context of semiclassical GR: Semiclassical effects would destroy any superluminal warp drive.}
However, in a more realistic situation, a warp drive would have to be created at a very low velocity in a given reference frame and then accelerated to superluminal speeds. One may then expect that the quantum state globally defined on such dynamical geometry would be automatically selected by the dynamics once suitable boundary conditions are provided (\eg, at early times). This is indeed the case for a gravitational collapse where it can be shown that, whenever a trapping horizon forms, the globally defined quantum state that is vacuum on \scrim~has to be thermal at \scrip~(at the Hawking temperature) and regular at the horizon. 
{In other words, the dynamics of the collapse avoids selecting a Boulware-like state, with its associated 
divergence at the horizon, ending up instead selecting the analogous for collapsing configurations of the Unruh vacuum state defined on eternal black holes, which leads to a perfectly regular RSET. Is the dynamics of the creation
of a warp drive, with its associated selection of the global vacuum state, able to avoid the presence of divergences
in the RSET?}
Indeed, in \cite{hiscock} it was  {already} noticed that an 
{Unruh-like state rather than a Boulware-like state should be expected to describe the quantum state characterizing a superluminal warp-drive creation}.

Here we settle this issue by explicitly considering the case of a warp drive which is created with zero velocity at early times and then accelerated up to some superluminal speed in a finite amount of time. This can be viewed as the warp-drive analogue of a semiclassical black hole collapse \cite{stresstensor}. {By restricting attention to warp drives in} $1+1$ dimensions (since this is the only case for which one can carry out a complete analytic treatment), we calculate the RSET in the warp-drive bubble and, using a specific model, we numerically estimate the energy density in the entire interior of the bubble.

After some transient effects generated when the horizons are created, in the center of the bubble there is a thermal flux of particles at the Hawking temperature corresponding to the surface gravity of the black horizon. However, the surface gravity can be shown to be inversely proportional to the thickness of the bubble walls which, as said, has to be of  {the} order of the Planck length if {some form of} QI holds. Hence, one has to conclude that an internal observer would soon find itself in the uncomfortable condition of being swamped by a thermal flux at the Planck temperature.

More importantly, the energy density measured by free-falling observers is negative on the white and black horizons (as expected for a vacuum polarization). While on the black one it is exponentially damped, on the white one it diverges exponentially to minus infinity.
Moreover, right-going null rays propagating in the bubble approach the Cauchy horizon (see end of Sec.~\ref{subsec:warpdrive}) at late times. As one should expect by analogy with other situation in which Cauchy horizons are present, as in Kerr--Newman black holes, the energy carried by such rays diverges, this time contributing with a positive sign, when approaching this horizon, thus representing an additional source of instability.

In Sec.~\ref{sect:radiation} we discuss the propagation of light rays in a dynamical warp-drive geometry in close analogy with the standard treatment for black holes.\footnote{A similar result was found by Gonz{\'a}lez-D{\'{\i}}az~\cite{gonzales2007}.} In Sec.~\ref{sect:RSET} we calculate the full RSET using the technique adopted in \cite{stresstensor} for black hole formation and look for its divergences and, in Sec.~\ref{sec:kink}, we present a specific model and we numerically compute the energy density seen by free-falling observers.

\section{Light-ray propagation}	%
\label{sect:radiation}		%

The causal structure of the dynamical warp drive discussed in Sec.~\ref{subsec:warpdrive} naturally leads to the expectation that some sort of Hawking radiation will be produced in a superluminal warp drive (as well as some transient particle emission). As sketched in Sec.~\ref{subsec:hawking} (see also~\cite{birreldavies}) for Hawking radiation, all the information about particle production is encoded in the way in which light rays propagate in a spacetime. That is, it is enough to know how light rays are bended in order to analyze the phenomenon of particle creation.
In the dynamical warp drive there is a single past null coordinate but three different future null coordinates associated with the final regions $\rm I$, $\rm II$ and $\rm III$, as described in Sec.~\ref{subsec:warpdrive}. From here on, we will be dealing exclusively with the connection between the past null coordinate $U$ at $\scri_L^-$ and the future null coordinate $u_{\rm II}$ at \scribub~in the interior of the bubble. Therefore, we will use $u$ to denote $u_{\rm II}$ whenever this does not lead to confusion. As discussed in \cite{particlecreation} the relation $U=p(u)$ encodes all the relevant information about particle production.

We want to study the features due to the two main properties of a dynamical warp-drive geometry, \ie, the spacetime is Minkowskian at early times and it is a warp drive at late times. In particular, we are now not interested in the transient features depending on how the transition between these two regimes is performed.\footnote{We shall discuss about transient terms in the Sec.~\ref{sec:kink}, where a numerical analysis allowing their investigation is performed.} Namely, we need only the behavior close to the horizons and at late times inside the whole bubble. It is clear from Fig.~\ref{fig:dynwd-pen} that, if one stays at constant $r$ inside the bubble and moves forward in time one crosses $u$ rays which pass closer and closer to the black horizon. Therefore, once we have determined the behavior of $p(u)$ close to the horizons,
we automatically have also the required information at late times in the whole bubble.

In general, the relation $U=p(u)$ is obtained by integrating the differential equation for the propagation of right-going light rays
\begin{equation}\label{eq:rays}
 \frac{\dd r}{\dd t}=c+\hat{v}(r,t).
\end{equation}
{Note that, while in Sec.~\ref{subsec:warpdrive} we considered a specific form of  $\hat{v}(r,t)$ [Eq.~\eqref{eq:dynamicfluidmetric}] in order to discuss the causal structure of the associated spacetime, here (and in what follows) the discussion holds for any $\hat{v}(r,t)$ that satisfies the requirements $\hat{v}(r,t)\to0$, for $t\to-\infty$  (sufficiently rapid for the spacetime to be asymptotically flat), and $\hat{v}(r, t)=\bar{v}$ after some finite time and within the warp-drive bubble.} 
Given the assumption that the spacetime settles down to a stationary warp-drive configuration, at late times the velocity profile will depend only on the $r$ coordinate. We can write, as in the stationary case Eq.~\eqref{eq:du}
\begin{equation}\label{eq:udef}
 \dd u=\dd t-\frac{\dd r}{c+\bar{v}(r)}.
\end{equation}
To find the required asymptotic relation one has to integrate this equation in the limit $r\to r_{1,2}$.
There, the velocity can be expanded as
\begin{equation}
 \bar{v}=-c\pm \kappa \left(r-r_{1,2}
 \right)+{\cal O}\left(\left(r-r_{1,2}
 \right)^2\right).
\end{equation}
Thus, we obtain
\begin{equation}
 u\simeq t\mp\frac{1}{\kappa}\ln\left|r-r_{1,2}
  \right|.\label{eq:uwhapprox}
\end{equation}
On the other hand, the coordinate $U$, obtained by integrating Eq.~\eqref{eq:rays} at early times, reduces to the Minkowski null coordinate
\begin{equation}
 U(t\to -\infty)=t-\frac{r}{c},
\end{equation}
and is regular in the whole spacetime, in particular, on the horizons. For instance, on a fixed $t$~slice in the stationary region, we can write $U$ as a regular function of $r$
\begin{equation}\label{eq:Uanalitic}
 U_\pm=\mathcal{U}_\pm\left(r-r_{1,2} \right),
\end{equation}
where we denoted with $U_+$ (respectively, $U_-$) the specific form of $U$ close to the black (respectively, white) horizon, and $\mathcal{U}\pm$ are analytic functions. Inserting Eq.~\eqref{eq:uwhapprox} in the above expression, at the same fixed time,
\begin{equation}
 U_\pm=p(u\to\pm \infty)=\mathcal{P}_\pm(e^{\mp\kappa u}),
\end{equation}
where $\mathcal{P}_\pm$ are again analytic functions. Note that the forms of these functions do not depend on the particular time slice chosen to perform the matching between Eqs.~\eqref{eq:uwhapprox} and \eqref{eq:Uanalitic}. In the proximity of the stationary horizons $u\to\pm\infty$, so $e^{\mp\kappa u}\to 0$ and the function $p$ can be expanded around the horizons. Up to the first order we get
\begin{equation}\label{eq:Uu}
 U=p(u \to \pm \infty)= U_{\substack{\rm BH \\ \rm WH}} \mp A_{\pm} e^{\mp\kappa u}+{\cal O}\left(e^{\mp2\kappa u}\right),
\end{equation}
where $A_{\pm}$ are positive constants.

This is indeed the asymptotic behavior one would expect in the presence of trapping horizons and, for the black hole case, it is the standard relation between $u$ and $U$. In fact, it leads to the conclusion that an observer at \scrip~will detect Hawking radiation with temperature $\kappa /2\pi$.
It is important to note that the result is completely general. The asymptotic behavior of $U=p(u)$ for large absolute values of $u$,
which is the only feature of $p(u)$ relevant for the present analysis (see Sec.~\ref{sec:kink} for a complete numerical solution),
does not depend on the specific velocity profile adopted. It is only necessary that it interpolates from Minkowski spacetime at early times to a stationary warp-drive geometry at late times.

While Eq.~\eqref{eq:Uu} is exactly of the expected form, its implications for particle production are not as straightforward as in the black hole case. In fact, in the warp-drive geometry the late-time modes labeled by $u$ will not be standard plane waves in an asymptotically flat region of spacetime as they will be characterized by the strange form given in Eq.~\eqref{eq:uwhapprox}.
Of course, if the surface gravity $\kappa$ is large enough so that the typical wavelength of the emitted radiation is much smaller than the bubble size, then the plane-wave approximation is fine and in the center of the bubble, at late times, one will measure standard Hawking radiation at temperature $\Th$. Nonetheless, in the general case, even if the calculation for the Bogoliubov coefficient is the standard one \cite{birreldavies}, the particles created will not be standard plane waves. The physics associated with the particle production by the white horizon is even less clear. This is why, in the next section, we shall consider the behavior of the RSET to get more significant information.

\begin{figure}[tb]
\centering
 \includegraphics{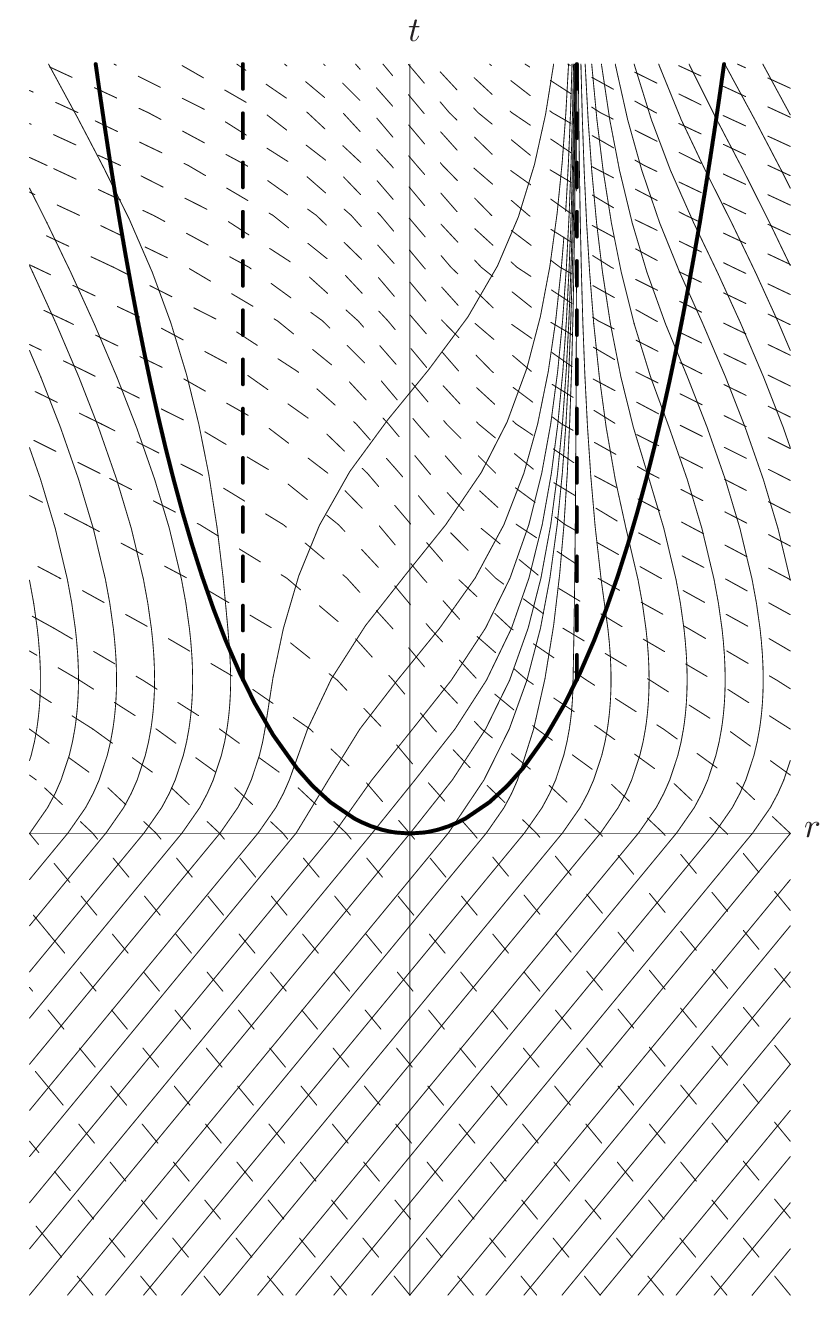}
 \caption{Light rays propagating rightward (solid lines) and leftward (dashed lines)  in the plane $(t,r)$ in a warp-drive spacetime with velocity profile of Eq.~\eqref{eq:ppwarpvelocity}. The out region in which the geometry is a stationary warp drive is at $r<\pm\text{arccosh}(t+1)$ (heavy solid lines). At $t<0$ the metric is Minkowskian. The horizons at $r_1$ and $r_2$ (heavy dashed lines) are formed at $t_{\rm H}=1$. Please refer to Sec.~\ref{sec:kink} for details.}
 \label{fig:lightrayssimp}
\end{figure}

In order to do so, we also need the relation between ingoing and outgoing left-going rays. In fact, these modes are excited too when the warp drive forms, even if we do not have a thermal particle production as for right-going modes. Left-going rays are the solution of the following differential equation [see Eq.~\eqref{eq:rays}]:
\begin{equation}\label{eq:raysleft}
 \frac{\dd r}{\dd t}=-c+\hat{v}(r,t).
\end{equation}
Looking at Fig.~\ref{fig:dynwd-pen}, we note that left-going rays do not see the horizons, that is, they cross them from $\scri^-_R$ to $\scri^+_L$. As a consequence, both the past and the future null coordinates $W$ and $w$ are defined at the asymptotic region outside the bubble. However, after the geometry inside the bubble has settled down to its final stationary form, we can define, just for convenience, a different coordinate $\tilde w$ inside the bubble, as in Eq.~\eqref{eq:dw}:
\begin{equation}\label{eq:wdef}
 \dd\tilde w=\dd t+\frac{\dd r}{c-\bar{v}(r)}.
\end{equation}
Note that $w$ and $\tilde w$ may or may not coincide depending on how fast the metric in the external region settles down to its final stationary form (refer to Fig.~\ref{fig:lightrayssimp} and Sec.~\ref{sec:kink} for an example in which they do not coincide).

$W$ is obtained in the {usual} way, by integrating Eq.~\eqref{eq:raysleft} at early times, when the spacetime is Minkowski
\begin{equation}
 W(t\to -\infty)=t+\frac{r}{c}.
\end{equation}
The relation $W=q(\tilde{w})$ can be found explicitly for specific cases.
The important point is that one can prove that this relation is always regular, so that it cannot give place to any phenomenon like Hawking radiation. Of course, one can choose a very unusual way to interpolate from Minkowski to the warp drive, such that a lot of particles are created in this sector, but this is not a general feature of dynamical warp drives. If we use a regular enough transition, only transient effects are present in this sector. In Sec.~\ref{sec:kink} we shall show that it is possible to find such a transition.

As an example, in Fig.~\ref{fig:lightrayssimp} we plot both right-going and left-going light rays propagating in the particular dynamic warp-drive spacetime analyzed in Sec.~\ref{sec:kink}.

%

\sectionmark{RSET}				%
\section{Renormalized Stress-Energy Tensor}	%
\label{sect:RSET}				%
\sectionmark{RSET}				%

For the calculation of the RSET inside the warp-drive bubble we use the method used in \cite{stresstensor} for a 
collapsing configuration to form a black hole.
In null coordinates $U$ and $W$, affine on \scrim, the metric can be written as
\begin{equation}\label{eq:metricUW}
 \dd s^2=-C(U,W)\,\dd U\dd W.
\end{equation}
As we described in the previous section, in the out region (the region in which the metric is stationary, \ie, $\bar{v}$ depends only on $r$) we can also use a different set of null coordinates $u, \tilde w$.
The coordinate $u$ is affine on $\hor_C^+$ and $\tilde w$ is the coordinate defined in~Eq.~\eqref{eq:wdef}. In these coordinates the metric is expressed as
\begin{equation}\label{eq:metricuw}
 \dd s^2=-\bar{C}(u,\tilde w)\, \dd u \dd\tilde w,
\end{equation}
which implies
\begin{equation}\label{eq:transfC}
 C(U,W) = \frac{\bar{C}(u,\tilde w)}{\dot{p}(u)\dot{q}(\tilde w)},
\end{equation}
and
\begin{equation}\label{eq:transdiff}
 U=p(u), \qquad W=q(\tilde w).
\end{equation}
These transformations are such that $\bar{C}$ has precisely the form of the future stationary warp-drive geometry,
\ie, it depends only on $r$ through $u,\tilde w$.

For concreteness let us refer to the RSET associated with having a single quantum massless scalar field living on the
spacetime. In this case the RSET components have the following form \cite{birreldavies}:
\begin{align}
 T_{UU} &= -\frac{1}{12\pi}C^{1/2}\partial_U^2 C^{-1/2},\label{eq:TUU}\\
 T_{WW} &= -\frac{1}{12\pi}C^{1/2}\partial_W^2 C^{-1/2},\label{eq:TWW}\\
 T_{UW} &= T_{WU}
 =\frac{1}{96\pi} C R.\label{eq:TUWR}
\end{align}
Qualitatively, things would not change if there were other fields present in the theory. The only modification will be 
that the previous expressions will get multiplied by a specific numerical factor. It is clear that in the in region (where the spacetime is Minkowskian) the RSET is trivially zero.

Now, for a metric in the form of Eq.~\eqref{eq:metricUW}, the curvature can be calculated as in~\cite{birreldavies}:
\begin{equation}
 R=\Box\ln|C|={(-g)}^{-1/2}\partial_\mu\left[{(-g)}^{1/2}g^{\mu\nu}\partial_\nu\ln|C|\right]
	=-\frac{4}{C}\partial_U\partial_W\ln|C|.
\end{equation}
Replacing this result in Eq.~\eqref{eq:TUWR} we obtain
\begin{equation}\label{eq:TUW}
 T_{UW}=T_{WU}=-\frac{1}{24\pi}\partial_U\partial_W\ln|C|.
\end{equation}
Using transformations~\eqref{eq:transfC} and \eqref{eq:transdiff} in Eqs.~\eqref{eq:TUU}, \eqref{eq:TWW}, and \eqref{eq:TUW} we obtain
\begin{align}
 T_{UU} &= -\frac{1}{12\pi}\frac{1}{\dot{p}^2}\left[\bar{C}^{1/2}\partial_u^2 \bar{C}^{-1/2}-\dot{p}^{1/2}\partial_u^2 \dot{p}^{-1/2}\right],\label{eq:TUUbar}\\
 T_{WW} &= -\frac{1}{12\pi}\frac{1}{\dot{q}^2}\left[\bar{C}^{1/2}\partial_{\tilde{w}}^2 \bar{C}^{-1/2}-\dot{q}^{1/2}\partial_{\tilde{w}}^2 \dot{q}^{-1/2}\right],\label{eq:TWWbar}\\
 T_{UW} &= T_{WU}= -\frac{1}{24\pi}\frac{1}{\dot{p}\dot{q}}\partial_u\partial_{\tilde{w}}\ln|\bar{C}|.\label{eq:TUWbar}
\end{align}

We can express the derivatives with respect to $u$ and $\tilde w$ in terms of derivatives with respect to $r$ and $t$. Note that the following expressions are obtained from Eqs.~\eqref{eq:udef} and \eqref{eq:wdef}, so they are valid only when the velocity profile depends only on the $r$ coordinate. In this fashion, we can study the stress-energy tensor at the end of the creation of the warp drive (or at the end of a collapse if we are studying a black hole).
\begin{equation}\label{eq:rtfromuw}
\begin{pmatrix}
\partial_r \\ 
\partial_t 
\end{pmatrix}
=
\begin{pmatrix}
 u_r & \tilde{w}_r\\
 u_t & \tilde{w}_t
\end{pmatrix}
\begin{pmatrix}
\partial_u \\
\partial_{\tilde{w}}
\end{pmatrix}
=
\begin{pmatrix}
  -1/(c+\bar{v}) & 1/(c-\bar{v})\\
  1 & 1
\end{pmatrix}
\begin{pmatrix}
\partial_u \\ 
\partial_{\tilde{w}}
\end{pmatrix}.
\end{equation}

To proceed, one must compute the four components of the derivative matrix. These are directly read from Eqs.~\eqref{eq:udef} and \eqref{eq:wdef}:
\begin{equation}\label{eq:uwderivs}
\begin{aligned}
  u_r &= -\frac{1}{c+\bar{v}(r)},	&\quad		\tilde{w}_r &=\frac{1}{c-\bar{v}(r)},\\
  u_t &= 1,				&\quad		\tilde{w}_t &=1.
\end{aligned}
\end{equation}

Note that the definition of $U$ and $W$ {can be extended} in the out region by following the light rays coming from the in region. 
For instance, one can {naturally} define $U(t,r)$ in the out region {as} $U(t,r)=p\left(u(t,r)\right)$ and in analogous ways for the other null coordinate. This means that it makes sense to take derivatives of the null coordinates with respect to $r$ and $t$ in all the spacetime. We write here the derivatives of $U$ and $W$ in the out region:
\begin{equation}\label{eq:UWderivspq}
\begin{aligned}
  U_r &=\dot{p}(u)u_r = -\frac{\dot{p}(u)}{c+\bar{v}(r)},	&\quad		W_r &=\dot{q}(\tilde{w})\tilde{w}_r =\frac{\dot{q}(\tilde{w})}{c-\bar{v}(r)},\\
  U_t &=\dot{p}(u)u_t = \dot{p}(u),				&\quad		 W_t &=\dot{q}(\tilde{w})\tilde{w}_t =\dot{q}(\tilde{w}).
\end{aligned}
\end{equation}

By inverting the derivative matrix appearing in Eq.~\eqref{eq:rtfromuw}, we therefore obtain the required result:
\begin{align}
 \partial_u &= -\frac{c^2-\bar{v}^2}{2c}\partial_r+\frac{c+\bar{v}}{2c}\partial_t,\\
 \partial_{\tilde{w}} &= \frac{c^2-\bar{v}^2}{2c}\partial_r+\frac{c-\bar{v}}{2c}\partial_t.
\end{align}

We are interested in calculating the RSET inside the bubble when the two horizons have been formed and when the configuration has settled down to a stationary warp drive. Since in this region the velocity depends only on $r$, we can replace the derivatives in Eqs.~\eqref{eq:TUUbar}, \eqref{eq:TWWbar} and \eqref{eq:TUWbar} with
\begin{align}
 \partial_u &\rightarrow -\frac{1-\bar{v}^2}{2}\partial_r = -\frac{\bar{C}}{2}\partial_r,\\
 \partial_{\tilde{w}} &\rightarrow  \frac{1-\bar{v}^2}{2}\partial_r = \frac{\bar{C}}{2}\partial_r.
\end{align}
Here, we have put $c=1$, for simplicity, and we have used
$\bar{C}=1-\bar{v}^2$, that can be obtained by writing the line element as
\begin{equation}
  \dd s^2=-\bar{C}(u,\tilde{w})\,\dd u\dd\tilde{w} 
	=-\bar{C}(u,\tilde{w})\left(u_t \dd t + u_r \dd r\right)\left(\tilde{w}_t \dd t + \tilde{w}_r \dd r\right)
   	=\frac{\bar{C}(u,\tilde{w})}{c^2-\bar{v}(r)^2}\left\{c^2 \dd t^2-\left[\dd r-\bar{v}(r) \dd t\right]^2\right\},
\end{equation}
and comparing it with Eq.~\eqref{eq:dynamicfluidmetric}, so that
\begin{equation}
 \bar{C}(u,\tilde{w})=c^2-\bar{v}(r)^2.\label{eq:barC}
\end{equation}

Moreover we indicate with $'$ the differentiation with respect to $r$. After some calculations we obtain
\begin{align}
 \bar{C}^{1/2}\partial_u^2 \bar{C}^{-1/2} &=\bar{C}^{1/2}\partial_{\tilde{w}}^2 \bar{C}^{-1/2}=\frac{1}{16}\left[{\left(\bar{C}'\right)}^2-2\bar{C}\bar{C}''\right],\\
 \dot{p}^{1/2}\partial_u^2 \dot{p}^{-1/2} &=\frac{1}{4\dot{p}^2}\left[3\ddot{p}^2-2\dot{p}\,\dddot{p}\right],\\
 \dot{q}^{1/2}\partial_{\tilde{w}}^2 \dot{q}^{-1/2} &=\frac{1}{4\dot{q}^2}\left[3\ddot{q}^2-2\dot{q}\,\dddot{q}\right],\\
 \partial_u\partial_{\tilde{w}}\ln|\bar{C}| &=-\frac{1}{4}\bar{C}\ddot{\bar{C}}.
\end{align}
Using again $\bar{C}=1-\bar{v}^2$ we get the final result:
\begin{align}
 T_{UU} &= -\frac{1}{48\pi}\frac{1}{\dot{p}^2}\left[\bar{v}'\,^2+\left(1-\bar{v}^2\right)\bar{v}\bar{v}''-\frac{3\ddot{p}^2-2\dot{p}\,\dddot{p}}{\dot{p}^2}\right],\\
 T_{WW} &= -\frac{1}{48\pi}\frac{1}{\dot{q}^2}\left[\bar{v}'\,^2+\left(1-\bar{v}^2\right)\bar{v}\bar{v}''-\frac{3\ddot{q}^2-2\dot{q}\,\dddot{q}}{\dot{q}^2}\right],\\
 T_{UW} &= T_{WU}=-\frac{1}{48\pi}\frac{1}{\dot{p}\dot{q}}\left(1-\bar{v}^2\right)\left[\bar{v}'\,^2+\bar{v}\bar{v}''\right].
\end{align}
One can check that these quantities do not diverge at the horizons, just like in \cite{stresstensor}. However we want to look at the energy density inside the bubble and try to understand whether it remains small or not as time increases.

In particular, it is interesting to look at the energy measured by a set of free-falling observers, whose four velocity is simply $u_{c}^\mu=(1,\bar{v})$ in $(t,r)$ components.
These observers measure an energy density $\rho$:
\begin{multline}\label{eq:rhowarpdrive}
 \rho=T_{\mu\nu}u_c^\mu u_c^\nu=T_{tt}+2\bar{v}T_{tr}+\bar{v}^2T_{rr} =U_t^2 T_{UU}+2 U_t W_t T_{UW} +W_t^2 T_{WW}\\
 +2\bar{v}\left[U_t U_r T_{UU}+\left(U_t W_r + W_t U_r \right)T_{UW} +W_t W_r T_{WW}\right]+\bar{v}^2\left[U_r^2 T_{UU}+2 U_r W_r T_{UW} +W_r^2 T_{WW}\right]\\
 =\left(U_t+\bar{v} U_r\right)^2T_{UU}+2\left(U_t+\bar{v} U_r\right)\left(W_t+\bar{v} W_r\right)T_{UW}+\left(W_t+\bar{v} W_r\right)^2T_{WW}\\
 =\frac{\dot{p}^2}{\left(1+\bar{v}\right)^2}T_{UU}+2\frac{\dot{p}\dot{q}}{1-\bar{v}^2}T_{UW}+\frac{\dot{q}^2}{\left(1-\bar{v}\right)^2}T_{WW}\\
=-\frac{1}{48\pi}\left[\frac{2\left(\bar{v}^4-\bar{v}^2+2\right)}{\left(1-\bar{v}^2\right)^2}\bar{v}'\,^2+\frac{4\bar{v}}{1-\bar{v}^2}\bar{v}''-\frac{f(u)}{\left(1+\bar{v}\right)^2}-\frac{g(\tilde{w})}{\left(1-\bar{v}\right)^2} \right],
\end{multline}
where we have defined
\begin{align}
 f(u)&\equiv\frac{3\ddot{p}^2(u)-2\dot{p}(u)\,\dddot{p}(u)}{\dot{p}^2(u)},\label{eq:fu}\\
 g(\tilde{w})&\equiv\frac{3\ddot{q}^2(\tilde{w})-2\dot{q}(\tilde{w})\,\dddot{q}(\tilde{w})}{\dot{q}^2(\tilde{w})}.
\end{align}

We want to study what happens to a spaceship placed at rest in the center of the bubble to investigate whether a warp drive can be used as a transportation device. Moreover we want to check whether the components of the RSET in a regular coordinate system are regular at the horizons. Freely falling observers are at rest in the center of the bubble but left
moving otherwise.

Looking at the above expressions for any component of the RSET or at the energy density $\rho$, we see that they can be split as a sum of three terms, one purely static, depending only on the $r$ coordinate through the shift velocity $\bar{v}$, and two dynamic pieces depending also on the $u$ and $\tilde{w}$ coordinates, respectively. They correspond to energy traveling on right-going and left-going light rays, respectively, eventually red-/blueshifted by a term depending on $r$.
Being interested in studying the energy density measured by free-falling observers, we write
\begin{equation}
 \rho=\rho_{\rm st}+\rho_{{\rm dyn}-u}+\rho_{{\rm dyn}-\tilde{w}},\\
\end{equation}
where we defined static terms (labeled by subscript $\rm st$) and dynamic terms (labeled by subscript $\rm dyn$) for each component. Right-going and left-going terms are respectively labeled by $u$ and $\tilde{w}$.
\begin{gather}
 \rho_{\rm st}\equiv-\frac{1}{24\pi}\left[\frac{\left(\bar{v}^4-\bar{v}^2+2\right)}{\left(1-\bar{v}^2\right)^2}\bar{v}'\,^2+\frac{2\bar{v}}{1-\bar{v}^2}\bar{v}''\right],\label{eq:rhocs}\\
 \rho_{{\rm dyn}-u}\equiv\frac{1}{48\pi}\frac{f(u)}{\left(1+\bar{v}\right)^2},\label{eq:rhodu}\\
 \rho_{{\rm dyn}-\tilde{w}}\equiv\frac{1}{48\pi}\frac{g(\tilde{w})}{\left(1-\bar{v}\right)^2}.\label{eq:rhodw}
\end{gather}
Let us start with $\rho_{{\rm dyn}-\tilde{w}}$. Its denominator is bounded for each $r$ because the shift velocity is negative, and its numerator is vanishing with time. It is easy to convince oneself that all the contributions to the RSET coming from the $\tilde{w}$ sector are not universal but depend exclusively on the specific interpolation between the early Minkowski spacetime and the final warp-drive spacetime. It is always possible to choose an interpolation so that all these contributions vanish at late times (see Sec.~\ref{sec:kink}).

From now on, we neglect the dynamic term $\rho_{{\rm  dyn}-\tilde{w}}$ and study
\begin{equation}
 \rho\simeq\rho_{\rm  st}+\rho_{{\rm dyn}-u}.\\
\end{equation}
%

\subsection{RSET at the center of the warp-drive bubble}	%

We shall now study the behavior of the RSET in the center of the bubble at late times. In this point $\bar{v}(r=0)=\bar{v}'(r=0)=0$ and the static term $\rho_{\rm st}$ vanishes.
By integrating Eq.~\eqref{eq:udef} in the stationary region, one can show that $u(t,r)$ is linear in $t$ so that, for fixed $r$, it will acquire arbitrarily large positive values (see Fig.~\ref{fig:dynwd-pen}).
The dynamic term in Eq.~\eqref{eq:rhodu} can be evaluated by using a late-time expansion of $p(u)$
in a Taylor series in $e^{\mp\kappa u}$. In fact we found that [see Eq.~\eqref{eq:Uu}]
\begin{equation}
 U_\pm=p(u\to\pm \infty)=\mathcal{P}_\pm(e^{\mp\kappa u}),
\end{equation}
where $\mathcal{P}_\pm$ are analytic functions. We now expand it up to the third order:
\begin{equation}\label{eq:Uulong}
 U=p(u \to \pm \infty)= U_{\substack{\rm BH \\ \rm WH}}+A_{1\pm} e^{\mp\kappa u}
	+\frac{A_{2\pm}}{2} e^{\mp 2\kappa u}+\frac{A_{3\pm}}{6}e^{\mp 3\kappa u}+{\cal O}\left(e^{\mp 4\kappa u}\right),
\end{equation}
where the upper (lower) signs correspond to $u\to+\infty$ ($u\to -\infty$). For simplicity we have defined the coefficients $A_{1+}\equiv-A_+$ and $A_{1-}\equiv A_-$.
In order to calculate the stress-energy tensor we need to calculate Eq.~\eqref{eq:fu}
\begin{equation}
f(u)=\frac{3\ddot{p}^2-2\dot{p}\,\dddot{p}}{\dot{p}^2}.
\end{equation}
We have
\begin{gather}
 \dot{p}(u)=\mp\kappa A_{1\pm} e^{\mp\kappa u}\left[1+\frac{A_{2\pm}}{A_{1\pm}} e^{\mp\kappa u}+\frac{A_{3\pm}}{2A_{1\pm}}e^{\mp 2\kappa u}+{\cal O}\left(e^{\mp 3\kappa u}\right)\right],\label{eq:approxpu}\\
 \ddot{p}(u)=\kappa^2 A_{1\pm} e^{\mp\kappa u}\left[1 +\frac{2A_{2\pm}}{A_{1\pm}} e^{\mp\kappa u}+\frac{3 A_{3\pm}}{2A_{1\pm}}e^{\mp 2\kappa u}+{\cal O}\left(e^{\mp 3\kappa u}\right)\right],\\
 \dddot{p}(u)=\mp\kappa^3 A_{1\pm} e^{\mp\kappa u}\left[1 +\frac{4A_{2\pm}}{A_{1\pm}} e^{\mp\kappa u}+\frac{9 A_{3\pm}}{2A_{1\pm}}e^{\mp 2\kappa u}+{\cal O}\left(e^{\mp 3\kappa u}\right)\right].
\end{gather}
so that
\begin{multline}\label{eq:approxfu}
f(u)=\frac{\kappa^4 A_{1\pm}^2e^{\mp 2\kappa u}
		\left\{1+2{\left(A_{2\pm}/A_{1\pm}\right)}e^{\mp\kappa u}+\left[4{\left(A_{2\pm}/A_{1\pm}\right)}^2-A_{3\pm}/A_{1\pm}\right]e^{\mp 2\kappa u}+{\cal O}\left(e^{\mp 3\kappa u}\right)\right\}}
	{\kappa^2 A_{1\pm}^2e^{\mp 2\kappa u}
		\left\{1+2{\left(A_{2\pm}/A_{1\pm}\right)}e^{\mp\kappa u}+\left[{\left(A_{2\pm}/A_{1\pm}\right)}^2+A_{3\pm}/A_{1\pm}\right]e^{\mp 2\kappa u}+{\cal O}\left(e^{\mp 3\kappa u}\right)\right\}}\\
=\kappa^2\left\{1+\frac{\left[4{\left(A_{2\pm}/A_{1\pm}\right)}^2-A_{3\pm}/A_{1\pm}\right]}{1+2{\left(A_{2\pm}/A_{1\pm}\right)}e^{\mp\kappa u}}e^{\mp 2\kappa u}+{\cal O}\left(e^{\mp 3\kappa u}\right) \right\}\\
	\shoveright{\times\left\{1-\frac{\left[{\left(A_{2\pm}/A_{1\pm}\right)}^2+A_{3\pm}/A_{1\pm}\right]}{1+2{\left(A_{2\pm}/A_{1\pm}\right)}e^{\mp\kappa u}}e^{\mp 2\kappa u}+{\cal O}\left(e^{\mp 3\kappa u}\right) \right\}}\\
=\kappa^2\left\{1+\left[3{\left(\frac{A_{2\pm}}{A_{1\pm}}\right)}^2-2\frac{A_{3\pm}}{A_{1\pm}}\right]e^{\mp 2\kappa u}+{\cal O}\left(e^{\mp 3\kappa u}\right) \right\}.
\end{multline}

With this expansion it is easy to see that for $u\to+\infty$
\begin{equation}
 \rho \simeq \frac{\kappa^2}{48\pi}.
 \label{en-den}
\end{equation}
This result may be easily understood in the following way. The surface gravity of the black horizons is the velocity derivative evaluated at this horizon $\kappa=(d\bar{v}/dr)_{r=r_1}$. Moreover, the energy density of a scalar field at some finite temperature $T$ in $1+1$ 
dimensions is simply
\begin{equation}\label{eq:scalarfield}
 \rho_{T}=\int\frac{\omega}{\left(e^{\omega/T}+1\right)}\frac{\dd k}{2\pi}
	=\frac{\pi}{12}T^2.
\end{equation}
Defining the Hawking temperature in the usual way as $\Th\equiv \kappa/2\pi$, it is easy to see that Eq.~\eqref{en-den} can be rewritten exactly as $\rho=\left(\pi /12\right){\Th}^2$. 
Thus, the observer inside the warp-drive bubble will indeed measure thermal radiation at the temperature $\Th$.

\subsection{RSET at the bubble horizons}	%
\label{subsect:RSETH2}				%

Let us now move to study $\rho$ close to the horizons. Note that both $\rho_{\rm st}$ and $\rho_{{\rm dyn}-u}$ are divergent at the horizons ($r\to r_{1,2}$)
because of the $(1+\bar{v})$ factors in the denominator. Just like in \cite{stresstensor}, for a black hole, these divergences exactly cancel each other, but something different happens at the black and white horizon. Expanding the velocity up to the second order one gets
\begin{equation}\label{eq:vseries}
 \bar{v}_\pm(r)=-1\pm\kappa\left(r-r_{1,2}
 \right)+\frac{1}{2}\sigma \left(r-r_{1,2}
 \right)^2 + {\cal O}\left(\left(r-r_{1,2}
 \right)^3\right),
\end{equation}
where $\sigma$ is a constant.

Close to the horizons the static term of Eq.~\eqref{eq:rhocs} then becomes
\begin{equation}
 \rho_{\rm st}\left(r \simeq r_{1,2}
 \right)
 	=-\frac{1}{48\pi}\left[\frac{1}{\left(r-r_{1,2}
\right)^2}\mp\frac{\sigma}{\kappa \left(r-r_{1,2}
\right)}\right]+{\cal O}(1).
\end{equation}
Similarly one can expand the dynamic term $\rho_{{\rm dyn}-u}$ to the same order.
This involves determining the function $f(u)$ in the proximity of the horizons.
Using the expansion {of} Eq.~\eqref{eq:uwhapprox}, valid for points close to the horizons, into Eq.~\eqref{eq:approxfu}, we obtain
\begin{equation}
\lim_{r\to r_{1,2} }f(u) = \kappa^2\left\{1+\left[3{\left(\frac{A_{2\pm}}{A_{1\pm}}\right)}^2-2\frac{A_{3\pm}}{A_{1\pm}}\right]e^{\mp 2\kappa t}\left(r-r_{1,2}
\right)^2+{\cal O}\left(\left(r-r_{1,2}
\right)^3\right) \right\}.
\end{equation}
Inserting this expression in Eq.~\eqref{eq:rhodu} we obtain:
\begin{equation}
 \rho_{{\rm dyn}-u}\left(r \simeq r_{1,2}
\right)
	= \frac{1}{48\pi}\left[\frac{1}{\left(r-r_{1,2}
\right)^2}\mp\frac{\sigma}{\kappa\left(r-r_{1,2}
\right)}\right]+{\cal O}(1).
\end{equation}
It is now clear that the total $\rho$ is ${\cal O}(1)$ on the horizons and does not diverge (as expected from the Fulling--Sweeny--Wald theorem \cite{fsw}).

However, let us look to the subleading terms
\begin{equation}
 \rho\left(r \simeq r_{1,2}
\right)=\frac{1}{48\pi}\left[3{\left(\frac{A_{2\pm}}{A_{1\pm}}\right)}^2-2\frac{A_{3\pm}}{A_{1\pm}}\right]e^{\mp2\kappa t}
	+C_\pm+{\cal O}\left(r-r_{1,2}
\right),
\label{eq:sublead}
\end{equation}
where $C_\pm$ are constants. We can easily see that the behavior close to the black horizon is completely different from that close to the white horizon. In the former case the energy density as seen by a free-falling observer is damped exponentially with time.
On the white horizon, however, this energy density grows exponentially with time. {This means that, moving along $\hor_2^-$, $\rho$ is large and negative diverging while approaching the crossing point between $\hor_2^-$ and~\scribub}.

This asymptotic divergence is physical and not a matter of selection of coordinates. In a very short time after the white horizon is formed (of the order of $1/\kappa$), the backreaction of the RSET in this region of spacetime is no longer negligible {but rather very strong}.
This result does not violate the Fulling--Sweeny--Wald theorem~\cite{fsw}, which states that if the two-point function $G(x,x')$ of a scalar field has a Hadamard singularity in the coincidence limit $x\to x'$ on an open neighborhood of a Cauchy hypersurface, its Hadamard structure is preserved by time evolution in the Cauchy development. In fact, the above found divergence of the RSET takes place at the crossing point between the white horizon and the Cauchy horizon, that is on the border of the Cauchy development.

\subsection{RSET at late times approaching \texorpdfstring{\scribub}{the Cauchy horizon} }	%
\label{subsect:RSETHc}										%

In Sec.~\ref{subsect:RSETH2} we studied the RSET on the black and white horizons. However, it is interesting to analyze better its behavior close to the Cauchy horizon \scribub. As one can see from Fig.~\ref{fig:lightrayssimp}, every $u$ ray reaches values of $r$ very close to $r_2$, at  sufficiently late times. This means that, even for large positive values of $u$, some time exists after which the approximate behavior of all the $u$ rays is well described by
\begin{equation}
 u\simeq t+\frac{1}{\kappa}\ln{\left(r_2-r\right)},
\end{equation}
so that
\begin{equation}
 r_2-r\propto e^{-\kappa t}.
\end{equation}
Keeping in mind this point, let us study the term $\rho_{{\rm dyn}-u}$, at fixed $u$, when time increases. There is a time at which the $u$ ray will be close enough to $r_2$, such that the denominator in Eq.~\eqref{eq:rhodu} can be approximated by
\begin{equation}
 {\left(1+\bar{v}\right)}^2\simeq \kappa^2 {\left(r_2-r\right)}^2,
\end{equation}
which becomes, thanks to the previous result,
\begin{equation}
 {\left(1+\bar{v}\right)}^2\propto e^{-2\kappa t},
\end{equation}
such that, for $t\to+\infty$, $\rho$ diverges along every single $u$ ray as $e^{+2\kappa t}$, because of the blueshift factor $(1+{\bar{v}})^2$. 
Hence, $\rho$ will diverge on the whole line \scribub, which is a Cauchy horizon for the geometry (so also this result does not contradict the Fulling--Sweeny--Wald theorem~\cite{fsw}).  We then deduce that the warp-drive spacetime is again likely to become unstable in a very short time.

As a closing remark, it is perhaps important to stress that this divergence of the RSET on the Cauchy horizon is of different nature with respect to the one found in Sec.~\ref{subsect:RSETH2}.
In fact, the divergence of Sec.~\ref{subsect:RSETH2} is intrinsically due to the inevitable transient disturbances produced by the formation of the horizon.
In this sense it is a new and very effective instability present every time a white horizon is formed in some dynamical way.
On the contrary, the just found divergence on \scribub~can be seen as due to the well-known infinite blueshift suffered by light rays as they approach a Cauchy horizon, in this specific case as due to the accumulation of Hawking radiation produced by the black horizon.
In this sense it is not very different from the often claimed instability of inner horizons in Kerr--Newman black holes~\cite{Simpson:1973ua,poissonisrael,markovicpoisson}.

Of course the divergence and the appearance of the horizon \scribub~would be avoided if the superluminal warp drive were sustained just for a finite amount of time. In that case, no Cauchy horizons would arise and no actual infinity would be reached by the RSET.  However, the latter would still become huge in a very short time, increasing exponentially on a time scale $1/\kappa\approx \Delta/c$, where $\Delta$ is the thickness of the warp-drive bubble. Note that, in order to get a time scale of even $1$~s, one would need $\Delta\approx 3\times10^8$~m.

\section{Numerical analysis}	%
\label{sec:kink}		%

We present here a specific model of dynamical warp drive, which we used for numerical calculations~\cite{wdcrete}. It is particularly useful because it allows one to carry out an almost complete analytical treatment and to numerically compute the energy density $\rho$ of Eq.~\eqref{eq:rhowarpdrive}. We adapt here the method presented in~\cite{particlecreation,notrap} for stellar collapses to the creation of a warp drive from Minkowski spacetime.
We can choose a simplified piecewise velocity profile, which has the relevant properties of a dynamical warp drive, \ie, {it describes} a flat geometry at early times and coincides with {the metric in} Eq.~\eqref{eq:warpdrivemetric} after some finite time~$t>0$.

Using our velocity profile defined in Eq.~\eqref{eq:fluidvelocity}
\begin{equation}
 \bar{v}(r)=\alpha c\left[f(r)-1\right],
\end{equation}
we can define a dynamical profile by replacing $\hat{v}$ in Eq.~\eqref{eq:dynamicfluidmetric} by $\hat{v}_{\rm kink}$
\begin{equation}\label{eq:ppwarpvelocity}
 \hat{v}_{\rm kink}(r,t)=
 \left\{
  \begin{aligned}
    & \bar{v}(\xi(t)), \qquad&\text{if} \quad |r|\geq\xi(t),  \\
    & \bar{v}(r), \qquad&\text{if}\quad |r|<\xi(t),
  \end{aligned}
 \right.
\end{equation}
where $\xi(t)$ is a monotonically increasing function of $t$, such that $\xi(t)\to0$, for $t\to-\infty$, and $\xi(t_{\rm H})=r_2=-r_1$.

One may wonder whether defining a velocity profile with a kink, as in Eq. \eqref{eq:ppwarpvelocity},  may lead to unphysical phenomena. Indeed this computational trick induces some spurious effects, but these features are just transients and do not affect the results at late times.

\subsection{Right-going rays}	%

We apply the same procedure of \cite{particlecreation} to calculate the exact relation between the past null coordinate
$U$ in $\scri_L^-$ and the future null coordinate relevant at \scribub.
To find this relation one has to find the integral curves of the ray differential equation~\eqref{eq:rays}, which becomes at early times
\begin{equation}
 \frac{\dd r}{\dd t}=c
\end{equation}
and in a neighborhood of \scribub~($t\to +\infty$ and $r\to r_2$, see Fig.~\ref{fig:dynwd-pen})	
\begin{equation}\label{eq:rayslimit}
 \frac{\dd r}{\dd t}=c+\bar{v}(r)
 	=\kappa \left(r_2-r\right)+{\cal O}\left((r_2 -r)^2\right).
\end{equation}
Integrating the first equation we obtain the obvious result
\begin{equation}
 t=C+\frac{r}{c}
\end{equation}
and for the second one
\begin{equation}\label{eq:laterays}
 t=D-\frac{1}{\kappa }\ln\left(r_2-r\right).
\end{equation}
Following \cite{particlecreation} we identify initial events $P\equiv\left(r_i,t_i\right)$, with $r_i\sim ct_i$, final events $Q\equiv\left(r_f,t_f\right)$ with $r_f\sim r_2-e^{-\kappa t_f}$, and intermediate events $O\equiv\left(r_0,t_0\right)$.

Let us define
\begin{align}
    U &= \lim_{t_i\to-\infty}\left(t_i-\frac{r_i}{c}\right),\label{eq:Ulimit}\\
    u &= \lim_{t_f\to+\infty}\left[t_f+\frac{1}{\kappa }\ln\left(r_2-r_f\right)\right].\label{eq:ulimit}
\end{align}

Integrating Eq.~\eqref{eq:rays} between $P$ and an intermediate event $O$ in the in region of this spacetime (where the velocity profile depends only on $t$), we find
\begin{equation}\label{eq:PO1}
 r_0-r_i=\int_{t_i}^{t_0}\dd t\left[c+\bar{v}(\xi(t))\right].
\end{equation}
Using the definition of $U$ in Eq.~\eqref{eq:Ulimit} we obtain
\begin{equation}\label{eq:Utr}
 U=t_0-\frac{r_0}{c}+\frac{1}{c}\int_{-\infty}^{t_0}\dd t\,\bar{v}(\xi(t)).
\end{equation}
Then, integrating between an intermediate event $O$ now in the out region (where the velocity profile depends only on $r$) and $Q$ we find
\begin{equation}\label{eq:O2Q}
 t_f-t_0=\int_{r_0}^{r_f}\frac{\dd r}{c+\bar{v}(r)},
\end{equation}
and using the definition in Eq.~\eqref{eq:ulimit}
\begin{multline}
 u=t_0+\lim_{r_f\to r_2}\left[\frac{1}{\kappa }\ln\left(r_2-r_f\right) + \int_{r_0}^{r_f}\frac{\dd r}{c+\bar{v}(r)}\right]\\
=t_0+\frac{1}{\kappa}\ln\left[r_2-r_0\right]
-\frac{1}{\kappa}\ln\left[r_0-r_1\right]
+\frac{1}{\kappa}\ln\left[r_2-r_1\right]
+\lim_{r_f\to r_2} \int_{r_0}^{r_f}\dd r
\left[\frac{1}{c+\bar{v}(r)}-\frac{1}{\kappa \left(r_2-r\right)}-\frac{1}{\kappa \left(r-r_1\right)}\right].
\end{multline}
It is easy to see that the limit in the previous expression is finite for whatever $r_1<r_0<r_2$ so that we end up with the relation
\begin{equation}\label{eq:utr}
u=t_0+\frac{1}{\kappa}\ln\left[r_2-r_0\right]
-\frac{1}{\kappa}\ln\left[r_0-r_1\right]
+\frac{1}{\kappa}\ln\left[r_2-r_1\right]
+\int_{r_0}^{r_2}\dd r
\left[\frac{1}{c+\bar{v}(r)}-\frac{1}{\kappa \left(r_2-r\right)}-\frac{1}{\kappa \left(r-r_1\right)}\right].
\end{equation}

We now want to find the relation between $U$ and $u$. It is possible to find a particular form for the kink function $\xi$ (see Fig.~\ref{fig:lightrayssimp} and its caption for an example), such that all the rays cross either the 
left kink $r=-\xi(t)$ or the right kink $r=+\xi(t)$ only once. The event ``crossing the kink'' can be seen as belonging to both the in region and the out region. Therefore the relation $U=p(u)$ can be found by eliminating $t_0$ and $r_0$ between expressions \eqref{eq:Utr} and \eqref{eq:utr} taking into account that $r_0$ will be either $-\xi(t_0)$ or $+\xi(t_0)$. Figure~\ref{fig:lightrayssimp} shows an exact numerical calculation of ray propagation, performed with this method.

However, in order to study the late-time behavior of both particle production and the RSET, we only need the previous relation (see Sec.~\ref{sect:radiation}) for large and positive values of $u$ (the location of the black horizon) and for large and negative values of $u$ (the location of the white horizon). We want to show here that our particular example leads in fact to the general relation Eq.~\eqref{eq:Uu}. In this specific case, taking the limit for $|u|\to\infty$ corresponds to study the previous relation for light rays crossing the left kink at $r_0=-\xi(t_0)$, when $r_0$ is very close to $r_1$ [respectively, crossing the right kink at $r_0=+\xi(t_0)$, when $r_0$ is very close to $r_2$].
First note that, when $t\to t_{\rm H}$ (the time of the first appearance of the trapping horizons) the function $\xi$ can be expanded in the following form:
\begin{align}\label{eq:xi}
 \xi(t)&=r_2+\lambda\left(t-t_{\rm H}\right)+{\cal O}\left[{(t-t_{\rm H})}^2\right],\\
 -\xi(t)&=r_1-\lambda\left(t-t_{\rm H}\right)+{\cal O}\left[{(t-t_{\rm H})}^2\right],
\end{align}
with $\lambda$ a positive constant (remember that we have chosen for simplicity $r_1=-r_2$).
Defining
\begin{equation}
U_{\substack{\rm BH\\\rm WH}}
=t_{\rm H}\pm\frac{\xi(t_{\rm H})}{c}+\frac{1}{c}\int_{-\infty}^{t_{\rm H}}\dd t\,\bar{v}(\xi(t)),
\end{equation}
it is easy to see that
\begin{equation}
 U_\pm=U_{\substack{\rm BH\\\rm WH}}\mp \frac{\lambda}{c}\left(t_{\rm H}-t_0\right)+{\cal O}\left[{(t_{\rm H}-t_0)}^2\right],
\end{equation}
where with $U_+$ we denote the form of the function $U(t_0)$ for rays crossing the kink close to $r_1$ [respectively, with $U_-$ we denote the form of the function $U(t_0)$ for rays crossing the kink close to to $r_1$].
By expanding Eq.~\eqref{eq:utr} in the same limit $t_0\to t_{\rm H}$ and retaining only the dominant term we obtain
\begin{equation}
 u_\pm\simeq\mp\frac{1}{\kappa }\ln\left[\lambda\left(t_{\rm H}-t_0\right)\right],
\end{equation}
where we define $u_\pm$ in the same fashion as $U_\pm$.
Putting together the last two results we obtain (inside the bubble) the general relation Eq.~\eqref{eq:Uu}:
\begin{equation}
 U(u \to \pm \infty)=U_{\substack{\rm BH \\ \rm WH}} \mp A_{\pm} e^{\mp\kappa u}.
\end{equation}
%

\subsection{Left-going rays}	%

In Sec.~\ref{sect:RSET} we stated that it is possible to find a wide class of transitions from a Minkowskian geometry to a warp-drive one such that the contributions to the RSET due to left-going modes are just transient, vanishing at late times. We show here that this is the case for our specific model.

In order to do so, we shall need the relation between ingoing and outgoing left-going rays. Left-going rays are the solution of the differential equation~\eqref{eq:raysleft} which becomes at early times
\begin{equation}
 \frac{\dd r}{\dd t}=-c,
\end{equation}
while inside the bubble we cannot take any limit as in Eq.~\eqref{eq:rayslimit} because left-going rays are not confined inside the bubble, but escape through the black horizon, as we already noticed in Sec.~\ref{sect:radiation}. Integrating the equation at early times we obtain
\begin{equation}
 t=C-\frac{r}{c}.
\end{equation}
We identify initial events $P\equiv\left(r_i,t_i\right)$, with $r_i\sim ct_i$, final events $Q\equiv\left(r_f,t_f\right)$, and intermediate events $O\equiv\left(r_0,t_0\right)$ either in the in region (when the velocity profile depends only on $t$) or in the out region (when the velocity profile depends only on $r$).

Using $\tilde w$ defined as in Eq.~\eqref{eq:wdef}, choosing arbitrarily the constant of integration, we obtain
\begin{equation}\label{eq:w}
 \tilde w(t_0,r_0)\equiv t_0+\int_0^{r_0}\frac{\dd r}{c-\bar{v}(r)}.
\end{equation}

For $W$ we proceed as for $U$. We integrate Eq.~\eqref{eq:raysleft} between $P$ and an event $O$ in the in region
\begin{equation}\label{eq:PO1left}
 r_0-r_i=\int_{t_i}^{t_0}\dd t\left[-c+\bar{v}(\xi(t))\right].
\end{equation}
Defining $W$ as
\begin{align}
    W &= \lim_{t_i\to-\infty}\left(t_i+\frac{r_i}{c}\right),\label{eq:Wlimit}
\end{align}
we obtain
\begin{equation}\label{eq:Wtr}
 W=t_0+\frac{r_0}{c}-\frac{1}{c}\int_{-\infty}^{t_0}\dd t\,\bar{v}(\xi(t)).
\end{equation}
Now, the relation $W=q(\tilde w)$ can be found eliminating $t_0,r_0$ from Eqs.~\eqref{eq:wdef} and \eqref{eq:Wtr} in the limit where $(t_0,r_0)$ is on the kink.

We can now show that indeed the term $\rho_{{\rm dyn}-\tilde{w}}$ defined in Eq.~\eqref{eq:rhodw} is just a transient for this model. We see from Fig.~\ref{fig:lightrayssimp} that, as time grows, points inside the bubble are reached by $\tilde{w}$ rays that intersect the kink later and later. Using Eqs.~\eqref{eq:wdef} and \eqref{eq:Wtr} we can estimate $\dot{q}$, calculating the derivatives of $\tilde{w}$ and $W$ with respect to $t$ on the kink, at crossing time $t_c$:
\begin{multline}
 \dot{q}=\frac{\dd W}{\dd \tilde{w}}=\left.\frac{\dd W}{\dd t}\right|_{r=\xi(t_c)}\left(\left.\frac{\dd\tilde{w}}{\dd t}\right|_{r=\xi(t_c)}\right)^{-1}\\
	=\left(1+\frac{\dot\xi(t_c)}{c}-\frac{\bar{v}(\xi(t_c))}{c}\right)\left(1+\frac{\dot\xi(t_c)}{c-\bar{v}(\xi(t_c))}\right)^{-1}
 =1-\frac{\bar{v}(\xi(t_c))}{c}\to 1-\frac{\bar{v}(+\infty)}{c}.
\end{multline}
This is a constant value greater than $0$.
As a consequence, all the derivatives of $\dot{q}$ go to zero as time grows and $\rho_{{\rm dyn}-\tilde{w}}$ must go to zero in the same way. In conclusion, the dynamic term originated by the distortion of left-going rays can be different from zero when the horizon is created. However, this is only a transient term that is brought toward $\scri_L^+$, \ie, outside the bubble. That is, it is possible to create a transition region such that some radiation traveling leftward is produced only at the {onset} of the warp drive, but there is not any phenomenon like Hawking radiation originated in this way.

\subsection{Results}	%

In this section we present the result of the numerical calculation of $\rho$ inside the bubble. The velocity profile is given in Eq.~\eqref{eq:fluidvelocity} with $c=1$, $\alpha=2$ and $a=1$. With this choice, the surface gravity is $\kappa=\sqrt{3}/2$, the kink function is $\xi(t)=\text{arccosh}(t+1)$ and the horizons appear at $t_{\rm H}=1$.

\begin{figure}
 \centering
 \includegraphics[width=.45\textwidth]{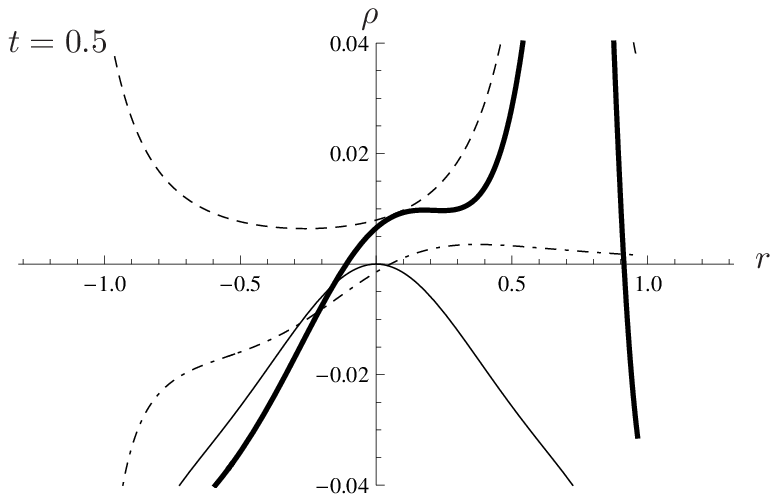}
 \hspace{.05\textwidth}
 \includegraphics[width=.45\textwidth]{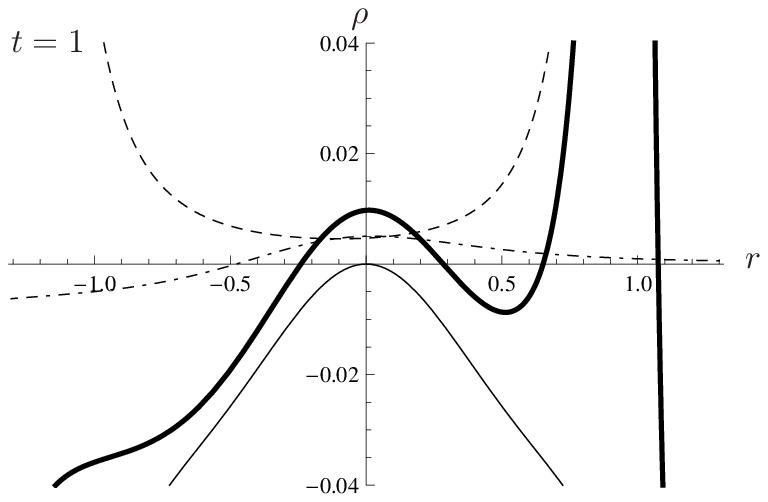}\\
 \vspace{.05\textwidth}
 \includegraphics[width=.45\textwidth]{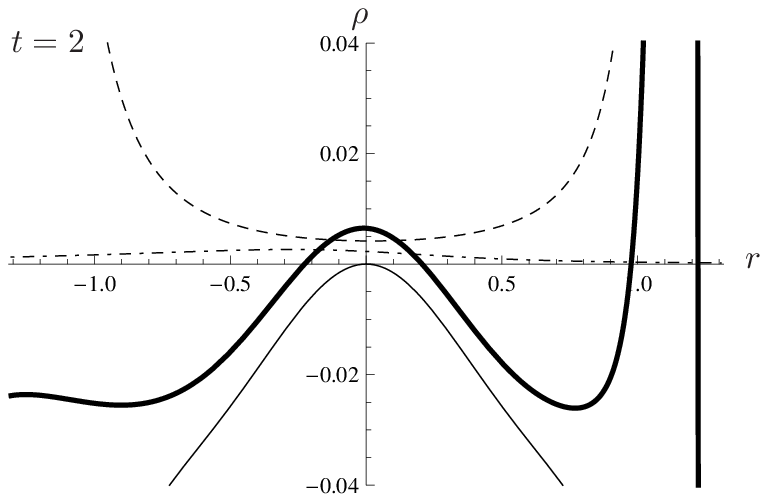}
 \hspace{.05\textwidth}
 \includegraphics[width=.45\textwidth]{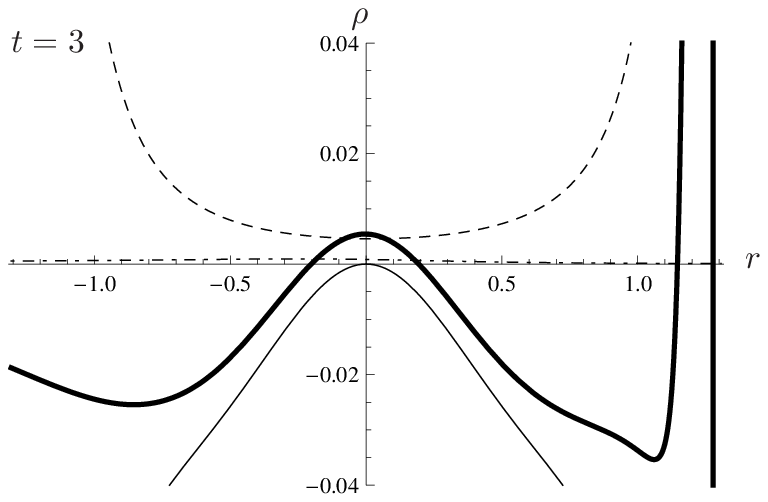}
 \caption{Energy density $\rho$ (thick solid line), $\rho_{{\rm st}}$ (solid line), $\rho_{{\rm dyn-}u}$ (dashed line) and $\rho_{{\rm dyn-}w}$ (dash-dotted line) as functions of $r$, $r_1<r<r_2$ at time $t=0.5,1,2,3$. Apparent horizons form at $t_{\rm H}=1$. }
 \label{fig:timev}
\end{figure}

In Fig.~\ref{fig:timev}, the energy density $\rho$ (thick solid line) of Eq.~\eqref{eq:rhowarpdrive} is plotted as a function of $r$ at different times ($t=0.5,1,2,3$) and  $r$ varies between $r_1$ and $r_2$, the locations of \hminus~and \hplus. The three terms in the right-hand-side of Eq.~\eqref{eq:rhowarpdrive}, $\rho_{\rm st}$, $\rho_{{\rm dyn-}u}$ and $\rho_{{\rm dyn-}w}$ are plotted, respectively with thin-solid, dashed and dash-dotted lines. In the first plot, all the components of $\rho$ vanish for $|r|\gtrsim 0.96$, because at time $t=0.5$ the kink is located at $r=\pm 0.96$ and the space is flat for $|r|>\xi(t)$.

The figure shows that the dynamic term $\rho_{{\rm dyn-}w}$ is transient, confirming the correctness of the approximation done in the analytical analysis, when this term was neglected to study the behavior of $\rho$ at late times. Note, however, that $\rho_{{\rm dyn-}w}$ gives an important contribution during the creation process of the warp drive.

In the center of the bubble ($r=0$) at late times, the energy density is $\rho\approx \kappa^2/48\pi\approx0.005$ and it is due only to the dynamic term $\rho_{{\rm dyn-}u}$, which can be naturally interpreted as Hawking radiation. For $t\lesssim 2$ the transient term $\rho_{{\rm dyn-}w}$ cannot be neglected. For instance, at $t=t_{\rm H}=1$, it causes the energy density inside the bubble to be about twice its late-time value.

\begin{figure}
 \centering
 \includegraphics[width=.45\textwidth]{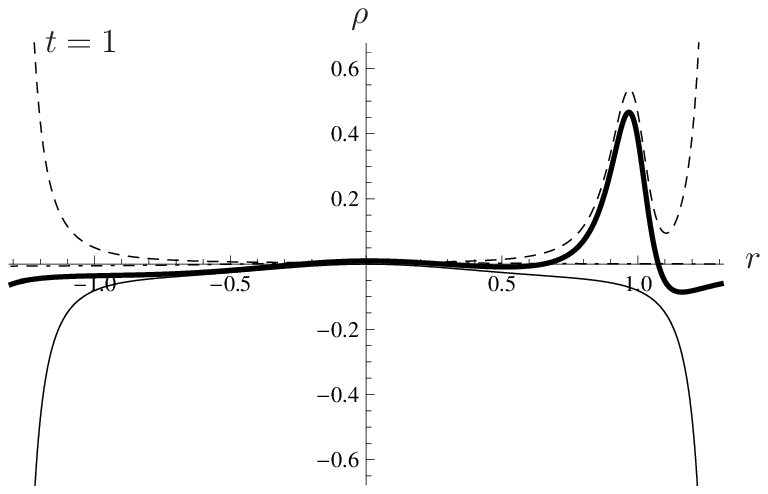}
 \hspace{.05\textwidth}
 \includegraphics[width=.45\textwidth]{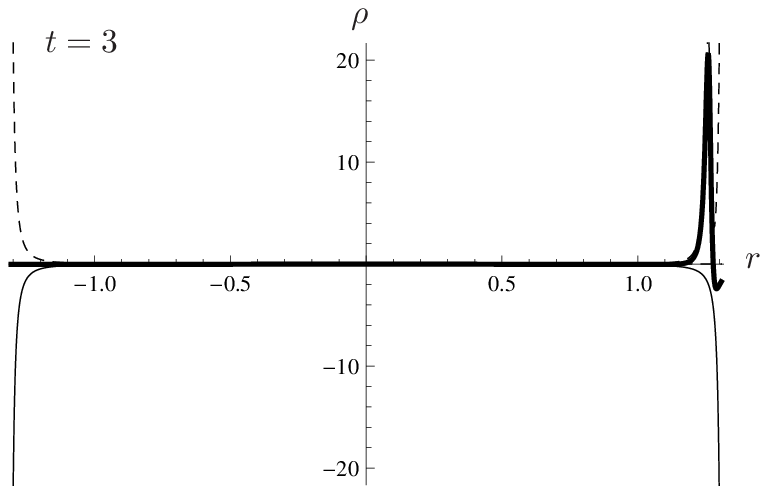}
 \caption{See caption of Fig.~\ref{fig:timev} for notation. Time slices at $t=1,3$. The scale of the $\rho$-axis is proportional to $e^{2\kappa t}$.}
 \label{fig:timevres}
\end{figure}

On the horizons $r=r_{1,2}$, both $\rho_{{\rm st}}$ and $\rho_{{\rm dyn-}u}$ are found to be divergent, but they cancel each other, leaving only a finite term, which vanishes with time on the black horizon ($r=r_1$), while it diverges to $-\infty$ on the white horizon ($r=r_2$), as shown in Eq.~\eqref{eq:sublead}. The claimed divergence on the Cauchy horizon is also found, even if it is difficult to recognize it from such plots. Looking at the Penrose diagram in Fig.~\ref{fig:dynwd-pen},
one indeed realizes that the location $(r=r_2$, $t=+\infty)$ does not represent a single point but a whole segment of the diagram. This means that one should disentangle the two effects (one at the Cauchy horizon, the other at the crossing point between \hminus~and~\scribub) which both take place at $r$ close to $r_2$. This can be done with a careful analysis: Exactly at $r=r_2$, as $t \to +\infty$, the value of $\rho$ goes towards $-\infty$; this contribution is superposed with a positive energy pulse whose peak-value grows towards $+\infty$ with time and whose center is located progressively closer and closer to $r_2$, approaching it as $r-r_2\propto e^{-\kappa t}$.
This can be better appreciated in Fig.~\ref{fig:timevres}, where the same quantities of Fig.~\ref{fig:timev} are plotted at time $t=2,3$, with different scales, proportional to $e^{2\kappa t}$ to compensate the exponential divergence of $\rho$.

\begin{figure}
 \centering
 \includegraphics[width=0.5\columnwidth]{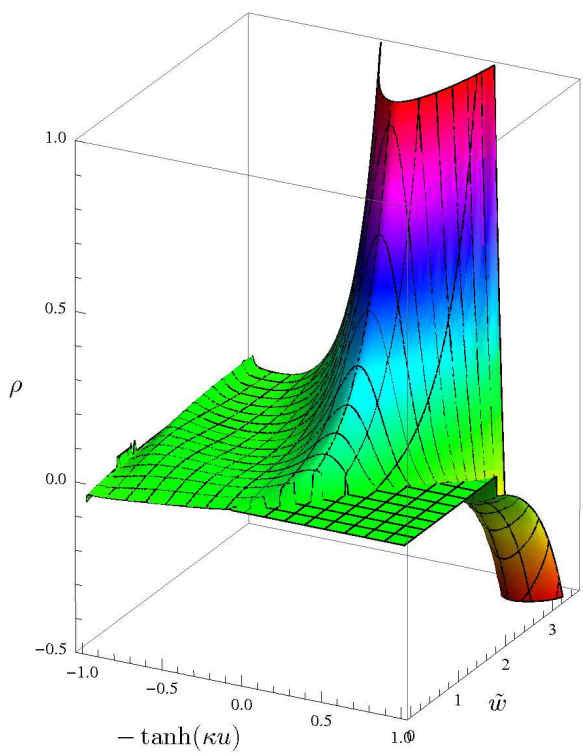}
 \caption{$\rho$ as a function of $-\tanh(\kappa u)$ and $\tilde w$.}
 \label{fig:rhouw}
\end{figure}

Since this degeneracy is due only to a bad choice of coordinates, it can be removed by using, for instance, the null coordinates $u$ and $\tilde w$. In Fig.~\ref{fig:rhouw}, $\rho$ is plotted as a function of $-\tanh(\kappa u)$ (in order to put the horizon $u=\pm\infty$ to finite positions) and $\tilde w$. Here, $\rho\to-\infty$ for $u\to-\infty$ ($-\tanh(\kappa u)=1$) and $\tilde w\to+\infty$, while $\rho\to+\infty$ for $u$ not too much negative and $\tilde w\to+\infty$. The former is the divergence on the white horizon, the latter is the divergence close to the Cauchy horizon: Whenever some energy travels on a $u$-ray, it is exponentially blueshifted at late times $\tilde w$ (late times $t$), when approaching
this horizon.

It is important to stress that these two divergences are of very different natures. In fact, the divergence at the crossing point between \hminus~and \scribub~is intrinsically due to the inevitable transient disturbances produced by the formation of the white horizon. In this sense it is a new and very effective instability.
On the contrary, the divergence on \scribub~is due to the well known infinite blueshift suffered by light rays as they approach a Cauchy horizon.
It is analogous to the often claimed instability of inner horizons in Kerr--Newman black holes~\cite{Simpson:1973ua,poissonisrael,markovicpoisson}.
In any case the backreaction of the RSET will doom the warp drive to be semiclassically unstable.

\section{Summary and discussion}	%
\label{sect:conclusion}			%

In a geometric optics approximation we have studied the propagation of light rays in dynamical geometries and found that the same exponential relation between affine coordinates on \scrim~and \scribub~is recovered at late times (large $u$) as in the case of black hole spacetimes. Given this relation, it is unavoidable the conclusion that indeed a Hawking flux will be observed by any observer inside the warp-drive bubble far from the black horizon. Indeed, the calculation shows explicitly the onset of such a flux. This radiation is produced at the black horizon and fills soon the interior of the bubble, traveling rightward at the speed of light. The central region of the warp drive behaves like the asymptotic region of a black hole: In both these regions the static contribution ($\rho_{\rm st}$) to the energy density vanishes so that the total energy density is due solely to the Hawking radiation generated at the black horizon.

When creating a warp drive one not only forms a black horizon but also a white one. Both are sources of right-going radiation. To understand better the nature of this radiation we have calculated the RSET in this geometry and, in particular, the energy density as measured by freely falling observers. In this way we recover that the RSET does not diverge at the horizons at any finite time. The singular behavior of the static terms (or vacuum polarization terms) of the RSET at the horizons is canceled by the leading contributions of the dynamical right-going terms, or what is equivalent, by the presence of Hawking radiation at both the horizons. 

{It is however easy to see that the behavior of the subleading terms of Eq.~\eqref{eq:sublead} is rather different between the black and the white horizons. The subleading term in the RSET associated with the formation of the black horizon does not produce any significant backreaction on the horizon itself. In fact, this term is just a transient which decays exponentially.\footnote{Furthermore, the analogy with a black hole originated by a star collapse of \cite{stresstensor} also allows us to infer that the warp drive must be created very rapidly in order to avoid a huge accumulation of vacuum polarization on the horizon and so a huge initial value of the energy density.}

The formation of a white horizon is also associated with a similar subleading term,} but this time it accumulates onto the white horizon itself. This causes the energy density $\rho$ seen by a free-falling observer to grow unboundedly with time on this horizon. The semiclassical backreaction of the RSET will make the superluminal warp drive become rapidly unstable, in a time scale of the order of $1/\kappa_2$, the inverse of the surface gravity of the white horizon. Indeed, if one trusted the QI \cite{pfenningford,broeck}, the wall thickness for $v_0\approx c$ would be $\Delta\lesssim 10^2\,L_{\rm P}$, and the surface gravity $\kappa\gtrsim10^{-2}\,{t_{\rm P}}^{-1}$, where $t_{\rm P}$ is the Planck time.\footnote{{In $1+1$ dimensions the warp-drive configuration is actually a vacuum solution of Einstein's equations. In this case, QI will not impose any conditions on the size of the wall thickness. However, as we will show, we expect our results to be valid also in $3+1$ dimensions. Therefore, we use the wall thickness of $3+1$ warp drives to obtain a realistic estimation for the Hawking temperature and the time scale of the exponential growing of the RSET.}}
This means that the time scale over which the backreaction of the RSET would become important is $\tau\sim1/\kappa\lesssim10^2\, t_{\rm P}$. Indeed, even forgetting about the QI, in order to get even a time scale $\tau\sim1 $ s for the growing rate of the RSET, one would need a wall as large as $3\times10^8$ m. Therefore, most probably one would be able to maintain a superluminal speed for just a very short interval of time. 
In addition to the above mentioned growing {term} on $\hor_2^{-}$, we have shown that there is also an unbounded accumulation of radiation on \scribub. Also this contribution will very rapidly lead to a significant backreaction on the superluminal warp drive and to some sort of semiclassical instability of the solution (that will most probably prevent the formation of the Cauchy horizons at late times).

Interestingly, the investigations of~\cite{MacherRP1} seem to imply that the above found asymptotic divergences of the RSET on the white horizon might disappear if Lorentz symmetry is broken at high energies. However, we shall show in Chap.~\ref{chap:warpdriveBEC} that, when the Lorentz breaking can be described through the introduction of a supersonic dispersion relation, another kind of instability appears, caused by a linear divergence of the radiation flux.

Even if the above described semiclassical instability may be avoided by some external action on the warp-drive bubble (or by some appropriate ultraviolet completion of the quantum field theory, like in~\cite{MacherRP1}), the QI lead to the conclusion  that the Hawking radiation in the center of the bubble will burn the internal observer with an excruciating temperature of $\Th\sim\kappa\gtrsim10^{-2}\,T_{\rm P}$, where $T_{\rm P}$ is the Planck temperature, about $10^{32}$ K. This would prevent the use of a superluminal warp drive for any kind of practical purpose. If we do not trust the QI, this high temperature might be avoided by making thicker walls. For instance, with $\Delta\sim 1$ m, one obtains a temperature of about $0.003$ K (roughly the temperature of radiation at a wavelength of $1$ m).\footnote{However, some very effective taming for the growing backreaction at the white horizon would be needed also in this case, given the previously estimated RSET growing rate.}

Finally, we comment on the fact that in this Thesis a 1+1 calculation was performed. In spherically symmetric spacetimes this might be seen as an $s$-wave approximation to the correct results. Unfortunately, this is not the case for the axisymmetric warp-drive geometry. However, the salient features of our results will be maintained in a full 3+1 calculation (most probably a numerical one), because they are still valid in a suitable open set of the horizons centered around the axis aligned with the direction of motion.

A suggestive interpretation of these results can be argued in connection with the so called chronology protection conjecture~\cite{hawking-chronology}. In fact, a time machine~\cite{everett} could be built through a couple of superluminal warp drives traveling in opposite directions. Thus, a protection mechanism seems to act at an early stage, forbidding the creation of a system which could be dangerous for causality.

In conclusion, we think that this work is casting strong doubts about the semiclassical stability of superluminal warp drives. Of course, all the aforementioned problems appear only when the velocity $v$ of the bubble becomes larger than the speed of light $c$, but they completely disappear when the bubble remains subluminal. Note that, even if $v$ is slightly smaller than $c$, no horizons form, no Hawking radiation is created, and neither strong temperature nor white horizon instability is found. The only remaining problem is that one would still need the presence of some amount of exotic matter to maintain the subluminal warp drive.

\chapter{Physical systems}	
\label{chap:physics}		

In this chapter we analyze three physical systems (water, Sec.~\ref{sec:water}, Bose--Einstein condensates (BECs), Sec.~\ref{sec:BECphysics}, and dielectric media, Sec.~\ref{sec:slowlight}) that are used as analogue models of gravity. It is not our intention to be exhaustive on all the possible systems proposed so far to realize the analogy~\cite{livrev}: phonons in weakly interacting BECs~\cite{BEC}, Fermi gases~\cite{fermi}, superfluid Helium~\cite{JacobsonVolovik}, slow light~\cite{slow_leonhardt,slow_reznik,slow_US}, non-linear electromagnetic waveguides~\cite{waveguide_SU,waveguide_leonhardt}, and ion rings~\cite{ions_prl,ions_long}.
Our choice of focussing only on three systems resides on two fundamental reasons. First, those three systems are the only ones for which experiments have been realized so far with quite interesting (albeit sometimes controversial) results. The second reason is that they are illustrative of the main issues that one faces in analogue gravity (AG), both on the experimental side (noise, stability of the background geometry\dots) and the theoretical side (modified dispersion relations at ultraviolet scales, quantization of perturbations, correlations between Hawking particles\dots).
In particular, before going into details of these three models, it is worth discussing how phenomena such as Hawking radiation are affected by short scale modifications that are present in any real physical system.

\sectionmark{UV behavior of the dispersion relation}		%
\section{Ultraviolet behavior of the dispersion relation}	%
\label{sec:UVdispersion}					%
\sectionmark{UV behavior of the dispersion relation}		%

As mentioned in Sec.~\ref{subsec:hawking}, one of the main concerns regarding Hawking radiation is the fact that quanta reaching \scrip\/ at late time come from a region very close to the horizon. Because of the infinite blueshift suffered by these modes when traced back towards the horizon, they would have entered the trans-Planckian regime (wavelength smaller than the Planck length). Since we expect standard general relativity and field theory to break down at such small scales, one might wonder whether such high energy modifications would affect (or possibly even kill) the particle spectrum at infinity. Seen from this perspective, analogue models are perfectly suitable for the investigation of these issues since they provide various short scale behaviors, depending on the specific considered system.

A quite general way of modeling the breaking of Lorentz symmetry at small scales is to modify the dispersion relation of fields (see~\cite{Mattingly} for a phenomenological approach to Lorentz violation). In this approach, the microphysics is parametrized by modifying the dispersion relation at high energy:
\begin{equation}
 \om^2=c^2 k^2+\Delta(k,K),
\end{equation}
where $K$ is some high energy scale, which is generally assumed to be the Planck scale and $\Delta(k,K)=O((k/K)^3)$. Assuming that measurements cannot be performed at very high energy, it is convenient to expand $\Delta$ in Taylor series around $k=0$. In some systems (for instance BECs, see~\ref{sec:BECphysics}), the Taylor series is finite, so that the expansion becomes exact.

Quite different physics appears according to the sign of the leading order term of the expansion. If it is positive (negative), the dispersion relation is called superluminal (subluminal) because the group velocity of the wave is larger (smaller) than the speed of light $c$. A similar terminology is used in the context of AG, referring to supersonic or subsonic dispersion relations, according to the sign of the difference between the speed of the waves at high frequency and the speed of sound.
Since we are going to use this concept in the context of analogue models of gravity, from now on we will talk about subsonic or supersonic dispersion relations, instead of subluminal or superluminal.

\subsection{Subsonic dispersion relation}	%
\label{subsec:UVsubluminal}			%

%
\begin{figure}
 \psfrag{om}[c][c]{$\Omega$}
 \psfrag{k}{$k$}
 \includegraphics[width=0.95\textwidth]{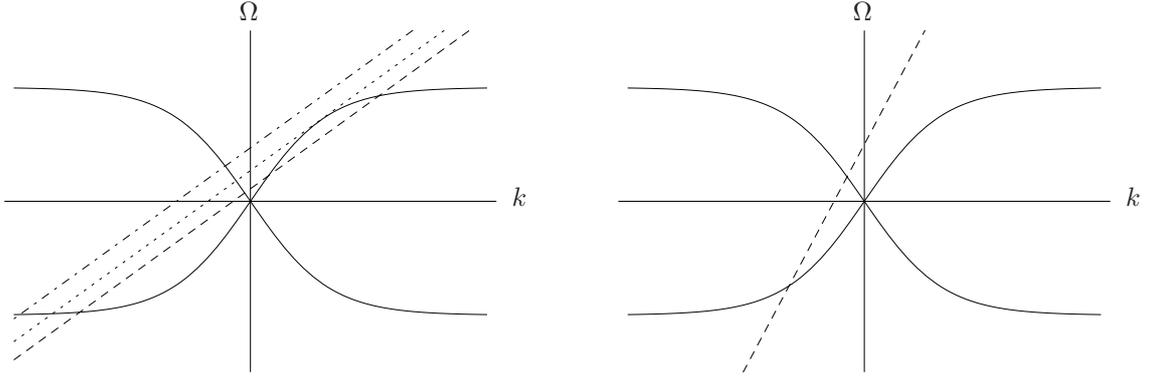}
 \caption{Graphical solution of the subsonic dispersion relation~\eqref{eq:subunruh} for subsonic flow (left panel) and supersonic flow (right panel). Solid lines: $\pm \Omega(k)$. Left panel, dashed, dotted, dotdashed lines: $\om-vk$ respectively for $\om<\ommax$, $\om=\ommax$, $\om>\ommax$. Right panel, dashed line: $(\om-vk)$. For $\om<\ommax$, one clearly sees that, in subsonic flows, four real roots exist.
The largest negative one, which is present also in supersonic flows, has negative group velocity (left-going) and represents the Hawking partner. The smallest positive one is the Hawking particle and has positive group velocity (right-going). The largest positive solution represents the incoming left-going particle which bounces at the turning point and it is converted into the Hawking particle.}
 \label{fig:dispersionsub}
\end{figure}

This was the case originally studied by Unruh~\cite{Unruh95} (see Fig.~\ref{fig:dispersionsub}):
\begin{equation}\label{eq:subunruh}
\omega-\bv\cdot\bk=cK\left(\tanh(k/K)^n\right)^{1/n}=ck\left\{1-\frac{1}{3n}\left(\frac{k}{K}\right)^{2n}+O\left[\left(\frac{k}{K}\right)^{4n}\right]\right\}\equiv \Omega(k),
\end{equation}
and by Corley and Jacobson~\cite{CJ96}
\begin{equation}\label{eq:subCJ}
 (\omega-\bv\cdot\bk)^2=c^2(k^2-k^4/K^2)=\Omega^2(k).
\end{equation}
Although the analytical forms of the dispersion relation differ, these two systems show a similar behavior, related to the fact that the wave velocity for large wavenumber is smaller than $c$. If one takes a flow profile with a single sonic point as in Sec.~\ref{subsec:schwarzschild} and traces back an outgoing mode (corresponding to the smallest positive solution in the left panel of Fig.~\ref{fig:dispersionsub}) from infinity towards the horizon, the wavenumber of that mode increases while approaching the horizon and, as a consequence of the subsonic dispersion relation, there is a point at which its group velocity $c_g<c$ equals the velocity of the fluid $v$. At that point (still going back in time), this mode appears as a superposition of two modes bouncing back from the horizon (corresponding to the largest positive and negative solutions in the left panel of Fig.~\ref{fig:dispersionsub}) with a wavelength increasing with the distance from the horizon. One of the two modes has negative norm (corresponding to the negative solution) with respect to the appropriate inner product~\cite{twofaces} and one has positive norm (corresponding to the largest negative solution in Fig.~\ref{fig:dispersionsub}).

The mixing between negative and positive norm modes is at the origin of Hawking particle creation, whose presence in this kind of systems has been confirmed both by analytical and numerical calculations~\cite{Unruh95,BMPS95,CJ96,Corley97,Saida,Tanaka99,UnruhSchu05,MacherRP1}.
Interestingly, those analyses show that the spectrum of Hawking radiation is thermal with a temperature $\Th$ [see Eq.~\eqref{eq:hawking_fluid}], when the surface gravity $\kappa$ is much smaller than $cK$. The spectrum remains Planckian up to a certain frequency $\omm$, which depends on $K$. 
Moreover, the classical counterpart of the described mode conversion has recently been observed in a wave tank experiment~\cite{Silke_exp}. We will describe this experiment and its result in Sec.~\ref{subsec:expwater}.

To conclude this section, note that modes propagating in these systems do not experience an infinite blueshift when they are close to the horizon.
However, the incoming waves are infinitely blueshifted when traced back in time, such that problems may arise in the determination of the proper initial conditions.

\subsection{Supersonic dispersion relation}	%
\label{subsec:UVsuperluminal}			%

\begin{figure}
 \psfrag{om}[c][c]{$\Omega$}
 \psfrag{k}{$k$}
 \includegraphics[width=0.95\textwidth]{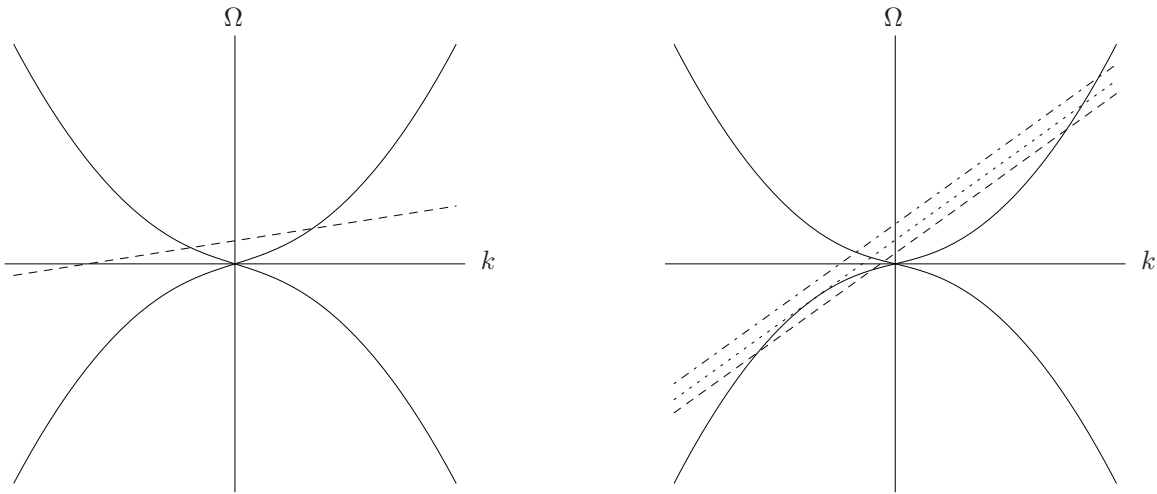}
 \caption{Graphical solution of the supersonic dispersion relation~\eqref{eq:dispersion1} of phonons in a BEC for subsonic flow (left panel) and supersonic flow (right panel). Solid lines: $\pm \Omega(k)$. Left panel, dashed line: $(\om-vk)$. Right panel, dashed, dotted, dotdashed lines: $\om-vk$ for $\om<\ommax$, $\om=\ommax$, $\om>\ommax$, respectively. For $\om<\ommax$, one clearly sees that, in supersonic flows, two extra (real) roots exist in the left lower quadrant.
The most negative one is called, in the text, $k^{(1)}_{\om}$ and the other one $k^{(2)}_{\om}$, so that $k^{(1)}_{\om} < k^{(2)}_{\om} < 0$. For details on the model, refer to Sec.~\ref{subsec:theobec} and Appendix~\ref{app:BEC}. In the Hawking process, the positive $k$ solution in the subsonic region (left panel) is the Hawking particle, while the negative solution $k^{(2)}_{\om}$ in the supersonic region is the negative norm left-going partner.}
 \label{fig:dispersion}
\end{figure}

The case of supersonic dispersion relation shows rather different features with respect to the case of subsonic dispersion. First, negative norm modes are now trapped in the supersonic region and cannot propagate outside the horizon as can be seen from the graphical solution of the typical supersonic dispersion relation of a BEC [see Eq.~\eqref{eq:dispersion1}] in Fig.~\ref{fig:dispersion}.

In this case Hawking radiation is still present because of negative norm modes.
If one traces back in time an outgoing mode in the supersonic region (positive solution of the dispersion relation in Fig.~\ref{fig:dispersion}, left panel, it gets blueshifted (but not infinitely) on the horizon and its group velocity increases, such that it remains always larger than the local speed of sound and actually can penetrate inside the horizon (positive solution of the dispersion relation in Fig.~\ref{fig:dispersion}, right panel). However, around the horizon, mode conversion takes place and the mode is decomposed not only in this right-going positive norm mode but also in a right-going negative norm mode (the largest negative solution $k^{(1)}_{\om}$ of Fig.~\ref{fig:dispersion}, right panel).

Playing the movie forward in time, right-going modes can actually cross the horizon from the interior region and escape to infinity. For a real black hole, this implies that, in the presence of superluminal dispersion relation, information may be extracted from the interior of the black hole and even from the singularity. This is a quite delicate issue, because, in order to obtain Hawking radiation from this system, one would need to impose boundary conditions on the singularity. However, whether Hawking radiation is present in the case of superluminal dispersion radiation is still a controversial issue. In~\cite{sensitivity_carlos}, the authors show that the radiation, although following a Planckian spectrum, dies off with time. Other results, instead, both analytical and numerical~\cite{carusotto1,carusotto2,MacherRP1,MacherBEC,Recati2009}, show that Hawking radiation is very robust also in the case of superluminal dispersion relation for frequencies $\om$ smaller than a certain cut off frequency $\omm$, for which there are only two real solutions of the dispersion relation also in the superluminal region (see Fig.~\ref{fig:dispersion}, right panel, dotdashed line).

The differences in those results are related with the choice of the correct quantum state. The results of~\cite{MacherRP1,MacherBEC,Recati2009} are obtained in a stationary geometry, by choosing a quantum state which is vacuum for ingoing modes (see Chap.~\ref{chap:hawking}). However, given the dispersive character of the theory, one might also choose a state with no incoming and outgoing particles in the subsonic region (a Boulware-like state~\cite{boulware}), corresponding to the exterior of the black hole~\cite{twofaces}. Of course, this would require a tuning of the initial occupation number of the incoming modes in the supersonic region (see discussion in Sec.~\ref{subsec:bogo}). Unfortunately, in a realistic case of a black hole collapse, it would be difficult to decide which of the two states would be selected by the dynamics of the system. Furthermore, in the case of the creation of a singularity (or more generally of a strong gravity region), it would be even more difficult to understand what are the proper boundary conditions on the modes coming out from this region.

Finally, let us mention that in the numerical simulation of~\cite{carusotto2}, the horizons are dynamically created in a BEC. This seems to exclude the possibility of a dynamical selection of the Boulware-like state. However, both the way in which the horizon is created in a BEC and the final velocity profile are completely different from what happens in a real collapsing star, ending up in a black hole. The possibility that superluminal dispersion relation might kill or heavily modify Hawking radiation in real astrophysics black holes cannot be totally excluded and further analyses are worthwhile in this direction.

\section{Surface waves in water}	%
\label{sec:water}			%

Water is probably the most common and popular physical system allowing the propagation of perturbations (\eg, surface waves). Its physics is widely studied and well understood from the point of view of classical fluid dynamics~\cite{Lamb}.
So it is natural to check if this system is suitable for the purposes of AG~\cite{Schutzhold:2002rf}.
Of course, one cannot hope to measure quantum effects in water waves on macroscopic scales. Moreover, the temperature at which water is liquid is order of magnitudes larger than the highest Hawking temperature that one may obtain with realistic flow profiles. So, why should we care about water? First, it is easy to control: Both the fluid velocity and the speed of surface waves are easily tunable by changing the depth of the basin.\footnote{Of course one cannot change in an arbitrary way both the depth of the basin and the velocity of the fluid because of mass conservation.} Second, to prove the existence of the Hawking effect, in principle, one would only need to measure the mixing between positive and negative norm modes (see Sec.~\ref{sec:phenomenon}). Once one measured that the $\beta$ coefficient governing the mixing between these modes followed a thermal distribution, the quantum spontaneous emission would be implied by quantum mechanics.

Some more comment on this point is worthwhile.
Whether one were satisfied with such an experimental result would of course depend on what would be defined as Hawking radiation and what would be considered as a proof for its existence. If one considers thermal spontaneous emission as the fundamental issue to be tested, water is not the right system at all. However, if one is satisfied with detecting mode conversion and measuring the proper thermal exponential factor in the $\beta$ coefficient, water is perfectly fine.

Finally, in water, as in any other real physical system, the local Lorentz invariance at the basis of the effective geometry is broken at scales comparable with the molecular one (see discussion in Sec.~\ref{subsec:hawking}). This makes the system suitable to study modifications that a possible breaking of the spacetime structure at small scales would produce on phenomena like Hawking radiation. In particular, we will show below that in water short-wavelength waves are slower than standard low frequency waves. This is just the opposite of what happens in a BEC (see Sec.~\ref{sec:BECphysics}). We shall discuss in details the differences and the analogies of those two different dispersion relations.

\subsection{Theoretical description}	%
\label{subsec:theowater}		%

\begin{figure}
 \psfrag{om}[c][c]{$\Omega$}
 \psfrag{k}{$k$}
 \includegraphics[width=0.95\textwidth]{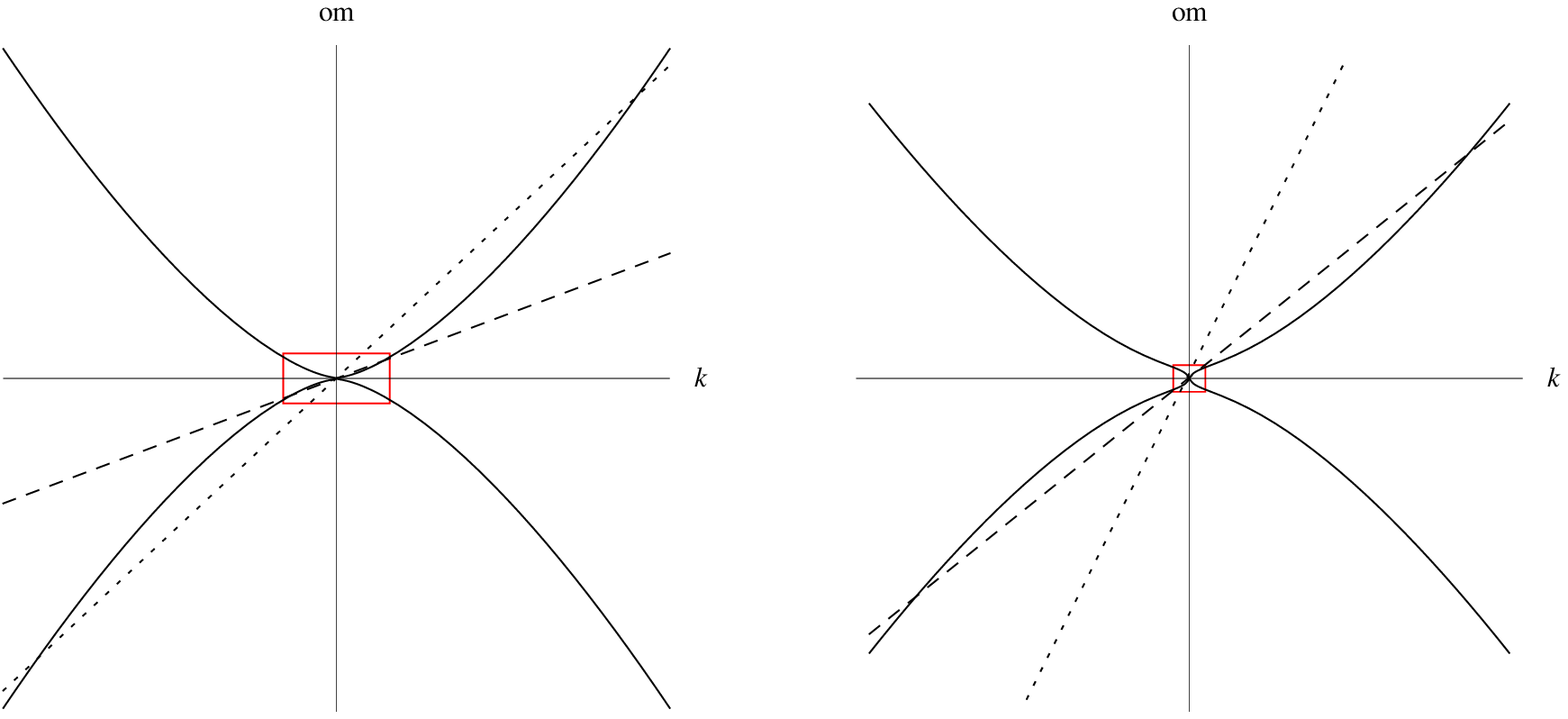}
 \centering
 \includegraphics[width=0.7\textwidth]{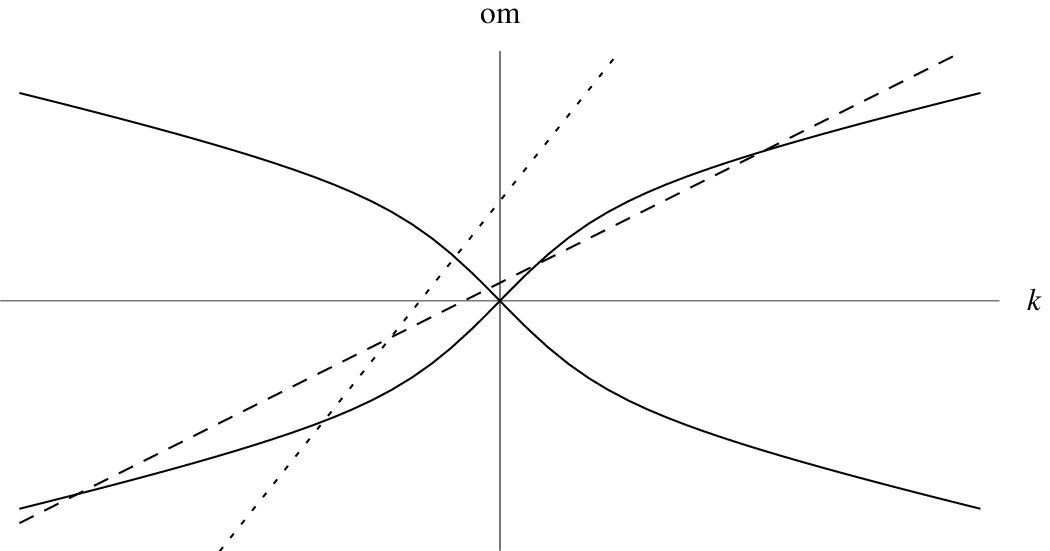}
 \caption{Graphical solution of the dispersion relation~\eqref{eq:waterdispersion} of surface waves in water. Solid lines: $\pm \Omega(k)$. Dashed and dotted lines: $(\om-vk)$ for subsonic and supersonic flows, respectively. The three plots are made at different scales. The region delimited by the red rectangle in the top left panel is zoomed in the top right panel and the region delimited by the red rectangle in the top right panel is zoomed in the bottom panel.
 When the flow is subsonic there can be up to 6 real $k$ solutions at a given frequency. When the flow is supersonic there are 4 real $k$ solutions. Note that, at large wavelength (bottom panel), the dispersion relation shows the same features as the subsonic dispersion relation of Fig.~\ref{fig:dispersionsub}, while, at very small wavelength (top left panel), it resembles the supersonic relation of Fig.~\ref{fig:dispersion}.
}
 \label{fig:dispersionwater}
\end{figure}

It is a well established fact (see~\cite{Philips}) that the dispersion relation of water waves is
\begin{equation}\label{eq:waterdispersion}
 \left(\om-\bv\cdot\bk\right)^2= \left(gk+\frac{\sigma k^3}{\rho}\right)\tanh(kh)=\Omega^2(k),
\end{equation}
where $h$ is the dept of the basin and $\sigma$ is the surface tension.
When $k>\sqrt{g\rho/\sigma}$ the cubic term dominates, the effects of surface tension are no longer negligible, and the wave is in the capillary regime. In Fig.~\ref{fig:dispersionwater} this dispersion relation is graphically solved. Because of the hyperbolic tangent and of the cubic term due to the surface tension, the dispersion relation is supersonic at very high momenta (compare the top left panel of Fig.~\ref{fig:dispersionwater} with Fig.~\ref{fig:dispersion}), while it is subsonic for low values of $k$ (compare the bottom panel of Fig.~\ref{fig:dispersionwater} with Fig.~\ref{fig:dispersionsub}).
Note that, in addition to the four solutions for subsonic flows, two extra solutions (capillary waves) appear at very high momentum. The analogue of Hawking radiation in this system is a classical mode conversion among modes in the subsonic region of the fluid, as shown in Sec.~\ref{subsec:UVsubluminal}. See~\cite{germainulf} for a detailed analysis on propagation and mode conversion of surface water waves in flows with horizons.

At first glance, Eq.~\eqref{eq:waterdispersion} is far from resembling a quadratic dispersion relation as needed when perturbations satisfy a hyperbolic equation as Eq.~\eqref{eq:KGfluid}. However, in the limit of low wavenumber
\begin{equation}\label{eq:waterconditions}
 k\ll\sqrt{\frac{g\rho}{\sigma}},\qquad
 k\ll\frac{1}{h},
\end{equation}
the dispersion relation is
\begin{equation}
 \om^2= gh\,k^2+O(k^4).
\end{equation}
In that limit all the waves have the same speed
\begin{equation}
 c=\sqrt{gh},
\end{equation}
depending on the depth of the basin.

The first condition of Eq.~\eqref{eq:waterconditions} requires that waves are not in the capillary regime. The second condition relates the wavelength of the wave to the height of the fluid. In order for surface waves to satisfy a quadratic dispersion relation, the depth of the basin must be much smaller than the wavelength (shallow water regime). In the opposite limit (deep water), when $k$ satisfy
\begin{equation}
 \frac{1}{h}\ll k\ll\sqrt{\frac{g\rho}{\sigma}},
\end{equation}
the dispersion relation is
\begin{equation}
 \om^2\approx gk
\end{equation}
and both the phase ($c_p$) and the group ($c_g$) velocity of the wave depend on its wavenumber
\begin{equation}
 c_p=\sqrt{\frac{g}{k}},\qquad c_g=\frac{c_p}{2}.
\end{equation}
As a consequence, a proper geometry cannot be defined.\footnote{Nevertheless, one can define a so-called \emph{rainbow} geometry, characterized by a $k$-dependent metric~\cite{rainbow}.}

In the shallow water regime, it is instead possible to show~\cite{Schutzhold:2002rf} that the perturbation velocity potential $\phi_1$ [see Eq.~\eqref{eq:vpotential}] satisfies Eq.~\eqref{eq:KGfluid}, in $d=2$ spatial dimension (because they are surface waves). The metric is exactly the same as in Eq.~\eqref{eq:fluidmetric}, but now $\bv$ is replaced by the component of the velocity parallel to the water surface and the density $\rn$ is constant since water is incompressible:
\begin{equation}\label{eq:waterlineelement}\
 \dd s^2=\frac{1}{c^2}\left[-(c^2-v_\parallel^2)\,\dd t^2-2\bv_\parallel\cdot\dd\bx_\parallel\,\dd t+\dd\bx_\parallel\cdot\dd\bx_\parallel\right].
\end{equation}
%

\subsection{Experiments}	%
\label{subsec:expwater}		%

The existence of a phenomenon analogous to Hawking radiation in hydrodynamic systems was quite overlooked by fluid mechanics researchers till a few years ago. The first experiment studying a water flow with a horizon with the aim of observing Hawking-like phenomena is actually very recent~\cite{germain_exp}. That experiment has the undoubted merit of having put forward an interesting phenomenology to the attention of the hydrodynamic community~\cite{germainulf}. However, for a series of reasons, it was not performed in the proper regime to detect the classical analogue of the Hawking phenomenon.

Nevertheless, this first experiment motivated a second one with the direct participation of Unruh~\cite{Silke_exp}. 
This second experiment was set in a much more accurate way, designed just to measure the classical Hawking mechanism.\footnote{In~\cite{germain_exp}, it was used a too large flume tank, built for commercial purposes. This did probably not allow to reach suitable settings in the short available time.} Keeping the flow rate constant,
the analogue of a white horizon was realized by changing the depth of the flume with a wing shaped obstacle to prevent flow separation. To avoid spurious effects due to the interaction of the fluid with the wall, the height of the waves was measured in the center of the tank (in~\cite{germain_exp}, it was measured on the wall) by illuminating the free surface of the water with a green laser light sheet. Water contained a fluorescent dye to create a sharp intensity maximum that was eventually detected by a high-resolution monochrome camera.

Analyzing the supersonic dispersion relation in Fig.~\ref{fig:dispersionwater}, bottom panel, in a subsonic flow (dashed line) corresponding to the exterior of a white hole, one sees that the positive norm mode with smallest positive $k$ (the solution in the linear regime of the dispersion relation) propagates in the opposite direction with respect to the flow. This implies that, in a white hole configuration, this mode propagates toward the horizon. At the same frequency, there are other two roots on the same branch of the dispersion relation, both with large absolute value of $k$, but opposite sign and opposite norm. Moreover, they are both propagating in the same direction of the flow, \ie, they travel away from the horizon.

In~\cite{Silke_exp}, the low momentum wave was artificially created in the supersonic region. Then, it propagated toward the white horizon where it was blocked and converted into the other two high-momentum waves, propagating in the opposite direction.
The conserved norms $|\alpha_\om|^2$ and $|\beta_\om|^2$ of the positive and negative outgoing waves, respectively, can be computed from the measured amplitudes of these waves. When mode conversion is due to a classical stimulated Hawking emission, the ratio between the two norms is
\begin{equation}\label{eq:expratiosilke}
 \frac{|\beta_\om|^2}{|\alpha_\om|^2}=\ee^{-2\pi\om/\kappa},
\end{equation}
where $\kappa$ is the surface gravity of the horizon.

\begin{figure}
 \includegraphics[width=0.48\textwidth]{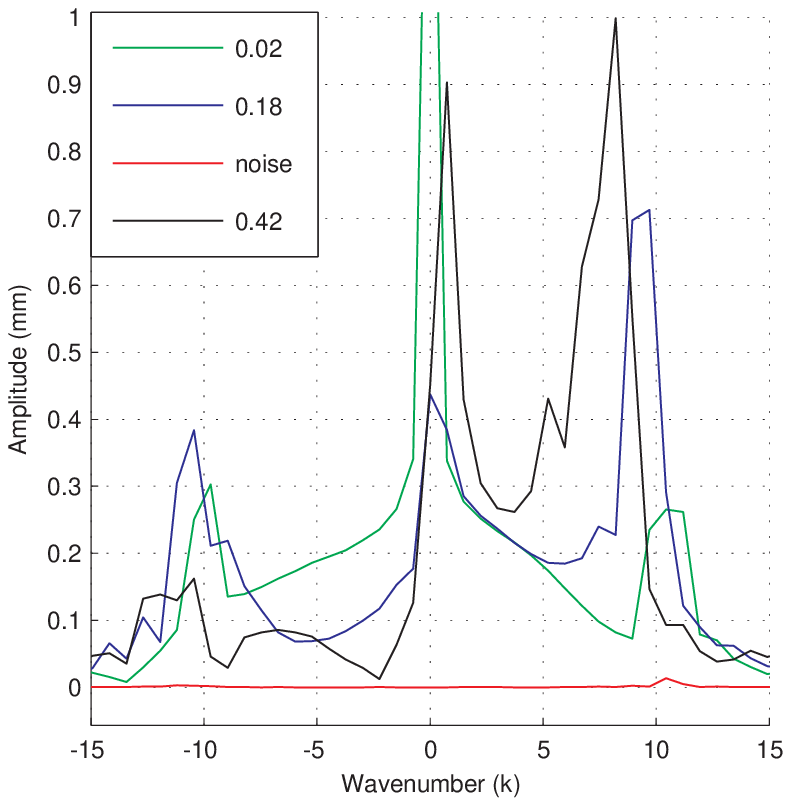}
 \includegraphics[width=0.48\textwidth]{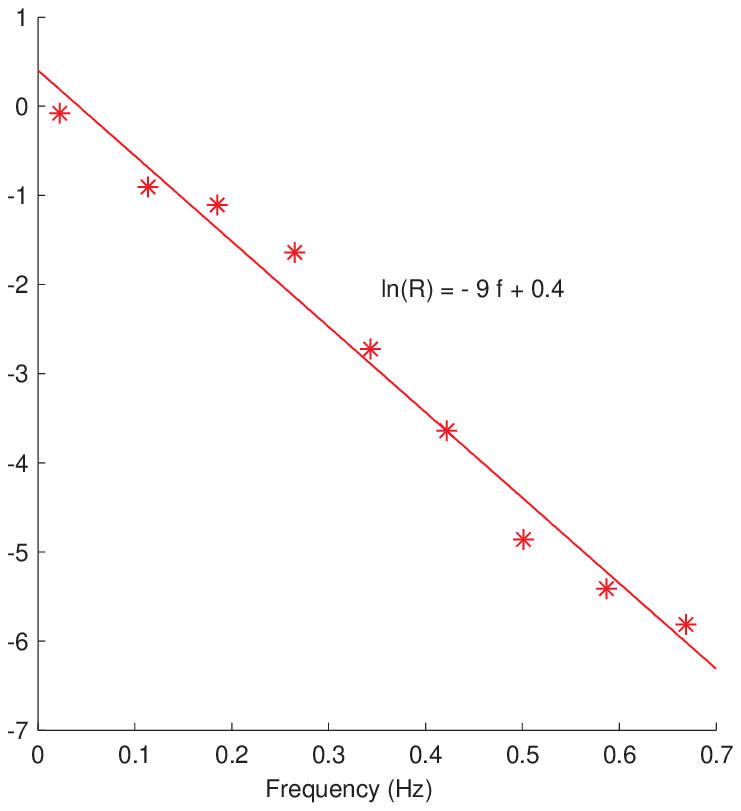}
 \caption{Left panel: Fourier transform in space of the wave amplitude for incoming waves of different frequencies. Right panel: ratio between the norm of the negative and positive norm outgoing modes [see Eq.~\eqref{eq:expratiosilke}] (From~\cite{Silke_exp}).}
 \label{fig:silkeresults}
\end{figure}

In Fig.~\ref{fig:silkeresults}, we report the result of the experiment. In the left panel, the authors plot the Fourier transform in $k$ of the surface of the water, at fixed frequency $\omega$. The central peak, at very low momentum represents the incoming linear wave. The right and the left peaks represent the amplitudes of the two outgoing waves with large momenta (in the non-linear regime of the dispersion relation). In the right panel, they plot the ratio of Eq.~\eqref{eq:expratiosilke}, that shows a nice exponential behavior. The fitted intercept is close to 0 and the slope corresponds to an amazingly low Hawking temperature $\Th=6\times10^{-12}\,\mbox{K}$, which dispels any doubt about the possibility of measuring the spontaneous emission in water, but nicely confirms the thermal behavior of Hawking emission.

\section{Bose--Einstein condensates}	%
\label{sec:BECphysics}			%

One of the main motivations to investigate analogue models of gravity is the possibility of simulating quantum field theory phenomena in curved geometries, such as Hawking radiation or cosmological particle production. As discussed in the previous section, classical systems like water are suitable to detect the presence of negative norm modes and the mode conversion at the basis of those phenomena. However, it would be interesting to detect also the quantum spontaneous emission that would exactly correspond, for instance, to Hawking radiation.

A system in which this aim could be achieved must satisfy the following properties:
\begin{itemize}
 \item high degree of quantum coherence,
 \item very low temperature to minimize the noise, since the expected signal is very faint,
 \item low speed of sound to create a supersonic region with velocities that can be handled in a laboratory.
\end{itemize}
All those requirements are satisfied by BECs, that have actually been one of the most studied analogue systems of gravity. They were the first quantum system to be successfully proposed~\cite{BEC,Garay} and then extensively investigated in~\cite{Barcelo:2001cp,Barcelo:2003wu,Barcelo:2001ca,expandinguniverse_ff,Fedichev:2003bv,Fedichev:2003dj} for various flow configurations corresponding to different geometries.

A lot of efforts have subsequently been devoted both to analytical and numerical investigations on the phenomenology of Hawking radiation~\cite{ulf,Gardiner,carusotto1,carusotto2,Carusottowhite,Recati2009,MacherBEC,bhlasers,smearhor} and cosmological particle production in BECs~\cite{Lidsey:2003ze,Weinfurtner:2004mu,expandinguniverse_piyush}. Analogy with particle production in expanding universes~\cite{Calzetta:2002zz} was used to explain the Bosenova experiment~\cite{Donley} (see Sec.~\ref{subsec:universe}). BECs have also been used as toy models to study quantum gravity phenomenology~\cite{kapusta81,two-componentBEC,two-componentBEC2,gravdynam,cosmobec} and a new system has been proposed~\cite{relbec} with this purpose.
Finally, for the first time, a sonic horizon was realized in a BEC~\cite{technion}, opening the way for a future experimental measure of spontaneous emission of Hawking particles.

In this section we present the basics of the theoretical description of BECs (a rigorous and detailed analysis is given in Appendix~\ref{app:BEC}) and of the most relevant numerical and experimental results. In Part~\ref{part:BECphenomenology} we will enter in the details of our investigation of Hawking radiation in BECs and of the instabilities arising in particular configurations. In Part~\ref{part:BECemergent} we will show how these systems can be used as tools for the investigation of emergent gravity issues.

\subsection{Theoretical description}	%
\label{subsec:theobec}			%

A dilute gas of bosons~\cite{Dalfovo} can be described by a quantum field $\hP$ satisfying the bosonic commutation relation
\begin{equation}\label{eq:comm1}
 \comm{\hP(t,\bx)}{\hPd(t,\bx')}=\delta^3(\bx-\bx'),
\end{equation}
and whose equation of motion
\begin{equation}\label{eq:hPmotion}
 \ii\hbar\pdt\hP(t,\bx)=\left[-\frac{\hbar^2}{2m}\nx^2 + V(\bx)  + g \hPd\hP\right]\hP
\end{equation}
is generated by the following Hamiltonian
\begin{equation}\label{eq:Hsc1}
 \hH = \int\! \dtx \left[\frac{\hbar^2}{2m} \nx \hPd \, \nx\hP + V  \hPd\hP  + \frac{g}{2} 
 \hPd\hPd\hP\hP    \right],
\end{equation}
where $m$ is the atom mass, $V$ the external potential and $g$ the effective coupling.

Condensation takes place when a macroscopic fraction of bosons is in the ground state. In this situation, the condensed part is described by a function $\Pn$, while $\hP-\Pn$ is a quantum field representing fluctuations on top of the condensate. To simplify the algebra, it is convenient to define the relative fluctuation operator:
\begin{equation}\label{eq:defphi1}
 \hp\equiv\frac{\hP-\Pn}{\Pn},
\end{equation}
which satisfies the following commutation relation
\begin{equation}\label{eq:commphi1}
 \comm{\hp(t,\bx)}{\hpd(t,\bx')}=\frac{1}{\rnbx}\delta(\bx-\bx'),
\end{equation}
where $\rn\equiv|\Pn|^2$ is the density of the condensate. Note that the above decomposition assumes a spontaneous breaking of the $U(1)$ symmetry of the system. In fact, in this analysis the condensate has a well-defined phase, since it is characterazed by the classical field $\Pn$, which is the expectation value of the field operator $\hP$ on a coherent state. As a consequence, such a state does not have a well-defined number of particles $N$. Thus, this approach is not fully appropriate to describe systems with a fixed $N$. Nevertheless, as shown by a fully rigorous treatment~\cite{dumcastin}, it becomes accurate in the limit $N\to\infty$, especially for a steady state at very low temperature.

Linearizing the equation of motion~\eqref{eq:hPmotion} with respect to $\hp$, one obtains the Gross--Pitaevskii equation for the condensed part $\Pn$
\begin{equation}
 \ii\hbar\pdt\Pn=\left[-\hbm\nx^2+V+g\rn\right]\Pn
\label{eq:GP1}
\end{equation}
and the Bogoliubov--de Gennes equation for the fluctuation field
\begin{equation}
 \ii\hbar\pdt\hp = \left[T_\rho-\ii \hbar\bv\!\cdot\!\nx + m c^2 \right]
\hp + m c^2 \hpd,\label{eq:BdG1}
\end{equation}
where $c\equiv\sqrt{g\rn/m}$ is the speed of sound and
\begin{equation}\label{eq:Trho1}
 T_\rho\equiv -\frac{\hbar^2}{2m} \frac {1}{\rn}\nx\!\cdot\!\rn\nx
\end{equation}
is a second order differential operator.

This equation is linear, it couples $\hp$ with $\hpd$, and, except for the $T_\rho$ term, it contains only first derivatives with respect to $\bx$ and $t$. This makes it very similar to Dirac's equation. In fact, $T_\rho$ can be neglected when the typical scale of perturbations is smaller than the healing length (see Appendix~\ref{app:BEC})
\begin{equation}\label{eq:healing1}
 \xi\equiv\frac{\hbar}{\sqrt{2}mc},
\end{equation}
which is nothing but the acoustic Compton wavelength of the atoms, so that Eq.~\eqref{eq:BdG1} really contains only first order derivatives. In that case, in the same way as the Dirac equation leads to the Klein--Gordon equation for an electron,\footnote{The converse is of course not true. The solutions of the Dirac equation are a subset of those of the Klein--Gordon equation.} $\hpd$ can be eliminated from Eq.~\eqref{eq:BdG1}. One then obtains a hyperbolic differential equation, that can be cast in the typical form of Eq.~\eqref{eq:KGfluid} for a scalar field in a curved geometry [see Eq.~\eqref{eq:KGhp}], with a fluid metric $g_{\mu\nu}$ as in Eq.~\eqref{eq:fluidmetric}, where $c$ is the propagation speed of phonons and $\bv$ is the velocity of the condensate.

When the wavelength of the phonons becomes comparable with the healing length, $T_\rho$ cannot be neglected anymore and Eq~\eqref{eq:BdG1} yields an exact quartic supersonic dispersion relation (see Fig.~\ref{fig:dispersion}):
\begin{equation}\label{eq:dispersion1}
 (\omega-vk)^2=\Omega^2(k)=c^2 k^2\left(1+\frac{k^2\xi^2}{2}\right)=c^2k^2+\frac{c^4 k^4}{\Lambda^2},
\end{equation}
where $\Lambda=2m c^2/\hbar$ is the ``healing frequency'' of the condensate.
The discussion of Sec.~\ref{subsec:UVsuperluminal} is therefore exactly valid for such a system. The simplicity of this dispersion relation makes BECs particularly suitable for the analysis of the robustness of particle creation phenomena against ultraviolet Lorentz breaking and for a clean comparison of theoretical predictions with experiments.

\subsection{Numerical results and experiments}	%
\label{subsec:becexp}				%

Since, as we saw in the previous section, phonons in condensates can be described by a modified Klein--Gordon equation in a curved spacetime, it is immediate to see that phenomena analogous to Hawking radiation or cosmological particle production must show up, once the appropriate flows (horizon, variation in time of $v$ or $c$) are set up.

In the following chapters the spectral properties of Hawking-like emission are deeply investigated from a theoretical standpoint. For the sake of completeness and as a convincing argument in favor of the possibility of experimentally observing this phonon emission in a real experiment, we present here the numerical results of~\cite{carusotto2}, where the authors claims to have numerically ``observed'' the set up of a stationary emission with the same features as Hawking radiation, in a one-dimensional condensate where an acoustic horizon is formed at a finite time.

\subsubsection{Hawking partners and correlations}	%
\label{subsec:correlations_carusotto}			%

The simple but clever idea at the basis of~\cite{carusotto2} is presented in~\cite{carusotto1}. Starting from the consideration that the faintness of Hawking radiation would probably be hidden by noise ({\it in primis}\/ thermal noise), they looked for other features that could be a strong signature of Hawking radiation. In fact, because of the thermal nature of the spectrum, if one just measured the flux of phonons, the signal would be likely confused with other spurious effects. The way out is noting that Hawking radiation has a very peculiar feature, being originated by a particle pair production mechanism. For each Hawking particle emitted outside the horizon there must be a partner propagating inside the horizon. While in a real black hole one cannot access the region inside the horizon, in an analogue model this is possible and one can detect both the Hawking particle and its partner. Thus, a strong signature of Hawking radiation would be the detection of strongly correlated signals on both sides of the horizon. Since in BECs one can easily measure the density of the fluid, the presence of Hawking radiation can be inferred by the presence of characteristic correlations in the density profile. In the particular case in which an acoustic black hole is build with a constant-velocity one-dimensional flow by varying the speed of sound $c$, it was shown in~\cite{carusotto1} that the expected correlation function should be
\begin{equation}\label{eq:correlationsCarusotto}
 G^{(2)}(x,x')\simeq 
 -\frac{\rho\,\kappa^2\,\sqrt{\xi_{\rm sub}\xi_{\rm sup}}}{16\pi(c_{\rm sub}-v)(v-c_{\rm sup})}\,
\cosh^{-2}\!\left[\frac{\kappa}{2}\left(\frac{x}{c_{\rm sub}-v}+\frac{x'}{v-c_{\rm sup}}\right)
\right],
\end{equation}
where, as usual, $\kappa$ is the surface gravity, $v$ is the velocity of the fluid, $c$ the speed of sound, and $\xi$ the healing length. Subscripts sup and sup denote whether the subscripted quantity is measured in the supersonic or in the subsonic region. This work has also stimulated further investigations on quantum correlation across a black horizon in presence of modified dispersion relation~\cite{unruhcorrelations,From2010} in a more general framework. Moreover, this technique has been recently applied to an expanding BEC simulating cosmological metrics~\cite{angus}.

\begin{figure}
 \includegraphics[width=0.5\textwidth]{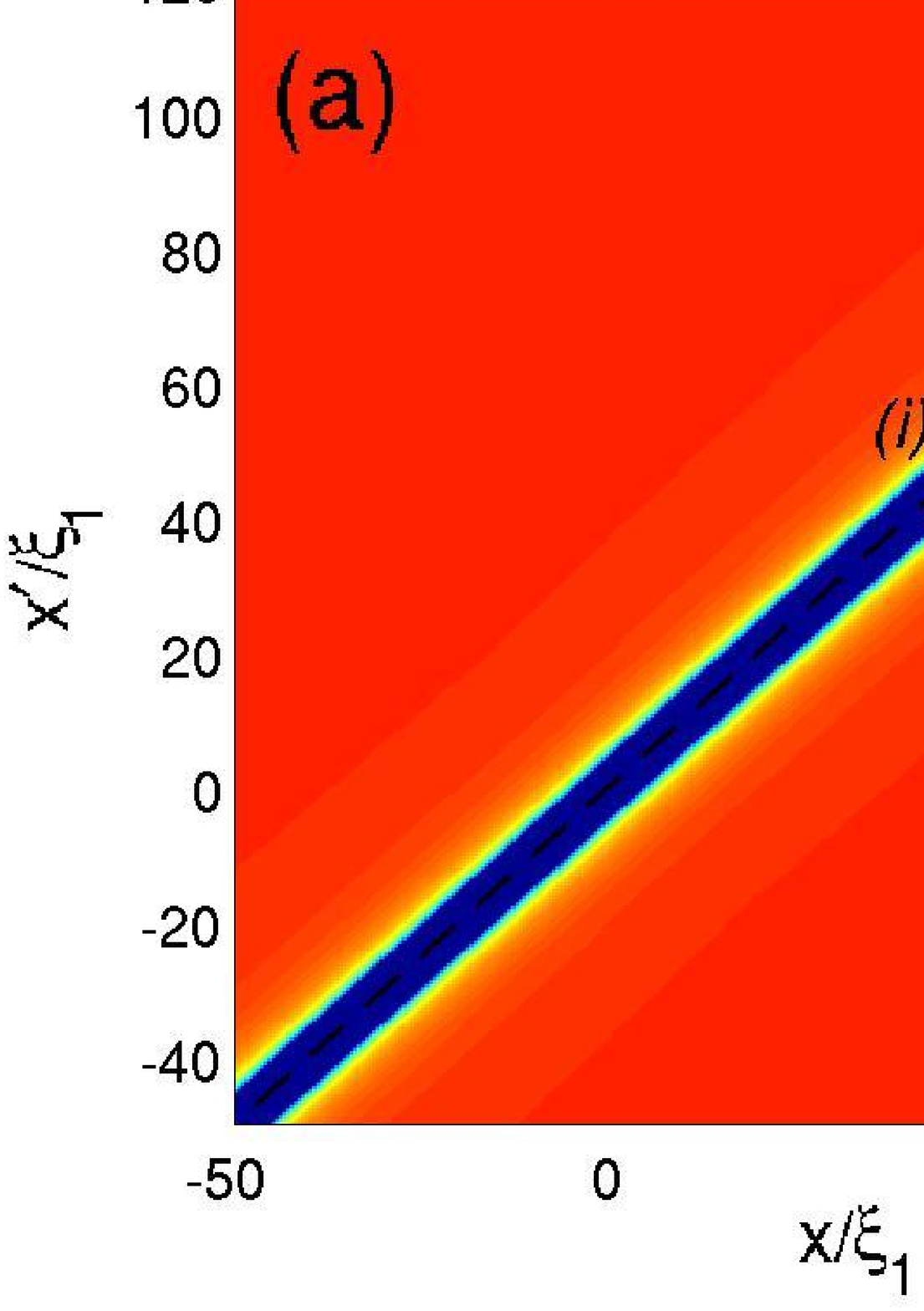}
 \includegraphics[width=0.5\textwidth]{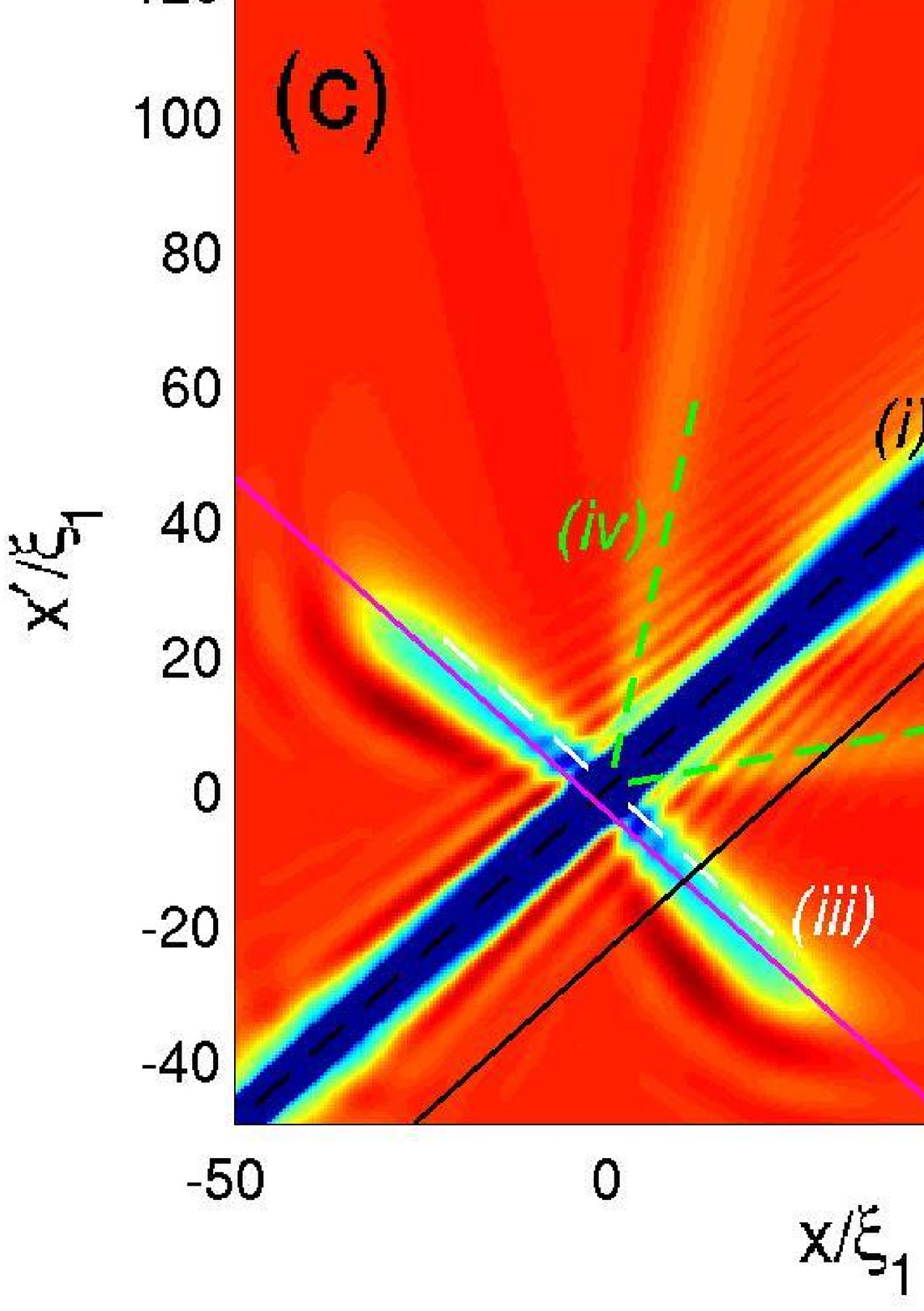}
 \caption{Rescaled density correlation $\rho\xi_{\rm sub}[G^{(2)}(x,x')-1]]$ as a function of $x$ and $x'$ at the initial time and after the formation of the horizon (From~\cite{carusotto2}). The feature labelled by (iii) is the signature of Hawking radiation.}
 \label{fig:correlationscarusotto}
\end{figure}

The above prediction was numerically checked in~\cite{carusotto2} through a numerical simulation in which the fluid was kept at constant velocity $v$ and the speed of sound $c$ was progressively changed until an acoustic horizon was formed. In Fig.~\ref{fig:correlationscarusotto} the rescaled density correlation $\rho\xi_{\rm sub}[G^{(2)}(x,x')-1]$ is plotted as a function of $x$ and $x'$ before (left panel) and after the formation of the horizon (right panel).
When the horizon is not formed, the only present feature (i) is a very strong anticorrelation line for $x=x'$, related to the repulsive interaction of the atoms. When the velocity of sound $c$ is changed, until it becomes smaller than the velocity of the fluid for $x>0$ (the interior of the black hole), some transient features related to dynamical Casimir effect show up~\cite{carusottocasimir}. In addition, two fringes [labelled by (iii) and (iv)] appear when the horizon is formed and they stabilize when no more change in the speed of sound is operated. The former one (iii) indicates that there are correlations between pairs of points located on opposite sides of the horizon. Furthermore, the slope of this correlation line is given by the ratio between the speed of propagation of the two signals. It was checked~\cite{carusotto2} that this slope is compatible with the speed of propagation of the Hawking particle (positive solution of the dispersion relation in Fig.~\ref{fig:dispersion}, left panel) and its partner (negative solution named $k^{(2)}$ in the caption of Fig.~\ref{fig:dispersion}). The latter feature, labelled by (iv), is a weaker emission of pairs, where both particles are emitted in the supersonic region ($x>0$). They correspond to $k^{(2)}$ and to the negative solution with smallest absolute value of Fig.~\ref{fig:dispersion}, right panel, on the ``left'' branch of the dispersion relation. Furthermore, there is a third feature in between (iii) and (iv) that the authors failed to observe in~\cite{carusotto2} but that was reported in~\cite{Carusottowhite}, representing the correlation between outgoing particles in the subsonic region (the same positive solution of Fig.~\ref{fig:dispersion}, left panel, which also corresponds to Hawking particles) and outgoing particles in the supersonic region corresponding to the less negative solution of the dispersion relation on the ``left'' branch.
When the speed of sound is varied but without reaching the formation of the horizon, the transient dynamical Casimir effect remains, but the three features related to pair production disappear.

Furthermore, a scenario in which the condensate is not at zero temperature was also considered in~\cite{carusotto2}. Interestingly, the signature of Hawking radiation in the correlation pattern is not hidden by the presence of a thermal bath, since thermal phonons are completely uncorrelated, but the signal is even reinforced because thermal phonons stimulate the emission of Hawking particles.

The results that we have just summarized show that the detection of Hawking radiation is indeed possible thanks to its correlation pattern, once a suitable flow with a horizon is sustained for a long enough time. Moreover, the first BEC flow with horizons was realized in 2009 at Technion~\cite{technion} but no measurement of correlations has been made yet. However, we will not probably have to wait for a long time before having the first observations of acoustic Hawking particle creation.

\subsubsection{The first analogue black-hole in a BEC}	%
\label{subsec:results_bec}				%

From an experimental point of view, according to~\cite{technion}, the main challenge is the realization of a supersonic flow. In fact, when a fluid becomes faster than sound, it can emit phonons due to Landau instability. However, by momentum conservation, this can happen only if there is some impurity, providing the required momentum in the opposite direction with respect to the fluid. The authors of~\cite{technion} managed to realize an experimental apparatus which does not supply much momentum in this direction, allowing supersonic flow during the time scale of the experiment ($20~\mbox{ms}$). At the same time, this free flow, required to prevent the Landau instability, also prevents the formation of quantized vortexes that usually form at flow speeds much lower than the speed of sound.

\begin{figure}
 \centering
 \includegraphics[width=.65\textwidth]{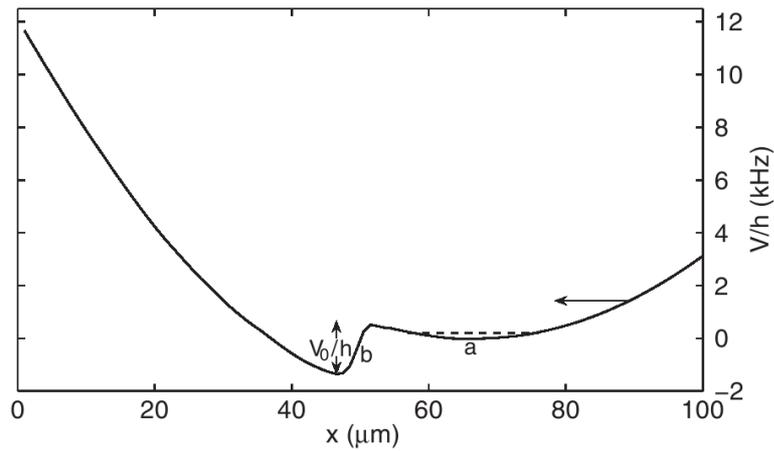}
 \caption{Potential used in the Technion experiment at a time before the minimum of the harmonic potential (a) has reached the steplike potential (b) (From~\cite{technion}). The harmonic potential moves leftward and the dashed line represents the chemical potential of the condensate. When the condensate crosses the steplike potential its density decreases and its velocity increases.}
 \label{fig:techpotential}
\end{figure}

In the experiment the black horizon is obtained by changing the velocity of the fluid (and, as a consequence, also its density and the local speed of sound) with a harmonic potential, created with a magnetic trap, and a steplike potential, created with a suitable laser beam (see Fig.~\ref{fig:techpotential}). The harmonic potential is accelerated so that the condensate passes over the steplike potential. After the step the density of the condensate is smaller so that, by continuity, its velocity increases.
This is a quite different way of obtaining a flow with a supersonic and a subsonic region separated by an acoustic horizon with respect to that proposed in~\cite{carusotto1} or in~\cite{expandinguniverse_blv,Barcelo:2003wu}, for the simulation of cosmological metrics. In those papers, the acoustic geometry is changed not by playing with the velocity of the fluid but by changing the scattering length [see Eq.~\eqref{eq:scatteringlength}] via a Feshbach resonance, by tuning an external magnetic field.

Finally, we remark that, after the condensate has sped up at a velocity larger than the speed of sound, it eventually slows down, becoming again subsonic. The point at which the velocity of the fluid crosses the speed of sound for the second time is the acoustic analogue of a white horizon. Even if the white horizon is made as smoother as possible, it seems very hard to get rid of it.
Unfortunately, it is well known that white holes are unstable at both classical~\cite{eardley} and quantum level~\cite{zns_zh,zns_jetp,wald_white}. However, even if white holes have been claimed to be unstable also in BECs~\cite{whiteholeinstability,ulf}, more recent results show that they should be stabilized by the modified dispersion relation~\cite{carlos_stability,MacherBEC}. In the Technion experiment, in fact, no white hole instability has been observed.

Furthermore, even if the white hole were stable, another instability (black hole laser~\cite{cj}) would arise in a system coupling a white and a black horizon in the same way as this experiment did. We will diffusely discuss about this effect in Chap.~\ref{chap:bhlaser}. As we shall show, this instability can be interpreted as an exponential amplification of Hawking radiation due to the presence of two horizons acting as a resonant cavity. In fact, we show that for a short time after the creation of the horizon, the radiation emitted by a white-hole--black-hole system is indistinguishable from standard Hawking radiation. Then, at later times, the instability shows up (see Fig.~\ref{fig:fluxes}). As shown in Sec.~\ref{sec:technion}, the duration of the Technion experiment was too short (probably by an order of magnitude) for this instability to grow.

\section{Refractive index perturbations}	%
\label{sec:slowlight}				%

The last model that we briefly consider is very different from the previous two, since the underlying physical system is not a fluid but a transparent medium, where one can manage to control the speed of light by varying the refractive index~\cite{waveguide_leonhardt,waveguide_leonhardt_app}. By using this technique, it has recently been reported the detection of phonons in an experimental set up with a black-hole--white-hole configuration~\cite{fiber_exp}. However, the interpretation of the observed emission as Hawking radiation is still controversial~\cite{fiber_comment,fiber_reply}.

The idea at the basis of this model~\cite{waveguide_leonhardt} is that one can control the refractive index $n$ in a dielectric medium by sending a laser pulse with large intensity $I$. In this way a refractive index perturbation (RIP) $\delta n=n_2\,I$ propagates in the dielectric medium with a speed of $v<c$, where $n_2$ is called nonlinear Kerr index. As a consequence, inside the pulse, the speed of light decreases because of the increasing of the refractive index.

If one sends after the RIP a probe pulse of different frequency with larger group velocity and propagating in the same direction, this pulse will get closer and closer to the RIP but it will not be able to overtake it. In fact, when the probe comes closer to the RIP, it feels the higher dispersive index and it is consequently slowed down, until a certain point, where it travels at the same velocity of the RIP.
Going to the RIP reference frame, the probe pulse initially travels towards the RIP, then it slows down till a point where its velocity vanishes. This points acts as a white horizon preventing light from travelling inside the RIP. Analogously, there is a point at the front of the pulse that behaves as a black horizon, not allowing light to escape from the inside of the RIP. In~\cite{waveguide_leonhardt} it is shown that the Hawking temperature corresponding to such a horizon can be of the order of $10^3\,\mbox{K}$, making this signal easier to be detected with respect to the expected signals in other analogue models.

The implementation of such an idea has recently been realized in a transparent dielectric medium (fused silica glass) using ultrashort laser pulse filaments to create a traveling RIP. Since the refractive index $n$ depends on frequency, a region where the velocity of light is smaller than the group velocity of the RIP $v$ exists only for a window of frequencies. In practice, only those frequencies $\om_1<\om<\om_2$ for which 
\begin{equation}\label{eq:phasehor}
 \frac{1}{n_0(\om)+\delta n}<\frac{v}{c}<\frac{1}{n_0(\om)}
\end{equation}
can feel the presence of the two horizons (see~\cite{fiber_theo_short,fiber_theo_long} for a theoretical analysis of this system).

The main challenge of this experiment is the suppression or elimination of spurious effects. \v{C}erenkov emission, Rayleigh scattering and some other effects are minimized by looking in a direction perpendicular to the propagation of the RIP. The main problem according to the authors is fluorescence. Because of that, in the experiment, the window of frequencies that feels the horizon was chosen in a range between $800$ and $900\,\mbox{nm}$, where there is no fluorescence emission.

As a final result, they actually observed some radiation whose spectrum is cut off exactly at $\om_1$ and $\om_2$.
However, whether this is Hawking radiation or not is an open debate~\cite{fiber_comment,fiber_reply}. As pointed out by~\cite{fiber_comment}, the first problematic issue is the definition of Hawking radiation itself. If one requires the thermality of the spectrum as a fundamental condition, then dielectric systems with frequency dependent refractive index are not suitable at all, because the spectrum is cut off and strongly modified close to $\om_1$ and $\om_2$, given that only a narrow window of frequencies can see the horizons. Furthermore, also the link between the surface gravity and the Hawking temperature is broken, just because the temperature cannot be defined.
Moreover, there are at least other two issues that are not very clear. First, Hawking radiation should not be emitted parallely to the horizon but in the orthogonal direction (that is in the direction of RIP propagation). Second, and even more importantly, there are no frequencies for which the group velocity of light equals the velocity of the RIP. This means that properly speaking there is no horizon, neither in a narrow window of frequencies. In fact, wave packets centered around an arbitrary frequency can cross RIP. This is because Eq.~\eqref{eq:phasehor} does not really define a horizon but just what is called a ``phase velocity horizon''.

Even if there is almost no doubt that the observed radiation is somehow connected with mixing of modes with negative and positive norm, it seems unlikely that the origin of this emission is what is usually called Hawking radiation. Note, however, that the presence of a horizon is not a strictly fundamental requirement for particle creation in stationary geometries in the presence of modified dispersion relations. In Sec.~\ref{sec:asymmetric} and in Chap.~\ref{chap:warpdriveBEC}, we in fact show that particles can be spontaneously created in supersonic BEC flows without horizons. This interesting phenomenon is now under deeper investigation~\cite{4x4}. A second possibility is that the observed emission be related to a sort of dynamical Casimir effect, as numerically observed in~\cite{carusotto2,carusottocasimir} even when the horizon is not formed.

\cleardoublepage
\part{A test field for QFT in CS: Bose--Einstein~condensates}
\label{part:BECphenomenology}

\chapter{Hawking radiation in~Bose--Einstein~condensates}			
\label{chap:hawking}								
\chaptermark{Hawking radiation in BECs}						

As broadly discussed in the first part of this Thesis, in the hydrodynamic approximation, \ie\/ for long wavelengths, the propagation of sound waves in a moving fluid that crosses the speed of sound is analogous to that of light in a black hole metric~\cite{unruh}.
However, we also stressed that Hawking radiation relies on short wavelength modes~\cite{TJ91,93,Primer}.
Therefore the dispersion of sound waves must be taken into account when computing the spectrum emitted by an acoustic black hole.
To this end, Unruh wrote a dispersive wave equation in a supersonic flow~\cite{Unruh95}.
Through a numerical analysis, he then showed that the spectrum was robust, provided the dispersive scale $\xi$, the ``healing length'', is much smaller than the surface gravity scale $1/\kappa$ which fixes the Hawking temperature $\Th = \kappa/2\pi$ (in units where $\hbar = k_{\rm B} = c = 1$).

Although this insensitivity was then confirmed by analytical~\cite{BMPS95,Corley97,Tanaka99,UnruhSchu05} and numerical~\cite{CJ96,UnruhTrieste,MacherRP1,MacherBEC} methods, {\it deviations} with respect to the standard flux are much harder to characterize, and so far there is no consensus on what are the relevant parameters determining those deviations.

To address this question, in this chapter we consider what happens in Bose--Einstein condensates (BECs) with velocity profiles characterized by {several} scales, and numerically compute the spectrum by varying them separately (see Secs.~\ref{sec:asymmetric} and~\ref{sec:broadhor}).
We show that, whenever $\omm/\Th \gtrsim 10$, where $\omm$ is a critical frequency~\cite{MacherRP1,MacherBEC} (see Sec.~\ref{subsec:UVsuperluminal} for its physical interpretation) that scales with $1/\xi$ but also depends on the asymptotic flow velocity, the flux remains Planckian to high accuracy, {\it even} when the temperature completely differs from the Hawking temperature $\Th$. In fact, we show that the temperature is determined by the average of the velocity gradient across the horizon over a critical length fixed by $\xi$ and $\kappa$.
In~\cite{cfp}, we analytically explain the origin of that length.
We checked that these properties hold for large classes of flows.
They cease to be valid only if the flow induces nonadiabatic effects that interfere with the Hawking effect, thereby introducing oscillations in the spectrum.
This is illustrated by considering {\it undulations} similar to those appearing in {white} holes' flows~\cite{Carusottowhite,Silke_exp}.

\sectionmark{Flows with single horizon}			%
\section{One-dimensional flows with single horizon}	%
\label{sec:1dflow}					%
\sectionmark{Flows with single horizon}			%

\subsection{The metric}	%

In this chapter we consider elongated condensates that are stationary flowing along the longitudinal direction $x$. We also assume that the transverse dimensions are small enough that relevant phonon excitations are longitudinal. In that case, one effectively deals with a 1+1 dimensional field (see discussion in Sec.~\ref{sec:generalmetric} about acoustic geometries in $1+1$ dimensions) propagating in a curved spacetime described by a metric of the form
\begin{equation}\label{eq:metric}
 \dd s^2=-c^2 \dd t^2+(\dd x-v \, \dd t)^2,
\end{equation}
where, as usual, $v$ is the flow velocity and $c$ the speed of sound.

For stationary flows, $c$ and $v$ only depend on $x$.
Assuming that the condensate flows from right to left ($v<0$), a sonic horizon is present where $w = c+v$ crosses $0$.
It corresponds to a black hole horizon when $\partial_x w > 0$, and a white one when $\partial_x w < 0$.
We shall call it a Killing horizon since the norm of the Killing field $\partial_t$ vanishes when $w=0$~\cite{livrev}.
Its location is taken at $x=0$. 

When ignoring short distance dispersion, the spectrum of the upstream phonons spontaneously emitted from a black hole horizon is simple,
and strictly corresponds to the Hawking radiation~\cite{unruh,Unruh95}.
It follows a Planck law and the temperature is fixed by the Killing surface gravity~\eqref{eq:kappa}
\begin{equation}\label{eq:temprelnew}
\kk \equiv \partial_x (c+v) \vert_{x = 0}.
\end{equation}
Note that, for relativistic fields, $\kk$ gives the late time decay rate of the frequency of outgoing modes, as seen by asymptotic observers, see Eq.~(32.8) in~\cite{MisnerThorneWheeler} for classical waves, and Eq.~(3.44) in~\cite{Primer} for Hawking radiation. As discussed in Sec.~\ref{subsec:hawking}, $\kk$ fixes the temperature to be exactly $\Th= \kk/ 2 \pi $, in units where $\hbar=k_B=1$, but
independently of $\chor$, the value of the sound speed at the horizon.

When including dispersion, these results are no longer true because the total amount of redshift saturates. Nevertheless, the $\kk$-decay law is recovered near the horizon where $w = c + v$ is linear in $x$. Moreover, when the healing length is much smaller than $\chor/\kk$, this guarantees that the spectrum and the correlations of the Hawking pairs are not significantly affected by dispersion~\cite{BMPS95,Rivista,From2010}.

We begin with a very simple velocity profile describing a geometry with a single black hole horizon by fixing
\begin{equation}\label{eq:velocity_simp}
 c(x)+v(x)= w(x) = \chor\,D\, \sign(x) \,
 \tanh^{1/n}\!\!\left[\left(\frac{\kappa|x|}{\chor D}\right)^{\!\!n}\right].
\end{equation}
Note that, for this velocity profile, the surface gravity is $\kk=\kappa$.
The quantity $D$ fixes the asymptotic value of $w$, whereas $n$ governs the smoothness of the transition from the region where $w$ is linear in $x$ to the asymptotic regime where it is constant. We shall work with $n =1$ or $2$ to have a well-defined range $\sim \chor D/\kappa$ where $w$ is linear, and yet avoid the nonadiabatic effects~\cite{CJ96,MacherRP1} when the transition is too sharp, \ie, when $n \geq 4$.
The metric~\eqref{eq:metric} is then completely fixed by introducing $q$,
\begin{equation}\label{eq:cv}
\begin{aligned}
 c(x)&=\chor+(1-q) w(x),\\
 v(x)&=-\chor+q \, w(x),
\end{aligned}
\end{equation}
which specifies how $c+v$ is shared between $c$ and $v$. 
In the simulations, when not explicitly specified, we work with $q=1/2$, because it minimizes the scattering between left and right moving phonons which is not related to the Hawking effect (see Fig. 11 in~\cite{MacherBEC}).
In fact, when $q=1/2$, the decoupling between right- and left-going waves is exact in the hydrodynamical limit when the quartic term in the dispersion relation~\eqref{eq:dispersion1} is negligible, that is for wavelength much longer than the condensate healing length. In this situation, in a WKB analysis when $k$ depends on $x$, the $x$-dependent generalization of Eq.~\eqref{eq:dispersion1} factorizes
\begin{equation}
 \left[\om-\left(v(x)+c(x)\right)k(x)\right]\left[\om-\left(v(x)-c(x)\right)k(x)\right]=0.
\end{equation}
Leftgoing waves, labelled in our notation by $v$, are described by the second branch
\begin{equation}
 k^v(x)=\frac{\om}{v(x)-c(x)},
\end{equation}
which, for $q=1/2$, reduces to
\begin{equation}
 k^v(x)=-\frac{\om}{2\chor}.
\end{equation}
This shows that $k^v$ does not depend on $x$, therefore being completely decoupled from the geometry. In this situation, left-going waves propagate as they were living in a flat geometry.

Hence working with $q=1/2$ will ease the identification of the other, more intrinsic, deviations with respect to the standard flux.
Similarly, to have well-defined asymptotic modes, we work with infinite condensates and with $c, \, v$ possessing well-defined values for $x \to \pm \infty$.

\subsection{Mode analysis and scattering matrix}	%
\label{subsec:bogo}					%

\begin{figure}
 \psfrag{om}[c][c]{$\Omega/\kappa$}
 \psfrag{k}{$\chor k/\kappa$}
 \includegraphics[width=0.95\textwidth]{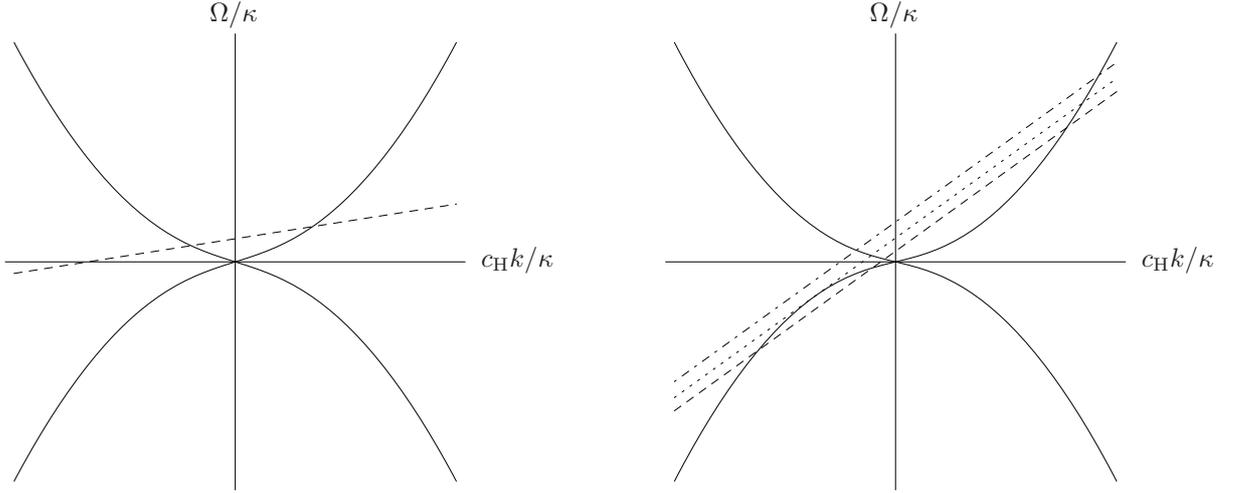}
 \caption{Graphical solution of the supersonic dispersion relation~\eqref{eq:dispersion1} of phonons in a BEC for subsonic flow (left panel) and supersonic flow (right panel). Solid lines: $\pm \Omega(k)/\kappa$. Left panel, dashed line: $(\om-vk)/\kappa$. Right panel, dashed, dotted, dotdashed lines: $\om-vk$ respectively for $\om<\ommax$, $\om=\ommax$, $\om>\ommax$. For $\om<\ommax$, one clearly sees that, in supersonic flows, two extra (real) roots exist in the left lower quadrant.
The most negative one is called, in the text, $k^{(1)}_{\om}$ and the other one $k^{(2)}_{\om}$, so that $k^{(1)}_{\om} < k^{(2)}_{\om} < 0$. For details on the model, refer to Sec.~\ref{subsec:theobec} and Appendix~\ref{app:BEC}. In the Hawking process, the positive $k$ solution in the subsonic region (left panel) is the Hawking particle, while the negative solution $k^{(2)}_{\om}$ in the supersonic region is the negative norm left-going partner.}
 \label{fig:dispersion_hawk}
\end{figure}

As discussed in Sec.~\ref{subsec:theobec} and broadly analyzed in Appendix~\ref{app:BEC}, perturbations in a BEC are described by a quantum field $\hp$ satisfying Eq.~\eqref{eq:BdG1}. In a stationary flow, this equation can be solved by Fourier analysis, expanding the field as
\begin{equation}\label{eq:phiexpansion1}
\hp(t,x) 
=\int_0^{+\infty}\!\dom\sum_{\alpha}\left[\phm{\om t}\pom\aom+\php{\om t}\vpoms\aomd\right],
\end{equation}
where $\alpha$ labels different solutions of the dispersion relation~\eqref{eq:dispersion1} (see Fig.~\ref{fig:dispersion_hawk} for its graphical solution), with the same conserved frequency $\omega$, and $\aom$ and $\aomd$ are standard destruction and creation operators, satisfying
\begin{equation}\label{eq:commaa}
 \comm{\aom}{\aomdp}=\delta_{\alpha\alpha'}\delta(\om-\omp).
\end{equation}
These rules follow from the commutation relation~\eqref{eq:commphi1} of $\hp$, when the modes are normalized through the appropriate non-positive-definite conserved scalar product (see Appendix~\ref{subsec:quantization})
\begin{equation}\label{eq:scalarphi1}
 \left\langle{\spin{\phi_1}{\varphi_1}}\right|\left.{\spin{\phi_2}{\varphi_2}}\right\rangle
 =\int\!\dtx\,\rnbx \left[\phi_1^{*}(t,\bx)\phi_2(t,\bx)-\varphi_1^{*}(t,\bx)\varphi_2(t,\bx)\right],
\end{equation}
in such a way that
\begin{equation}\label{eq:normalization}
 \left\langle{\spin{\phi_\om^\alpha}{\varphi_{\om}^{\alpha}}}\right|\left.{\spin{\phi_{\om'}^{\alpha'}}{\varphi_{\om'}^{\alpha'}}}\right\rangle=\delta_{\alpha\alpha'}\delta(\om-\omp).
\end{equation}
Note that the complex scalar field $\hp$ of Eq.~\eqref{eq:phiexpansion1} suitably describes quantized phonons, because phonons and anti-phonons must coincide. This property is rigorously derived in Appendix~\ref{subsec:quantization}, directly starting from the symmetries of Eq.~\eqref{eq:BdG1}.

Furthermore, in the WKB approximation, it is easy to show (see Appendix~\ref{subsec:mode1d}) that the modes $\pom$ and $\vpom$ can be written as
\begin{equation}\label{eq:WKBmodes}
\begin{aligned}
 \pom&=\sqrt{\frac{\partial k_\om^\alpha(x)}{\partial\om}}\frac{1}{\sqrt{1-D_{k_\om^\alpha}^2(x)}}\frac{\exp\left[\ii\int^x \dx' k_\om^\alpha(x')\right]}{\sqrt{2\pi\rn(x)}},\\
 \vpom&=\sqrt{\frac{\partial k_\om^\alpha(x)}{\partial\om}}\frac{D_{k_\om^\alpha}(x)}{\sqrt{1-D_{k_\om^\alpha}^2(x)}}\frac{\exp\left[\ii\int^x \dx' k_\om^\alpha(x')\right]}{\sqrt{2\pi\rn(x)}},
\end{aligned}
\end{equation}
where $k_\om^\alpha(x)$ are the solutions of the $x$-dependent dispersion relation
\begin{equation}\label{eq:dispersionWKB1}
 (\om-v(x)k(x))^2=\Omega^2(k(x))=c^2(x) k^2(x)+\frac{\hbar^2 k^4(x)}{4 m^2}, 
\end{equation}
and
\begin{equation}\label{eq:dkWKB}
 D_k(x)=\frac{1}{mc^2(x)}\left[\hbar\sqrt{c^2(x)k^2(x)+\frac{\hbar^2k^4(x)}{4m^2}} - \frac{\hbar^2k^2(x)}{2m} - mc^2(x)\right].
\end{equation}

In general, the {\it number} of modes $\alpha$ sharing the same frequency $\om$ depends on the asymptotic properties of $w = c + v$. This is important as it explains why, when there is dispersion, there is no flux above a critical frequency $\omm$ whose value is fixed by the asymptotic value $w$ (see Sec.~\ref{subsec:UVsuperluminal} for a qualitative discussion and Sec.~\ref{subsec:mode1d} for a rigorous analysis).

In homogeneous subsonic flows, $|v|< c$, for real $\omega$, two roots $k_\om$ of Eq.~\eqref{eq:dispersion1} are real, and describe the right and left moving modes, that we call $(\phi_\om^u,\varphi_\om^u)$ and $(\phi_\om^v,\varphi_\om^v)$, respectively.
The other two roots are complex, conjugated to each other, and describe modes that are asymptotically growing or decaying, say to left ({\it vice versa}\/ to right).
Only the first two correspond to perturbations that must be quantized, as the last two are not asymptotically bounded.
In homogeneous supersonic flows, the situation is quite different. For real $\omega$, there exists a critical frequency $\omm$ below which the four roots are real (see Fig.~\ref{fig:dispersion_hawk}) and above which two are real and two complex, as in subsonic flows.
The spectrum of asymptotically bounded modes (ABM) thus contains only two modes for $\om > \omm$ and four modes for $0 < \om < \omm$: the two standard right-going and left-going modes $\phi_\om^u$ and $\phi_\om^v$, plus two new propagating modes $(\phi_\om^{(1)},\varphi_\om^{(1)})$ and $(\phi_\om^{(2)},\varphi_\om^{(2)})$, with wavevectors $k_\om^{(1)}$ and $k_\om^{(2)}$, respectively.

Differently from the standard modes described by $(\phi_\om^u,\varphi_\om^u)$ and $(\phi_\om^v,\varphi_\om^v)$, those two new modes do not satisfy the normalization condition of Eq.~\eqref{eq:normalization}, because their norms are negative. This implies that also the corresponding destruction and creation operators $\hat a^{(i)}_\om$ and $\hat a^{(i)\dagger}_\om$ do not satisfy Eq.~\eqref{eq:commaa}, so that they are not suitable for entering the field expansion of Eq.~\eqref{eq:phiexpansion1}.
This problem is strictly related to the fact that the two solutions $k_\om^{(1)}$ and $k_\om^{(2)}$ of the dispersion relation have negative frequency $\Omega(k_\om^{(i)})$ in the frame comoving with the fluid (in Fig.~\ref{fig:dispersion_hawk}, right panel, they are in fact located in the left lower quadrant).
To go around this impasse, one can solve the mode equation for negative values $-\om$ of the Killing frequency (we define $\om$ to be always positive). In so doing, the two solutions $k_{-\om}^{(i)}=-k_{\om}^{(i)}$ are now located in the right upper quadrant, $\Omega(k_\om^{(i)})$ is positive, the norm of the corresponding modes $(\phi_{-\om}^{(i)},\varphi_{-\om}^{(i)})$ is also positive, and the destruction and creation operators $\hat a^{(i)}_{-\om}$ and $\hat a^{(i)\,\dagger}_{-\om}$ eventually satisfy the proper commutation relations~\eqref{eq:commaa}. Furthermore, because the modes $\phi_{-\om}^{(i)}$ have negative frequency, while $(\varphi_{-\om}^{(i)})^*$ have positive frequency, in a supersonic flow the field must be expanded as
\begin{multline}\label{eq:expsup1}
\hp_{\rm sup}=\int_0^{\infty}\!\dom\left\{ \ee^{-\ii\om t} \left[ \phi_\om^u \ha_\om^u+\phi_\om^v\ha_\om^v
+ \theta(\ommax - \om)
\sum_{i=1,2}(\varphi_{-\om}^{(i)})^*\,\ha_{-\om}^{{(i)}\dagger}\right] \right.
 \\
\left. + \ee^{+ \ii\om t} \left[ (\varphi_\om^u)^*\,\ha_\om^{u\, \dagger}+(\varphi_\om^v)^*\,\ha_\om^{v\, \dagger}
+ \theta(\ommax - \om)
\sum_{i=1,2}\phi_{-\om}^{(i)}\ha_{-\om}^{{(i)}}\right] \right\},
\end{multline}
where the Heaviside function $\theta(\ommax - \om)$ has been inserted because the two extra roots exist only for $\om<\omm$. Considering only positive frequency modes in the above expansion, with a slightly abuse of terminology, from now on we shall refer to $\phi^u_\om$ and $\phi^v_\om$ as positive norm modes, since their corresponding doublets $(\phi_\om^u,\varphi_\om^u)$ and $(\phi_\om^v,\varphi_\om^v)$ have positive norm. Analogously, we shall refer to $(\varphi_{-\om}^{(i)})^*$ as negative norm modes. In fact, the doublets $((\varphi_{-\om}^{(i)})^*,(\phi_{-\om}^{(i)})^*)$ have negative norm because, by construction, the doublets $(\phi_{-\om}^{(i)},\varphi_{-\om}^{(i)})$ have positive norm.

\begin{figure}
\psfrag{tp}[c][c]{Turning point}
\psfrag{lone}[c][c]{$\varphi_{-\om}^{*}$}
\psfrag{ltwo}[c][c]{$\varphi_{-\om}^{*}$}
\psfrag{rone}[c][c]{$\varphi_{-\om}^{*}$}
\psfrag{rtwo}[c][c]{$\varphi_{-\om}^{*}$}
\psfrag{ru}[c][c]{$\phi_\om^{u}$}
\psfrag{lu}[c][c]{$\phi_\om^{u}$}
\psfrag{rv}[c][c]{$\phi_\om^{v}$}
\psfrag{lv}[c][c]{$\phi_\om^{v}$}
 \includegraphics[width=0.48\textwidth]{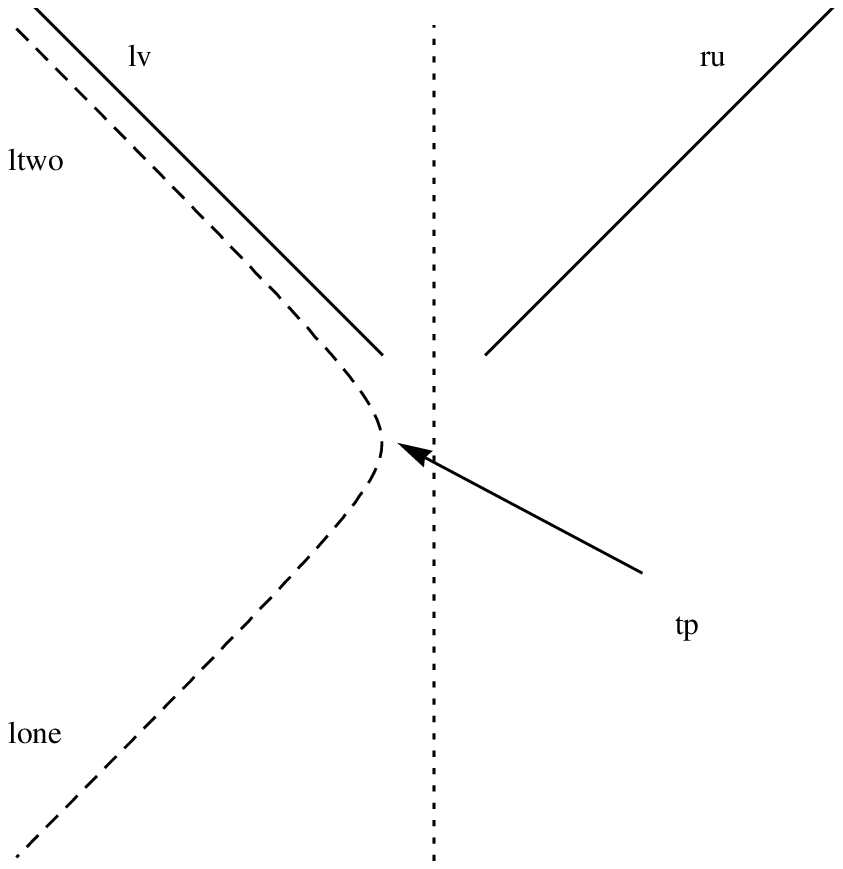}
 \includegraphics[width=0.48\textwidth]{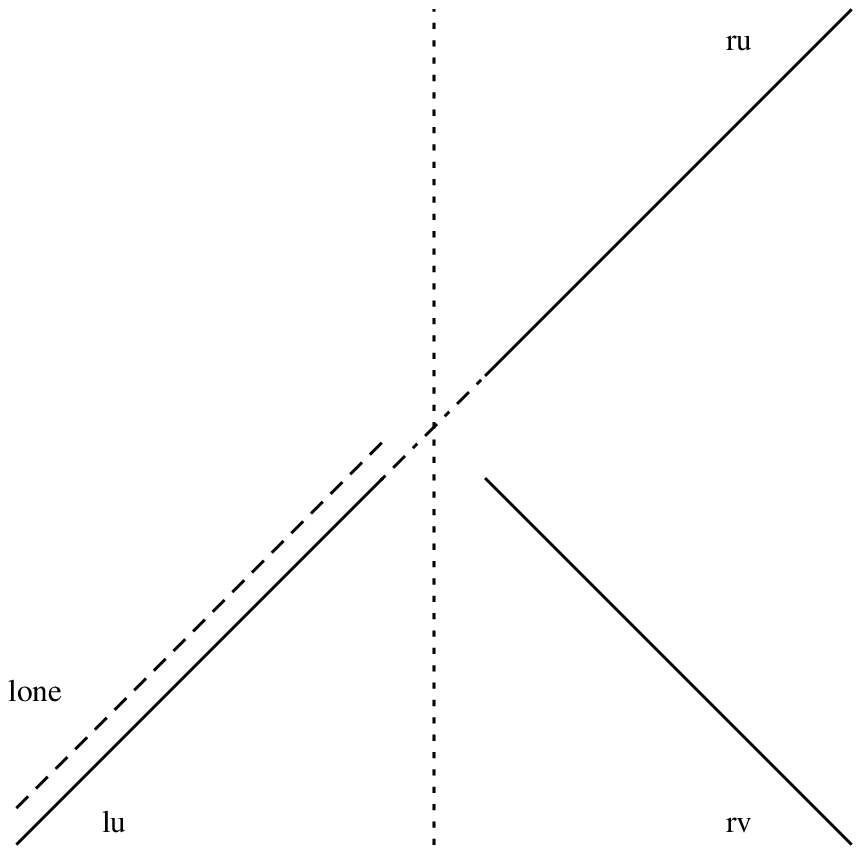}
 \caption{Left panel: graphical representation of the in mode $(\varphi_{-\om}^{\rm in})^*$ in a plane ($x$, $t$). It has only one incoming right-going branch $\varphi_{-\om}^{*}$ in the supersonic region on the left of the horizon, while it has 3 outgoing branches: two left-going ($\phi_\om^{v}$ ,$\varphi_{-\om}^{*}$) in the supersonic region and one right-going ($\phi_\om^{u}$) in the subsonic region. This modes arrives from $x\to-\infty$ with wavevector $k^{(1)}_\om$ and, at the turning point, it is partially reflected back into a mode with wavevector $k^{(2)}_\om$ and partially converted in the other out modes. Right panel: out mode $\phi_\om^{u,\rm out}$. It has one outgoing branch $\phi_\om^{u}$ in the subsonic region and 3 ingoing branches: one left-going ($\phi_\om^{v}$) in the subsonic region and two right-going ($\phi_\om^{u}$, $\varphi_{-\om}^{*}$) in the supersonic region. Positive (negative) norm modes are represented by solid (dashed) lines. The dotted vertical line is the horizon.}
 \label{fig:modes}
\end{figure}

In {\it inhomogeneous} flows that cross once the speed of sound and with two asymptotic flat regions, \ie, as those of Eq.~\eqref{eq:velocity_simp}, one gets a new situation: when $\om>\omm$, there are still only two modes, whereas for $\om<\omm$, there are now three independent globally defined ABM~\cite{MacherRP1}.
There are only three independent modes because the semiclassical trajectories associated with the two extra roots ($k^{(1)}_\om$ and $k^{(2)}_\om$, see Fig.~\ref{fig:dispersion_hawk}) necessarily possess a turning point for $x_{\rm tp} < 0$, in the supersonic region, since they cannot propagate in the subsonic region $x>0$. Hence these two roots describe the initial and final value of the momentum of the same mode (see Fig.~\ref{fig:modes}, left panel, dashed line).

The location of the turning point $x_{\rm tp}$ depends on $\om$ and its value is easily obtained by solving $v_{\rm gr} = 0$ [see Eq.~\eqref{eq:gr_vel}].
Using Eqs.~\eqref{eq:temprelnew} and~\eqref{eq:dispersion1}, for $\om \ll \Lambda$, one finds~\cite{CJ96}
\begin{equation}
\label{eq:tp}
-\frac{\kk  x_{\rm tp}(\om)}{\chor} \simeq \left(\frac{\om}{\Lambda}\right)^{2/3}.
\end{equation}
For $\om \to 0$ fixed $\Lambda$, or $\Lambda \to \infty$ fixed $\om$,
the turning point coincides with the Killing horizon at $x=0$, as expected
from the behavior of the semiclassical trajectories in the absence of dispersion.\footnote{When considering subsonic dispersion~\eqref{eq:subCJ} by flipping the sign of the $k^4$ term in Eq.~\eqref{eq:dispersion1},
one verifies that $x_{\rm tp}(\om)$ flips sign but keeps the same norm.}
When $\om \to \omm$, the turning point $x_{\rm tp}(\om) \to -\infty$.
Hence, the threshold frequency $\omm$ is fixed by the {\it asymptotic} value of $w(x \to - \infty)$.
In our velocity flows of Eq.~\eqref{eq:velocity_simp}, one has
\begin{equation}\label{eq:omm}
\omm = \Lambda \, f(D).
\end{equation}

As discussed above, in backgrounds containing one sonic horizon, there are two or three ABM, depending on whether $\om$ is larger or smaller than $\omm$.
Thus, $\hp$ is can be expanded as [see also Eq.~\eqref{eq:expsup1}]
\begin{multline}\label{eq:expansion_bh}
 \hp_{\rm BH}=\int_0^{\infty}\!\dom\left\{ \ee^{-\ii\om t} \left[ \phi_\om^u \ha_\om^u+\phi_\om^v\ha_\om^v
+ \theta(\ommax - \om)
(\varphi_{-\om})^*\,\ha_{-\om}^{\dagger}\right] \right.
 \\
\left. + \ee^{+ \ii\om t} \left[ (\varphi_\om^u)^*\,\ha_\om^{u\, \dagger}+(\varphi_\om^v)^*\,\ha_\om^{v\, \dagger}
+ \theta(\ommax - \om)
\phi_{-\om}\ha_{-\om}\right] \right\}.
\end{multline}
As for homogeneous profiles, the modes $\phi_\om^u$ and $\phi_\om^v$ have positive norm and they are associated with the usual real roots of the dispersion relation~\eqref{eq:dispersionWKB1} (see Fig.~\ref{fig:dispersion_hawk}). They describe, respectively, propagating right-$u$ and left-$v$ moving waves with respect to the condensed atoms.
On the contrary, $(\varphi_{-\om})^*$ has negative norm. It is associated with the extra roots that exist in supersonic flows for $\om < \omm$, and describes phonons that are trapped inside the horizon ($x<0$)
in the supersonic region.
These modes describe the negative frequency partners of the outgoing Hawking phonons represented by $\phi_\om^u$ and $\ha_\om^u$.

In our infinite condensates, the above modes become superpositions of plane waves both in the left and right asymptotic regions.
We can thus construct without ambiguity the in- and out-mode bases.
The in modes are such that each of them contains only one asymptotic branch with group velocity directed towards the horizon.
The definition of out modes is analogous and based on the criterion that each out mode contains only one asymptotic branch with group velocity directed away from the horizon.
For instance, the right-going negative norm ingoing mode (Fig.~\ref{fig:modes}, left panel) is the wave $(\varphi_{-\om}^{\rm in})^*$, whose projection on the left-going mode $\phi_\om^{v}$ vanishes in the right asymptotic region and whose projection on the right-going mode $\phi_\om^u$ vanishes in the left asymptotic region. The right-going positive norm $u$ out mode (Fig.~\ref{fig:modes}, right panel) is the wave $\phi_\om^{u,{\rm out}}$ whose projections on the left-going modes $\phi_\om^v$ and $\varphi_{-\om}^*$ vanish in the left asymptotic region.

It is clear from the above discussion and Eq.~\eqref{eq:expsup1} that, for $\om>\omm$, there exist only two positive norm modes and, therefore, in- and out-mode bases are linearly related by a trivial (elastic) $2 \times 2$ transformation
\begin{equation}\label{eq:el_scatt}
\begin{aligned}
 \phi_\om^{u,{\rm in}} &= {\cal T}_\om \phi_\om^{u,\rm out} + {\cal R}_\om \phi_\om^{v,\rm out},\\
 \phi_\om^{v,{\rm in}} &= \tilde {\cal R}_\om \phi_\om^{u,\rm out} + \tilde {\cal T}_\om  \,\phi_\om^{v,\rm out},
\end{aligned}
\end{equation}
with the condition imposed by unitarity
\begin{align}
 |{\cal T}_\om|^2+|{\cal R}_\om|^2=1,\\
 {\cal R}_\om\tilde {\cal T}_\om^*+{\cal T}_\om \tilde {\cal R}_\om^*=0.
\end{align}
Instead, when  $\om < \omm$, the in and out bases contain three modes
which mix with each other in a nontrivial way by a $3 \times 3 $ Bogoliubov transformation~\cite{MacherBEC}
\begin{equation}\label{eq:bog_transf}
\begin{aligned}
 \phi_\om^{u,{\rm in}} &= \alpha_\om \phi_\om^{u,\rm out} + \beta_{-\om} \left(\varphi_{-\om}^{\rm out}\right)^*
 				+ \tilde A_\om \phi_\om^{v,\rm out},\\
  (\varphi_{-\om}^{{\rm in}})^* &= \beta_\om \phi_\om^{u,\rm out}+ \alpha_{-\om} \left(\varphi_{-\om}^{\rm out}\right)^*
 				+ \tilde B_\om\phi_\om^{v,\rm out},\\
 \phi_\om^{v,{\rm in}} &= A_\om \phi_\om^{u,\rm out} + B_{-\om} \left(\varphi_{-\om}^{\rm out}\right)^*
 				+ \alpha_\om^v \phi_\om^{v,\rm out},
\end{aligned}
\end{equation}
where we denote with
\begin{itemize}
 \item $\alpha$: the projections of the in modes on the corresponding out modes,
 \item $\beta$: the projections of the right-going positive norm modes $u$ on the negative norm mode (and {\it vice versa}),
 \item $A$: the projections of right-going positive norm mode $u$ on the left-going positive norm mode $v$ (and {\it vice versa}),
 \item $B$: the projections of left-going positive norm mode $v$ on the negative norm mode (and {\it vice versa}).
\end{itemize}
The standard normalization of the modes yields relations such as (from the first equation)
\begin{equation}
 |\alpha_\om|^2 - |\beta_{-\om}|^2 + |\tilde A_\om|^2 = 1,
\end{equation}
where the minus sign in front of the $\beta$ coefficient comes from the normalization of the negative norm mode $(\varphi_{-\om})^*$.%
\footnote{More correctly, we should say that the doublet $((\varphi_{-\om})^*,(\phi_{-\om})^*)$ has negative norm because, by construction, $(\phi_{-\om},\varphi_{-\om})$ has positive norm. See discussion on page~\pageref{eq:expsup1}.}

Given a quantum state $\ket{\Xi}$, which is an eigenstate of the three initial number operators
\begin{equation}
 \hat n_\om^{i,\rm in}\equiv\ha_\om^{i,{\rm in}\,\dagger}\ha_\om^{i,{\rm in}},
\end{equation}
the final occupation numbers
\begin{equation}
n_\om^{i,\rm out}\equiv \langle\Xi| \ha_\om^{i,{\rm out}\dagger}\ha_\om^{i,\rm out} |\Xi\rangle
\end{equation}
can be expressed in terms of initial ones
\begin{equation}
n_\om^{i,\rm in}\equiv \langle\Xi| \ha_\om^{i,{\rm in}\dagger}\ha_\om^{i,\rm in} |\Xi\rangle,
\end{equation}
by using the coefficients of the scattering matrix~\eqref{eq:bog_transf}:
\begin{equation}\label{eq:finaloccupationnumbers}
 \begin{aligned}
 &
  n_\om^{u,\rm out} = n_\om^{u,\rm in} + |A_\om|^2 ( n_\om^{v,\rm in} - n_\om^{u,\rm in} )
  			+|\beta_\om|^2 ( 1 + n_\om^{u,\rm in} + n_{-\om}^{\rm in} ),
  \\
  &
  n_\om^{v,\rm out} = n_{\om}^{v,\rm in} + |\tilde A_\om|^2 ( n_\om^{u,\rm in} - n_\om^{v,\rm in} )
  			+ |\tilde B_\om|^2 ( 1 + n_\om^{v,\rm in} + n_{-\om}^{\rm in} ),
  \\
  &
  n_{-\om}^{\rm out} = n_{-\om}^{u,\rm in}+ |\beta_{-\om}|^2 ( 1 + n_\om^{u,\rm in} + n_{-\om}^{u,\rm in} )
  			+ |B_{-\om}|^2 ( 1 + n_\om^{v,\rm in} + n_{-\om}^{\rm in} ).
 \end{aligned}
\end{equation}
The first term in every equation is the initial occupation number of the corresponding mode. In the first two equations, the second term is due to the elastic scattering between right- and left-going modes. The third term, which is due to the presence of negative norm modes, is the sum of spontaneous (represented by the 1 in the parenthesis) and stimulated emission involving the initial occupation number of the partner and the species itself. In the third equation the elastic scattering term is missing, but there are two spontaneous/stimulated emission terms related to the two pair-creation channels.

When the initial state is vacuum,the final occupation numbers are
\begin{gather}
 n_\om^{u,\rm out} 
= |\beta_\om|^2,\label{eq:occvacuum}\\
  n_\om^{v,{\rm out}} 
= |\tilde B_\om|^2,\label{eq:occvacuum1}\\
  n_{-\om}^{\rm out}
= |\beta_{- \om}|^2 + |B_{-\om}|^2=
   n_\om^{u,\rm out} +  n_\om^{v.\rm out}.\label{eq:occvacuum2}
\end{gather}
From these relations one sees that the norm of $\beta_\om$ fixes the occupation number of outgoing phonons spontaneously produced by the scattering of the quantum field near the black hole horizon. In the standard analysis without dispersion, $n_\om^{u,\rm out}$ is Planckian and describes the Hawking effect~\cite{Primer}.
In that case, for massless conformally invariant fields, $n_\om^{v,\rm out}$, the number of left moving phonons spontaneously produced, identically vanishes.
In the presence of dispersion, one generally finds $ n_\om^{v,\rm out} \ll n_\om^{u,\rm out}$. Hence the scattering of a dispersive field on a sonic horizon basically consists of a more general and slightly modified version of the standard Hawking effect. This is true when $\omm \gtrsim 2 \kappa$. For larger values of $\kappa$, the deviations become large, but the structure of Eq.~\eqref{eq:bog_transf} and the meaning of its coefficients remain unchanged~\cite{MacherBEC,Recati2009}.

Note that, if we choose a quantum state $\ket{\chi}$ which is not an eigenstate of the initial occupation operators $\hat n_\om^{i,\rm in}$, Eq.~\eqref{eq:finaloccupationnumbers} is not complete and other terms must be added. For instance, if
\begin{equation}
 \ket{\chi}=\frac{1}{\sqrt{2}}\left(\ha_\om^{u,{\rm in}\,\dagger}+\ha_\om^{v,{\rm in}\,\dagger}\right)\ket{0_{\rm in}}.
\end{equation}
the final occupation number of the $u$ mode would be
\begin{equation}
 \frac{1}{2}\left|\alpha_\om+A_\om\right|^2+\beta^2=\frac{1}{2}+\frac{3}{2}|\beta|^2+\re({\alpha_\om A_\om^*}),
\end{equation}
while, Eq.~\eqref{eq:finaloccupationnumbers} would yield a wrong result
\begin{equation}
 \frac{1}{2}+\frac{3}{2}|\beta|^2,
\end{equation}
where the cross term $\re({\alpha_\om A_\om^*})$ is missing.

In that case, Eq.~\eqref{eq:finaloccupationnumbers} becomes rather complicated and it is difficult to have a grasp on the expected particle content of the out modes. Let us imagine what would happen in the case in which the horizon is not always present but it is created at some finite time as in~\cite{sensitivity_carlos}. In that case, at early times, there are only two ABM for each frequency $\omega$, because the negative norm modes are not present, since the condensate is everywhere supersonic. If the initial state was the Fock vacuum for these two modes, once the horizon was formed, one would reasonably expect that this quantum state would look as ``almost'' vacuum for the destruction operators $\ha_\om^{u,{\rm in}}$, $\ha_\om^{v,{\rm in}}$. More difficult would be to determine whether the state would also be vacuum for $\ha_{-\om}^{{\rm in}}$. It is of course reasonable to assume that, if the horizon was formed adiabatically, the state would not contain negative frequency particles. If this were the case, the results of~\cite{MacherBEC} for a stationary geometry with vacuum initial condition might be applicable also to a realistic scenario when the horizon is created at a finite time.

In~\cite{carusotto2}, a numerical simulation shows that this is indeed what happens when the state is initially vacuum and the horizon is formed. After some transient radiation, the emission settles down to a stationary regime, coinciding with what is expected from calculations in eternal black holes (see Sec.~\ref{subsec:becexp}).
Probably, even in a real experiment where both the initial state are not so under control and the formation of the horizon cannot be made arbitrary slow, things would not differ so much from this situation. The initial state would almost coincide with the eternal in vacuum state and standard Hawking radiation would be propably detected, after some transient period, in the same fashion as in numerical simulations~\cite{carusotto2}.

However, it is meaningful to ask what would happen in a real black hole if one allowed for modified superluminal dispersion relation.
In general relativity a quantum state that is vacuum before the star collapse at late times is equivalent, at late times, to the Unruh state~\cite{Unruh:1976db} of an eternal black hole. In our language, this corresponds to say that the quantum state which is vacuum before the star collapse would be equivalent to the vacuum state described above ($n_\om^{u,\rm in}=n_\om^{v,\rm in}=n_{-\om}^{\rm in}=0$) for a stationary geometry. Would this be also true for superluminal modifications to the dispersion relation? What would the corresponding occupation number $n_{-\om}^{\rm in}$ be? Would this quantum state show Hawking radiation at late times?
In general relativity the Unruh state is the only possible one. If the collapse tried to select another state, vacuum polarization and energy carried by Hawking radiation would not cancel each other on the horizon, leading there to a divergence of the stress energy tensor, so that the back reaction would likely prevent the formation of the horizon~\cite{stresstensor}. So the Unruh state is the only possible state for real black holes and it must be stable against perturbations of initial data, since other possibilities would eventually yield divergences on the horizon.

In the presence of superluminal dispersion relation the situation is completely different. First, the divergence on the horizon is tamed by the modifications of the dispersion relation. In fact, modes traced back in time toward the horizon are still blueshifted but not infinitely, because when they reach a certain wavenumber they can cross the horizon and propagate inside the horizon. This leads to the second difference: Playing now the movie forward in time, modes can escape from the horizon carrying information about the interior and, in the case of a real black hole, about the singularity. As a consequence one should put conditions on the singularity (or on a very strong curvature region, assuming that the superluminal dispersion relation tamed also the singularity) in order to predict what one would observe at late times far from the black hole. In any case, it is not clear at all why the quantum state should appear as the equivalent of the Unruh state [that is the state considered above in Eq.~\eqref{eq:occvacuum}, \eqref{eq:occvacuum1}, and~\eqref{eq:occvacuum2}] and, for instance, not as the equivalent of the Boulware state~\cite{boulware}, with no emission of particle, or something in between.

\subsection{Observables}	%
\label{subsec:observables}	%

In what follows we study, for a stationary geometry, the properties of $n_\omega^{\rm out}$ of Eq.~\eqref{eq:occvacuum}, the mean occupation number of phonons of frequency $\omega$ spontaneously emitted to the right region, \ie, against the flow, when the quantum state is vacuum for ingoing modes ($n_\om^{i,\rm in}=0$).
The figures of this chapter have been obtained by numerically solving Eq.~\eqref{eq:BdG1} for Fourier modes of Eq.~\eqref{eq:phiexpansion1} [see Eq.~\eqref{eq:phi}] with the code described in Appendix~\ref{app:code}.

\begin{figure}
\centering
 \includegraphics[width=0.48\textwidth]{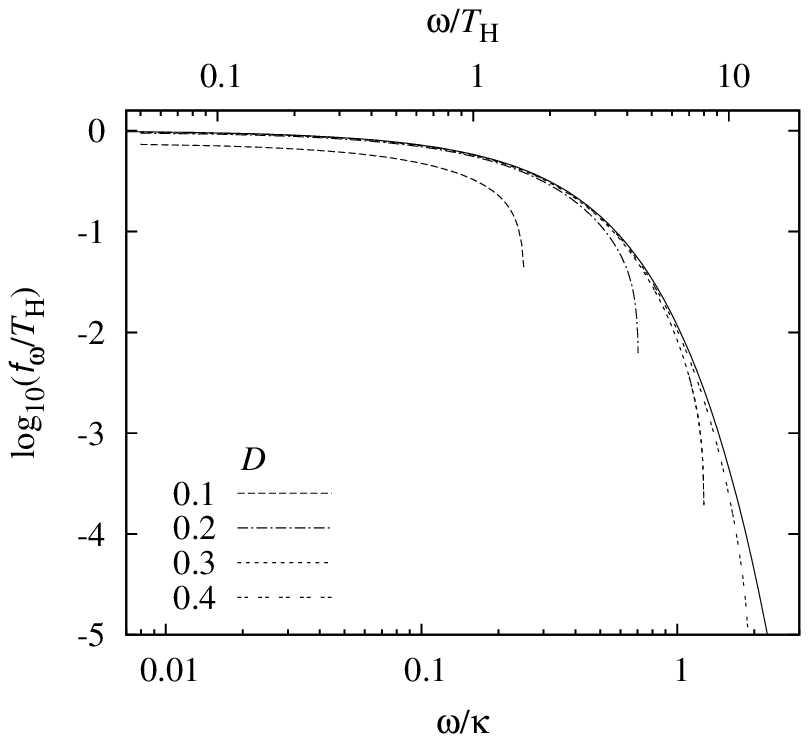}
 \hspace{0.02\textwidth}
 \includegraphics[width=0.48\textwidth]{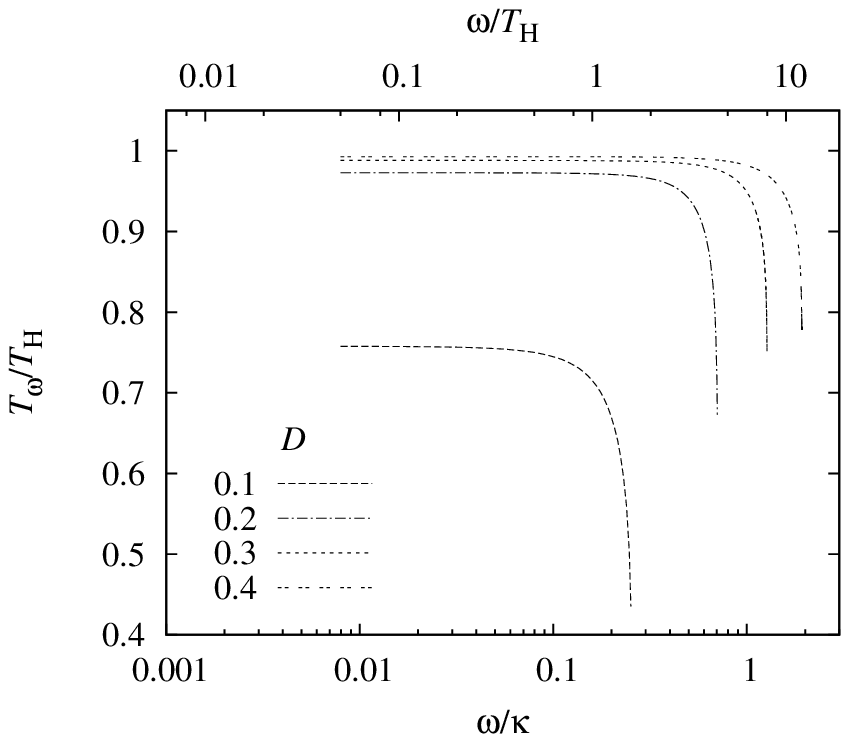}
\caption{$\log_{10}(f_\om/\Th)$ (left panel) and $T_\om/\Th$ of Eq.~\eqref{eq:Tom} (right panel)
versus $\om/\kappa$ for different values of $D$.
The solid line in the left panel represents the Planck flux at temperature $\Th=\kk/2\pi$.
When $D > 0.1$, the phonon flux hardly differs from it, while
for $D=0.1$, $\omm\ll 10 \, T_{\rm H}$ and the flux is weaker.
In the right panel, for all values of $D$, $T_\om$ hardly varies until $\om \to \omm$ of Eq.~\eqref{eq:omm}, where the flux vanishes.
For $\omm \gtrsim 10 \, T_{\rm H}$, as can be seen from the horizontal upper scale, these spectra are, to a high precision, Planckian.
In both panels, the fixed parameters are $\Lambda/\kappa = 15$, $n=\nt=1$.
\label{fig:tom_fom_nopert}
}
\end{figure}

In Fig.~\ref{fig:tom_fom_nopert}, left panel, we present the energy flux
\begin{equation}\label{eq:flux}
f_\om \equiv \om \, n_\om^{u,\rm out},
\end{equation}
as a function of $\om$, for different values of $D$, keeping fixed all the other parameters.
When working with a relativistic massless field, one would obtain a Planck spectrum with a temperature $\Th = \kk/2\pi$ (solid line).
When $D$ is sufficiently large ($D\gtrsim0.1$), the phonon flux does not differ from the Planck flux at the corresponding temperature. However, for $D=0.1$, $\omm$ is much smaller than the Hawking temperature, such that this cut off occurs at a frequency much lower than $T_{\rm H}$. In that case the flux is much weaker than the corresponding Planckian one.
However, the spectrum is remarkably close to the Planckian spectrum when $\omm$ is sufficiently large and, even when it differs from it, it changes very smoothly with $D$, becoming weaker but with the same shape. This is in accordance with the 
{\it robustness} of the flux already recognized in~\cite{Unruh95,BMPS95,CJ96}.

Since comparing energy fluxes $f_\om$ is not convenient, we introduce the temperature function, implicitly defined by
 \begin{equation}\label{eq:Tom}
  n_\om^{u,\rm out} \equiv \frac{1}{\exp(\om/ T_\om)-1}.
 \end{equation}
As such, $T_\om$ is simply another way to express $n_\om^{u,\rm out}$, but it presents the great advantage of being constant whenever the flux is Planckian.

In fact, as can be seen in Fig.~\ref{fig:tom_fom_nopert}, right panel, $T_\om$ remains remarkably constant, until $\om$ approaches the critical frequency $\omm$ of Eq.~\eqref{eq:omm}.
At that frequency, the signal drops down as it must do, since the Bogoliubov transformation \eqref{eq:bog_transf} becomes trivial, because the negative norm mode $\varphi_{-\om}$ no longer exists above $\omm$.
This cutoff effect relies on the asymptotic value of the flow $v+c$, and has been studied in detail in~\cite{MacherRP1,MacherBEC}. In what follows we focus instead on the deviations of the spectrum which depend on the near horizon properties of $v+c$.
Furthermore, from Fig.~\ref{fig:tom_fom_nopert}, it is even clearer what we observed for the flux $f_\om$: The spectrum is close to a Planckian one even for very small values of $D$ (and, as a consequence, of $\omm$) but with a much lower value of the temperature. This robustness of the thermal properties of the spectrum was completely unexpected and can actually be properly investigated only with numerical techniques. In Sec.~\ref{sec:broadhor}, we will show that the robustness of the thermality of the spectrum is a very general property and its temperature, although possibly very different from the standard Hawking temperature, can be estimated in a simple way as an average of the standard decay rate~\eqref{eq:temprelnew} over a small region around the horizon [see Eq.~\eqref{eq:kav}].

\begin{figure}
\centering
 \includegraphics[width=0.48\textwidth]{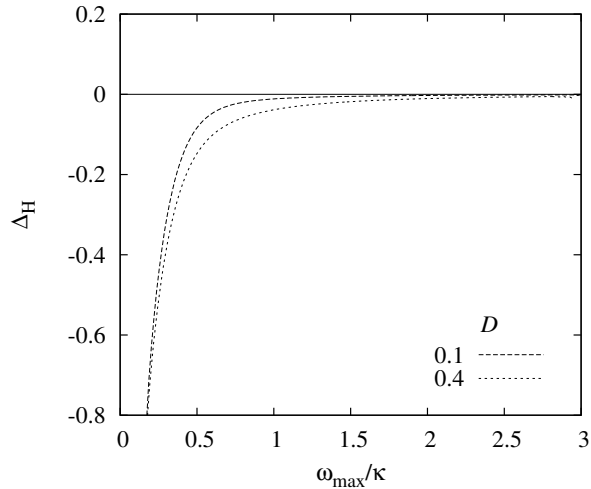}
\caption{ 
$\dH$ of Eq.~\eqref{eq:dH} as a function of $\omm/\kappa$ for two values of $D$.
When $\omm/\kappa$ is large enough, deviations from Planckianity are small and, more importantly,
governed by the ratio $\omm/\kappa$ only and not independently by $D$ or $\Lambda$.
\label{fig:deltaH_nopert}}
\end{figure}

Since one of our main goals is to identify the parameters governing the deviations from Planckianity, it is convenient to define
\begin{equation}\label{eq:dH}
 \dH\equiv\frac{f_{\omega=\Th}-f^{\Th}_{\omega=\Th}}{f^{\Th}_{\omega=\Th}},
\end{equation}
where $f_\om$ is the energy flux defined in Eq.~\eqref{eq:flux} and $f^{\Th}_\om$ is the thermal flux at temperature $\Th$. This quantity parametrizes the deviation of the observed flux $f$ at a reference frequency $\om=\Th$, with respect to the standard thermal spectrum $f_{\Th}$ at temperature $\Th$.
In Fig.~\ref{fig:deltaH_nopert}, $\dH$ is plotted as a function of $\omm/\kappa$ for two values of $D$. As shown in Eq.~\eqref{eq:omm}, $\omm$ depends both on $\Lambda$ and $D$, so, for each curve, $\omm/\kappa$ is varied by varying only $\Lambda$ and keeping $D$ constant. When $\omm/\kappa\gtrsim2$, that is $\omm$ is an order of magnitude larger than the Hawking temperature $\Th$, there are no substantial deviations of the observed flux from the Planckian one. Furthermore the two curves merge together, signaling that those small deviations are governed only by the ratio $\omm/\kappa$ and not independently by $D$ or $\Lambda$. This means that, as far as this analysis allow to discriminate, all the effects of dispersion are encoded in a single parameter $\omm$.

All the above results~\cite{MacherBEC} were obtained using the symmetrical profiles of Eq.~\eqref{eq:velocity_simp}. One may therefore wonder whether the observed robustness of the spectrum is not partially due to the symmetry of the flow with respect to the horizon, namely $w(-x) = - w(x)$.
Furthermore, using a velocity profile as simple as that of Eq.~\eqref{eq:velocity_simp}, some difficulties appear in the characterization of the deviations of the observed spectrum from a Planckian one at temperature $\Th$.
In fact, these deviations have different origins and are governed by different parameters.
To investigate these two issues, in the following sections we shall consider flows which are highly asymmetric and with short scales perturbations.

\section{Asymmetric velocity profiles}	%
\label{sec:asymmetric}			%

To investigate whether the robustness of the spectrum observed in~\cite{MacherBEC} was due or not to the symmetry of the flow with respect to the horizon, we study here a more general class of flows with respect to those of Eq.~\eqref{eq:velocity_simp}, namely
\begin{equation}
\frac{w(x)}{\chor } = \Dta + D \, \tanh^{1/n}\!\!\left[\left(\frac{\kappa x}{D \chor}\right)^{\!\!n}\right]
\label{vdparam}
\end{equation}
and, for the sake of simplicity, we investigate only the case $n=1$.
In particular, we shall study the spectrum in the interesting limiting cases where the horizon disappears because the flux becomes either sub or supersonic.

\begin{figure}
 \centering
 \psfrag{w}[c][c]{$w(x)/\chor$}
 \psfrag{x}[c][c]{$\kappa x/\chor$}
 \psfrag{dm}[c][r]{$D-\Dta$}
 \psfrag{dp}[c][l]{$D+\Dta$}
 \includegraphics[width=0.95\textwidth]{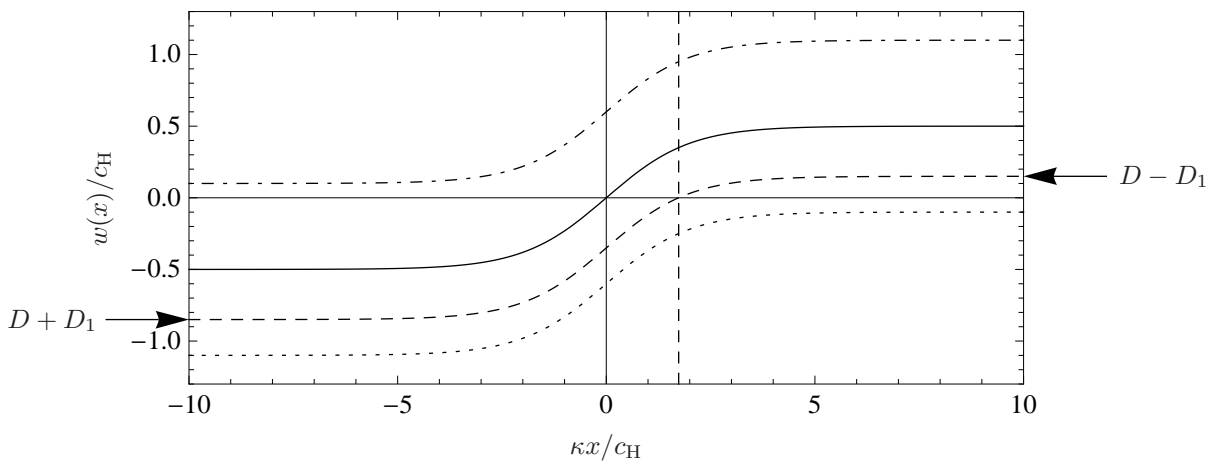}
 \caption{\label{fig:profile}
Four velocity profiles $w=c+v$ of Eq. (\ref{vdparam}) with the same height $D$ but four values of $\Dta$. For $ \Dta= 0$ (solid line), one recovers the symmetric case of~\cite{MacherRP1} with the Killing horizon (solid vertical line) at $x=0$.
For $-D<\Dta<0$ (dashed line), the horizon (dashed vertical line) is shifted to $x>0$. For illustrative purposes, we indicate on the plot the values of $D+\Dta=D-|\Dta|$ and $D-\Dta=D+|\Dta|$.
In the last two cases, there is no horizon: for $\Dta<-D$ (dotted line), the flow is everywhere supersonic, and for $\Dta>D$ (dotdashed line), it is subsonic.}
\end{figure}

The new parameter $\Dta$, which was taken to be $0$ in~\cite{MacherRP1,MacherBEC}, fixes the asymmetry between sub and supersonic regimes. When $\Dta > D$, the flow is everywhere subsonic and, when $\Dta <- D $, it is supersonic, see Fig.~\ref{fig:profile}.
Instead, when $|\Dta|<D$, there is a sonic (Killing) horizon where $w=0$, localized at
\begin{equation}
 x_{\rm H}=-\frac{D\chor}{\kappa}\,\mbox{arctanh}\!\left(\frac{\Dta}{D}\right).
\end{equation}
It separates the sub ($x>x_{\rm H}$) from the supersonic region ($x < x_{\rm H}$).
In such a flow, when ignoring dispersion, the spectrum of upstream phonons spontaneously emitted from the horizon
would be very simple, and would strictly correspond to Hawking radiation~\cite{unruh,Unruh95}. It would follow a Planck law at Hawking temperature $\Th={\kk}/{2\pi}$, where the surface gravity $\kk$ is
\begin{equation}\label{eq:temprelD2}
\kk  \equiv \partial_x (c+v) \vert_{x = x_H}
= \kappa 
\left[1-\left(\frac{\Dta}{D}\right)^{\!\!2}\right].
\end{equation}
Note that, as for the symmetric velocity profiles of Eq.~\eqref{eq:velocity_simp}, $\kappa$ still fixes the maximum slope of $w(x)$, but it cannot now be interpreted as the surface gravity of the horizon, that is determined by $\kk$.

\begin{figure}
\centering
\includegraphics[width=.48\textwidth]{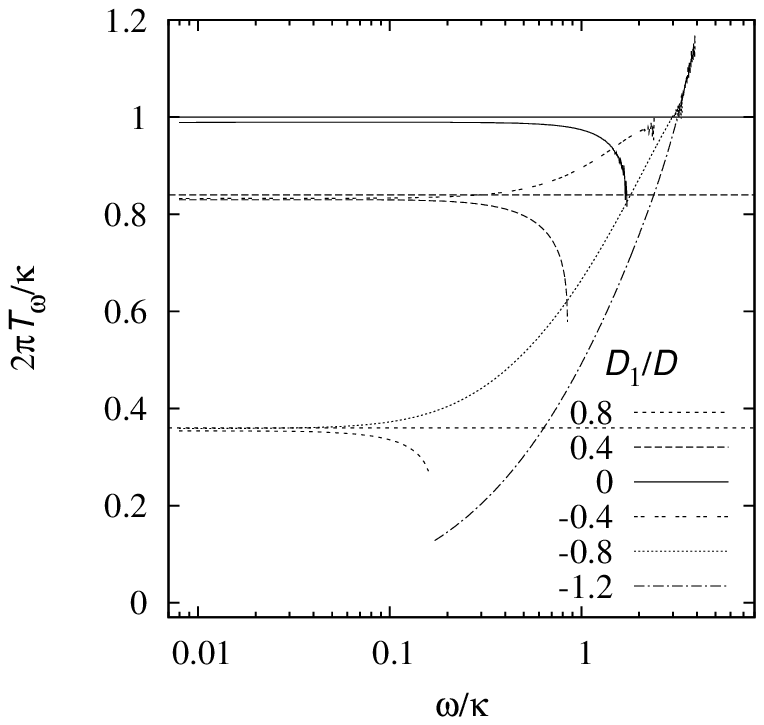}
\includegraphics[width=.48\textwidth]{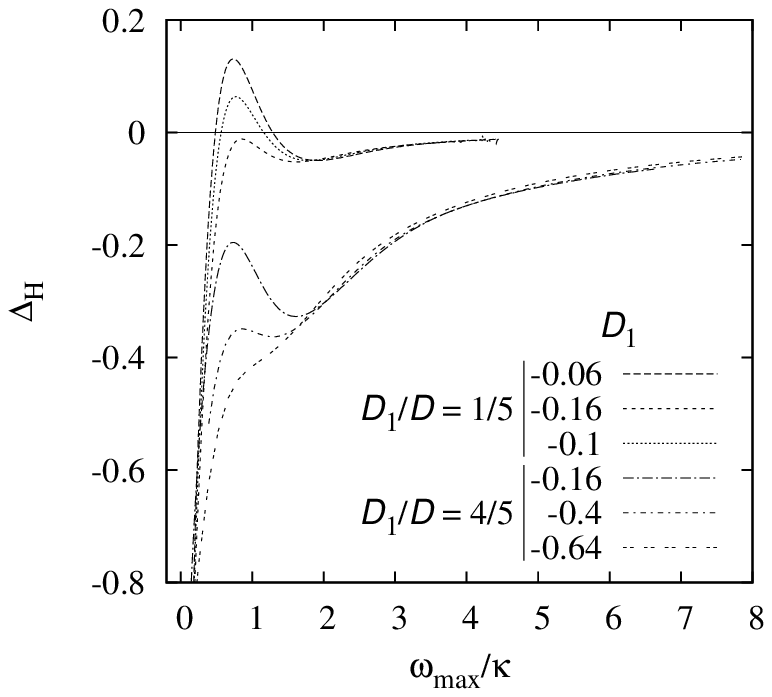}
\caption{\label{fig:tom}Left panel: $T_\om$ for 6 values of $\Dta/D$,
with $D=0.5$ and $\Lambda/\kappa=10$ fixed. The horizontal lines represent $\Th(\Dta)$ of Eq.~\eqref{eq:temprelD2}, respectively for $|\Dta|=0,0.2,0.4$.
For $\Dta<-D$ the curve is truncated at $\ommin$ as explained in the text.
Right panel: $\dH$ as a function of $\omm/\kappa$ for two values of $\Dta/D$ and  various $\Dta$.
When $\omm/\kappa$ is large enough, deviations from Planckianity are small and, more importantly, 
governed by the ratio $\Dta/D$ only.}
\end{figure}

In Fig.~\ref{fig:tom}, $T_\om$ and $\Th$ (horizontal lines) are plotted for various values of $\Dta$. They are both normalized using the temperature scale $\kappa/2\pi$, determined by $\kappa$, in order to better appreciate the scaling of $\kk$ with $\Dta$, which is given in Eq.~\eqref{eq:temprelD2}.
We distinguish four regimes:
\begin{itemize}
 \item For $|\Dta|$ sufficiently smaller than $D$, the spectrum is Planckian until $\om$ approaches $\omm$, where it vanishes.
 For $\Dta > 0$, $\omm$ decreases since $\vert w(x = - \infty)\vert$ does so, and conversely for $\Dta < 0$.
 However for both signs, at low frequency $T_\om$ closely follows $\kk/2\pi$ of \eqref{eq:temprelD2}.
 \item When $|\Dta|\to D$, $\kk$ of \eqref{eq:temprelD2} drops down to $0$ and the sonic horizon disappears.
 In this critical regime, deviations from Planckianity appear even at low frequency.
 \item For $\Dta>D$, the flow is everywhere subsonic. There is no particle production because there are no negative norm modes with $\om > 0$.
 \item For $\Dta< - D$, the flow is supersonic. A new critical frequency $\ommin<\omm$ appears.
 For $0< \om<\ommin$, there are now $4$ asymptotic in and out modes, and the scattering matrix is $4\times4$, and the treatment of Sec.~\ref{subsec:bogo} does not apply.
 However, for $\ommin<\om<\omm$ there are only 3 modes, this frequency band can be handled by this analysis.
 In this range, as can be seen in the left panel, $n_\om^{u,\rm out}$ is close to the case where $\Dta > - D$, even though the spectrum is not at all Planckian. 
\end{itemize}

Furthermore, we checked that the {\it shape} of the spectrum is hardly changed when varying the ``healing frequency'' $\Lambda$.
Namely, spectra calculated with different values of $\xi$ almost coincide when $T_\om$ is plotted versus the rescaled quantity $\om/\omm(\Lambda)$.

Finally, to identify the parameters governing the deviation from Planckianity,
$\dH$, defined in Eq.~\eqref{eq:dH}, is plotted as a function of $\omm(\Lambda)/\kappa$ in Fig.~\ref{fig:tom} (right panel).
Various values of $\Dta$ are used, while the ratio $\Dta/D$ is kept constant respectively for the three lower curves ($\Dta/D=-1/5$), and for the three upper curves ($\Dta/D=4/5$).
As expected, $\dH$ goes to 0 for large values of $\omm/\kappa$. What we learn here is that the leading deviations from Planckianity do not depend separately on $\Dta$ or $D$ but only on their ratio.

To conclude, we stress that the thermal spectrum at the na\"ive surface gravity~\eqref{eq:temprelD2} 
provides a reliable approximation of the actual phonon spectrum in all cases but the critical ones where the sonic horizon disappears. In these extreme cases, the spectrum is no longer thermal and its properties are governed by the dispersive properties of field.

\sectionmark{Broadening the horizon}				%
\section{Short-scale perturbations: Broadening the horizon}	%
\label{sec:broadhor}						%
\sectionmark{Broadening the horizon}				%

As discussed at the beginning of this chapter, the robustness of Hawking radiation against modifications of the dispersion relation is well established. However, it is not well known for which flow configurations this is true, since only very smooth and symmetric flow profiles have been studied so far, with the exception of the step-like profile of~\cite{Recati2009}. In this case, somehow unexpectedly, the predicted spectrum is thermal even if the Hawking temperature cannot be computed because the surface gravity is virtually infinite because of the non-smooth profile. To understand why this is possible, we consider velocity profiles with quite sharp perturbations on very small scales and, varying the velocity profile, we study the deviations of the spectrum from Planckianity and, in the case the spectrum is thermal, the deviation of its temperature from Hawking one.

\subsection{Profiles}	%
\label{sec:pertflows}	%

As already mentioned, it is difficult to characterize the deviations from the Planck spectrum by using a velocity profile described by such few parameters as that of Eq.~\eqref{eq:velocity_simp}.

To this aim, we introduce flow profiles governed by {\it several} scales of the form
\begin{equation}\label{eq:velocity_dec}
 c(x)+v(x)= w(x) = \wB(x) + \wt(x),
\end{equation}
where $\wB$ is a reference background profile and $\wt$ a perturbation of smaller amplitude: $0 \leq | \wt(x)| \lesssim |\wB(x)|$.

\begin{figure}
\centering
 {\includegraphics[width=0.48\textwidth]{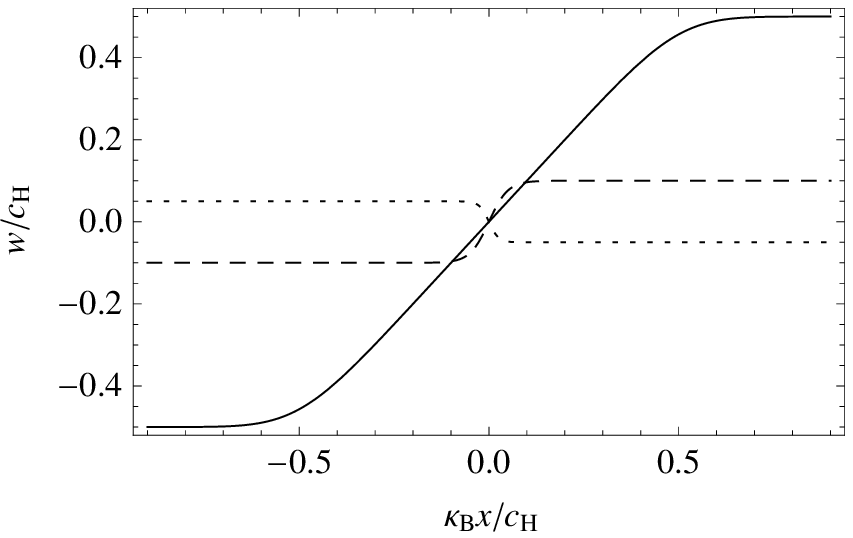}}\\[10pt]
 {\includegraphics[width=0.48\textwidth]{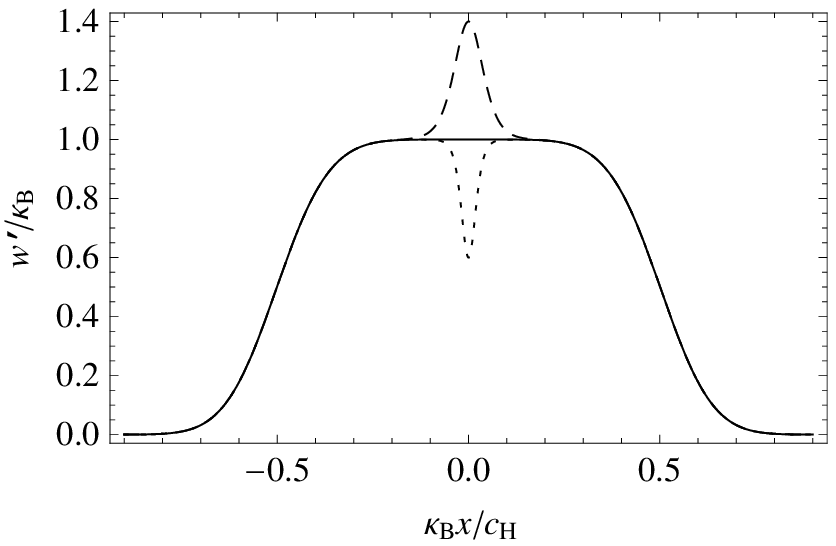}}
 \hspace{0.02\textwidth}
 {\includegraphics[width=0.48\textwidth]{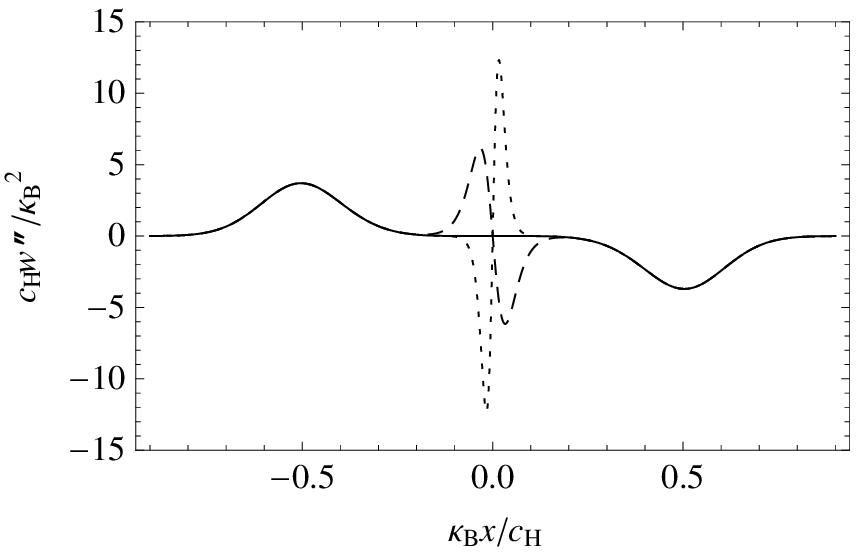}}
 \caption{Top panel: background flow $\wB$ (solid line) for $n=3$, $D=0.5$ and two perturbations $\wt$ (amplified by a factor of 5, dashed and dotted lines), with $\nt=1$, and $(\kt/\ka, \Dt)$ equal to $(0.4, 0.02)$ and $(-0.4, 0.01)$.
 Bottom panels: first and second derivative of $\wB$ (solid line) and of $w=\wB+\wt$ (dashed and dotted lines) for the same $\wt$. Narrower perturbations induce higher modifications of derivatives of $w$.
}
\label{fig:w}
\end{figure}

The background profile is still given by Eq.~\eqref{eq:velocity_simp}
\begin{equation}\label{eq:velocityB}
 \frac{\wB(x)}{\chor} =  D\,\sign(x)\, \tanh^{1/n}\!\!\left[\left(\frac{\ka|x|}{\chor D}\right)^{\!\!n}\right].
\end{equation}
We shall consider two classes of perturbations, symmetric ones with respect to the horizon, \ie, $\wt(x)= - \wt(-x)$,
and then, in Sec.~\ref{subsec:assym}, more general ones without that symmetry. The symmetric ones are similar to $\wB$:
\begin{equation}\label{eq:velocity2}
\frac{\wt(x)}{\chor} = 
\Dt\,\sign(x)
\, \tanh^{1/{\nt}}\!\!\left[\left(\frac{\kt|x|}{\chor \Dt}\right)^{\!\!\nt}\right].
\end{equation}
The condition $0 \leq | \wt| \lesssim |\wB|$ is simply $0 \leq \Dt \lesssim D$.
In Fig.~\ref{fig:w}, we plot $w(x)$, its first and second derivatives
for different values of $\kt/\ka$ and $\Dt$.

The smoothness parameter $\nt$ will range from $0.5 $ to $4$.
We shall see that it only affects the flux marginally. In fact the most important quantity is $\Dt$.
Its role is to fix 
\begin{equation}\label{eq:xlin}
|x_{\rm lin}| \simeq  \frac{\chor \Dt}{\kt},
\end{equation}
the interval of $x$ over which $w(x)$ is linear.
In this respect, it should be noticed that the surface gravity is equal to
\begin{equation}\label{eq:temprel}
\kk \equiv \partial_x (c+v) \vert_{x = 0} = \ka + \kt,
\end{equation}
{\it irrespectively} of the value of $\Dt$, and that of $\nt$. 
For relativistic fields, $\Dt$ would play no role since the temperature is $\kk/2\pi$ (when ignoring gray body factors, a correct approximation in our settings when $q=1/2$).
For dispersive fields instead, $\Dt$ plays a crucial role.

\subsection{Planckian character}	%
\label{sec:general}			%

\begin{figure}
\centering
 \includegraphics[width=0.48\textwidth]{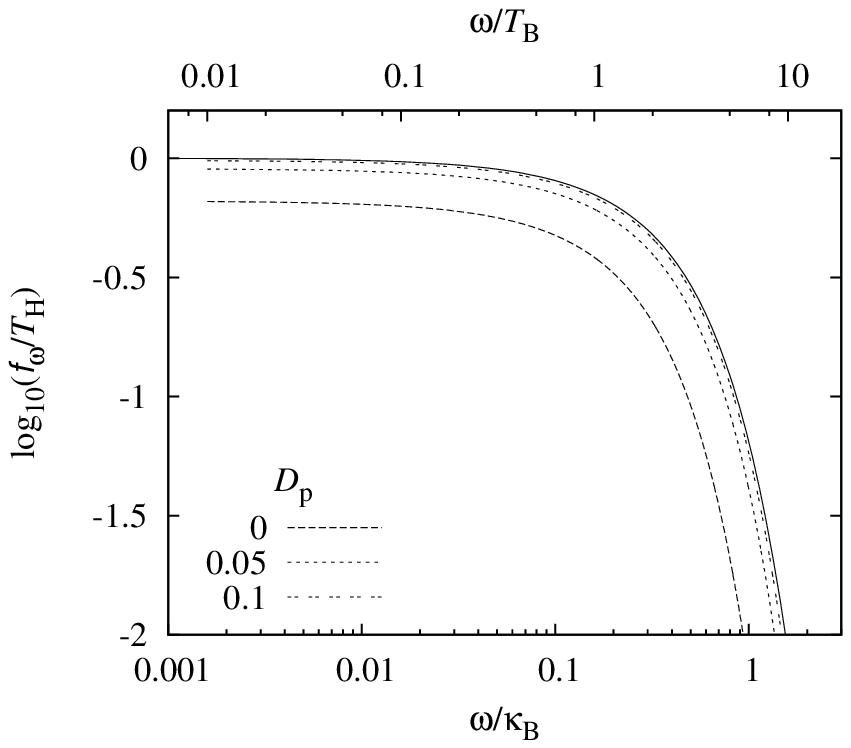}
 \hspace{0.02\textwidth}
 \includegraphics[width=0.48\textwidth]{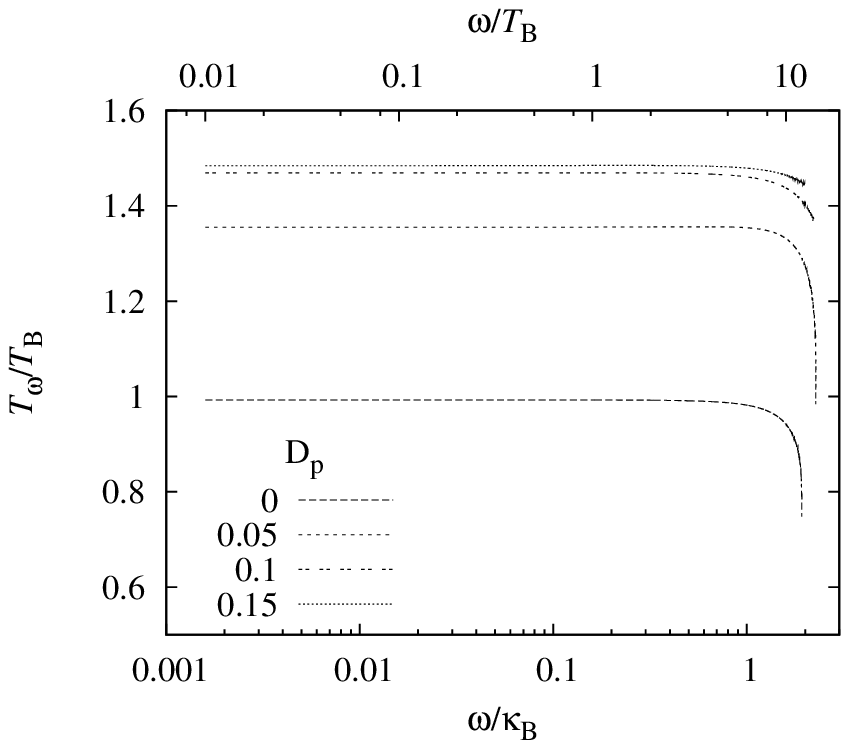}
\caption{$\log_{10}(f_\om/\Th)$ (left panel) and $T_\om/\TB$ of Eq.~\eqref{eq:Tom} (right panel) versus $\omok$ for different values of $\Dt$, but the same surface gravity $\kk= 1.5 \ka$. 
The solid line in the left panel represents the Planck flux at temperature $\Th=\kk/2\pi$.
When $\Dt= 0.1$, the phonon flux hardly differs from it.
For smaller values of $\Dt$, the flux is weaker, and for $\Dt \to 0$, it is fixed by the background surface gravity $\ka$.
In the right panel, $\TB=\ka/2\pi$ is the Hawking temperature of the background flow.
For all values of $\Dt$, $T_\om$ hardly varies until $\om \to \omm$ of Eq.~\eqref{eq:omm},
where the flux vanishes.
Since $\omm \gtrsim 10 \, \TB$, as can be seen from the horizontal upper scale,
these spectra are, to a high precision, Planckian.
In both panels, the fixed parameters are $\Lambda/\ka = 15$, $D = 0.4$, $\kt/\ka=0.5$, $n=\nt=1$.
\label{fig:tom_fom}
}
\end{figure}

In Fig.~\ref{fig:tom_fom}, left panel, we present the energy flux of Eq.~\eqref{eq:flux} as a function of $\om$ for different values of the $\Dt$, keeping fixed all the other parameters.
When working with a relativistic massless field, one would obtain a Planck spectrum with a temperature $\Th = \kk/2\pi$, irrespectively of the value of $\Dt$.
To ease the comparison with the dispersive fluxes, we have represented this flux by a solid line.
Working with the Bogoliubov--de~Gennes equation~\eqref{eq:BdG1}, we see instead that the flux varies with $\Dt$.
In fact it  increases monotonically. 
Moreover, when $\Dt \to 0$, we verified that it coincides with the flux one would obtain in the reference flow $\wB$.
We also see that when $\Dt$ is sufficiently large, the flux saturates and agrees to a high accuracy with the Planck spectrum at temperature $\Th$, in accordance with the {\it robustness}~\cite{Unruh95,BMPS95,CJ96} of the flux, as discussed in Sec.~\ref{subsec:observables}.

The Planckian character of the spectrum is evident from Fig.~\ref{fig:tom_fom} (right panel), in which we plot
$T_\om/\TB$, where
\begin{equation}\label{eq:TB}
 \TB\equiv\frac{\ka}{2\pi}
\end{equation}
is the Hawking temperature for an unperturbed flow profile of Eq.~\eqref{eq:velocityB}.
Note that $T_\om$ remains remarkably constant, until $\om$ approaches the critical frequency $\omm$ of Eq.~\eqref{eq:omm}.
In addition to the observables defined in Sec.~\ref{subsec:observables}, to quantify the small deviations from a Planck spectrum at temperature
\begin{equation}
  T_0\equiv\lim_{\om\to0}T_\om,
\end{equation}
we study the relative deviation
\begin{equation}\label{eq:run}
 \run\equiv \frac{T_0-T_{\om}}{T_0}.
\end{equation}
For all values of $\Dt$ we found that
\begin{equation}\label{eq:run2}
  |\runT| \lesssim 3.\times 10^{-4}.
 \end{equation}
Note that $\runT \neq 0$ is not due to numerical errors, but it really characterizes the deviations.
Equation~\eqref{eq:run2} implies that, to a high accuracy, the spectrum remains Planckian even when $T_0$ completely differs from $\Th$. (In the present case, it differs by $33 \%$, since $\kk = 1.5\, \ka$.)
The smallness of $\runT$ also implies that the parameters that govern the deviations from Planckianity {\it differ} from those governing the shift $T_0-\Th$.
Hence when studying the robustness of the Hawking flux against introducing dispersion,
one must differentiate these two types of deviations.

These are the first important results of~\cite{smearhor}.
They generalize what was observed in~\cite{MacherRP1,MacherBEC} to a much wider class of flows: First, that the thermality is well preserved whenever $\omm/\TB \gtrsim 10$, and, second, that the relative temperature shift $T_0/\Th - 1$ is much larger than $\runT$.
In the sequel we work with $\omm/\TB \gtrsim 10$ and study how the low frequency temperature $T_0$ depends on $\kt$, $\Dt, \nt$ and $\Lambda$.

\subsection{The critical value of \texorpdfstring{$\Dt$}{Dp}}	%
\label{sec:T0}							%

%
\begin{figure}
\centering
  \includegraphics[width=0.48\textwidth]{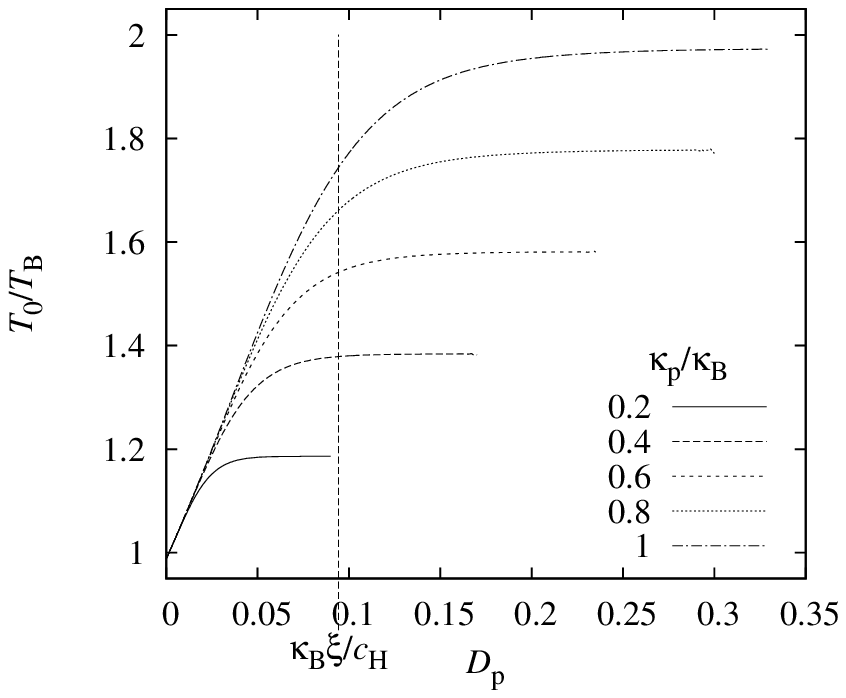}
  \hspace{0.02\textwidth}
  \includegraphics[width=0.48\textwidth]{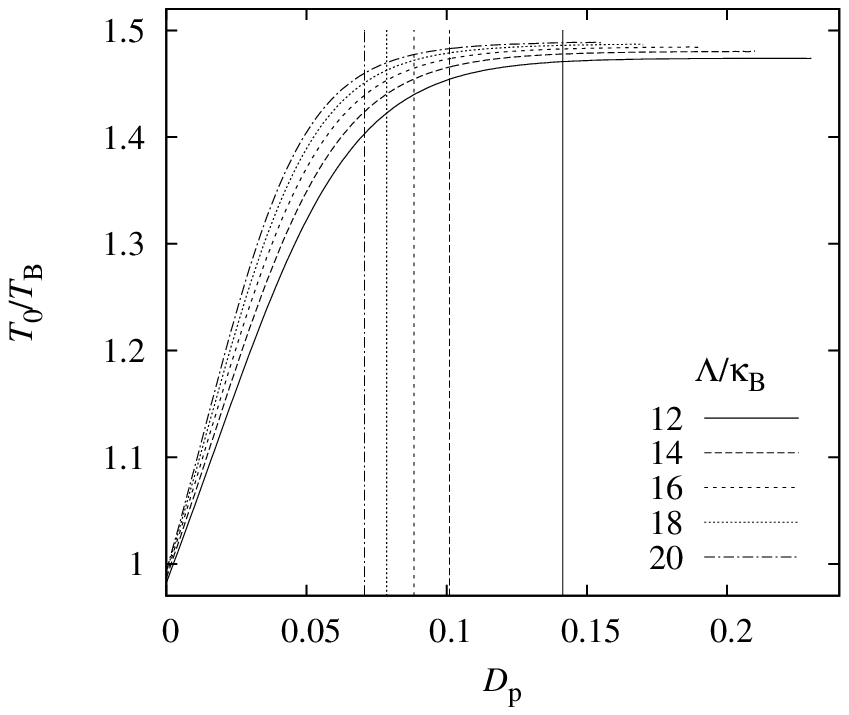}
   \caption{Left panel: 
$T_0/\TB$ versus $\Dt$ for $\kt/\ka$ from $0.2$ to $1$, and with $\Lambda/\ka = 15, D = 0.3$, $n=\nt=1$.
Along each curve the surface gravity $\kk$ of Eq.~\eqref{eq:temprel} is {\it constant}.
It fixes the temperature {\it only} for $D \gg \dcrit$.
The vertical line gives the value of the healing length $\xi$.
Right panel: $T_0/\TB$ versus $\Dt$ for $\Lambda/\ka$ from $12$ to $20$, and with $\kt/\ka = 0.5, D = 0.3$, $n=\nt=1$.
The vertical lines give the values of the healing length $\xi = \sqrt{2}\chor/\Lambda$.
One can see that $\dcrit$ of Eq.~\eqref{eq:dcrit1} is {\it not} (directly) related to $\xi$.
}
\label{fig:T0_k3}
\end{figure}

In Fig.~\ref{fig:tom_fom}, right panel, we saw that $T_0$ starts from $\ka/2\pi$ for $\Dt\to0$, and asymptotes to $\Th = \kk/2\pi$ for $\Dt$ larger than some critical value $\dcrit$. (We have checked that this is also the case when $\kt < 0$, and this down
to $\kt = - \ka$, in which case the surface gravity $\kk$ vanishes.)
To understand what fixes $\dcrit$, we first consider series of flows $\wB + \wt(\Dt)$ for different values of $\kt$ at fixed $\ka$.
For each series, the surface gravity $\kk = \ka + \kt$ is thus constant.
The resulting temperatures $T_0(\Dt)$ are presented in Fig.~\ref{fig:T0_k3}, left panel.
This figure reveals many interesting features.
For small values of $\Dt$, $T_0$  increases linearly with $\Dt$ in a manner essentially {\it independent} of $\kt$.
In addition, for each series, when $\Dt$ is large enough, $T_0$ saturates at the standard result $\Th = \kk/2\pi$.

Given the linear behavior of $T_0$ for small $\Dt$ and the saturation for large values, \ie,
\begin{equation}\label{eq:limits}
\begin{aligned}
 2\pi T_0(\Dt\ll\dcrit)&\sim \ka+s\, \Dt, \\
 2\pi T_0(\Dt\gg\dcrit)&\sim \ka+\kt=\kk,
\end{aligned}
\end{equation}
we define $\dcrit$ as the value of $\Dt$ where these intercept, and obtain
\begin{equation}\label{eq:dcrit}
 \dcrit = \frac{\kt}{s}.
\end{equation}
To understand what fixes the slope $s$, we consider series of flows with different values of the ultraviolet dispersion scale $\Lambda$.
The results are presented in Fig.~\ref{fig:T0_k3}, right panel.
It is numerically easy to see that $s\propto \Lambda^{2/3}$.
Hence we get
\begin{equation}\label{eq:dcrit1}
 \dcrit\propto \kt\, \Lambda^{-2/3}.
\end{equation}

We now present a simple argument telling how $\dcrit$ should scale with the various parameters.
In~\cite{MacherBEC}, by considering $\wB$ of Eq.~\eqref{eq:velocityB}, it was {\it observed} that the leading deviations due to dispersion are governed by inverse powers of $\ommok$.
For these unperturbed flows, $\omm$ of Eq.~\eqref{eq:omm}, even though initially defined as the frequency such that the turning point $x_{\rm tp}(\omega)$ of Eq.~\eqref{eq:tp} is pushed to $-\infty$, also corresponds to the frequency such that the turning point is at the edge of the domain $\sim \chor D/\ka$, where $\wB$ is linear in $x$.

When working with $w= \wB + \wt$ of Eq.~\eqref{eq:velocity2} and with $\Dt \ll D$, the linear domain is now given by $x_{\rm lin}$ of Eq.~\eqref{eq:xlin}.
Hence we expect that the deviations will be now governed by inverse powers of $\ommt/\kk$, where $\ommt$ is the new critical frequency fixed by $x_{\rm lin}$.
The value of $\ommt$ is such that the turning point $| x_{\rm tp}(\omega)|$ of Eq.~\eqref{eq:tp} equals $x_{\rm lin}$.
Using Eq.~\eqref{eq:xlin}, $| x_{\rm tp}(\ommt)| = x_{\rm lin}$ gives
\begin{equation}\label{eq:omm2}
 \ommt \simeq\Lambda\left( \frac{\kk}{\kt} \Dt\right)^{3/2}.
\end{equation}
Since the deviations from the standard result should be small when $\ommt/\kappa_K\gg1$, the critical value of $\Dt$ below which the temperature of the emitted radiation is no longer given by $\Th$ should scale as
\begin{equation}
 \dcrit \sim \frac{\kt}{(\kk\, \Lambda^2 )^{1/3}}.\label{eq:dcrittheo}
\end{equation}
A rigorous study~\cite{cfp} of the connection formulas~\cite{BMPS95,Corley97,Tanaka99,UnruhSchu05} confirms the result of this simple reasoning.

Equation~\eqref{eq:dcrittheo} is in agreement with Eq.~\eqref{eq:dcrit1} for $\kt \ll \ka$.
In the following section, we shall numerically validate it for all values of $\kt$.
Before proceeding, we checked that $\dcrit$ does not significantly depend on the background quantities $n$, $D$ and $\ka$, as one could have expected.
The weak, subleading, dependence on $\nt$ is studied in Sec.~\ref{sec:n2}.

\subsection{The averaged surface gravity}	%
\label{sec:bark}				%

So far we have some understanding of the Planckian character of the flux in the robust regime when $\Dt > \dcrit$, and of the value of $\dcrit$~\cite{cfp}.
Instead, what fixes $T_\om$ outside this regime, for $\Dt < \dcrit$, is {\it terra incognita}.

We shall now numerically establish that, to a good approximation when $\omm/\TB \gtrsim 10$, the temperature is determined by the average of the gradient $dw/dx$ over a width that we parametrize by $d_{\xi}=\dxl + \dxr $, where $\dxl$ and $\dxr $ are respectively the width on the left and on the right calculated from the Killing horizon at $x=0$.

In fact, given that the dispersion of Eq.~\eqref{eq:dispersion1} defines a ``healing length'' $\xi = \sqrt{2}c/\Lambda$, it is not surprising that $\xi$ in turns defines a minimal resolution length, such that details of $w(x)$ smaller than $d_{\xi}$ are not ``seen'' by the phonon field.
In other words, because of dispersion, it is as if we were dealing with a horizon of width $d_{\xi}$.\footnote{\label{foot:hor}This important result should be opposed to the possibility discussed in~\cite{UnruhSchu05,08} according to which $T_\om$ could be governed by a local, $\om$-dependent, function of $w$, \eg\/ by the value of  $\partial_x w$ evaluated at the turning point of Eq.~\eqref{eq:tp}.
In fact our numerical results indicate that it is a {\it non-local} quantity that fixes the temperature, and in a manner essentially independent of $\om$. This is reminiscent of the ideas that quantum gravitational effects might blur the horizon~\cite{quantum_metric_fluctRP,Beyond_RP} and that dispersion might regulate the black hole entanglement entropy~\cite{ent_ent_TJRP}.}
Moreover, when $\dxl \neq \dxr $, this means that the center of the effective horizon is {\it displaced} with respect to the Killing horizon to $\bar x_{\rm hor}= (\dxr - \dxl)/2$.

To test the idea that the temperature be determined by an average ``surface gravity,'' we introduce
\begin{eqnarray}\label{eq:kav}
 \bar\kappa\equiv \frac{1}{d_{\xi}} \int_{-\dxl}^{\dxr} \dd x \, \frac{\dd w(x)}{\dd x}=\frac{w(\dxr) -w ( -\dxl) }{d_{\xi}}.
\end{eqnarray}
Using Eqs.~\eqref{eq:velocityB} and~\eqref{eq:velocity2}, we get
\begin{eqnarray}\label{eq:kavs}
 \bar\kappa=\ka+ \frac{\wt(\dxr)  + \wt(\dxl)}{d_{\xi}} 
+O\left[\left(\frac{\ka d_{\xi}}{D\chor}\right)^{2n}\right],
\end{eqnarray}
where we assume that $d_\xi \ll D/\ka$, so that the average of the background term $\wB$ is approximately $\ka$, plus corrections of the order of $(\ka d_{\xi}/{D\chor})^{2n}$.

\begin{figure}
\centering
 \includegraphics[width=0.48\textwidth]{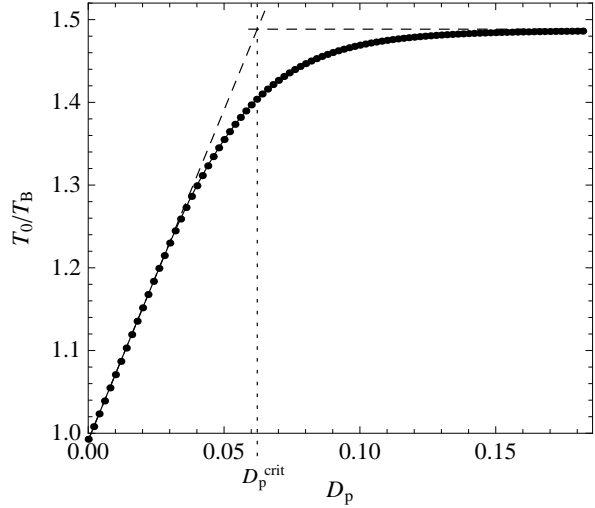}
 \caption{Dots: $T_0/\TB$ versus $\Dt$ computed for $\Lambda/\ka=15$, $D=0.4$, $\nt=n=1$, $\kt/\ka=0.5$.
 Solid line: best fit curve $\bar \kappa/2\pi\TB$ obtained by using $\bar\kappa$ of Eq.~\eqref{eq:kavs}.
 The perfect agreement establishes that the temperature $T_0$ equals $\bar \kappa/2\pi$.
 Dashed lines: asymptotic behaviors for $\Dt\ll\dcrit$ and $\Dt\gg\dcrit$ [see Eqs.~\eqref{eq:limits} and~\eqref{eq:dcrit}].}
 \label{fig:fitbark}
\end{figure}

We proceed as follows: We compute $T_0$ for about 200 values of $\Dt$ for some fixed values of the perturbed flow parameters $(\nt, \kt)$ and the dispersive scale $\Lambda$, \ie, along  series as those represented in Fig.~\ref{fig:T0_k3}.
Out of this set we extract five quantities, namely, $\ka^{(\rm fit)}$, $\nt^{(\rm fit)}$, $\kt^{(\rm fit)}$, $\dxl$, and $\dxr$, by fitting $\bar \kappa(\ka,\nt,\kt,\dxl, \dxr;\Dt)/2\pi$ of Eq.~\eqref{eq:kavs} with a nonlinear least-squares method.
The agreement between the couples $T_0(\Dt)$ and the fitted function $\bar \kappa(\Dt)/2\pi$ is really striking---see Fig.~\ref{fig:fitbark} for an example.

Moreover, the fitted values $\ka^{(\rm fit)}$ and $\kt^{(\rm fit)}$ are in very good agreement (less than 1.5\% and 1\%, respectively) with their values used in the flow $w$.
This is a necessary check to validate that it is the average of $\partial_x w$ that governs the temperature.
However, the agreement between $\nt^{(\rm fit)}$ and $\nt$ is less good. This indicates that the actual temperature is not {\it exactly} given by the average of Eq.~\eqref{eq:kav}.\footnote{We notice that different averaging procedures could have been considered in the place of Eq.~\eqref{eq:kav}, and they would have given slightly different results. At this point, we are not in the position to distinguish among them. What is important is that the scaling laws we shall derive in Sec.~\ref{sec:dxi} are independent of the particular choice.}
Even though this is a subdominant effect, it is interesting to see under which conditions a better agreement between $\nt^{(\rm fit)}$ and $\nt$ is reached.

\subsection{Subleading effects in \texorpdfstring{$\nt$}{np}}	%
\label{sec:n2}							%

The parameter $\nt$ of Eq.~\eqref{eq:velocity2} determines the smoothness of the transition between the region where $\wt(x)$ is linear, with constant derivative $\kt$, and the region where it is flat, equal to $\Dt \chor$. Larger values of $\nt$ make that transition sharper.
Therefore higher derivatives of $w$ will become very large at the end of the linear-regime region, near $x_{\rm lin}$ of Eq.~\eqref{eq:xlin}.
As a consequence, nonadiabatic effects become more important, thereby generating larger deviations between $T_0$ and $\bar \kappa/2\pi$ of Eq.~\eqref{eq:kavs}.

\begin{figure}
  \centering
  \includegraphics[width=0.48\textwidth]{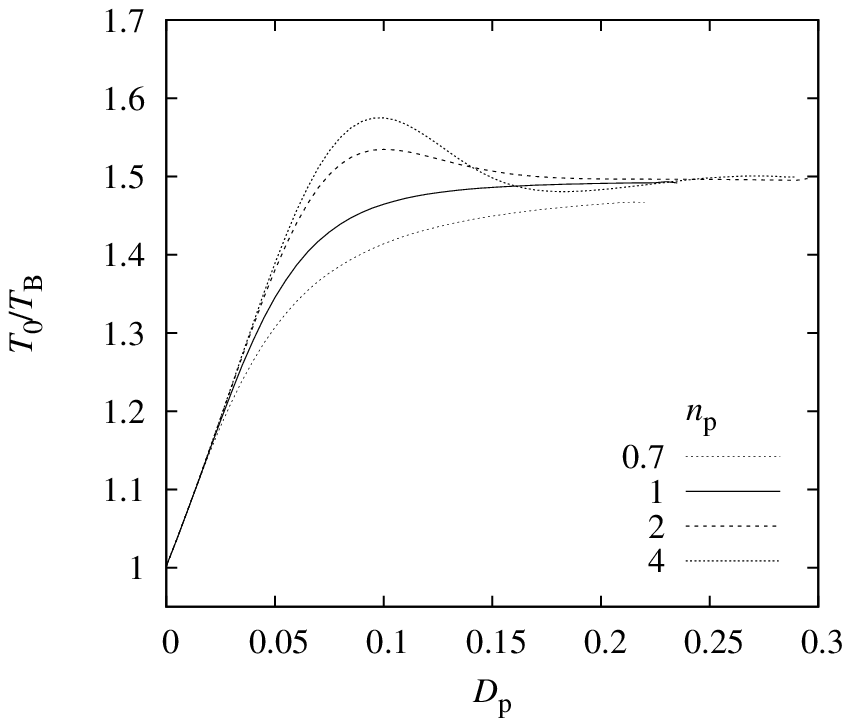}
  \hspace{0.02\textwidth}
  \includegraphics[width=0.48\textwidth]{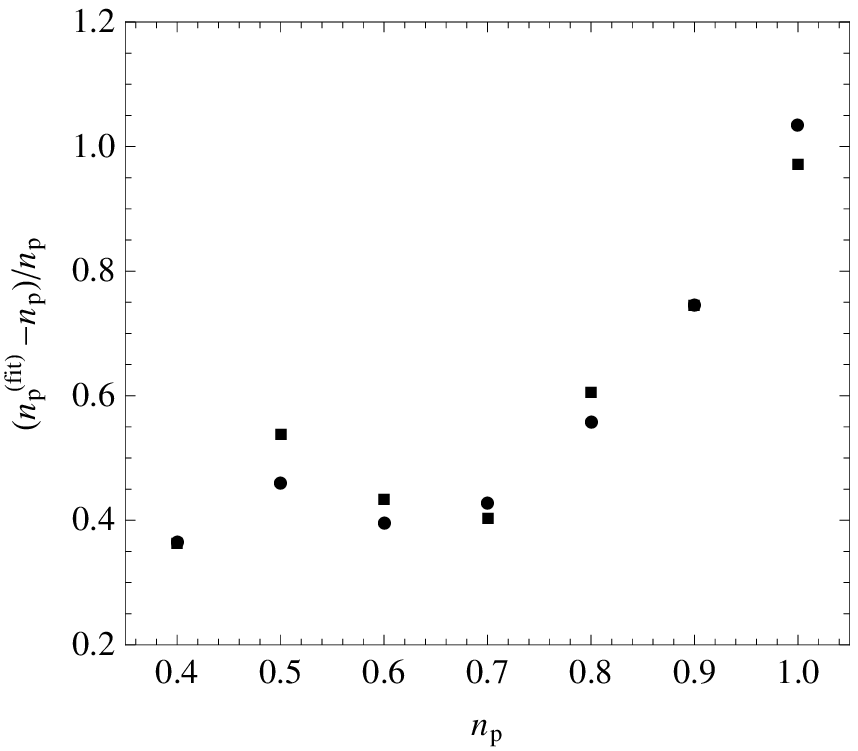}
  \caption{Left panel: $T_0/\TB$ versus $\Dt$ for different values of $\nt$ of  Eq.~\eqref{eq:velocity2} with 
   $\kt/\ka = 0.5$, $D = 0.4$, $n=2$. $T_0$ monotonically approaches $\Th$ for $\nt\leq1$. When $\nt>1$ instead, it exceeds $\Th$ near $\dcrit$ and then approaches it with oscillations.  Right panel: the relative difference $(\nt^{\rm (fit)}-\nt)/\nt$ versus $\nt$. The difference decreases for smaller $\nt$, in a rather insensitive way with respect to $\Lambda/\ka$ (dots, $\Lambda/\ka=15$; squares, $\Lambda/\ka=20$).}
  \label{fig:n2}
\end{figure}

To investigate these deviations, we compare $T_0(\Dt)$ for various values of $\nt$.
As it appears from Fig.~\ref{fig:n2}, left panel, $s$ and $\dcrit$ do not significantly depend on $\nt$.
However, the transition region between the linear regime and the constant regime depends on $\nt$. The transition is indeed monotonic for small values of $\nt$ whereas, for $\nt > 1$, $T_0$ oscillates around $\Th = \kk/2 \pi$. 
The monotonic function of Eq.~\eqref{eq:kav} will of course never describe these oscillations.
However, for $\nt\leq1$, one can compare $\nt^{(\rm fit)}$ with $\nt$. 
The result is shown in Fig.~\ref{fig:n2}, right panel. The dots and the squares are computed using different values of $\Lambda/\ka$, respectively 15 and 20.
One sees that $\nt^{\rm fit}$ is always larger than $\nt$. 
One also sees that the discrepancy becomes smaller when $\nt$ decreases, as expected, since nonadiabatic effects are smaller for smoother profiles, hence smaller $\nt$.
One should not overestimate the relevance of this discrepancy because we found that the value of $\nt^{(\rm fit)}$ is correlated in the fitting procedure to the difference $\dxr - \dxl$.
For instance, when imposing $\dxr = \dxl$, $\nt^{(\rm fit)} - \nt$ is reduced by a factor of about $2$.

The important conclusion is the following.
When nonadiabatic effects are small, the temperature of the phonon flux is well approximated by $1/2\pi$ times the average of $\partial_x w$ over a certain width $d_\xi$ [see Eq.~\eqref{eq:kav}].

\subsection{The scaling laws of the width \texorpdfstring{$d_\xi$}{d}}	%
\label{sec:dxi}								%

Unlike $\ka$, $\kt, \nt, \Dt$ which characterize the flow $w$ and $\Lambda$ which fixes the healing length, $d_{\xi}$ is not a parameter entering the mode equation~\eqref{eq:eqphi}, which is directly derived from the Bogoliubov--de Gennes equation~\eqref{eq:BdG1}.
Therefore the value obtained in a fit cannot
be compared with some a priori known value.
In fact $d_\xi$ should be conceived as an {\it  emergent}, effective, length scale.

Before using numerical results to determine how $d_\xi$ scales when changing $\kk$, $\kt$ and $\Lambda$, we point out that we can predict how it should do, as $d_\xi$ is closely related to $\dcrit$ of Eq.~\eqref{eq:dcrittheo}.
Indeed, taking the limits of Eq.~\eqref{eq:kavs} for small and large $\Dt$ we obtain
\begin{equation}\begin{aligned}
 \bar\kappa\left(\Dt\ll 
\frac{\kt d_\xi}{\chor}
\right)&\sim\ka+\frac{2\Dt \chor}{d_\xi},\\
 \bar\kappa\left(\Dt\gg 
\frac{\kt d_\xi}{\chor}\right)&\sim 
\kk,
\end{aligned}\end{equation}
and hence, using Eq.~\eqref{eq:limits} and~\eqref{eq:dcrit}, we get
\begin{equation}\label{eq:hdcrit}
 d_\xi = \frac{2\chor\dcrit}{\kt}.
\end{equation}
In other words, the critical value of $\Dt$ separating the two regimes of $T_0(\Dt)$ shown in Fig.~\ref{fig:fitbark} strictly corresponds to the width $d_\xi$ one should use in Eq.~\eqref{eq:kav} to follow $T_0(\Dt)$.
Using Eq.~\eqref{eq:dcrittheo}, we get
\begin{equation}\label{eq:h}
 d_\xi = d \times \frac{\chor}{(\kk \, \Lambda^2)^{1/3}} ,
\end{equation}
where $d$ is a constant. In terms of the healing length $\xi = \sqrt{2}\chor/\Lambda$, this gives
\begin{equation}\label{eq:hh}
 \frac{d_\xi}{\xi} = d\times\left(\frac{2\xi}{d_K}\right)^{-1/3},
\end{equation}
where $d_K\equiv \chor/\kk$ is the surface gravity length.
Hence, when $\xi \ll d_K$, $d_\xi$ is significantly larger than $\xi$.

The prediction of Eq.~\eqref{eq:h} can be numerically validated by a {linear}\footnote{When the function is linear in the parameters, a close-form solution of the $\chi^2$ minimization problem is available. This method is therefore preferable to non-linear ones.} least-square fit of 
\begin{equation}\label{eq:hfit}
 \log\left(d_\xi \frac{\ka}{\chor}\right)
 =\log(d)+a\log\!\left(\frac{\Lambda}{\ka}\right)+b\log\!\left(\frac{\kt}{\ka}\right)
 +c\log\!\left(1+\frac{\kt}{\ka}\right).
\end{equation}
We used a grid of 110 values of $(\kt,\Lambda)$. 
The corresponding values of $d_\xi(\kt,\Lambda)$ are extracted from series of about 200 couples $(T_0,\Dt)$ using the procedure described after Eq.~\eqref{eq:kav}. $\kt/\ka$ ranges from 0.1 to 1, with step 0.1 and $\Lambda/\ka$ ranges from 10 to 20 with unitary step. The other parameters are $D=0.4$, $n=\nt=1$, $q=0.5$.\footnote{Because of numerical noise, for some series ($\Lambda$,$\kt$), it was not possible to obtain a sufficiently large number of points $(T_0,\Dt)$. In particular it is difficult to obtain $T_0$ for large values of $\Dt$. We therefore decided to use only the series for which we obtained $T_0$ for every value of $\Dt$ from $0$ to $0.1$ (namely 100 points, with step 0.001).
In this way, there are points with $\Dt\gtrsim\dcrit$ in the series $(T_0,\Dt)$ used to determine $d_{\xi}$. This condition is satisfied by 71 series over 110.}
The result of the fit is
\begin{equation}
\label{eq:hfitparams}
\begin{aligned}
 &a=-0.627,\quad &b=0.024,\\
 &c=-0.33,\quad &d=1.58,
 \end{aligned}
\end{equation}
where the numerical error is on the last figure. The exponent $a$ differs from about $5\%$ from its expected value $-0.667$. The exponent $b$ is also close to its expected value 0. Finally, $c$ perfectly matches its theoretical estimate. Furthermore, we notice that the unknown constant $d$ introduced in Eq.~\eqref{eq:h} is of the order of 1.

In brief, for the considered class of flows, we have established that, to some accuracy,
\begin{itemize}
\item the spectrum is Planckian, 
\item the temperature equals $\bar \kappa/2\pi$ for a width $d_\xi$, and
\item $d_\xi$ scales according to Eq.~\eqref{eq:h} with $d \approx 1.6$.
\end{itemize}
From our fits we also found that the broadened horizon is slightly displaced with respect to the Killing horizon by a shift $|\bar{x}_{\rm hor}|\approx0.1d_\xi$.
However, because of the symmetry of Eq.~\eqref{eq:kavs}, we cannot know the sign of $\bar{x}_{\rm hor}$.
To determine it, it is necessary to consider asymmetric flows.

\subsection{Asymmetric flow profiles}	%
\label{subsec:assym}			%

In this section, we consider profiles  of the form
\begin{equation}\label{eq:velocity2gen}
\wtl(x) = \wt(x-\delta) - \wt(-\delta),
\end{equation}
which are simply $\wt$ of Eq.~\eqref{eq:velocity2} shifted from the horizon by $\delta$.
We have subtracted the constant term $\wt(-\delta)$ so that $w_B + \wtl$ still vanishes at $x=0$.
\begin{figure}
\centering
  \includegraphics[width=0.48\textwidth]{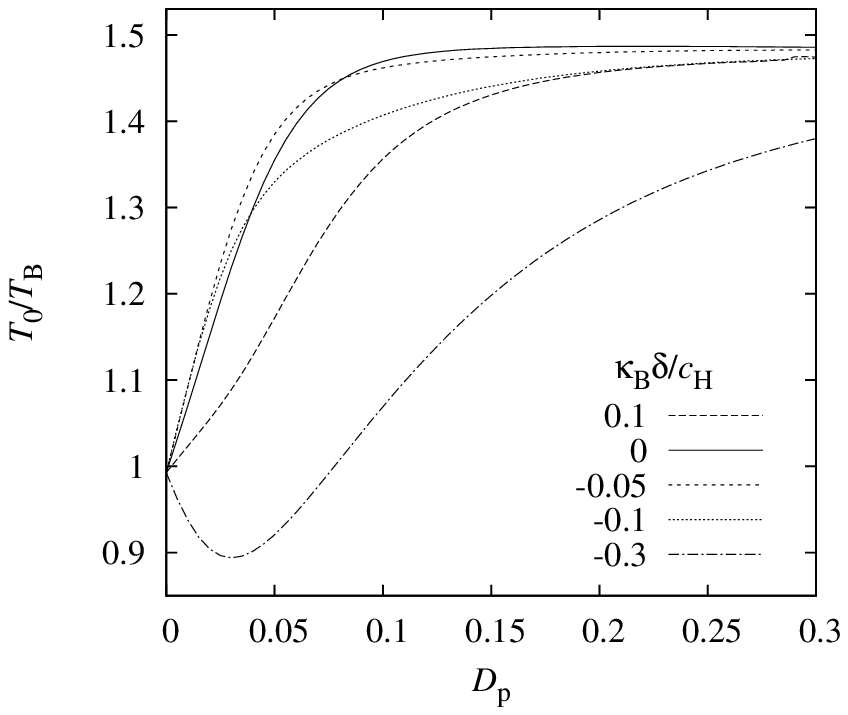}
  \hspace{0.02\textwidth}
   \includegraphics[width=0.48\textwidth]{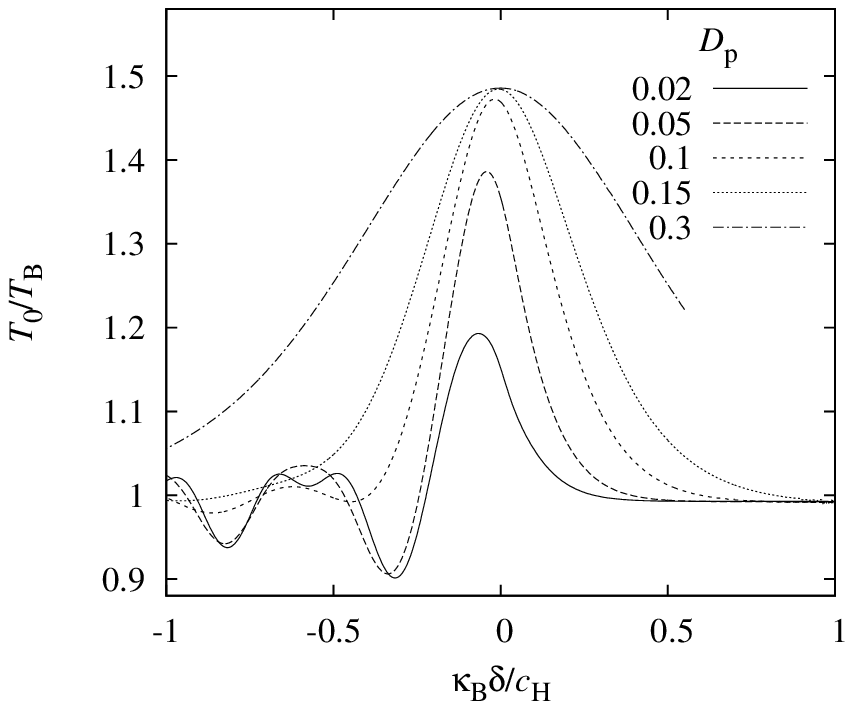}
   \caption{On the left panel, $T_0/\TB$ as a function of $\Dt$ for various values of $\delta$, and, on the right, $T_0/\TB$ as a function of $\delta$ with $\Dt$ from $\Dt=0.02$ to $\Dt=0.3$. $\Lambda/\ka = 15$, $D = 0.4$, $\kt/\ka=0.5$, $n=\nt=1$.
   On the left panel, for values of $\delta \geq -0.05$, $T_0(\Dt)$ behaves as in Fig.~\ref{fig:fitbark}, whereas, for more negative values of $\delta$, $T_0$ behaves differently.
   On the right panel, the asymmetry of $T_0$ under $\delta \to - \delta$ is clearly visible for $\Dt < \dcrit\approx0.06$.}
  \label{fig:LS}
\end{figure}
In Fig.~\ref{fig:LS}, $T_0$ is plotted as a function of $\Dt$ for different $\delta$ (left plot), and as a function of $\delta$ for different $\Dt$ (right plot).
The behavior of these curves is quite complicated. To understand it, we first consider {\it smooth} perturbations, \ie\/ perturbations with $\Dt>\dcrit\approx0.06$.

From the right panel of Fig.~\ref{fig:LS}, when $\Dt>\dcrit$, but for both signs of $\delta$, we see that $T_0$ monotonically interpolates from $\Th=\kk/2\pi$, when $\delta$ is small and the perturbation is close to the horizon, to $\TB=\ka/2\pi$, when $\delta\to\infty$. In that case, the perturbation is far from the horizon and thus no longer contributes to $\kk$ which equals its background value $\ka$.
When considering instead well localized perturbations, that is values of $\Dt$ smaller than $\dcrit$, we see that $T_0$ oscillates, but only for $\delta$ sufficiently negative, \ie\/ when the perturbation is sufficiently displaced in the region where the {\it decaying mode} oscillates.\footnote{We have adopted this formulation because, for subluminal dispersion, we expect that the sign of $\bar x_{\rm hor}$ will be the opposite, and this in virtue of the ``symmetry'' between the behavior of modes for sub and superluminal dispersion~\cite{cfp}.
Indeed, for (sub) superluminal dispersion the turning point is in the (sub) super-sonic region, see Eq.~\eqref{eq:tp}.
In both cases, the mode decays on the other side of the turning point~\cite{CJ96,MacherRP1}.}
The origin of these oscillations can be understood as follows.
The scattering on the well localized bump in $\partial_x w(x)$ interferes with the scattering on the horizon, thereby producing oscillatory behaviors.\footnote{This is very similar to what has been found in inflation. When introducing a sharp modification of mode propagation before horizon exit, there are interferences between this localized scattering and the standard mode amplification which produce oscillations in the power spectrum, see~\cite{CNP} for references and a critical analysis on the generic character of this.}

It is quite clear that the average of Eq.~\eqref{eq:kav} cannot describe these oscillations.
However this should {\it not} be conceived as limiting the validity of the observation that the temperature is given by this average. Indeed, these oscillations result from interferences between two spatially separated scatterings. In these circumstances it is meaningless to apply Eq.~\eqref{eq:kav}, which is based on the near horizon properties of $w$.

In addition, when oscillations are found for $T_0$ as a function of $\Dt$ or $\delta$, the Planck character of the spectrum is {\it also} lost.
\begin{figure}
\centering
\includegraphics[width=0.48\textwidth]{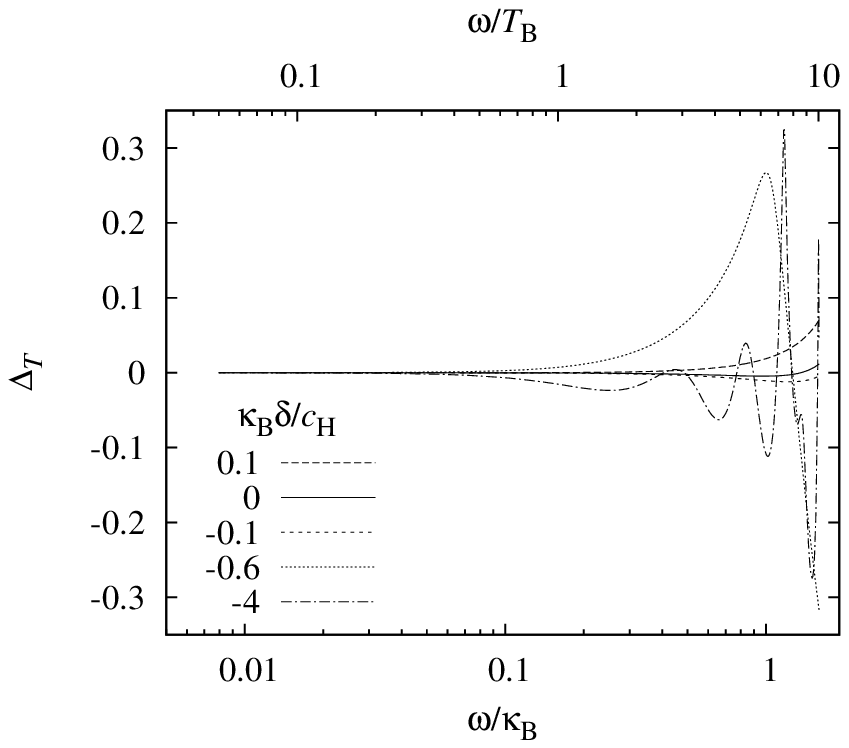}
\hspace{0.02\textwidth}
\includegraphics[width=0.48\textwidth]{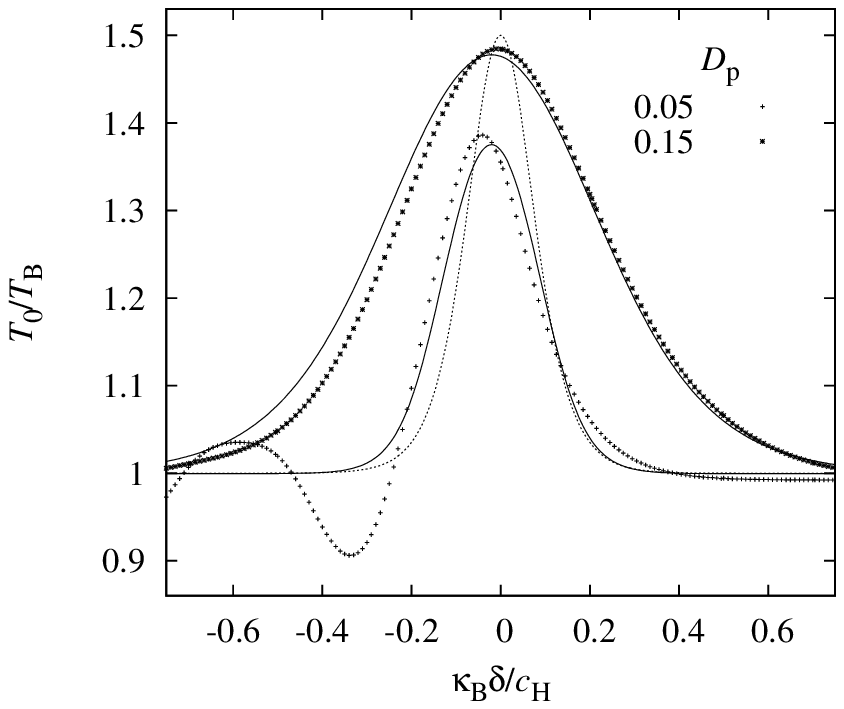}
\caption{Left panel: The parameter $\run$ of Eq.~\eqref{eq:run} versus $\omok$ for different values of $\delta$.
For $\delta=0$, continuous line, the perturbation is symmetric with respect to the horizon, and $\run$ is extremely small in agreement with Eq.~\eqref{eq:run2}.
When $\delta > 0$ and $(-\delta) \leq d_\xi \approx 0.2 \, \chor/\ka $, $\run$ stays very small and no oscillations are found.
Instead when $-\delta \gg d_\xi$, oscillations of large amplitudes appear even for low frequencies, \ie, $\om \ll \omm \approx 13 \,\TB$, thereby indicating that the spectrum is no longer Planckian. The fixed parameters are $\Lambda/\ka = 15$, $D = 0.4$, $\Dt=0.02$, $\kt/\ka=0.5$, $n=\nt=1$. Right panel: Numerical data (dots) for the series $T_0(\delta)/\TB$ with $\Dt=0.15\approx 2.5\dcrit$ and $0.05\approx0.8\dcrit$ of Fig.~\ref{fig:LS}, right plot. The solid lines are the values of the fitted function of Eq.~\eqref{eq:kav}.
The dotted line represents the surface gravity $\kk(\delta)/2\pi$ for $\Dt=0.05$.
The agreement between the temperature and $\bar \kappa/2\pi$ is quite good in the domains with no interference.
The lowering of the peak and its shift to the left for small values of $\Dt$ are rather well reproduced.
None of these effects are found when using the surface gravity $\kk(\delta)$.
}
\label{fig:LS_run_fit}
\end{figure}
Indeed, the parameter $\run$ of Eq.~\eqref{eq:run} characterizing deviations from Planckianity also displays oscillations (of relative amplitude $\sim 20 \%$) (see Fig.~\ref{fig:LS_run_fit}, left panel).
As can be seen, the number of peaks increases with increasing values of $-\delta$.
This is to be expected since more ``room'' is available to fit resonating modes. In fact this phenomenon can be viewed as a precursor of the black hole laser effect~\cite{cj}.
If the perturbation were strong enough to behave as a white horizon, the resonating modes would start growing exponentially~\cite{cp,bhlasers} as shown in Chap.~\ref{chap:bhlaser}. In that case complex-frequency modes appear when the distance between the horizons is sufficiently large, and their number increases with that distance.

To complete this analysis of the flows of Eq.~\eqref{eq:velocity2gen}, we verified that $\bar \kappa$ of Eq.~\eqref{eq:kav} still approximately governs the temperature.
In Fig.~\ref{fig:LS_run_fit} (right panel), both the numerical data (dots) and the fitted function (solid line) are reported for the
series $T_0(\delta)$, with $\Dt=0.05$ and $0.15$, of Fig.~\ref{fig:LS} (right panel). This function is obtained as a two-variable function of $\delta$ and $\Dt$ by using the numerical data of $T_0$. The fitting region is restricted to those values of $\delta$ for which there are no oscillations and for three values of $\Dt$, namely $\Dt=0.05,0.1,0.15$.
As a reference, the dotted line represents $\kk(\delta)/2\pi$, \ie\/ the temperature one would obtain without dispersion, for $\Dt=0.05$.
As a check, we also fitted  the values of $\ka$, $\kt$ and $\nt$ and compared them with the corresponding constants used in the numerical analysis. The relative difference is very small, respectively, around $0.1\%$, $1\%$ and $5\%$.

As one can see in Fig.~\ref{fig:LS_run_fit} (right panel), the agreement between the actual temperature $T_0$ and the average surface gravity $\bar \kappa/2\pi$ is quite good in the parameter region where there are no interference effects. 
One can also see that for $\Dt = 0.05$, $\bar \kappa/2\pi$ of Eq.~\eqref{eq:kav} follows $T_0$ much more closely than the 
surface gravity $\kk(\delta)/2\pi$.

In summary, the analysis of asymmetrical perturbations first indicates that the shift $\bar x_{\rm hor}$ is always negative. That is, the broadened horizon is displaced towards the region where the {decaying mode} oscillates. Moreover we find that the value of the shift does not depend directly on $\delta$.
By this we mean that $\bar x_{\rm hor}$ depends on $\delta$ only because $d_\xi$ of Eq.~\eqref{eq:h} is a function of $\delta$ through $\kk(\delta)$.
Second, for sufficiently smooth perturbations we found that the temperature is still approximately given by the average of Eq.~\eqref{eq:kav}.
Finally, when the perturbation is sharp and sufficiently deep inside the supersonic region, we found that the spectrum is affected by interferences and explained why it is so.

\subsection{Undulations}	%
\label{sec:oscillation}		%

Supersonic flows, possessing a sonic horizon with a negative {surface gravity} $\kk$, behave like the time reverse of black hole horizons: long wavelength modes are blueshifted when scattered on such a horizon. Thanks to dispersion, the blueshift effect saturates
and, as a result, the scattering matrix is well-defined. Moreover, for each $\om$, this matrix is closely related to Eq.~\eqref{eq:bog_transf} evaluated in the corresponding black hole flow~\cite{MacherRP1,MacherBEC}.
In spite of this correspondence, it has been recently observed that white hole flows display a specific phenomenon related to the limit $\om \to 0$~\cite{Carusottowhite,Silke_exp}.
Indeed, these flows produce a zero-frequency mode, a stationary undulation, which has a finite wavelength $k_Z$ and a high amplitude.

\begin{figure}
 \centering
 \psfrag{x}[c][c]{$x$}
 \psfrag{w}[c][c]{$\wt(x)$}
 \psfrag{mA}[c][c]{$-A$}
 \psfrag{A}[c][c]{$A$}
 \psfrag{md}[c][c]{$-\Delta$}
 \psfrag{d}[c][c]{$\Delta$}
 \psfrag{kz}[c][c]{$2\pi/k_Z$}
 \psfrag{th}[c][c]{$\theta/k_Z$}
 \includegraphics[width=0.8\textwidth]{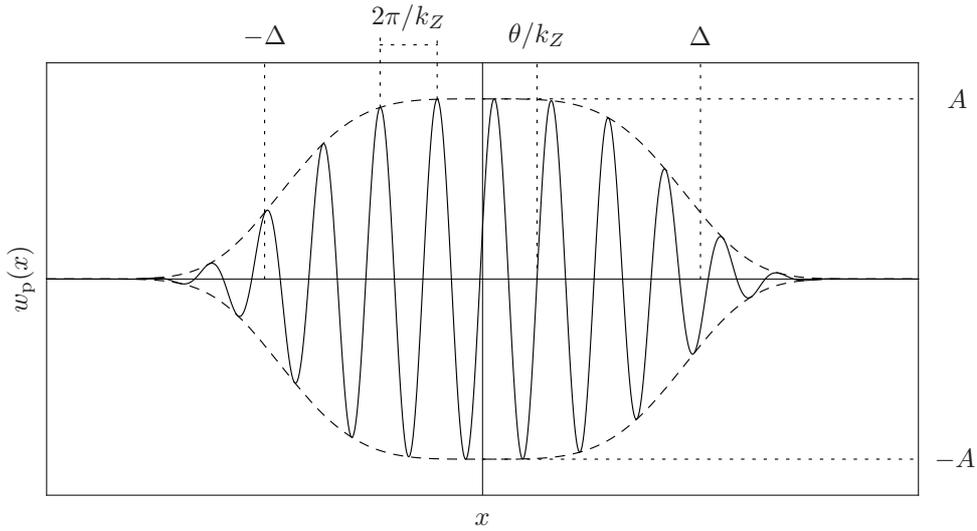}
 \caption{Velocity profile perturbation $\wt$ defined in Eq.~\eqref{eq:periodic} (solid line) and its exponential envelope $\exp[-(x/\Delta)^4)]$ (dashed lines). The geometrical meaning of the various parameters is evident from this graphical representation.}
 \label{fig:sinexp}
\end{figure}

To schematically describe such undulation and study its impact on the emitted spectrum, we consider the following perturbations: 
\begin{equation}\label{eq:periodic}
\frac{\wt(x)}{\chor} = A\,  \sin\left(k_Z x-\theta\right) \, \ee^{-(x/\Delta)^4}.
\end{equation}
$A$ is the amplitude of the undulation, and $\theta$ fixes its phase on the horizon at $x=0$. The new length $\Delta$ is taken to be larger than $1/k_Z$, and comparable to $D \chor/\ka$ which characterizes the linear regime of the background flow $\wB$ (see Fig.~\ref{fig:sinexp} for a graphical representation).

\begin{figure}
\centering
  \includegraphics[width=0.48\textwidth]{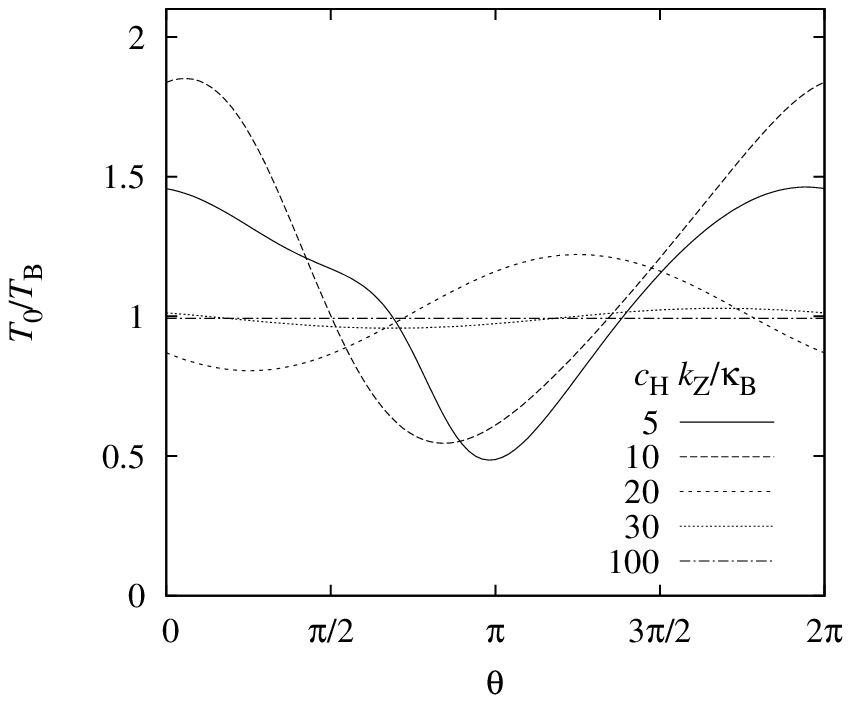}
   \caption{$T_0/\TB$ as a function of $\theta$ for Eq.~\eqref{eq:periodic}, varying $k_Z$ and $A$ so that $A\, k_Z=0.5\ka$ stays fixed. $\Delta=2 D \chor/\ka$, $\Lambda/\ka = 15$, $D = 0.4$, $n=1$.
   For very high values of $k_Z$, \ie\/ $k_Z = 100$ and $30$ $\ka/\chor$, the temperature is approximately $\ka/2\pi$, \ie\/ equal to $\bar \kappa/2\pi$ of Eq.~\eqref{eq:kav}.
   For small values of $k_Z$, \ie\/ $k_Z =  5 \, \ka/\chor$, $T_0$
   also approximately follows $\bar \kappa(\theta)/2\pi\sim \kk(\theta)/2\pi$ since the undulation wavelength is larger than $d_\xi$. In the intermediate regime instead, for $k_Z = 10$ and $20 \, \ka/\chor$, the temperature  widely oscillates and no longer follows $\bar \kappa/2\pi$.}
\label{fig:T0_delta}
\end{figure}

In Fig.~\ref{fig:T0_delta}, $T_0$ is plotted versus $\theta$ for different values of $k_Z$, keeping fixed $A\, k_Z$ so that the surface gravity of Eq.~\eqref{eq:temprel} does not change with $k_Z$. Indeed one gets $\kk(\theta) = \ka+ A\,k_Z \cos(\theta)$.
When $k_Z \gg 1/d_\xi$, as expected from our previous results, the temperature is very close to $\ka/2\pi$, that of the background flow $\wB$.
Hence, independently of the value of $\theta$, the perturbation is completely washed out by dispersion.
On the contrary, when the wavelength is larger than the width of the horizon, \ie\/ when $k_Z < 2\pi/d_\xi \approx 30 \ka/\chor$, we expect that $2\pi\, T_0$ will be approximately given by $\kk(\theta)$ given above.
This can be observed for $k_Z$ equal to $5\, \ka/\chor$.
One notices that the curve is not exactly symmetric around $\theta=\pi$.
But this is also expected from the former analysis where we saw that moving the perturbation in the supersonic or in the subsonic region affects differently the spectrum.

The novelties are found for intermediate values of $k_Z$ (say, between $10\ka/\chor$ and $20\ka/\chor$). For these values, $\delta_Z = 2\pi/k_Z$ is comparable to $d_\xi$. Hence the undulation can enter into resonance with the scattering on the horizon, thereby producing a kind of parametric amplification of the flux.
This can be observed for $\theta = 0 $ and $k_Z = 10\ka/\chor$ where the temperature is larger than $\kk/2\pi = 1.5 \TB$.
The fact that this resonance occurs for wavelengths $\sim d_\xi$, reinforces the fact that the effective width of the horizon is indeed $d_\xi$.

\section{Summary and discussion}	%
\label{sec:conclusion_broadhor}		%

By considering series of supersonic flows consisting of local perturbations $\wt$ defined on the top of a background flow $\wB$, we were able to identify the parameters that fix the spectral properties of Bogoliubov phonons emitted by a (black hole)
sonic horizon.

We showed that the flux remains Planckian to high accuracy, see Eq.~\eqref{eq:run2}, even when the temperature strongly differs from $\Th = \kk/2\pi$, the standard value fixed by the surface gravity, see Fig.~\ref{fig:tom_fom}, right panel.
This result implies two things.  In spite of the fact that each mode of frequency $\om$ has its own turning point of Eq.~\eqref{eq:tp}, the Planckianity first implies that the temperature function $T_\om$ of Eq.~\eqref{eq:Tom} is {\it not} determined by some function of $w(x)$ evaluated at an $\om$-dependent location, as $T_\om$ is common to all low frequencies with respect to $\omm$ of Eq.~\eqref{eq:omm}.
Second, it implies that the parameters governing the small deviations from Planckianity differ from those governing the difference between the low frequency temperature $T_0$ and $\Th$.
In fact, whereas the difference $T_0 - \Th$ is governed by the {\it near horizon} properties, deviations from Planckianity are governed by inverse powers of $\omm/T_0$, where $\omm$ is fixed by $D_{\rm asympt} = D + \Dt$, characterizing the {\it global} properties of the supersonic transition region, such as the asymptotic velocity excess.

Moreover, in contradistinction to the relativistic case where the temperature is locally fixed by $\kk$ of Eq.~\eqref{eq:temprel}, we show that the temperature $T_0$ is determined, to a high accuracy when $\omm/\Th \gtrsim 10$, by a {\it nonlocal} quantity: the spatial
average across the horizon of $\partial_x w$ over a width $d_\xi$, as if the horizon were broadened.\footnote{This is in agreement with what was found in~\cite{cp} when considering the black hole laser effect~\cite{cj}. In that case, the effect disappears when the distance $\delta_{\rm H}$ between the black and the white horizon becomes too small, because the dispersive field is unable to ``resolve'' the two horizons. When using linear profiles for both horizons, one finds that the critical value of $\delta_{\rm H}$ is about $d_\xi$.}
Hence, whenever $w(x)$ hardly varies over $d_\xi$ in the near horizon region, $\bar \kappa$ of Eq.~\eqref{eq:kav} equals the surface gravity $\kk$, and the standard temperature is found.
Instead, when $w$ varies on scales shorter than $d_\xi$, $T_0$ is no longer the standard one, but it is well approximated by $\bar \kappa/2\pi$ for all frequencies $\om \ll \omm$.
This result is reinforced by the fact that the effective horizon width $d_\xi$ scales according to a well-defined law given in
Eq.~\eqref{eq:h}.
Under this new perspective, the findings of~\cite{Recati2009}, where the emitted spectrum is still Planckian are not surprising. In fact the average $\bar\kappa$ is well defined and finite, even in the extremal case in which the surface gravity $\kk$ diverges and the Hawking temperature $\Th$ is not definite.

Furthermore, the above results are basically valid for all flows. They cease to be valid under the same restricted conditions, namely when the flow contains a sharp perturbation well localized far enough from the horizon.
Then the perturbation induces a scattering that interferes with the Hawking effect, thereby engendering oscillations.
These are similar to those found in inflationary spectra, and are also attributable to interferences between two scatterings.
It is clear that the local average of Eq.~\eqref{eq:kav} cannot describe these oscillations which are {nonlocal} on a scale much larger than the width $d_\xi$.
Hence we can say that when $\omm/\Th \gtrsim 10$ and when the scattering essentially arises near the horizon, the spectrum is near Planckian, and with a temperature approximately given by $\bar \kappa$ of Eq.~\eqref{eq:kav}.
Finally, to mimic what is found in white hole flows~\cite{Carusottowhite,Silke_exp}, we studied profiles containing an undulation on the top of a smooth profile. By varying its wavelength, we have found three different behaviors that match the above analysis.

To conclude, we stress that, even if these results on acoustic black holes in realistic BEC flows are quite solid, further analyses are required in order to apply those arguments to real black holes, with modified superluminal dispersion relation modelled on the phononic one. As discussed in Sec.~\ref{subsec:bogo}, the situation is complicated by the fact that a black hole is originated by a star collapse which might not naturally select an Unruh-like state, because the superluminal dispersion relation would tame any divergence on the horizons for non-Unruh states. Moreover, the presence of a singularity or of a strong curvature region would not be hidden beyond the horizon because there would exist modes that would be able to cross the horizon outward. Thus, fixing the boundary conditions in this strong curvature region would be quite problematic. There is no {\it a priori}\/ reason why, after the collapse, the initial vacuum state should appear as if the modes coming out from the singularity were in their vacuum state.

\chapter{Warp drives in Bose--Einstein condensates}	
\chaptermark{Warp drives in BECs}			
\label{chap:warpdriveBEC}				

One of the main objectives of this Thesis is to show the importance of analogue gravity as a set of tools for the investigation of interesting problems both in condensed matter and in general relativity.
The analysis of this chapter is a significant example of this twofold usefulness of analogue models of gravity.
In fact, we extend here the analysis of Chap.~\ref{chap:hawking} to Bose--Einstein condensates (BECs) with two horizons, separating three regions: one internal subsonic region and two external supersonic ones. The coupling of a black and a white acoustic horizon introduces new interesting and nontrivial effects with respect to the single horizon scenario. First, this system is mildly unstable, showing a linear divergence (in time) of the detection rate measured by a detector coupled with density fluctuations. Second, in this framework, one can study the limiting case in which the two horizons collapse into a single point. In that case, the horizons and the subsonic region  disappear. Interestingly, some particle production is still present, confirming the first results of~\cite{robustness}, described in Sec.~\ref{sec:asymmetric}.

Furthermore, the acoustic metrics associated with this class of flows have the same features of the warp-drive spacetimes investigated in Chap.~\ref{chap:warpdrive}. However, in the present system, the phonon field is described by a dispersion relation that naturally shows supersonic modifications at large frequencies. Thus, this analysis allows to learn some lessons about how the instability found in Chap.~\ref{chap:warpdrive} would be affected by trans-Planckian physics in the case a superluminal modification of the dispersion relation of quantum fields took place at trans-Planckian frequencies.
This sort of studies is of fundamental importance because, as broadly discussed in Chap.~\ref{chap:warpdrive}, the possibility of violating causality (\eg\/ of realizing time travel) is strictly related to spacetime shortcuts allowing faster-than-light travel~\cite{everett}.
In spacetimes described by analogue metrics closed timelike curves are forbidden because these geometries inherit the property of stable causality from the underlying Newtonian ``world'' endowed with an absolute time. Nevertheless, it is interesting to understand if chronology is protected by some more fundamental mechanism even in spacetimes that do not emerge from a Galilean invariant system.
In particular, the result of the present chapter is in favor of the instability of dynamical warp drives, even in the presence of superluminal dispersion relations. However, this instability is only linear and it is therefore much milder than the exponential growth of the renormalized stress-energy tensor for scalar fields with relativistic dispersion relation.

\section{Mode analysis and scattering matrix}	%
\label{sec:scatteringwarpdrive}			%

\begin{figure}
 \centering
 \psfrag{bh}[c][c]{BH}
 \psfrag{wh}[c][c]{WH}
 \psfrag{w}[c][c]{$w(x)/\chor$}
 \psfrag{x}[c][c]{$\kappa x/\chor$}
 \psfrag{tl}[c][c]{$2L$}
 \psfrag{one}[c][c]{I}
 \psfrag{two}[c][c]{II}
 \psfrag{thr}[c][c]{III}
 \includegraphics[width=0.85\textwidth]{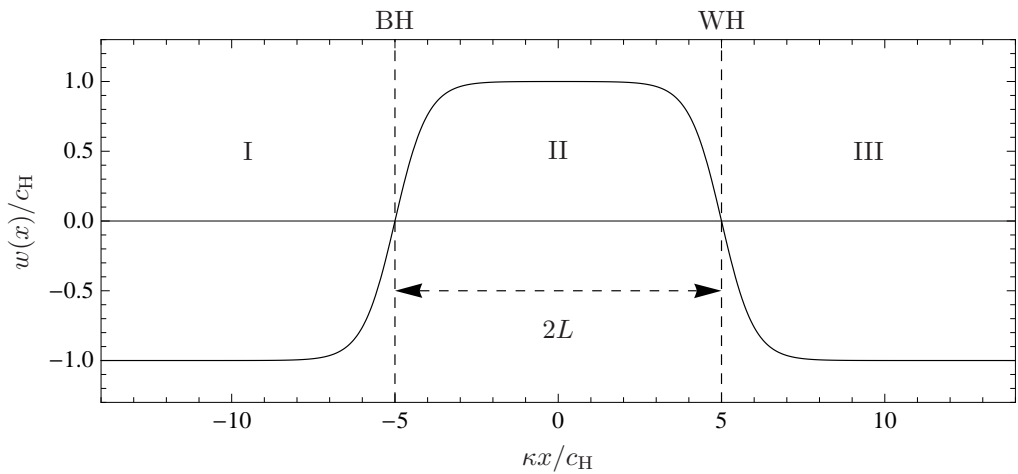}
 \caption{\label{fig:profile_4x4}
 Velocity profile $w=c+v$ of Eq. \eqref{eq:velocity4x4}. The dashed lines denote the horizons. In the central region II the flow is subsonic ($-v<c$), in region I and III it is supersonic ($-v>c$).}
\end{figure}

In this chapter we work with a specific class of flows, showing the same qualitative features of the warp-drive geometries investigated in Chap.~\ref{chap:warpdrive}. Thus, at the same time, we shall investigate an interesting set up in BECs and we shall shed some light on how a superluminal modification of the dispersion relation affects the results of Chap.~\ref{chap:warpdrive}.
For concreteness, let us take a $1+1$ dimensional (see discussion in Sec.~\ref{sec:generalmetric} about acoustic geometries in $1+1$ dimensions) left-moving fluid ($v<0$) with a velocity profile with a black and a white acoustic horizons (see Fig.~\ref{fig:profile_4x4}), obtained by properly combining two profiles of the form of Eq.~\eqref{eq:velocity_simp}:
\begin{equation}\label{eq:velocity4x4}
 c(x)+v(x)=w(x)=-\chor D\, \sign(x^2-L^2) 
 \tanh^{1/n}\!\!\left[\left(\frac{\kb|x+L|}{\chor D}\right)^{\!\!n}\right] \tanh^{1/n}\!\!\left[\left(\frac{\kw|x-L|}{\chor D}\right)^{\!\!n}\right].
\end{equation}
We define regions I, II, and III as the three regions with $x<-L$, $-L<x<L$, and $x>L$, respectively. In region II $w>0$, \ie, the fluid is subsonic. On the contrary, in region I and III $w<0$ and the fluid is supersonic.
From the point of view of an observer sitting in region II, $x=-L$ and $x=L$ appear as a black and a white horizon, respectively. Accordingly, we have called $\kb$ and $\kw$ the corresponding surface gravities. $D$ is a parameter defining the asymptotic value of the velocity profile. Finally, the velocity of the fluid $v$ and the speed of sound $c$ are fixed by introducing a new constant parameter $q$, determining how $w$ is shared between $v$ and $c$ [see Eq.~\eqref{eq:cv}]. To minimize the scattering between right- and left-going modes we work with $q=1/2$ [see comment after Eq.~\eqref{eq:cv}].

In Sec.~\ref{subsec:UVsuperluminal} and in Chap.~\ref{chap:hawking} we broadly described the properties of the various solutions of a dispersion relation with supersonic modifications (see Fig.~\ref{fig:dispersion}). When the flow is supersonic, we found four propagating modes with real wavenumber, while, when the flow is subsonic, in addition to the two standard solutions with real $k$, there are two roots with complex wavevector, respectively corresponding to decaying and growing modes at $x\to+\infty$ ({\it vice versa}\/ for $x\to-\infty$).
Furthermore, in an acoustic single black-hole geometry, the presence of those two extra-roots (real in the supersonic region and complex in the subsonic one) generates three globally defined asymptotically bounded modes (ABM) for each value of the Killing frequency $\omega$.

In the present situation, where the flow is subsonic in the compact region II and supersonic in the two asymptotic external regions I and III, the number of ABM sharing the same conserved frequency $\om$ is 4. Indeed, in regions I and III, there are 4 propagating modes with real wavenumber. Moreover, all the four solutions of the dispersion relation in the subsonic region II are physically acceptable. In fact, also the two modes with complex wavenumber are bounded, because of the compactness of region II.

Starting from the mode expansion of $\hp$ in the supersonic regions
\begin{multline}\label{eq:expsup2}
 \hp_{\rm sup}=\int_0^{\infty}\!\dom\left\{ \ee^{-\ii\om t} \left[ \phi_\om^u \ha_\om^u+\phi_\om^v\ha_\om^v
+ \theta(\ommax - \om)
\sum_{i=1,2}(\varphi_{-\om}^{(i)})^*\,\ha_{-\om}^{{(i)}\dagger}\right] \right.
 \\
\left. + \ee^{+ \ii\om t} \left[ (\varphi_\om^u)^*\,\ha_\om^{u\, \dagger}+(\varphi_\om^v)^*\,\ha_\om^{v\, \dagger}
+ \theta(\ommax - \om)
\sum_{i=1,2}\phi_{-\om}^{(i)}\ha_{-\om}^{{(i)}}\right] \right\},
\end{multline}
we can identify the four globally defined modes in the black-hole--white-hole geometry.
Using the same notation of Chap.~\ref{chap:hawking}, $\phi_\om^u$ and $\phi_\om^v$ represent the two standard positive-norm right- and left-going waves (see discussion in Sec.~\ref{subsec:bogo}). Analogously,  $(\varphi_{-\om}^{(1)})^*$ and $(\varphi_{-\om}^{(2)})^*$ are the two extra modes with negative norm and wavevectors $\kone$ and $\ktwo$, which are the two extra solutions of the dispersion relation~\eqref{eq:dispersion1} with the most negative and the less negative values of $k$, respectively (see Fig.~\ref{fig:dispersion}). Accordingly to these definitions, $(\varphi_{-\om}^{(1)})^*$ has positive group velocity and $(\varphi_{-\om}^{(2)})^*$ has negative group velocity.

\begin{figure}
\psfrag{lone}[c][c]{$(\varphi_{-\om}^{(1),\rm I})^*$}
\psfrag{ltwo}[c][c]{$(\varphi_{-\om}^{(2),\rm I})^*$}
\psfrag{rone}[c][c]{$(\varphi_{-\om}^{(1),\rm III})^*$}
\psfrag{rtwo}[c][c]{$(\varphi_{-\om}^{(2),\rm III})^*$}
\psfrag{ru}[c][c]{$\phi_\om^{u,\rm III}$}
\psfrag{lu}[c][c]{$\phi_\om^{u,\rm I}$}
\psfrag{rv}[c][c]{$\phi_\om^{v,\rm III}$}
\psfrag{lv}[c][c]{$\phi_\om^{v,\rm I}$}
\includegraphics[width=0.48\textwidth]{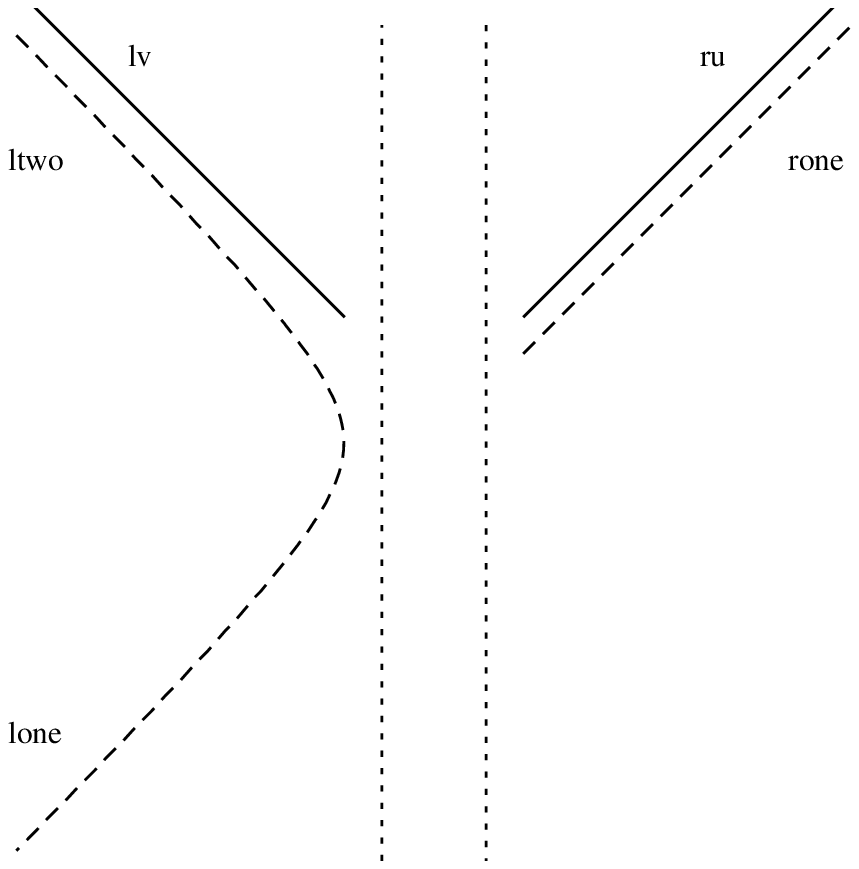}
 \includegraphics[width=0.48\textwidth]{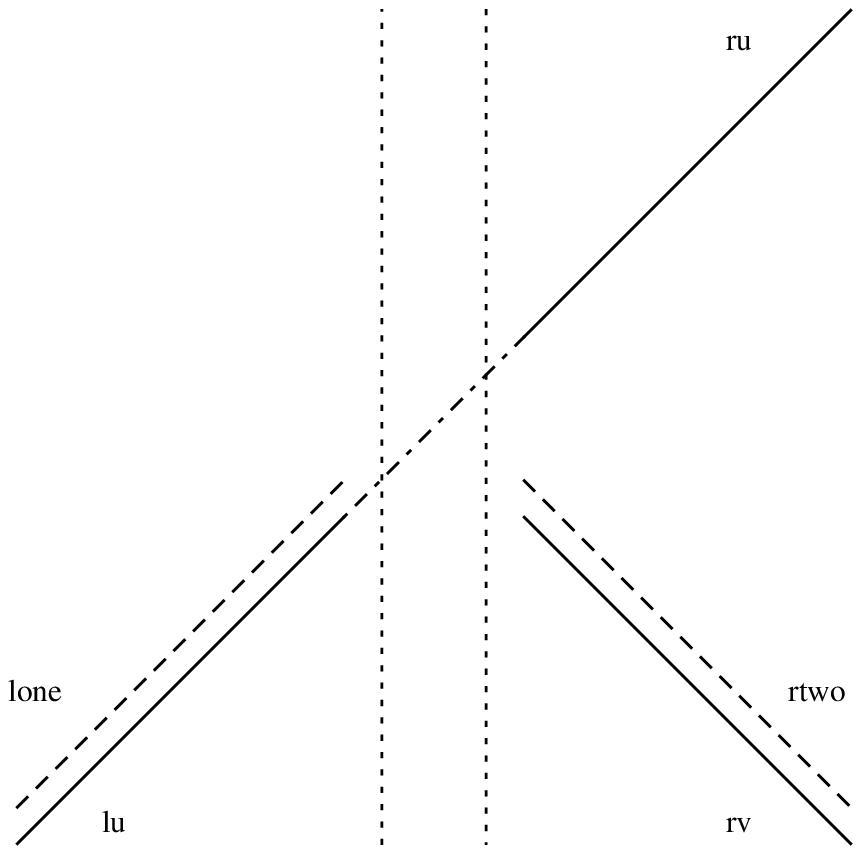}
 \caption{Left panel: graphical representation of the in mode $(\varphi_{-\om}^{(1),\rm in})^*$ in the plane ($x$, $t$). $(\varphi_{-\om}^{(1),\rm in})^*$ has only one incoming branch $(\varphi_{-\om}^{(1),\rm I})^*$, in region I, and 4 outgoing branches, two left-going in region I and two right-going in region III. Right panel: out mode $\phi_\om^{u,\rm out}$, with a single outgoing branch $\phi_\om^{u,\rm III}$, in region III, and 4 ingoing branches, two right-going in region I and two left-going in region III. Positive (negative) norm modes are represented by solid (dashed) lines. The dotted vertical lines are the horizons.}
 \label{fig:mode4x4}
\end{figure}

Along the lines of what we did for geometries with a single horizon in Sec.~\ref{subsec:bogo}, it is possible to define a basis of incoming modes, having only one asymptotic branch with group velocity directed toward region II, either from region I or III, and a basis of outgoing modes, having only one asymptotic branch propagating away from the horizons.
For instance, $(\varphi_{-\om}^{(1),\rm in})^*$ is the mode with unitary projection on the right-going branch $(\varphi_{-\om}^{(1),\rm I})^*$ in the asymptotic region I, while its projections on the right-going $\phi_\om^{u,\rm I}$ in region I and on the left-going  $\phi_\om^{v,\rm III}$ and $(\varphi_{-\om}^{(2),\rm III})^*$ in region III all vanish (see Fig.~\ref{fig:mode4x4}, left panel).
As a further example, the projection of the out mode $\phi_\om^{u,\rm out}$ on the right-going branch $\phi_\om^{u,\rm III}$ in the asymptotic region III is 1, while its projections on the right-going $(\varphi_{-\om}^{(1),\rm III})^*$ in region III and on the left-going $\phi_\om^{v,\rm I}$ and $(\varphi_{-\om}^{(2),\rm I})^*$ in region I all vanish (see Fig.~\ref{fig:mode4x4}, right panel).

Since the globally defined ABM are 4, the scattering matrix between in and out modes must be $4\times4$.  With standard notation, we denote with
\begin{itemize}
 \item $\alpha$: the projections of in modes on the corresponding out modes,
 \item $\beta$: the projections of the right-going positive norm mode $u$ on the negative norm modes (and {\it vice versa}),
 \item $A$: the projections of the right-going positive norm mode $u$ on the left-going positive norm mode $v$ (and {\it vice versa}), or the projections of one negative norm mode on the other one (and {\it vice versa}).
 \item $B$: the projections of the left-going positive norm mode $v$ on the negative norm modes (and {\it vice versa}).
\end{itemize}
With this notation, the in modes are related to the out modes through the following expressions
\begin{equation}
\begin{aligned}
\phi^{u,\rm in }_{\om}
	&= \alpha_{\om}^u \, \phi^{u,\rm out}_{\om}
	+ \beta^{(1)}_{-\om}\left(\varphi^{(1),\rm out}_{-\om}\right)^*
	+ \beta^{(2)}_{-\om}   \left(\varphi^{(2),\rm out}_{-\om}\right)^* 
	+ A_{\om} \,  \phi^{v,\rm out}_{\om},
\\
\left(\varphi^{(1),\rm in }_{-\om}\right)^*
	&= \beta^{(1)}_{\om}\, \phi^{u,\rm out}_{\om}
	+ \alpha_{-\om}^{(1)} \left(\varphi^{(1),\rm out}_{-\om}\right)^*
	+ A_{-\om}  \left(\varphi^{(2), out}_{-\om}\right)^*
	+ \tilde B^{(1)}_{\om}\, \phi^{v,\rm out}_{\om}
\\
\left(\varphi^{(2),\rm in }_{-\om}\right)^*
	&= \beta^{(2)}_{\om}\, \phi^{u,\rm out}_{\om}
	+ \tilde A_{-\om}  \left(\varphi^{(1),\rm out}_{-\om}\right)^*
	+ \alpha_{-\om}^{(2)} \left(\varphi^{(2), out}_{-\om}\right)^*
	+ \tilde B^{(2)}_{\om}\, \phi^{v,\rm out}_{\om}	
\\
\phi^{v,\rm in }_{\om}
	&= \alpha^v_{\om} \, \phi^{v,\rm out}_{\om}
	+ B^{(1)}_{-\om}   \left(\varphi^{(1),\rm out}_{-\om}\right)^*
	+ B^{(2)}_{-\om}   \left(\varphi^{(2),\rm out}_{-\om}\right)^*
	+ \tilde A_{\om}\,  \phi^{u,\rm out}_{\om}.
\label{eq:bogo4x4}
\end{aligned}
\end{equation}
The standard mode normalization [see Eq.~\eqref{eq:scalarphi1}] yields relations such as [from the first equation of~\eqref{eq:bogo4x4}]
\begin{equation}
 |\alpha_\om|^2 - |\beta_{-\om}^{(1)}|^2 - |\beta_{-\om}^{(2)}|^2 + |\tilde A_\om|^2 = 1,
\end{equation}
where the minus signs come from the normalizations of the negative norm modes $(\varphi^{(i)}_{-\om})^*$.

Starting with the in vacuum, and using Eq.~\eqref{eq:bogo4x4} to express the out operators in terms of in ones, the mean occupation numbers of the four types of phonons are
\begin{align}
n^{u,\rm out}_{\om}
&=\langle 0_{\rm in} | \hat a^{u, \rm out \dagger}_\om \hat a^{u,\rm out}_\om  | 0_{\rm in} \rangle 
= | \beta^{(1)}_{\om}|^2 + | \beta^{(2)}_{\om} |^2, 
\\ 
n^{v,{\rm out}}_{\om} 
& =\langle 0_{\rm in} | \hat a^{v, \rm out \dagger}_\om \hat a^{v, \rm out}_\om | 0_{\rm in} \rangle 
= | \tilde B^{(1)}_{\om} |^2 + | \tilde B^{(2)}_{\om} |^2, 
\\
n^{(i),{\rm out}}_{-\om} 
&= \langle 0_{\rm in} | \hat a^{(i),\rm out \dagger}_{-\om } \hat a^{(i),\rm out}_{-\om} | 0_{\rm in} \rangle 
= |\beta^{(i)}_{-\om} |^2 +    |B^{(i)}_{-\om} |^2,
\label{fouroccn}
\end{align}
and the final occupation numbers are related by
\begin{equation}
  n^{u,{\rm out}}_{\om} +  n^{v,{\rm out}}_{\om} =
 n^{(1),{\rm out}}_{-\om} +  n^{(2),{\rm out}}_{-\om} .
\end{equation}
With respect to the case of a geometry with a single horizon, there are two possible channels for the production of Hawking-like particles, given by the mode conversion of the two in modes with negative norm: the right-going $(\varphi^{(1),\rm in }_{-\om})^*$, coming from region I, and the left-going $(\varphi^{(2),\rm in }_{-\om})^*$, coming from region III. The projections of these modes on the positive norm out mode $\phi^{u,\rm out}_{\om}$ give the final occupation number of the mode $\phi^{u,\rm out}_{\om}$ and represent the contributions of two different creation processes. Indeed, let us remark that $\beta^{(2)}_{\om}$ is related to a single scattering on the white horizon ($x=L$), while $\beta^{(1)}_{\om}$ is related to a double scattering process, first on the black horizon ($x=-L$) and eventually on the white horizon.
Because of their different origins, those two terms contribute very differently to the total spectrum of emitted particles, as shown in the following section.

\section{Numerical analysis}	%
\label{sec:results4x4}		%

\subsection{Results}		%
\label{subsec:results}		%

We present here the results of a numerical computation of the coefficients of the $4\times4$ scattering matrix [Eq.~\eqref{eq:bogo4x4}], performed by a modified version of the code used for the analysis of Chap.~\ref{chap:hawking} (see Appendix~\ref{appsec:4x4}).
Since the $\beta_\om$ coefficient of a single black hole geometry is now generalized by the two $\beta_\om^{(1)}$ and $\beta_\om^{(2)}$, we focus only on these two coefficients, because they provide the most relevant physical information about the spontaneous emission of particles in a black-hole--white-hole geometry.
As described in the previous section, they indeed represent the rate of mode conversion into the positive norm outgoing mode $\phi^{u,\rm out}_{\om}$ of the incoming negative norm modes $(\varphi^{(i),\rm in }_{-\om})^*$.

\begin{figure}
\centering
 \includegraphics[width=0.48\textwidth]{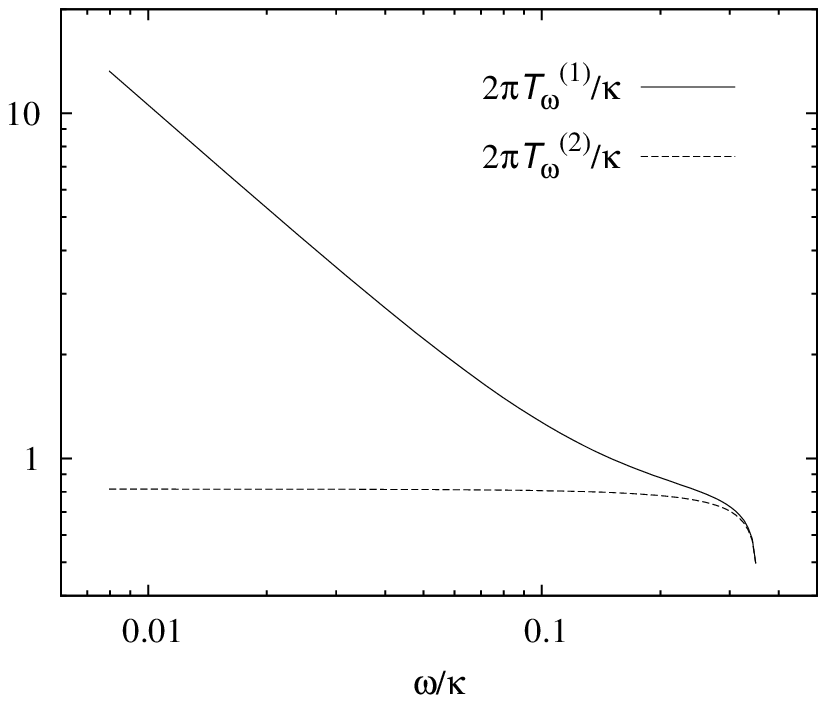}
 \hspace{0.02\textwidth}
 \includegraphics[width=0.48\textwidth]{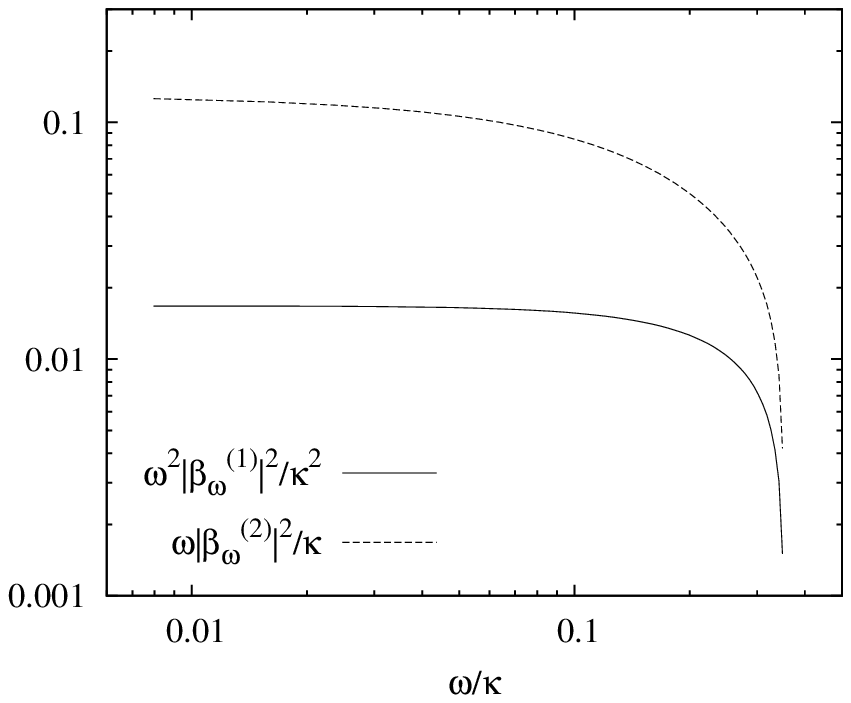}
\caption{Left panel: $T_\om^{(1)}$ (solid line) and $T_\om^{(2)}$ (dashed line)
versus $\om/\kappa$ for a velocity profile as in Eq.~\eqref{eq:velocity4x4}.
The contribution $\beta_\om^{(2)}$ to the final occupation number of positive norm out modes $u$ is Planckian for $\om<\omm$, while the contribution of $\beta^{(1)}_\om$ is infrared divergent.
Right panel: $\om^2|\beta_\om^{(1)}|^2$ (solid line) and $\om|\beta_\om^{(2)}|^2$ (dashed line). $|\beta_\om^{(2)}|^2$ behaves like a standard $\beta$-coefficient in a single horizon geometry, while $|\beta_\om^{(1)}|^2$ diverges like $1/\om^2$.
In both panels, the fixed parameters are $\Lambda/\kappa = 2$, $n=1$, $D=0.5$, $\kw=\kb=\kappa$, $L\kappa/\chor=4$.
\label{fig:beta_warpdrive}}
\end{figure}

In Chap.~\ref{chap:hawking} we proved that $|\beta_\om|^2$ follows a Planckian distribution up to a certain frequency $\ommax$, by plotting the effective temperature $T_\om$ in Fig.~\ref{fig:tom_fom_nopert}, which in fact remains constant when $\om<\ommax$.
In analogy with the definition of $T_\om$ given in Eq.~\eqref{eq:Tom}, we shall investigate the behavior of two quantities $T_\om^{(1)}$ and $T_\om^{(2)}$, implicitly defined by
\begin{equation}
 |\beta_{\om}^{(i)}|^2\equiv\frac{1}{\ee^{\om/T_\om^{(i)}}-1}.
\end{equation}
In the left panel of Fig.~\ref{fig:beta_warpdrive}, we plot $T_\om^{(1)}$ (solid line) and $T_\om^{(2)}$ (dashed line) as functions of the frequency $\om$.
While $T_\om^{(2)}$ is constant up to $\omm$, thus showing the very same behavior of $T_\om$ (see Fig.~\ref{fig:tom_fom_nopert}, right panel), $T_\om^{(1)}$ shows an infrared divergence.

These two different behaviors are better spotted in the right panel of the same figure, where we plot $\om^2|\beta_\om^{(1)}|^2$ (solid line) and $\om|\beta_\om^{(2)}|^2$ (dashed line).
In the limit of $\om\to0$, $\om|\beta_\om^{(2)}|^2$ goes to a constant for $\om\to0$,
accordingly to the asymptotic behavior of the Planckian distribution
\begin{equation}\label{eq:planck0freq}
 |\beta_{\om}|^2=\frac{1}{\ee^{\om/\Th}-1}\sim\frac{\Th}{\om}
\end{equation}
(in units where $\hbar=k_{\rm B}=1$), for a constant temperature parameter $T_\om^{(2)}$.
Differently, $|\beta_\om^{(1)}|^2$ diverges as $\om^{-2}$, as proved by the constancy of $\om^2|\beta_\om^{(1)}|^2$ at low frequencies.

\begin{figure}
\centering
 \includegraphics[width=0.48\textwidth]{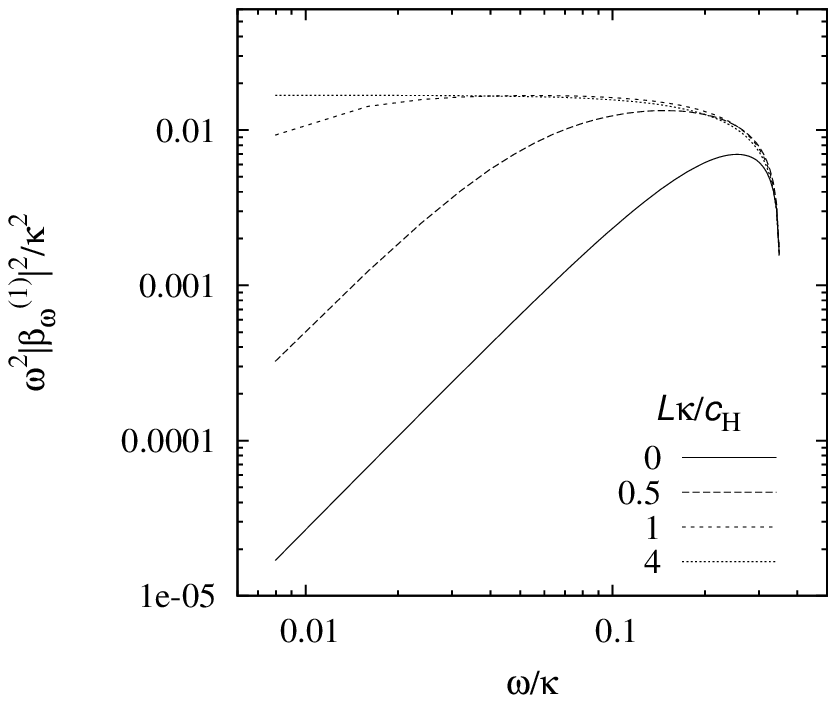}
 \hspace{0.02\textwidth}
 \includegraphics[width=0.48\textwidth]{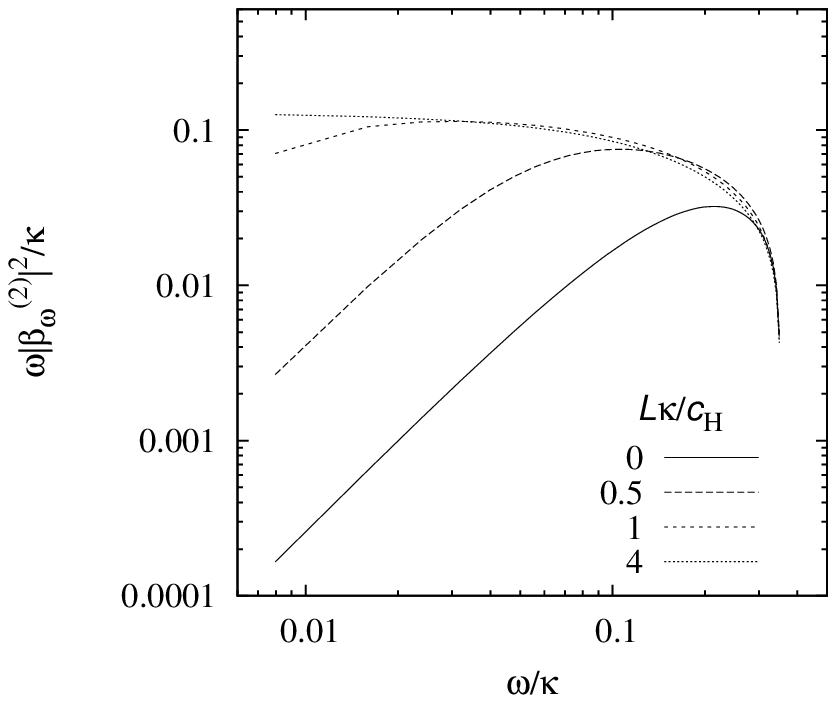}
\caption{$\om^2|\beta_\om^{(1)}|^2$ (left panel) and $\om|\beta_\om^{(2)}|^2$ (right panel), for various values of $L\kappa/\chor$. Interestingly, particle production is present even when $L=0$ and no horizon is present.
In both panels, the fixed parameters are $\Lambda/\kappa = 2$, $n=1$, $D=0.5$, $\kw=\kb=\kappa$.
\label{fig:beta_warpdrive_L}
}
\end{figure}

Finally, in Fig.~\ref{fig:beta_warpdrive_L}, we compare $\om^2|\beta_\om^{(1)}|^2$ (left panel) and $\om|\beta_\om^{(2)}|^2$ (right panel), for various values of the distance $2L$ between the horizons. Note that, for small values of $L$, the subsonic region between the horizons tends to disappear and the spectrum get strongly modified at low frequencies. This is somehow natural, because low frequencies correspond to long wavelengths. In this case the spectrum is no more Planckian even for the $\beta^{(2)}_\om$-channel. Nevertheless, some radiation is still present, even in the limiting case $L=0$, when there is no horizon. In fact, the presence of two supersonic regions, connected by a non flat compact region around $x=0$ seems sufficient to trigger particle production. This result is analogous to what we discovered in~\cite{robustness} and reviewed in Sec.~\ref{sec:asymmetric}, for asymmetric supersonic velocity profiles without horizons (see curves with negative $\Dta/D$ in the left panel of Fig.~\ref{fig:tom}).

\subsection{Interpretation}	%
\label{subsec:interpretation}	%

The main result of this chapter is the different scaling of the two $\beta$-coefficients:
\begin{itemize}
 \item $|\beta_\omega^{(1)}|^2\sim{1}/\omega^2$,
 \item $|\beta_\omega^{(2)}|^2\sim{1}/\omega$.
\end{itemize}
We already noticed in the end of Sec.~\ref{sec:scatteringwarpdrive} that $\beta_\omega^{(1)}$ is related to a double scattering on both the horizons. Because of that, $\beta_\omega^{(1)}$ is made by the sum of the products of the black hole $\alpha_\om$ times the white hole $\beta_\om$ and {\it vice versa}. Given that each of those coefficients diverges as $\omega^{-1}$, their products go as $\omega^{-2}$.
On the contrary, $\beta_\om^{(2)}$ is related to a single scattering on the white horizon. So it must behave as $\om^{-1}$ just like the $\beta$-coefficient in a single horizon geometry.

In this section, we investigate the consequences of the above scalings of the $\beta$-coefficients on the flux of emitted particles in a dynamical scenario where the horizons are created at finite time $t_{\rm H}=0$.%
\footnote{This is the situation investigated in Chap.~\ref{chap:warpdrive} for a relativistic scalar field in a dynamic warp-drive geometry.}
It is well known that a spectrum behaving as $\om^{-1}$ at low frequency gives rise to a constant emission of particles in time. The question addressed here is what is the effect of the $\om^{-2}$ scaling of $|\beta_\omega^{(2)}|^2$.
Let us imagine to put a particle detector coupled to the density perturbation $\hr=\hat\rho-\rn$ in region III, far from the horizons and after their formation.
Using Fermi's golden rule~\cite{fermi}, we compute the probability of detecting a quantum of frequency $\om$ per unit of time as a function of the time $T$ elapsed after the formation of the horizons.
By definition of $\hp$ in Eq.~\eqref{eq:defphi1}, $\hr$ is given by
\begin{equation}
 \hr=\rn(\hp+\hpd)= \rn \, \hch,
\end{equation}
where $\hch\equiv\hp+\hpd$ is a Hermitian field. Its expansion in terms of in modes $\chi_\om^{i,\rm in}$ and in operators $\hat a_\om^{i,\rm in}$ is [see Eq.~\eqref{eq:expsup2}]

\begin{multline}\label{eq:chiexpansion_wd}
 \hch(t,x)=\int_0^{+\infty}\dd\om\,\phm{\om t}\left\{\chi_\om^{u,\rm in} \ha_\om^{u,\rm in}+\chi_\om^{v,\rm in}\ha_\om^{v,\rm in}\right.\\ \left.
+ \theta(\ommax - \om)\left[ (\chi_{-\om}^{(1),\rm in})^*\ha_{-\om}^{(1),\rm in\dagger}+(\chi_{-\om}^{(2),\rm in})^*\ha_{-\om}^{(2),\rm in\dagger}\right]\right\}+\mbox{h.c.},
\end{multline}
where
\begin{equation}\label{eq:chimodes_wd}
 \com\equiv\pom+\vpom.
\end{equation}
In the vacuum state $\ket{0_{\rm in}}$, defined before the formation of the horizons at $t_{\rm H}=0$, the two-point function is
\begin{multline}\label{eq:gp}
 G^+(x,t;x',t')=\bra{0_{\rm in}}{\hch(t,x) \, \hch(t',x')}\ket{0_{\rm in}} 
\\= 
\int_0^{+\infty}\!\dom\,\phm{\om(t-t')}\left[\chi_\om^{u,\rm in}(x)(\chi_\om^{u,\rm in}(x'))^*+\chi_\om^{v,\rm in}(x)(\chi_\om^{v,\rm in}(x'))^*\right]\\
+\int_0^{+\ommax}\!\dom\,\php{\om(t-t')}\left[(\chi_{-\om}^{(1),\rm in}(x))^*\chi_{-\om}^{(1),\rm in}(x')+(\chi_{-\om}^{(2),\rm in}(x))^*\chi_{-\om}^{(2),\rm in}(x')\right].
\end{multline}
Using $G^+$ one computes the number of quanta of frequency $\om_d$ counted by a detector moving along a world line $x(\tau)$~\cite{birreldavies}, from the creation of the horizons at $t_{\rm H}=0$,
\begin{equation}\label{eq:Pom}
 P_{\om_d}(T)=\int_0^T\dd\tau\int_0^T\dd\tau'\ee^{-\ii\om_d(\tau-\tau')}\rn(x(\tau))\rn(x(\tau'))\, G^+(x(\tau),\tau;x(\tau'),\tau').
\end{equation}
Placing now the detector at constant $x(\tau)=x_d$ and inserting $G^+$ of Eq.~\eqref{eq:gp} into Eq.~\eqref{eq:Pom}, the integral in the time coordinates yields a factor
\begin{equation}
 \int_0^T\dd\tau\int_0^T\dd\tau'\ee^{-\ii(\om_d\pm\om)(\tau-\tau')}=\frac{4\sin^2[(\om_d\pm\om)T/2]}{(\om_d\pm\om)^2}\sim 2\pi\,T\,\delta(\om_d\pm\om),
\end{equation}
where the last approximation is valid for large $T$, the $+$ sign holds for positive norm $u$ and $v$ modes, and the $-$ sign holds for the two negative norm modes.
As a consequence, since the integration over frequencies is performed only on positive values of $\om$, only those terms of $G^+$ related to negative norm modes contribute to the final result.

Aiming to measure the production of Hawking quanta in region III, far from the horizons, we rewrite the in modes of Eq.~\eqref{eq:gp} in term of out modes. Furthermore, we look only at right-going modes in the right asymptotic region and assume that the detector can distinguish positive norm modes $\chi_\om^{u,\rm out}$ from negative norm modes $\chi_{-\om}^{(1),\rm out}$. Under these assumptions, $P_{\om_d}$ reduces to
\begin{equation}
 P_{\om_d}(T)=\rn(x_d)^2\int_0^{\ommax}\!\dd\om\,\frac{4\sin^2[(\om_d-\om)T/2]}{(\om_d-\om)^2}\left(|\beta_\om^{(1)}|^2+|\beta_\om^{(2)}|^2\right)|\chi_{\om}^{u,\rm out}(x)|^2.
\end{equation}
Using the expression for $\chi$ in term of $\phi$ and $\varphi$ given in Eq.~\eqref{eq:chimodes_wd} and the expressions for the WKB modes of Eq.~\eqref{eq:WKBmodes}
\begin{equation}\label{eq:Pom2}
 P_{\om_d}(T)=\frac{\rn(x_d)}{2\pi}\int_0^{\ommax}\!\dd\om\,\frac{4\sin^2[(\om_d-\om)T/2]}{(\om_d-\om)^2}\left(|\beta_\om^{(1)}|^2+|\beta_\om^{(2)}|^2\right)\frac{1+D_{\ku}(x_d)}{[1-D_{\ku}(x_d)]v_g(\ku,x_d)},
\end{equation}
where $D_k$ is given in Eq.~\eqref{eq:dkWKB} and $v_g=\partial\om/\partial k$ is the group velocity. Note that, by repeating the same computation for a single horizon, one would obtain exactly the same expression with a unique $|\beta_\om|^2$, instead of $(|\beta_\om^{(1)}|^2+|\beta_\om^{(2)}|^2)$.

For large $T$
\begin{equation}
 \frac{4\sin^2[(\om_d-\om)T/2]}{(\om_d-\om)^2}\sim2\pi\, T\,\delta(\om_d-\om)
\end{equation}
and Eq.~\eqref{eq:Pom2} becomes
\begin{equation}\label{eq:Pom3}
 P_{\om_d}(T)=\rn(x_d)
  T
 \left(|\beta_{\om_d}^{(1)}|^2+|\beta_{\om_d}^{(2)}|^2\right)\frac{1+D_{k_{\om_d}^u}(x_d)}{[1-D_{k_{\om_d}^u}(x_d)]v_g(k_{\om_d}^u,x_d)}.
\end{equation}
Finally, the rate of detection of particles with frequency $\om_d$ is obtained by differentiating the above expression with respect to $T$:
\begin{equation}\label{eq:Rom}
 R_{\om_d}=\rn(x_d)\left(|\beta_{\om_d}^{(1)}|^2+|\beta_{\om_d}^{(2)}|^2\right)\frac{1+D_{k_{\om_d}^u}(x_d)}{[1-D_{k_{\om_d}^u}(x_d)]v_g(k_{\om_d}^u,x_d)}.
\end{equation}
To compute the total detection rate, this expression must be integrated over the detection frequency $\om_d$. 
Since the horizons are created at a finite time $t_{\rm H}=0$, no particles with frequency $\om$ can be detected before a time $1/\om$. This provides an infrared cutoff $1/T$:
\begin{equation}
 R_{\rm tot}(T) = \int_{1/T}^{\ommax}\dd\om_d\, R_{\om_d}.
\end{equation}
To estimate this integral, note that $v$ and $c$ are constant in the limit $x_d\to+\infty$, corresponding to a detector placed in the supersonic region III very far from the white hole horizon. Moreover the integral is dominated by low frequencies because of the infrared divergence of $|\beta^{(i)}_{\om_d}|^2$. Thus, $D_k$ and $v_g$ can be approximated by their asymptotic behaviors for $\om_d\to0$.
In this limit, the momentum $\ku$ of the right-going mode $\phi_\om^u$ goes to a constant value $k_0^u$ (see the right panel of Fig.~\ref{fig:dispersion_hawk}). This value might be straightforwardly computed from the dispersion relation~\eqref{eq:dispersionWKB1}, using the asymptotic values of $v$ and $c$ from Eq.~\eqref{eq:velocity4x4}. However, to proceed with the present analysis, it is enough to note that, when $\om_d\to0$, the group velocity $v_g$ of the mode $\phi_\om^u$ does not vanish and $D_{k_{\om_d}^u}$ assumes a finite value different from 1. As a consequence,
\begin{equation}
 \frac{1+D_{k_{\om_d}^u}}{[1-D_{k_{\om_d}^u}]v_g(k_{\om_d}^u)}
\end{equation}
goes to a constant, and Eq.~\eqref{eq:Rom} yields
\begin{equation}
   R_{\rm tot}(T)\propto \int_{1/T}^{\ommax}\dd\om_d\left(|\beta_{\om_d}^{(1)}|^2+|\beta_{\om_d}^{(2)}|^2\right).
\end{equation}
Using the leading behavior of $|\beta^{(1)}_{\om}|^2\propto\om^{-2}$, we eventually obtain
\begin{equation}
R_{{\rm tot}}(T)\propto T,\label{eq:Rtot}
\end{equation}
\ie, the flux of density perturbations increases linearly with time. This linear divergence has the same origin as the logarithmic divergence found in a white-hole configuration in a BEC~\cite[Eq.~(32)]{Carusottowhite}.
Both those divergences are related to the presence of a zero-frequency mode with non-vanishing momentum $k_0^u$, which can propagate in the supersonic region away from a white-hole horizon (region III of a warp-drive geometry corresponds indeed to the interior of a white hole).
However, in the present situation the divergence is stronger (linear instead of logarithmic) because of the double scattering, first on the black horizon and then on the white one, yielding $|\beta_\om^{(1)}|^2\propto \om^{-2}$. Note that the properties of modes that scatter only on the white hole are encoded by $|\beta_\om^{(2)}|^2\propto \om^{-1}$, as in geometries with a single white hole~\cite{Carusottowhite}.

With respect to the exponential divergence found in Chap.~\ref{chap:warpdrive} for a warp-drive geometry in the case of relativistic dispersion relation, the linear divergence of Eq.~\eqref{eq:Rtot} is extremely milder. However, this result implies that warp-drive spacetimes cannot be stabilized by the presence of a superluminal dispersion relation.

As a consistency check, we show that standard results are recovered when this analysis is applied to a single black-hole configuration, where $|\beta_\om|^2\propto\om^{-1}$.
In this situation, the outgoing mode $\phi_\om^u$ propagates in the subsonic region (see left panel of Fig.~\ref{fig:dispersion_hawk}). Its wavenumber $\ku$ therefore goes to 0, when $\om\to0$, and
\begin{equation}
 \frac{1+D_{\ku}}{[1-D_{\ku}]v_g(\ku)}\propto\om.
\end{equation}
Repeating the above analysis under these assumptions, one obtains that $R^{\rm BH}_{\om_d}(T)$ is constant, \ie, the flux of emitted particles is stationary.

\section{Summary and discussion}	%
\label{sec:conclustions4x4}		%

We computed the flux of Hawking phonons in a BEC flow with two sonic horizons, corresponding to an acoustic black hole and an acoustic white hole, respectively, such that the flow is subsonic in between the horizons and supersonic elsewhere.
This configuration exactly reproduces a warp-drive geometry like that studied in Chap.~\ref{chap:warpdrive}, the main difference being the presence of a non-relativistic supersonic dispersion relation.

We found that the flux of outgoing phonons is given by the sum of two contributions. The first one is due to mode conversion of the negative norm mode coming from the left asymptotic region, which crosses both the black and the white horizons, before eventually reaching the right asymptotic region. The second one is originated by the mode conversion of the negative norm mode coming from the right asymptotic region, which scatters only on the white horizon.
The latter term contributes to the final occupation number of positive norm outgoing modes in a standard way, \ie, the associated $|\beta_\om^{(2)}|^2$ has a Planckian distribution with a cut off at $\ommax$, the scale given by dispersion. In particular, it behaves as $\om^{-1}$ for $\om\to0$. The former one, being related to the scattering across both the horizons, is described by a coefficient $|\beta_\om^{(1)}|^2$ which behaves as $\om^{-2}$ instead, \ie, as the square of $|\beta_\om|^2$ for a single-horizon.

This has important consequences on physical observables measured in this system. In fact, if a density perturbation detector is coupled to the phonon field, it will measure a flux of phonons which is finite at any finite time, but diverges linearly with time. Since the back reaction of phonons will become not negligible, a more accurate analysis taking into account non-linear effects on the background is needed to predict the evolution of the system. Another aspect of this geometry deserving further investigations~\cite{4x4} is the description of the physical observables measured by an observer at rest in the subsonic internal region. Moreover, it would be interesting to compute the correlation pattern in such a geometry. In this case one should find a divergence similar to that found in a single white-hole geometry~\cite{Carusottowhite}. However, in the present situation the divergence would be linear instead of logarithmic.

A second important result of this analysis is that particle production is still present even in the limiting situation where the horizons collapse into a single point. This is somehow unexpected because in that case neither the horizons nor the subsonic region are present. The only relevant feature needed for particle creation is indeed the presence of two supersonic regions, where negative norm modes propagate, and of a curved region connecting them, where mode conversion takes place. This result is in agreement with what found in~\cite{robustness} and reviewed in Sec.~\ref{sec:asymmetric}, where particle production was observed in a supersonic flow.

Finally, we remark that the analogy between this simple model in a BEC and a warp-drive geometry in general relativity provides valuable information on the stability of a warp-drive bubble. In Chap.~\ref{chap:warpdrive} we showed that such a geometry is semiclassicaly unstable when a quantum scalar field with relativistic dispersion relation lives in that spacetime. In particular, we found that the renormalized stress-energy tensor of this field diverges exponentially with time close to the white horizon and the Cauchy horizon. Here, we showed that the presence of a modified dispersion relation partially tames such a divergence. However, when coupling a detector to the phonon field, the flux of detected phonons diverges linearly with time.
Thus, superluminal warp-drive spacetimes remain unstable for superluminal modifications of the dispersion relation.

\chapter{Black hole lasers in Bose--Einstein condensates}	%
\chaptermark{Black hole lasers in BECs}				%
\label{chap:bhlaser}						%

Quite recently, a near-stationary supersonic flow engendering a pair of horizons (a black hole one followed by a white hole one) was realized in a Bose--Einstein condensate (BEC)~\cite{technion} (see Sec.~\ref{subsec:becexp}).
In such a background, because of the anomalous dispersion of Bogoliubov phonons, one expects to get a kind of laser effect~\cite{cj,lf,cp} due to a self-amplification of the Hawking radiation.

In this chapter, we provide the theoretical basis of this effect starting from the Bogoliubov--de Gennes (BdG) equation~\eqref{eq:BdG1}, by extending to flows containing two horizons the treatment of Chap.~\ref{chap:hawking}, that was applied only to a single sonic horizon.
Following~\cite{cp} we then establish that the laser effect is governed by a discrete set of complex frequency modes that correspond to the resonant modes of the cavity formed by the region bordered by the two horizons.
In a classical description, this discrete set governs the dynamical instability of the flow.
The second aim is to compute the spectrum of the complex frequency modes
by theoretical and numerical methods.
The excellent agreement of the results~\cite{bhlasers,cfp} validates both the concepts and the semi-classical methods used in the theoretical approach.
Finally we consider two observables: the mean number of emitted phonons, and the two-point function of the phonon field
that governs the density--density correlation pattern~\cite{carusotto1,carusotto2}. At sufficiently late time,
both observables are governed by a single mode, the most unstable one.
In this late time regime, their behaviors are identical to those one would obtain using a classical description
of the density perturbations.
At early time instead, when starting from vacuum configurations, correspondence with the quantum phonon flux emitted by a black hole horizon~\cite{MacherBEC} is established.

The propagation of phonons in flows containing two horizons has already been considered. However, toroidal configurations with periodic boundary conditions were generally used~\cite{Garay,Gardiner}.
In that case, the spectrum is very complicated because it results from a combination of two effects: the discreteness of the wave vectors defined on the torus interferes with that associated with modes that are trapped in the supersonic region.
As a result, not only the analysis is difficult, but the relationships with the Hawking effect and the black hole laser effect~\cite{cj,lf,cp} are hard to draw.
On the contrary, when dealing with continuous wave vectors, the analysis of the complex frequency mode is simpler, and the relationship with Hawking radiation is easily made.

\section{Theoretical analysis}	%
\label{sec:theory}		%

\subsection{Black-hole--white-hole geometries}	%
\label{subsec:bhlasers}				%

As in Chaps.~\ref{chap:hawking} and~\ref{chap:warpdriveBEC}, we work with elongated condensates that can be effectively considered as 1-dimensional, so that we can write the metric line element as in Eq.~\eqref{eq:metric} (see discussion in Sec.~\ref{sec:generalmetric} about acoustic geometries in $1+1$ dimensions).
In this language, a black-hole--white-hole geometry is obtained when $v$ crosses twice $c$.
Assuming that the fluid flows from right to left ($v<0$), the locations of the white hole horizon and that of the black hole are $x_{\rm W}=-L$ and $x_{\rm B}=L$, respectively.
These divide the $x$-axis in three regions, which we call I on the left of the white horizon ($x<-L$), II between the two horizons ($-L<x<L$) and III on the right of the black horizon ($x>L$). The flow is supersonic in the internal region II and subsonic otherwise. This is the opposite case of what considered in Chap.~\ref{chap:warpdriveBEC}, where the velocity was subsonic inside.

\begin{figure}
 \psfrag{q}[c][c]{\small $q$}
 \psfrag{d}[c][c]{\small $D$}
 \psfrag{x}[c][c]{\small $\kappa x/\chor$}
 \psfrag{vc}[c][c]{\small $|v|/\chor,\,c/\chor)$}
 \begin{flushright}
  \includegraphics[width=\textwidth]{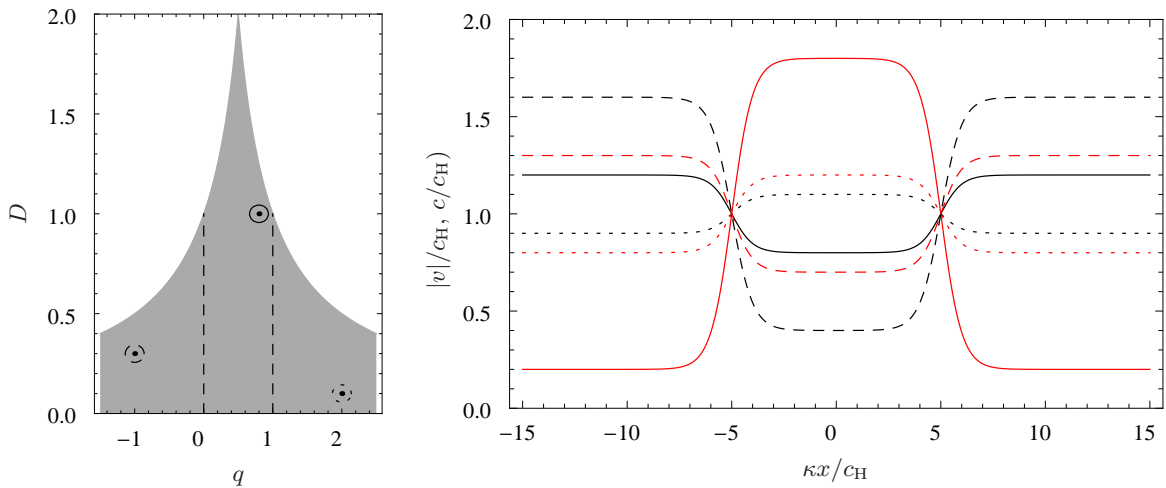}
 \end{flushright}
 \caption{Left panel: allowed values for $q$ and $D$ giving rise to a black-hole--white-hole geometry.
Right panel: Velocity profiles $c(x)/\chor$ (black lines) and $|v(x)|/\chor$ (red lines) for three different values of $(q,D)$: $q=0.8$, $D=1$ (solid lines), $q=-1$, $D=0.3$ (dashed lines) and $q=2$, $D=0.1$ (dotted lines). These values are also reported in the left plot. All profiles with $\kw=\kb=\kappa$.}
 \label{fig:qD}
\end{figure}

A stationary velocity profile satisfying the above requirement is
\begin{equation}\label{eq:velocity}
 c(x)+v(x)=w(x)=\chor D\, \sign(x^2-L^2)\,
 \tanh^{1/n}\!\!\left[\left(\frac{\kw|x+L|}{\chor D}\right)^{\!\!n}\right] \tanh^{1/n}\!\!\left[\left(\frac{\kb|x-L|}{\chor D}\right)^{\!\!n}\right],
\end{equation}
which generalize Eq.~\eqref{eq:velocity_simp} to two horizons. $\chor$ is the sound speed at the horizons, $2L$ is the distance between them, $D$ determines the size of the near-horizon regions where the metric is not flat, $n$ controls the sharpness of the transition to the flat regions, and $\kw$ and $\kb$ control the surface gravities [see Eq.~\eqref{eq:kappa}].
The metric is then completely fixed by introducing an extra parameter $q$, as in Eq.~\eqref{eq:cv}, which specifies how $c+v$ is shared between $c$ and $v$.
When restricting to the cases where
\begin{equation}
 c(x)>0,\quad v(x)<0,\quad D>0,
\end{equation}
the range of $D$ and $q$ is limited.
The allowed values are graphically illustrated by the shaded area in Fig.~\ref{fig:qD}, left panel.
Three velocity profiles corresponding to different couples $(q,D)$ are plotted in Fig.~\ref{fig:qD}, right panel.
%

\subsection{Modes in black-hole--white-hole geometries.}	%
\label{sec:laser}						%

To obtain the spectrum in condensates that cross twice the speed of sound, \ie\/ in flows defined in Eq.~\eqref{eq:velocity}, we need to take into account the nontrivial propagation in the non-homogeneous regions localized near the two horizons, and the fact that each subsonic region is now defined on the half line and no longer on the full real axis.
Then, starting from the general analysis of Sec.~\ref{subsec:mode1d}, we need to identify the number of globally defined independent modes, as done in Sec.~\ref{subsec:bogo} for flows with a single horizon.

It should be noted that this set does not depend on the particular form of the mode equation, but mainly resides on the behavior in the internal region II and in the asymptotic flat external regions I and III described in Sec.~\ref{subsec:bhlasers}.
Therefore, the result is the same when studying a phonon field in a BEC, as in the present analysis, or a generic scalar field with superluminal dispersion relation as in~\cite{cp}.
As explained in this reference, in geometries described by Eq.~\eqref{eq:velocity}, there is a continuous double set (right-going and left-going waves) of real frequency modes, and a discrete set of complex frequency modes.
Here we give a more mathematical argument yielding the same result.

In flows~\eqref{eq:velocity}, two asymptotically flat and subsonic regions play the most important role in defining the set of modes, for both real and complex frequencies $\lambda$.
For $\lambda= \om$ real, there are three bounded modes: the two usual propagating modes associated with the real roots of Eq.~\eqref{eq:dispersion1} plus the mode that asymptotically decays. This mode should now be taken into account as it is bounded in the subsonic region where it is defined.
We will call the two sets of three modes $\phi_\om^{i,{\rm I}}$ and $\phi_\om^{i,{\rm III}}$, respectively for the left (I) and right (III) asymptotic region, where the superscript $i$ takes three values: $ u, v, CJ$ to characterize the asymptotic right-going $u$-mode, the left-going $v$-mode and the decaying $CJ$-mode, named after the paper of Corley and Jacobson~\cite{CJ96}.
In the internal region II, we have four modes that we call $\phi_\om^{j,{\rm II}}$ with $j=1,2,3,4$.
Since this region is compact, all the four roots of Eq.~\eqref{eq:dispersion1} should be taken into account.

Therefore, any globally defined solution of frequency $\om$ can be expanded in two forms by referring to its left, or right, asymptotic behavior:
\begin{align}
\phi_\om&=\sum_i L_\om^i \, \phi_\om^{i,{\rm I}} , \label{eq:phileft} \\
\phi_\om&=\sum_i R_\om^i \, \phi_\om^{i,{\rm III}}.\label{eq:phiright}
\end{align}
Moreover, each partial wave $\phi_\om^{i,{\rm I}}$ or $\phi_\om^{i,{\rm III}}$ can be written in term of the $\phi_\om^{j,{\rm II}}$:
\begin{align}
\phi_\om^{i,{\rm I}}&=\sum_j M_{R,\om}^{ij} \, \phi_\om^{j,{\rm II}}, \label{eq:phicenterleft}\\
 \phi_\om^{i,{\rm III}}&=\sum_j M_{L,\om}^{ij} \, \phi_\om^{j,{\rm II}}.\label{eq:phicenterright}
\end{align}
Putting together the above equations, one obtains 
\begin{equation}\label{eq:system}
 \sum_i M_{R,\om}^{ij}R_\om^i=\sum_i M_{L,\om}^{ij}L_\om^i.
\end{equation}
This system of four equations in six unknowns (the coefficients $R_\om^i$ and $L_\om^i$) has a two-dimensional set of solutions, corresponding to two linearly independent modes.
For instance, one can choose as independent solutions, $\phi^{u, \rm in}_{\om}$ and $\phi^{v, \rm in}_{\om}$, the two right-going and left-going in-modes, defined as the solution of Eq.~\eqref{eq:system} with respectively $L_\om^u=1,R_\om^v=0$ and $L_\om^u=0,R_\om^v=1$.
Similarly, one can chose the out-modes $\phi^{u, \rm out}_{\om}$ and $\phi^{v, \rm out}_{\om}$ defined respectively by $L_\om^v=0,R_\om^u=1$ and $L_\om^v=1,R_\om^u=0$.
Using the scalar product of Eq.~\eqref{eq:normphi} to normalize these modes, the two sets are related by~\cite{cp}
\begin{align}
\phi^{u, \rm in}_{\om} &= {\cal T}_{\om} \, \phi^{u, \rm out}_{\om} 
+ {\cal R}_{\om} \,  \phi^{v, \rm out}_{\om}, \\
\phi^{v, in}_{\om} &= \tilde {\cal T}_{\om}\,  \phi^{v, \rm out}_{\om} + \tilde
{\cal R}_{\om}\,  \phi^{u, \rm out}_{\om} .  
\label{RT}
\end{align}
Unitarity imposes $|{\cal T}_{\om}|^2 +|{\cal R}_{\om}|^2 = 1 = | \tilde {\cal T}_{\om}|^2 +|\tilde {\cal R}_{\om}|^2 $, and $R_\om \tilde {\cal T}_\om^* + {\cal T}_\om \tilde {\cal R}_\om^* = 0$ [see Eq.~\eqref{eq:el_scatt}].

Let us now move to the complex frequency case. Because all these expressions are analytic in $\lambda$, the above analysis applies as such when replacing $\om$ by $\lambda = \om + i \Gamma$, where $1/\Gamma$ is the time scale over which the unstable mode grows (decays), when $\Gamma$ is positive (negative).
The novel aspects only come through the condition that the modes be asymptotically bounded.
Indeed, $\Gamma >0$ implies that $k^u_\lambda$, the asymptotic wave number of the $u$ real root of Eq.~\eqref{eq:dispersion1} (more precisely of its analytical continuation in $\Gamma$), acquires a positive imaginary part. Thus the mode diverges at $x\to-\infty$, unless one puts $L_\lambda^u=0$.
Similarly, on the right side, the $v$ mode diverges at $x\to+\infty$, and imposing that it is bounded requires $R_\lambda^v=0$.
However, these two conditions imply that the system~\eqref{eq:system} has only the trivial solution $R_\lambda^i=L_\lambda^i=0$, {\it except} when its determinant vanishes. This condition defines an equation for the complex frequency $\lambda$ that has a finite set of solutions: $\{\la, a = 1, 2, ..., N\}$.

\subsection{Semi-classical treatment}	%
\label{sec:wkb}				%

Having established the condition that defines the complex frequencies $\la$, we now turn to the question of calculating them.
Since the above algebraic method does not depend on the specific form of the mode equation, the semi-classical treatment of~\cite{cp} applies to the present case. We just have to pay attention to the mode normalization, since we are now dealing with mode doublets $(\phi,\varphi)$ (see discussion in Sec.~\ref{subsec:bogo}).

When the background is not homogeneous but $v$ varies slowly with respect to the wavelength of the perturbation, the exact solutions are well approximated by their WKB approximation of Eqs.~\eqref{eq:phiWKB} and~\eqref{eq:vphiWKB}, and the wave numbers $k_\om^\alpha(x)$ acquire an $x$-dependence, being real roots of the dispersion relation~\eqref{eq:dispersionWKB} in a stationary inhomogeneous flow characterized by $v(x)$ and $c(x)$.
Such WKB equations are valid in the three regions I, II, III but not close to the horizons where the gradients of $v$ and $c$ cannot be neglected~\cite{cfp}.

The theoretical analysis can then be greatly simplified when taking into account the fact that the mixing between $u$ and $v$ modes is usually negligible~\cite{MacherBEC,MacherRP1}, in particular when $q=1/2$ [see discussion after Eq.~\eqref{eq:cv}].%
\footnote{The numerical analysis of the next section does not rely on this simplifying assumption and we shall discuss under which conditions it is actually reliable.}
In this approximation, the laser effect is completely due to the $u$-modes (for flows to the left $v< 0$).
Considering the $u$ in-mode in the internal region II, it can be written as a superposition of three WKB waves:
\begin{equation}\label{eq:expnov}
 \phi_\om^{u,{\rm in}}={\cal A}_\om \phi_\om^u + {\cal B}_\om^{(1)} {\varphi_{-\om}^{(1)}}^* + {\cal B}_\om^{(2)} {\varphi_{-\om}^{(2)}}^*,
\end{equation}
where $\varphi_{-\om}^{(i)}$ are the two negative frequency $\varphi^u$-modes, corresponding to the two extra roots $k_\om^{(i)}$, $\kone$ being the most negative one, see figure~\ref{fig:dispersion}.

In~\cite{ulf} it was shown that the propagation from one horizon to the other is efficiently described by a unitary scattering of a two-component vector
\begin{equation}
 \Phi=\spin{\phi_\om^u}{(\varphi_{-\om}^{{\rm trap}})^*},
\end{equation}
whose first component represents the $u$-wave $\phi_\om^u$ and the other represents the trapped wave $(\varphi_{-\om}^{{\rm trap}})^*$, described either by $(\varphi_{-\om}^{(1)})^*$ or $(\varphi_{-\om}^{(2)})^*$. The WKB propagation of this waves in the flat regions I, II and III, and their scattering at the horizons, where the WKB approximation breaks down, is described by an $S$ matrix that can be decomposed in four elementary matrices
\begin{equation}
 S=U_4 U_3 U_2 U_1,
\end{equation}
where
\begin{itemize}
 \item $U_1$ describes the scattering at the white hole horizon:
 \begin{equation}
 U_1=\begin{pmatrix}\alpha_\om & \alpha_\om z_\om \\ \tilde\alpha_\om z_\om^* & \tilde\alpha_\om\end{pmatrix}.
 \end{equation}
 The incoming rightgoing mode $\phi_\om^u$, propagating in region I, and the incoming leftgoing mode $(\varphi_{-\om}^{(2)})^*$, propagating in region II, are scattered by the white hole horizon on $\phi_\om^u$ and $(\varphi_{-\om}^{(1)})^*$, both propagating rightward in region II toward the black hole horizon.
 \item $U_2$ describes the propagation in region II between the white and the black hole horizon of the two rightgoing modes $\phi_\om^u$ and $(\varphi_{-\om}^{(1)})^*$:
 \begin{equation}
  U_2=\begin{pmatrix}\ee^{\ii S_\om^u} & 0 \\ 0 & \ee^{-\ii S_{-\om}^{(1)}}\end{pmatrix}.
 \end{equation}
 \item $U_3$ describes the scattering at the black hole horizon:
 \begin{equation}\label{eq:u3}
  U_3=\begin{pmatrix}\gamma_\om & \gamma_\om w_\om \\ \tilde\gamma_\om w_\om^* & \tilde\gamma_\om\end{pmatrix}.
 \end{equation}
 The modes $\phi_\om^u$ and $(\varphi_{-\om}^{(1)})^*$, both propagating rightward in region II, are scattered by the black hole horizon on the rightgoing $\phi_\om^u$, propagating in region III, and on the leftgoing $(\varphi_{-\om}^{(2)})^*$, propagating in region II back to the white hole horizon.
 \item $U_4$ describes the propagation back to the white hole horizon of $(\varphi_{-\om}^{(2)})^*$ and the free propagation of the outgoing $\phi_\om^u$, in region III on the right of the black horizon:
 \begin{equation}
  U_4=\begin{pmatrix}1 & 0 \\ 0 & \ee^{\ii S_{-\om}^{(2)}}\end{pmatrix}.
 \end{equation}
\end{itemize}
In the above matrices we used
\begin{align}
 S_\om^u&\equiv\int_{-L}^L\dx\,k^u_\om(x),\\
 S_{-\om}^{(i)}&\equiv\int_{-L_\om}^{R_\om}\dx\,\left[-k_\om^{(i)}(x)\right],\quad i=1,2,
\end{align}
which are the Hamilton--Jacobi actions governing the phase of the WKB modes, where $L_\om$ and $R_\om$ are the two turning points of the trapped mode. From the definition of $S$:
\begin{align}
S_{11} &=  \alpha_\om  \gamma_\om \, \ee^{\ii S^{u}_{\om}}
 \left[1 +  \frac{\tilde\alpha_\om}{\alpha_\om }z_\om^* w_\om  \, \ee^{-\ii(S^{u}_{\om}+S^{(1)}_{-\om})} \right],\\
S_{12} &= \alpha_\om  \gamma_\om \, \ee^{\ii S^{u}_{\om}}
\left[ z_\om + \frac{\tilde\alpha_\om}{\alpha_\om }w_\om\, \ee^{-\ii(S^{u}_{\om}+S^{(1)}_{-\om})} \right],\\
S_{21} &= \tilde\alpha_\om  \tilde\gamma_\om \, \ee^{-\ii(S^{(1)}_{-\om} -S^{(2)}_{-\om})}
\left[ z_\om^* + \frac{\alpha_\om}{\tilde\alpha_\om }w^*_\om\, \ee^{\ii(S^{u}_{\om}+S^{(1)}_{-\om})} \right],\\
S_{22} &=  \tilde\alpha_\om  \tilde\gamma_\om \, \ee^{-\ii(S^{(1)}_{-\om} -S^{(2)}_{-\om})}
 \left[1 +  \frac{\alpha_\om}{\tilde\alpha_\om }z_\om w^*_\om  \, \ee^{\ii(S^{u}_{\om}+S^{(1)}_{-\om})} \right]. 
\end{align}

Moreover, by unitarity,
\begin{equation}
 U_i^\dagger\sigma_3 U_i=\sigma_3,
\end{equation}
because of the specific form of the scalar product~\eqref{eq:scalar} and the fact that $\phi_\om^u$ and $(\varphi_{-\om}^{{\rm trap}})^*$ behave as the first and second component of a two-component vector $W_\om$, respectively [see Eq.~\eqref{eq:doublets}].
This implies that the parameters in $U_1$ and $U_3$ must satisfy
\begin{gather}
 |\alpha_\om|^2=|\tilde\alpha_\om|^2,\\
 |\gamma_\om|^2=|\tilde\gamma_\om|^2,\\
 |\alpha_\om|^2(1-|z_\om|^2)=|\gamma_\om|^2(1-|w_\om|^2)=1.
\end{gather}
%

\subsubsection{Real frequency}	%

The scattering matrix $S$ describes the propagation of the trapped modes $(\varphi_{-\om}^{(i)})^*$ back-and-forth between the horizons, and their interaction with $\phi_\om^u$. In order for $\Phi$ to represent a physical solution, it must be single-valued, that is $(\varphi_{-\om}^{(2)})$ must assume the same value before undergoing the scattering process described by $S$ and afterwords. Namely, fixing to 1 the amplitude of the incoming $u$-wave, if ${\cal B}_\om^{(2)}=b_\om$ is the initial amplitude of $(\varphi_{-\om}^{(2)})$ before the scattering, the following condition must hold~\cite{cp}:
\begin{equation}
 \spin{\ee^{\ii\theta_\om}}{b_\om}= S\spin{1}{b_\om},
\end{equation}
where $\theta_\om$ is real by unitarity.
From this matricial equation, one obtains
\begin{equation}\label{eq:btheta}
 b_\om= \frac{S_{21}}{1-S_{22}},\qquad \ee^{\ii\theta_\om}=-\frac{S_{11}}{S_{22}^*}\frac{1-S_{22}^*}{1-S_{22}}, 
\end{equation}
where, to obtain the second equations one has to use the unitarity of $S$.
The coefficients of Eq.~\eqref{eq:expnov} are
\begin{equation}\label{eq:abb}
 {\cal A_\om}=\alpha_\om(1+z_\om b_\om),\quad {\cal B}_\om^{(1)}=\tilde\alpha(z_\om^*+b_\om),\quad {\cal B}_\om^{(2)}=b_\om.
\end{equation}
%

\subsubsection{Complex frequency}	%

For complex frequency with $\Gamma>0$, one must impose that the incoming $u$ branch vanishes to keep the mode finite in the past ($t\to-\infty$). Moreover the trapped mode must again be single valued and we fix its amplitude to 1:
\begin{equation}
 \spin{\beta_a}{1}=S\spin{0}{1},
\end{equation}
which implies
\begin{equation}\label{eq:beta}
 \beta_a=S_{12},\quad S_{22}=1.
\end{equation}

Eqs.~\eqref{eq:btheta}, \eqref{eq:abb} and~\eqref{eq:beta} are the central equations for the analysis of the results of the numerical integration (see Sec.~\ref{sec:numerics}).
Even though these equations have been obtained through a semi-classical reasoning, their validity goes beyond that of the WKB approximation, as shall be established by the remarkable agreement between the relations one can derive from them and the numerical results.

\begin{figure}
 \centering
 \psfrag{om}[c][c]{\small $\omega/\ommax$}
 \psfrag{b2}[c][c]{\small $|{\cal B}_\om^{(2)}|^2$}
 \includegraphics[width=\textwidth]{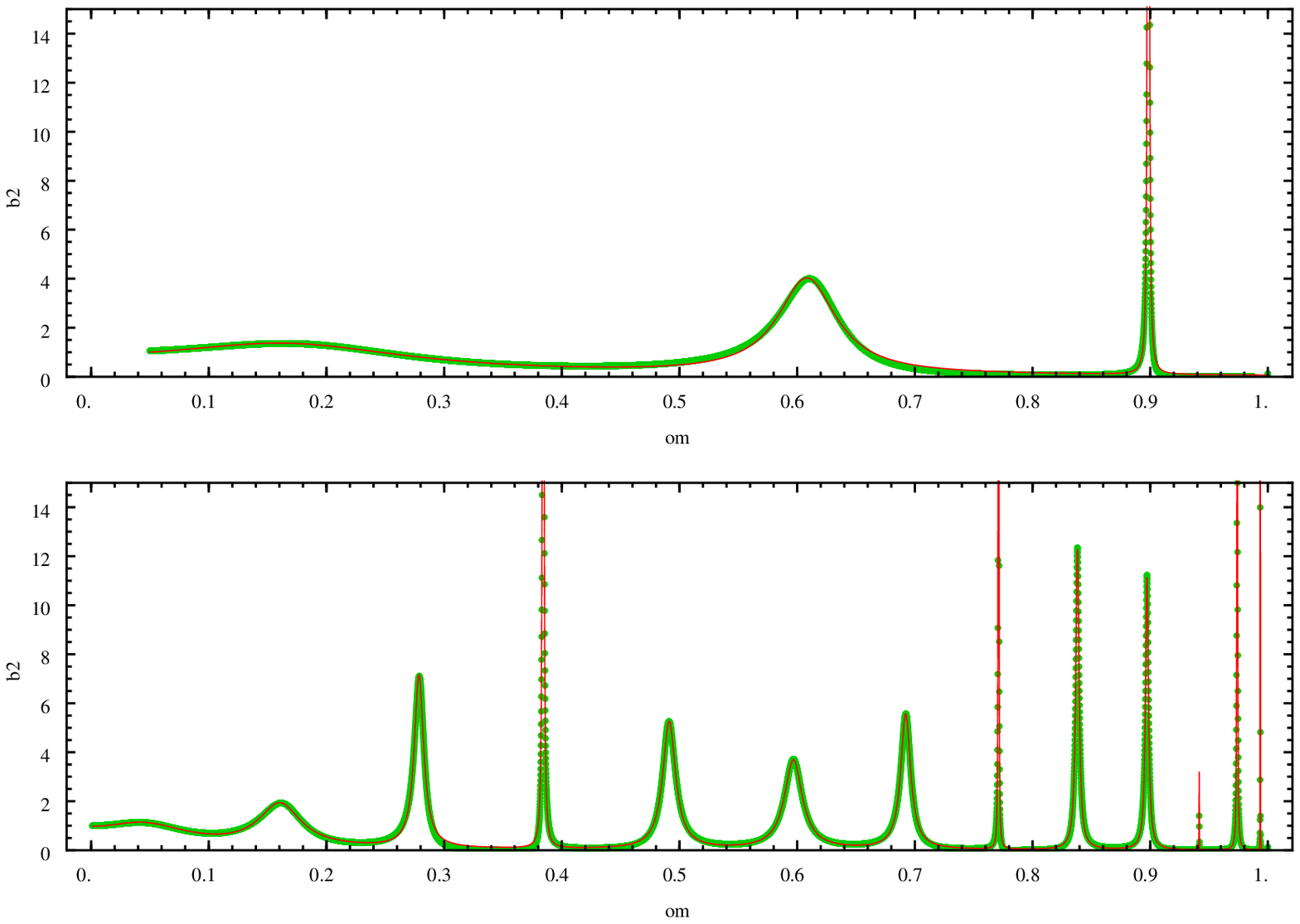}
 \caption{We represent $|{\cal B}_\om^{(2)}|^2$ as a function of $\omega/\ommax$ for two values of $L$, $L\kappa/\chor=6$ (upper panel), $L\kappa/\chor=25$ (lower panel), and for  $\kw=\kb=\kappa$,  $q=0.5$, $D=0.33$, $n=1$, $\Lambda/\kappa=2$, $\ommax/\kappa\approx0.195$. Green points: numerical simulation; red lines: fitted series of Lorentzians. Online movie \href{http://people.sissa.it/finazzi/bec_bhlasers/movies/eigenfrequencies.gif}{eigenfrequencies.gif} (available from~\urlnjp): Evolution of $|{\cal B}_\om^{(2)}|^2$ from $L\kappa/\chor=15$--25.}
 \label{fig:lorentzians}
\end{figure}

The main prediction concerns the relation between $\la$, solutions of Eq.~\eqref{eq:beta}, and the behavior of the coefficients of Eq.~\eqref{eq:abb}.
Indeed, note that if and only if $\la$ is a solution of $S_{22}=1$, it is also a pole of ${\cal B}_\om^{(2)}=b_\om=S_{21}/(1-S_{22})$. Therefore, when $\Gamma_a$ is small enough (this is always true in our numerical situations), the pole is close to the real axis and the behavior of ${\cal B}_\om^{(2)}$ for real frequencies $\om$ is dominated by the contribution of the pole:
\begin{equation}
 {\cal B}_\om^{(2)}=\frac{S_{21}}{1-S_{22}}\approx -\frac{\ii\Gamma_a}{\om-(\om_a+\ii\Gamma_a)}{\cal B}_{\om_a}^{(2)},
\end{equation}
from which
\begin{equation}\label{eq:lorentzian}
 |{\cal B}_\om^{(2)}|^2\approx \frac{\Gamma_a^2}{(\om-\om_a)^2+\Gamma_a^2}|{\cal B}_{\om_a}^{(2)}|^2.
\end{equation}
Thus, $|{\cal B}_\om^{(2)}|^2$ should be well described by a sum of Lorentzians characterized by the complex frequencies $\la$.
In Fig.~\ref{fig:lorentzians} we plot $|{\cal B}_\om^{(2)}|^2$ obtained from numerical simulations and the corresponding fitted sum of Lorentzians. The agreement is excellent.

We conclude this analysis by showing that a semi-classical treatment furnishes approximate expressions for the complex frequencies $\la$.
Assuming $|z_\om|^2$, $|w_\om^2|$ and $|z_\om w_\om|$ are much smaller than 1, one can expand the condition $S_{22}=1$ in powers of these quantities.
To zeroth order, one has $\lambda_a=\om_a$ real, with $\om_a$ solution of the Bohr--Sommerfeld (BS) condition~\cite{cp}
\begin{equation}\label{eq:bs}
 S_{-\om_a}^{(1)}-S_{-\om_a}^{(2)}-\arg(\tilde\alpha_{\om_a}\tilde\gamma_{\om_a})
 =\int_{-L_{\om_a}}^{R_{\om_a}}\dx\,\left[-k_{\om_a}^{(1)}(x)+k_{\om_a}^{(2)}(x)\right]
 -\arg(\tilde\alpha_{\om_a}\tilde\gamma_{\om_a})=2\pi \nbs,
\end{equation}
where $\nbs \equiv a = 1,2,\ldots,N$.
A semi-classical treatment gives $\arg(\tilde\alpha_{\om_a}\tilde\gamma_{\om_a})=-\pi$. A more refined estimate~\cite{cfp} gives
\begin{align}
 &\arg(\tilde\alpha_{\om_a}\tilde\gamma_{\om_a})\approx - \pi - f\left(\frac{\om_a}{\kb}\right) - f\left(\frac{\om_a}{\kw}\right),\label{eq:arg}\\
  &f(x)\equiv  \arg\Gamma_{\rm E}(\ii x) -  x\log(x) + x +\frac{\pi}{4},\label{eq:f}
\end{align}
where $\Gamma_{\rm E}$ is Euler's Gamma function.

To first order in $|z_{\om_a}|^2$, $|w_{\om_a}^2|$, $|z_{\om_a} w_{\om_a}|$ and $\delta\lambda_a=\delta\om_a+\ii\Gamma_a$, $S_{22}=1$ gives
\begin{align}
& \Gamma_a = \frac{1}{2 T^{\rm b}_{\om_a}}
\left(|z_{\om_a}|^2+|w_{\om_a}^2|+2|z_{\om_a} w_{\om_a}^*|\cos\vartheta_a\right)=\frac{1}{2 T^{\rm b}_{\om_a}}|S_{12}(\om_a)|^2,\label{eq:gamma}\\
& \delta\om_a = - \frac{1}{T_{\om_a}^{\rm b}}
|z_{\om_a} w_{\om_a}^*| \sin\vartheta_a,\label{eq:deltaom}\\
 &\vartheta_a=S_{\om_a}^u+ S_{-\om_a}^{(1)}+\arg\frac{z_{\om_a} w_{\om_a}^*\alpha_{\om_a}}{\tilde\alpha_{\om_a}},\label{eq:theta}\\
 &T_{\om_a}^{\rm b}=\frac{\partial}{\partial\om}\left[ S_{-\om}^{(2)} - S_{-\om}^{(1)} + \arg\tilde\alpha_{\om}\tilde\gamma_{\om} \right]_{\om=\om_a},\label{eq:tb}
\end{align}
where~\cite{cfp}
\begin{equation}
 \arg\frac{z_{\om_a} w_{\om_a}^*\alpha_{\om_a}}{\tilde\alpha_{\om_a}}\approx\frac{2\om_a}{3\kb}+\frac{2\om_a}{3\kw},
\end{equation}
and $T^{\rm b}_{\om_a}$ is the time for the trapped mode to make a full bounce. The phase $\vartheta_a$ modulates the frequency imaginary part $\Gamma_a$ and the first order correction to its real part $\delta\om_a$, as shall be seen in Sec.~\ref{sec:L}.

\subsection{Observables: Phonon fluxes and density--density correlation patterns}	%
\label{sec:correlations}								%

Having identified the set of eigenmodes, we now consider observable quantities.
We shall study both the phonon flux emitted by the black-hole--white-hole system, and the non-local density--density correlation pattern, as they illustrate very different aspects of the laser effect.

In what follows, we shall work in quantum settings because we assume that the state at the formation of the supersonic region is vacuum.%
\footnote{Refer to Appendix~\ref{app:BEC}, and in particular to Sec.~\ref{subsec:quantization} for a rigorous quantization of phonon in BECs and to Sec.~\ref{subsec:1D} for its application to one dimensional flows.}
In classical settings, the dynamical instability (the laser effect) would be present only if the initial density-perturbation possessed a non-vanishing overlap with some complex frequency mode, see section V.A. in~\cite{cp}.
For a perfectly adiabatic formation of the supersonic flow, the amplitudes associated with these modes would be zero.
However, any small deviation from adiabaticity will trigger the instability, thereby engendering a behavior very similar to that derived in quantum settings.
In fact, as we shall see, the main difference between the quantum and the classical treatment 
is at early times because in the quantum vacuum all complex frequency modes
contribute to the observables through the spontaneous excitation.

To compute observables, we first need to choose the quantum state.
Because of the instability, the energy~\eqref{eq:hamiltonianadd} is unbounded from below, and there is no clear definition of the vacuum. Nevertheless, if the formation  of the supersonic region at time $t_0=0$ is adiabatic, and if the temperature of the condensate is low enough, \ie\/ much smaller than $\kappa$, the Heisenberg state of the system can be approximated by the state $|0\rangle$, which is annihilated, at that initial time, by the destruction operators $\aom$, that are associated to the continuous set of real frequency modes, and $\dlpl$ and $\dlmi$, that are associated to the finite set of complex frequency modes [see Eq.~\eqref{eq:defd} for a rigorous definition]%
\footnote{We applied the standard reasoning to both the $\aom$ and the $\hat d$ operators because $\Gamma_a$, the imaginary part of the complex frequency $\la$, is smaller than a tenth of $\om_a = \re \la$. If the imaginary part $\la$ is large, there is no reason to assume that the complex frequency modes are in the vacuum state defined by Eq.~\eqref{eq:vacuum}. Nevertheless, a different choice of the initial state is crucial only for the description of the system soon after the creation of the horizons. At late time the exponential growth of the complex frequency modes (dynamical instability) will wash out any difference in the initial state.}
\begin{equation}\label{eq:vacuum}
 \aom|0\rangle=\dlpl|0\rangle=\dlmi|0\rangle=0.
\end{equation}
Since the operators $\dlpl$, $\dlmi$ are not stationary, the expectation values will not be stationary either, and will instead depend on the lapse of time since the formation at $t_0=0$. In what follows all expectation values will be computed in this state.

The two observables we shall study are related to the density fluctuation $\hr=\hat\rho-\rn$.
Given the definition of $\hp$ in Eq.~\eqref{eq:defphi1}, $\hr$ is given by
\begin{equation}
 \hr=\rn(\hp+\hpd)= \rn \, \hch,
\end{equation}
where $\hch\equiv\hp+\hpd$ is Hermitian. Its expansion in terms of operators $\hat a,\hat d$ is
\begin{equation}\label{eq:chiexpansion}
 \hch(t,x)=\int\!\!\dom\sum_{\alpha}\!\left[\phm{\om t}\com\aom+\mbox{h.c.}\right]
  +\sum_a\!\left[\clpl\dlpl+\clmi\dlmi+ \mbox{h.c.} \right],
\end{equation}
where
\begin{equation}\label{chimodes}
 \begin{aligned}
 &\com\equiv\pom+\vpom,\nonumber \\
&\chi_{a, \pm}(t,x)\equiv \phi_{a, \pm}(t, x)+ \varphi_{a, \pm}(t, x).
 \end{aligned}
\end{equation}
In the vacuum defined at $t_0=0$, the two-point function is
\begin{multline}
 \vev{\hch(t,x) \, \hch(t',x')} = \int\!\dom\sum_{\alpha}\phm{\om(t-t')}\com\comsxp
 \\
 +\sum_a\Big(\clpl\clplsxp+\clmi\clmisxp\Big).
\end{multline}
Because of the complex frequency modes, it is not a function of $t-t'$ only. This differs from the cases
of a single black hole (or white hole) horizon, where the two-point function is stationary in vacuum~\cite{MacherBEC}.

To extract more physical information, it is convenient to return to the frequency eigenmodes appearing in Eq.~\eqref{eq:phiexpansion}. Defining
\begin{equation}
 \sil\equiv\xl+\el,\quad\nl\equiv\psl+\zel,
\end{equation}
the two-point function becomes
\begin{multline}
 \vev{\hch(t,x) \, \hch(t',x')}
= \int\!\dom\sum_{\alpha}\phm{\om(t-t')}\com\comsxp
 \\
  +\sum_a\re\left[\ee^{\Gamma(t+t')}\phm{\om_a(t-t')}\sil\silsxp+\ee^{-\Gamma(t+t')}\phm{\om_a(t-t')}\nl\nlsxp\right]
 \\
 +\ii\sum_a\re\left[\phm{\la(t-t')}\sil\nlsxp-\phm{\las(t-t')}\nl\silsxp\right].
\end{multline}
The first line gives the (vacuum) contribution of the real frequency modes. Since these are only elastically scattered, see Eq.~\eqref{RT}, they stay in their ground state and contribute neither to the emitted fluxes nor to the pattern of density--density correlations. The third line contains a mixed contribution of the growing $\sil$ and decaying modes $\nl$. It does not contribute to the fluxes at any time because $\sil$ vanishes in region I, and $\nl$ does it in III, see the paragraph after Eq.~\eqref{RT}. Moreover, since it only depends on $t-t'$, it does not significantly contribute to the correlation pattern at late time.
Therefore, both fluxes and correlation patterns are governed by the growing mode contribution, the first term  of the second line in $\sil\silsxp$.

It is clear that at late times, \ie\/ $t-t_0 > 1/{\max}(\Gamma_{a})$, all observables are dominated by the mode with the largest $\Gamma_a$.
For instance, the equal time density--density correlation function is asymptotically given by
\begin{equation}\label{eq:densitycorrelation}
 \vev{\hr(t,x)\hr(t,x')} \sim \rnx\rnxp\ \times \ee^{2\Gamma_a t} \, \re\left[ \sil\silsxp \right].
\end{equation}
The real part of the frequency $\om_a$ drops out and, as a result, the locus of the maxima does not propagate with time.
This function is studied for various modes in Sec.~\ref{sec:modesandcorrelation}.
In case the initial state is not vacuum but a thermal state, the above two-point function would be multiplied by $2n_a+1$, where $n_a$ is the mean occupation number, which depends in a nontrivial way both on the temperature and on the blue shift effect due to the horizons (see equation (46) in~\cite{MacherBEC}).
Similar considerations apply to the quantity that is considered below.
Had we worked in classical settings, we would have obtained a coherent wave whose behavior in $x$ and $t$ is the same as that of $\vev{\hr(t,x)\hr(t',x')}$ at fixed $x',t'$, see equation (54) in~\cite{cp}. The main difference arises from the fact that the classical wave amplitude is fixed by initial conditions.

It is also interesting to study the onset of the instability for earlier time, \ie\/ $t-t_0 < 1/{\max}(\Gamma_{a})$.
To this end, we consider the following quantity:
\begin{align}
P_{\rm bh-wh}(\om, T) &\equiv \int_{0}^T dt'  \int_{0}^T dt  \,\ee^{-i\om(t-t')} \,
\vev{ \hch(t,x)\,  \hch(t',x)} ,
\nonumber \\
&\sim  \Sigma_a \, \ee^{\Gamma_a T}\, \vert \sil \vert^2 \, \frac{ 4 \vert \sin[(\om - \om_a - i \Gamma_a
)T/2] \vert^2}
{(\om - \om_a)^2 + (\Gamma_a)^2}.
\label{rate}
\end{align}
It governs the probability that a density fluctuation of frequency $\om$ be observed during the interval $[0,T]$ at some fixed location $x$ in the subsonic region III, see~\cite{cp}.

When $\Gamma_a \to 0$, when the discrete set labeled $a$ becomes continuous, and when $T$ is sufficiently large, $P_{\rm bh-wh}(\om, T)$ behaves as in Fermi's golden rule: $P_{\rm bh-wh}(\om, T)\propto  T\, \bar n_{\om}$, \ie\/ the lapse of time $T$ times $\bar n_{\om}$, the mean occupation number of frequency $\om$.
There can be prefactors that depend on the strength of the coupling, the amplitude of the mode $\phi_\om$ at $x$, if one deals with a derivative coupling.

It is more interesting to study $P_{\rm bh-wh}(\om, T)$ as a function of $T$ for a given black-hole--white-hole geometry, and to compare it with $P_{\rm bh}(\om, T)$, the corresponding quantity computed in the case where only the black hole horizon would be present. When $T \times {\max}(\Gamma_a) \ll 1$, the two observables behave in a very similar manner, even when only a few (say 10) complex frequency modes exist in the black-hole--white-hole case. This will be seen by a numerical analysis in Sec.~\ref{sec:growthofn}, but can be also understood from the properties of the growing modes $\sil$, see the central paragraph of section V.B.5 in~\cite{cp}.

For classical settings, the laser effect is present only when the initial density-perturbation profile is characterized by a non-vanishing amplitude of some complex frequency mode. Since for adiabatic formation of the horizons all the amplitudes associated with those modes are zero, no laser effect appears in this case. However, even a small deviation from perfect adiabaticity causes the presence of instabilities. The main difference in the quantum framework is therefore the possibility to make the system unstable through the spontaneous excitation of complex frequency modes.

\section{Numerical results}	%
\label{sec:numerics}		%

\subsection{The method}	%
\label{sec:nmethod}	%

In a few words, we describe how we proceeded to solve numerically the BdG equation~\eqref{eq:BdG1} in the flows described by Eq.~\eqref{eq:velocity}.
As in the case of a single black hole or white hole horizon, when working with a fixed frequency, the main task is to avoid the growing mode contaminating the oscillatory modes.
The way to get rid of this difficulty is by integrating from the subsonic region towards the horizon into the supersonic region where there is no growing mode~\cite{CJ96}.

Basically we solved separately the mode equation from region I to region II, and from region III to region II, using a code adapted from~\cite{MacherBEC} (refer to Appendix~\ref{app:code} for a detailed description of our code). In each case, as in that reference, we integrated the equations for initial conditions describing the three acceptable modes in regions I and III: the right-moving $u$-mode, the left-moving $v$-mode and the decaying mode.
Then for each of them we computed the amplitudes of the four oscillatory modes in the supersonic region II. The two globally defined modes are built using the procedure described in Sec.~\ref{sec:laser}, by matching the modes in region II.
To obtain the scattering from I to III, we eliminated the amplitudes in region II.

To obtain the correlation patterns of Figs.~\ref{fig:correlations}--\ref{fig:correlations2} we solved the mode equation with the corresponding complex frequency $\la$ that we had formerly computed.

\subsection{The discrete set of complex frequencies}	%
\label{sec:poles}					%

Two strategies can be used to determine the complex frequencies $\la = \om_a + i\Gamma_a$, solutions of Eq.~\eqref{eq:beta}.
They can be obtained either by solving directly the linear system~\eqref{eq:system}, or by determining the center and the width of the Lorentzians in $|{\cal B}_\om^{(2)}|^2$ of Eq.~\eqref{eq:lorentzian}.
We use the latter to localize them and the former to refine the results and study how they depend on the various parameters of the system.

To give a first idea, in Fig.~\ref{fig:lorentzians} we represent $|{\cal B}_\om^{(2)}|^2$ as a function of $\omega/\ommax$ (green points) for two different values of $L$, the distance between the horizons. A sum of Lorentzians functions (red line) is fitted to the data obtained from the numerical analysis [see Eq.~\eqref{eq:lorentzian}].
The quality of the fit confirms the correctness of the theoretical analysis of Sec.~\ref{sec:wkb} and of~\cite{cp}.
In the following sections, we study the dependence of $\la$ on both the parameters describing the geometry [see the velocity profile~\eqref{eq:velocity}] and the dispersive scale $\Lambda$ of Eq.~\eqref{eq:dispersion1}. The dependence on $n$ is not reported here, since it induces no significant change.%
\footnote{
The role of $n$ of Eq.~\eqref{eq:velocity} is to govern the smoothness of the transition from the near-horizon region to the flat asymptotic ones.
A smaller $n$ corresponds to smoother transition, while a larger $n$ corresponds to steeper transition and, consequently, to a larger almost-flat region between the horizons. The usefulness of $n$ is essentially technical: it allows one to control the numerical analysis when $L$, the distance between the horizons, is comparable to $D$, the width of the transition  between regions I--III of Sec.~\ref{subsec:bhlasers}.}
Before proceeding we make some comment about the properties and the validity of the BS condition~\eqref{eq:bs}.

\subsubsection{The validity of the semi-classical approximation}	%
\label{sec:Validity}							%

The action appearing in Eq.~\eqref{eq:bs} contains the {\it difference} of two wave vectors: $\ktwo -\kone $.
These have the same sign (negative for positive $\omega$), as they belong to the same $u$-branch of the dispersion relation (see Fig.~\ref{fig:dispersion}), but they have opposite group velocity: $\kone$ describes a right-going mode with respect to the lab, whereas $\ktwo$ a left-moving one.
This peculiarity give rise to an unusual phenomenon.
In the usual case, eigenmodes with small (large) frequency correspond to small (large) BS numbers $\nbs$, \ie\/ to modes with few (many) nodes. In the present case instead, a large $\nbs$ corresponds to low frequency modes and {\it vice versa}.
This can be understood from Eq.~\eqref{eq:bs}.
Because the wave vectors appeared subtracted, a small $\nbs$ implies $\kone$ close to $\ktwo$ and this happens for $\omega$ close to $\ommax$ (see Fig.~\ref{fig:dispersion}). On the contrary, a large $\nbs$ implies a large difference between $\kone$ and $\ktwo$, that is, $\omega$ close to 0.

\begin{figure}
 \centering
 \psfrag{om}[c][c]{\small $\omega/\ommax$}
 \psfrag{b2}[c][c]{\small $|{\cal B}_\om^{(2)}|^2$}
 \includegraphics[width=\textwidth]{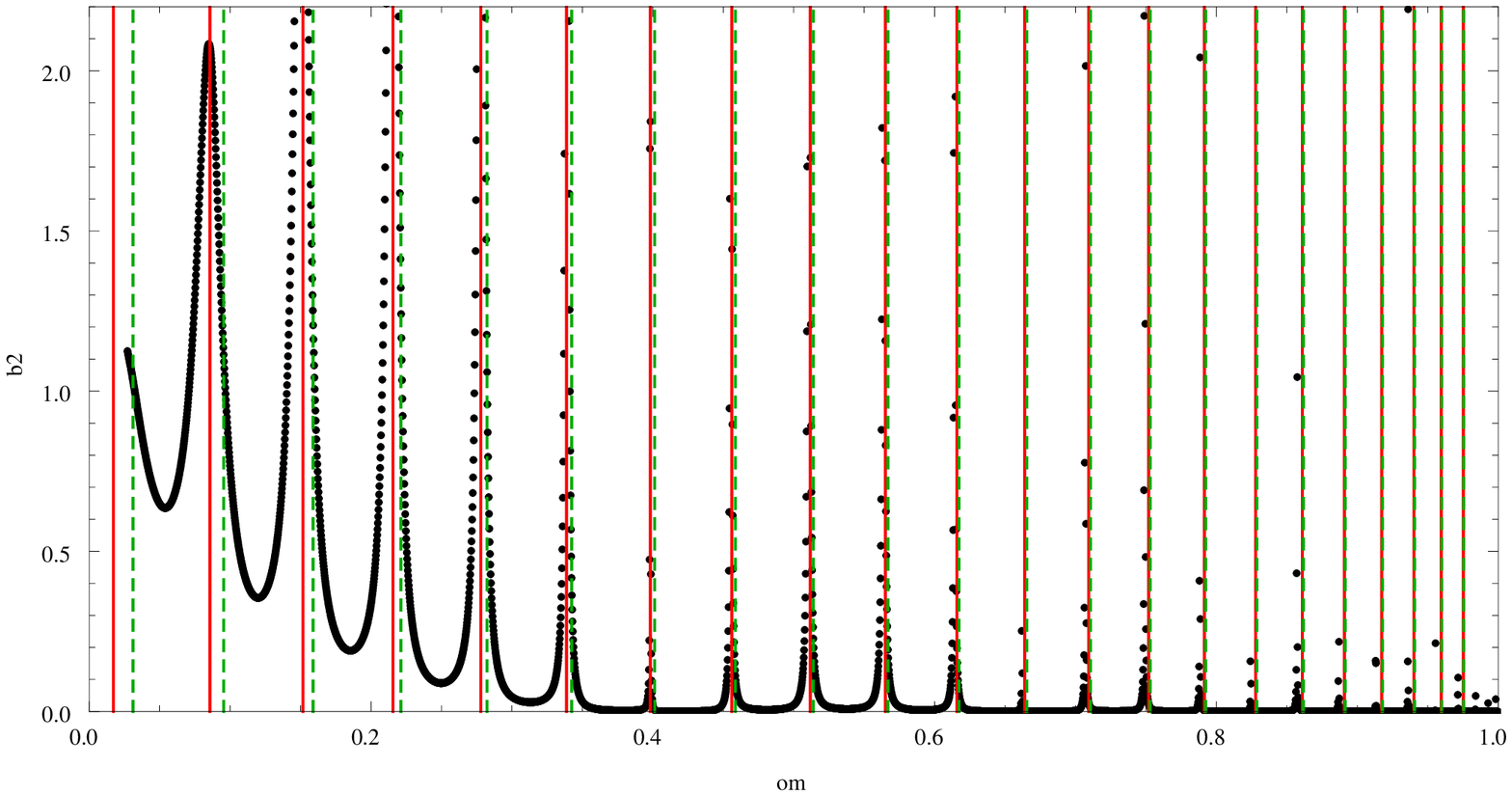}
 \includegraphics[width=\textwidth]{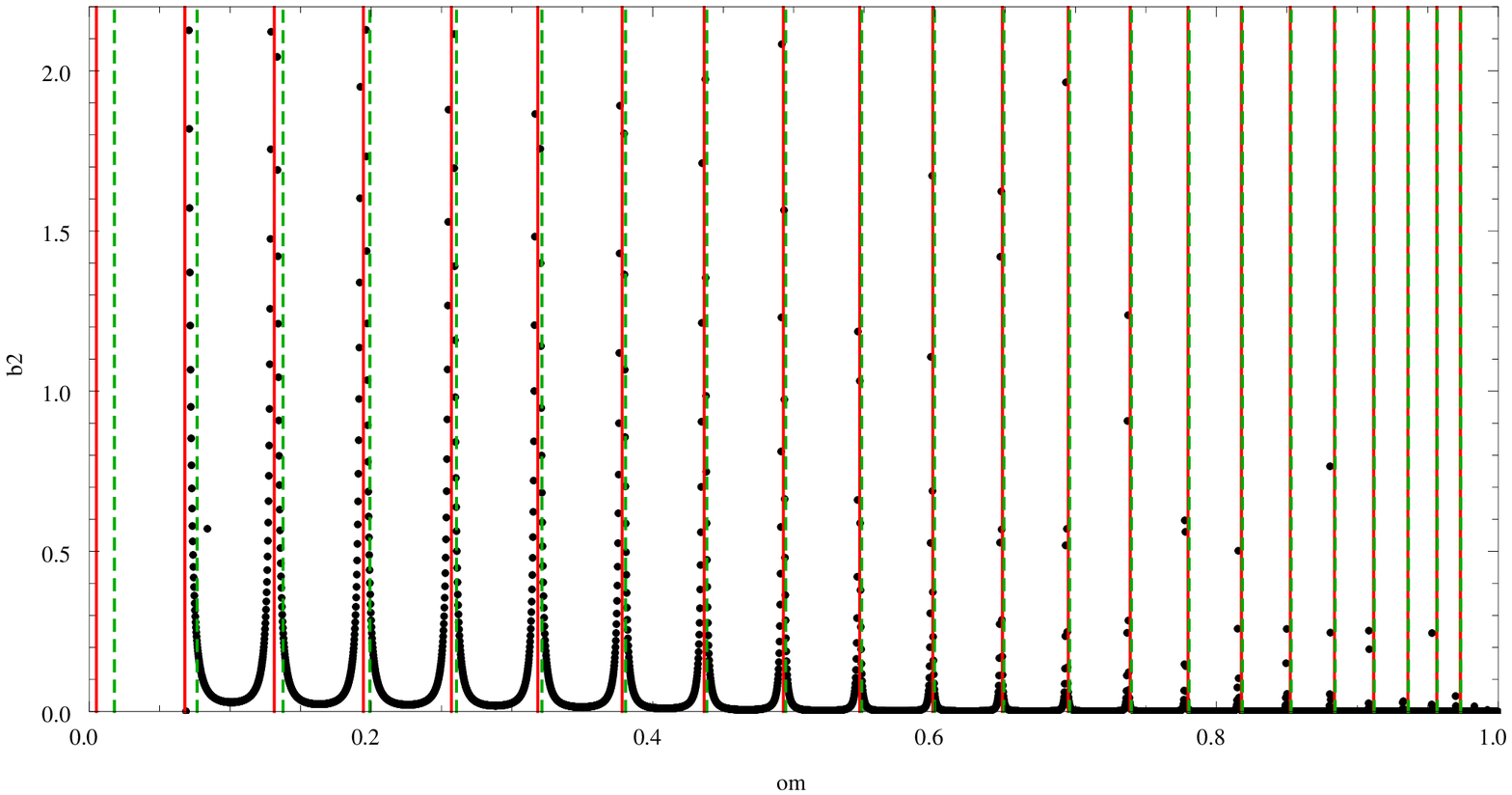}
 \caption{$|{\cal B}_\om^{(2)}|^2$ as a function of $\omega/\ommax$ for $\kw=\kb=\kappa$ and $\kw=0.5\kb=0.5\kappa$ (lower panel). In both plots $q=0.5$, $D=0.5$, $n=2$, $L\kappa/\chor=10$, $\Lambda/\kappa=8$, $\ommax/\kappa\approx1.408$ (upper panel). Points: numerical simulation; green dashed spikes: standard WKB approximation; red solid spikes: improved treatment of Eq.~\eqref{eq:arg}.}
 \label{fig:peakswkb}
\end{figure}

This has an unusual consequence. The BS condition is more reliable when the action is large, that is for $\nbs\gg1$. In the present case this happens for small $\omega/\ommax$. On the other hand the WKB approximation is expected to fail for small $\omega$, as can be verified by the fact that $f$ of Eq.~\eqref{eq:f} cannot be neglected when $\omega/\kappa \leq 1$.
Both these expectations are confirmed in Fig.~\ref{fig:peakswkb} where we compare the numerical results with the predictions obtained with the standard WKB approximation (green lines) and with the improved method (red lines), that is when using Eq.~\eqref{eq:arg}.
Firstly, the agreement between the BS condition and the numerical results is worse for $\om$ close to $\ommax$ and gets better when $\nbs$ increases.
Secondly, at low frequency, while the quality of the standard WKB prediction (green lines) becomes worse, the improved method continues to work very well, thereby establishing its validity.
Further studies about the agreement of the improved method and numerical results can be found in~\cite{cfp}.

%

\subsubsection{The birth of modes as a function of the distance between the horizons}	%
\label{sec:L}										%

\begin{figure}
 \centering
 \psfrag{oma}[c][c]{\scriptsize $\om_a/\ommax$}
 \psfrag{gamma}[c][c]{\scriptsize $\Gamma_a/\kappa$}
 \psfrag{L}[c][c]{\scriptsize $L\kappa/\chor$}
 \includegraphics[width=0.48\textwidth]{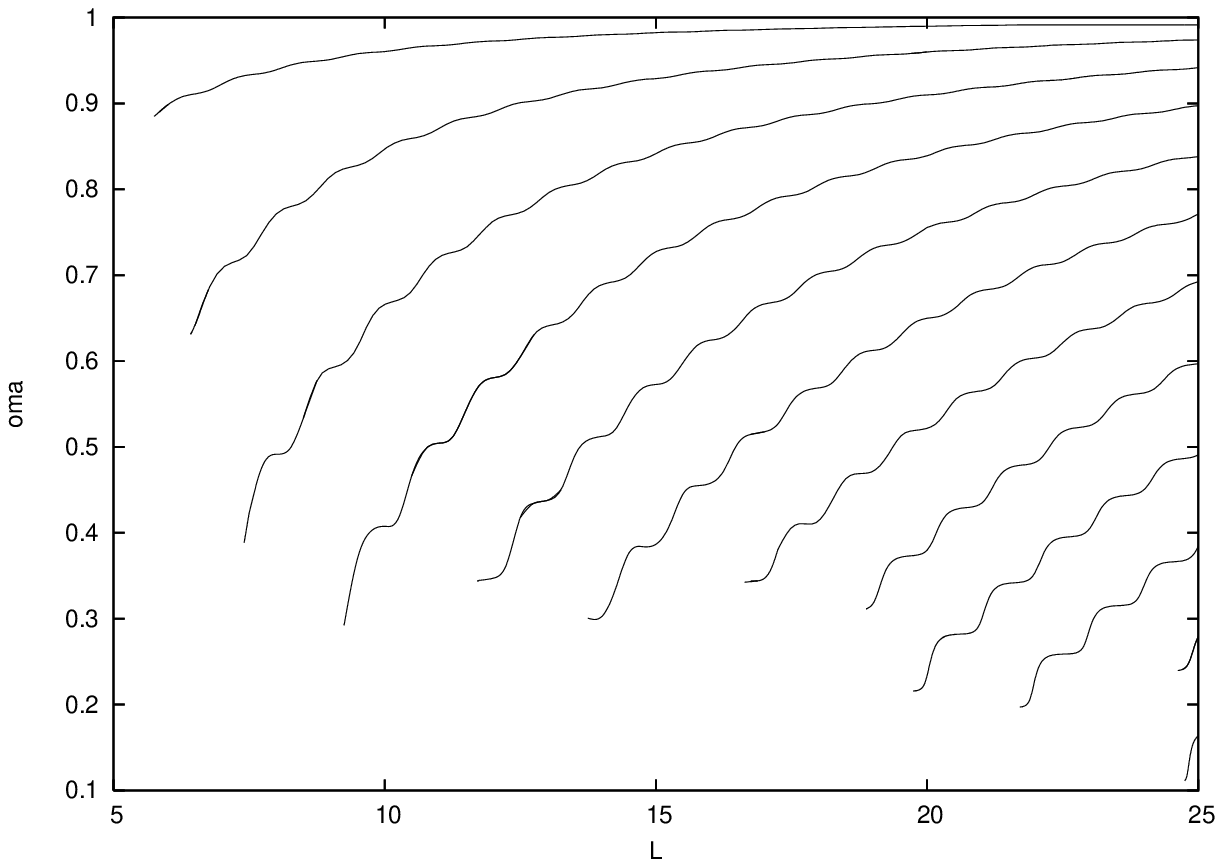}
 \includegraphics[width=0.48\textwidth]{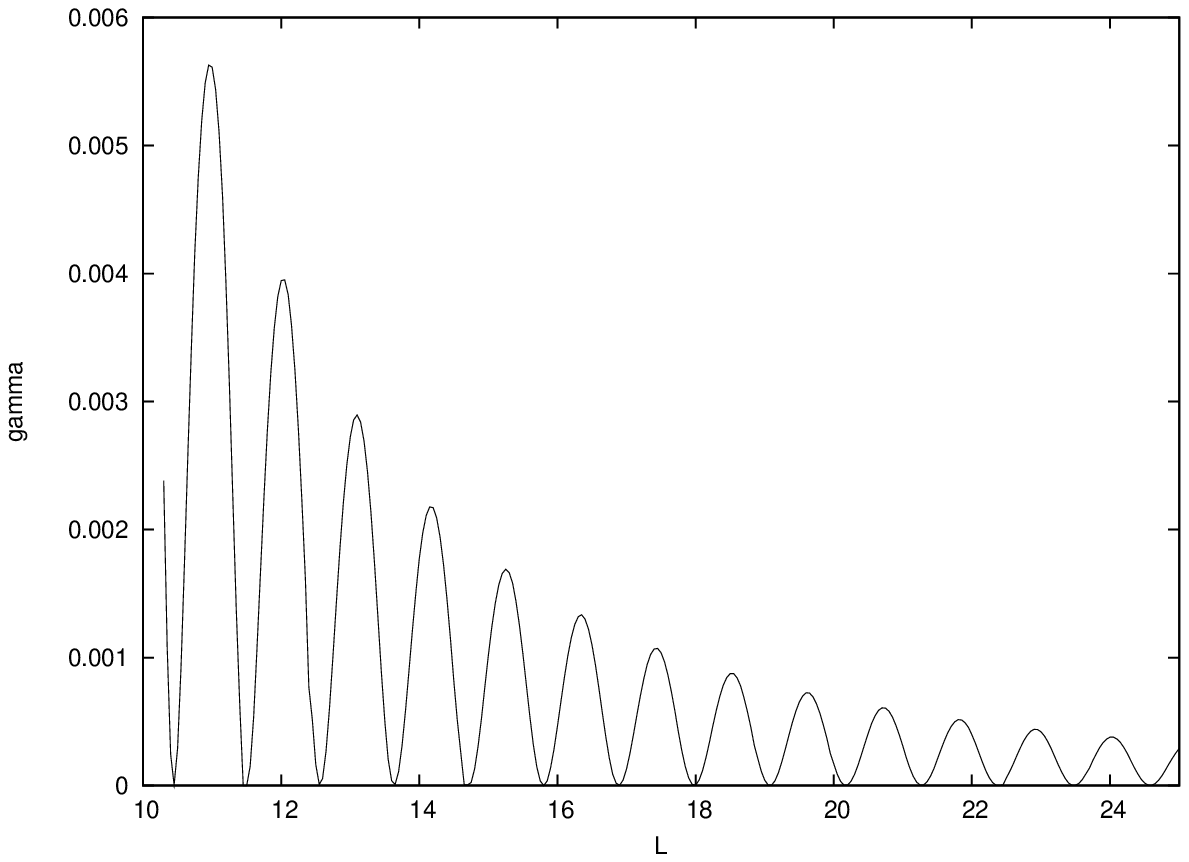}
 \caption{Left panel: The set of $\omega_a/\ommax$ as a function of $L$, from $\nbs$=1 (upper curve) to $\nbs=12$ (lower curve) for the same fixed parameters as in Fig.~\ref{fig:lorentzians}. At $\kappa L/\chor =25$, the values of $\omega_a/\ommax$ correspond to the peaks of the lower panel of that figure from right ($\nbs=1$) to left ($\nbs=12$).
The broad peak of Fig.~\ref{fig:lorentzians} near $\om/\ommax =0.05$ cannot be reproduced with this method because of numerical errors. For similar reasons, each curve $\om_a(L)$ ends for low $\omega$. Right panel: $\Gamma_a/\kappa$ for $\nbs=4$ as a function of $L$ for the same fixed parameters. The occurrence of zeros is due to the fact that $\kb=\kw$, as discussed in the text.}
 \label{fig:omega}
\end{figure}

In Fig.~\ref{fig:omega}, left panel, we plot $\om_a = \re(\la)$ for the entire spectrum of complex frequency modes as a function of the half-distance between the horizons $L$, all other quantities being fixed.
On the right panel, $\Gamma_a = \im(\lambda_a)$ is plotted as a function of $L$, for $\nbs$=4 and for the same values of the other parameters.
As conjectured in~\cite{cp}, when $L$ grows, all $\om_a(L)$ values increase, and when there is enough ``room'', a new eigenmode appears with $\omega_a \sim 0$. As a result, the eigenfrequencies become denser close to $\ommax$, because they neither cross it nor disappear.

It is interesting to consider more closely the birth of new modes near $\om=0$.
A rather difficult question to answer is the following: when extra modes appear, does the imaginary part of the frequency $\Gamma_a$ vanish, as found in figure~1 of \cite{Cardoso}, or not?
This question is physically relevant in that it governs the continuous character of the stability of the system: if $\Gamma_a$ vanishes, it implies that the birth of new modes leads to a continuous behavior in $L$ (as one can expect on the general ground that the properties of the eigenmodes, solutions of Eq.~\eqref{eq:BdG}, continuously depend on the parameters appearing in that equation).
From Fig.~\ref{fig:omega}, which is based on the roots of the determinant in Eq.~\eqref{eq:system}, no definite conclusion can be drawn since there is a loss of numerical precision for $\om \to 0$.
A more reliable method consists in studying the behavior of $|{\cal B}_\om^{(2)}|^2$ on the real frequency axis. We created a gif-animation (\href{http://people.sissa.it/finazzi/bec_bhlasers/movies/eigenfrequencies.gif}{eigenfrequencies.gif}, available from~\urlnjp) of $|{\cal B}_\om^{(2)}|^2$ as a function of $\om$ for increasing values of $L$ to illustrate the appearance of the new eigenfrequencies.
From this it is quite clear that when a new eigenmode appears, both the real part and the imaginary part of $\la$ vanish, as was also found in the appendix of \cite{Fullingbook}.

Another important feature confirming the theoretical analysis of Sec.~\ref{sec:wkb} is the presence of superposed oscillations on the overall trend of $\omega_a$. In fact, the latter is given by the solution of the BS equation~\eqref{eq:bs}, whereas the oscillations are due to the deviations governed by the sine in $\delta\om_a$ of Eq.~\eqref{eq:deltaom}.
For the imaginary part instead, there is no zeroth-order contribution, and the dominant contribution to the right plot of Fig.~\ref{fig:omega} is given in Eq.~\eqref{eq:gamma}. Those oscillations are related to the interference between the two rightgoing modes $\phi_\om^u$ and $(\varphi_{-\om}^{(1)})^*$ [see Eq.~\eqref{eq:theta}], propagating betweeen the horizons in region II.

\subsubsection{Asymmetric velocity profiles}	%
\label{sec:kw}					%

\begin{figure}
 \centering
 \psfrag{oma}[c][c]{\small $\om/\kb$}
 \psfrag{gamma}[c][c]{\small $\Gamma_a/\kb$}
 \psfrag{L}[c][c]{\small $L\kb/\chor$}
  \includegraphics[width=0.6\textwidth]{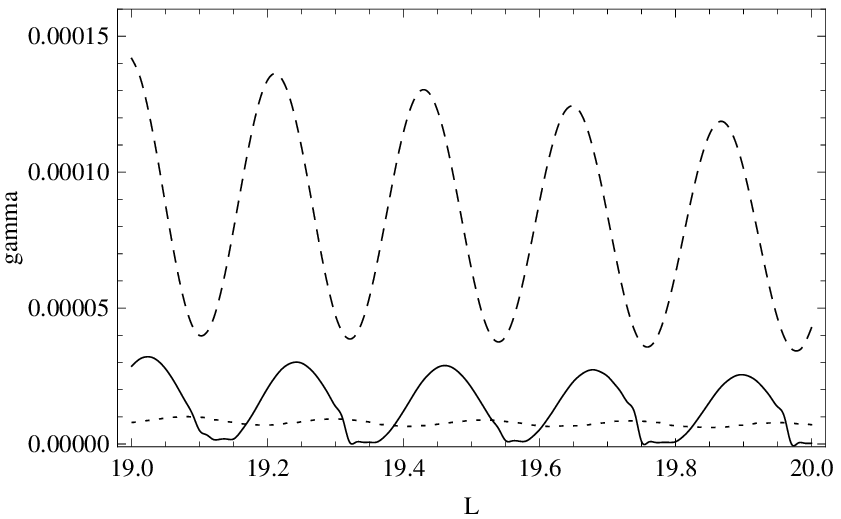}
 \caption{Left panel: $\Gamma_a/\kb$ as a function of $L\kb/\chor$ for $\kw=\kb$ (solid line), $\kw/\kb=0.5$ (dotted line) and $\kw/\kb=2$ (dashed line), for $\ommax/\kb=0.974$, $D=0.33$, $n=1$, $q=0.5$, $\Lambda/\kb=10$.
}
 \label{fig:kw}
\end{figure}

Using asymmetric velocity profiles, namely different surface gravities $\kw$ and $\kb$, does not lead to substantial modifications of the spectrum. 
The only observable change concerns $\Gamma_a$.
When $\kw=\kb$, $\Gamma_a$ vanishes for some particular combination of parameters, as can be seen in Fig.~\ref{fig:omega}, right panel. Eq.~\eqref{eq:gamma} can indeed be rewritten as
\begin{equation}
 \Gamma_a = \frac{1}{2 T_{\om_a}}\left[\left(|z_{\om_a}|-|w_{\om_a}|\right)^2+4|z_{\om_a} w_{\om_a}^*|\cos^2\left(\frac{\vartheta_a}{2}\right)\right],
\label{eq:Ga2}
\end{equation}
which shows that when $|z_{\om_a}|=|w_{\om_a}|$, $\Gamma_a$ vanishes when the cosine does, and this is because the scatterings at the black and the white horizons destructively interfere with each other.
When the two surface gravities are different, $\left(|z_{\om_a}|-|w_{\om_a}|\right)^2$ is non-zero and $\Gamma_a$ can no longer vanish (see Fig.~\ref{fig:kw}).
The non-zero offset is significant only for sufficiently large values of $\omega/\kappa$, which requires that $\ommax/\kappa$ be large enough, since all $\om_a < \ommax$.
Using the standard expressions $z_\om= e^{- \pi \om/\kw}$ and $w_\om= e^{- \pi \om/\kb}$, which furnish reliable estimates~\cite{MacherBEC}, the ratio between the offset and the amplitude of the oscillation is
\begin{equation}\label{eq:ratio}
 \frac{\left(|z_{\om_a}|-|w_{\om_a}|\right)^2}{4|z_{\om_a} w_{\om_a}^*|}= \sinh^2(\pi \om_a(\kb-\kw)/2).
\end{equation}
Even though this function increases in  $\omega$ when $\kw\neq\kb$, the consequences of this are never important because when the ratio is large, the corresponding mode will not significantly contribute to the instability since $\Gamma_a \ll 1$, as can be seen from Eq.~\eqref{eq:Ga2}.

\subsubsection{The \texorpdfstring{$u$--$v$}{u-v} mixing and the parameter \texorpdfstring{$q$}{q}}	%
\label{sec:Tq}												%

\begin{figure}
 \centering
 \psfrag{q}[c][c]{\small $q$}
 \psfrag{T2}[c][c]{\small $|T|^2$}
  \includegraphics[width=0.6\textwidth]{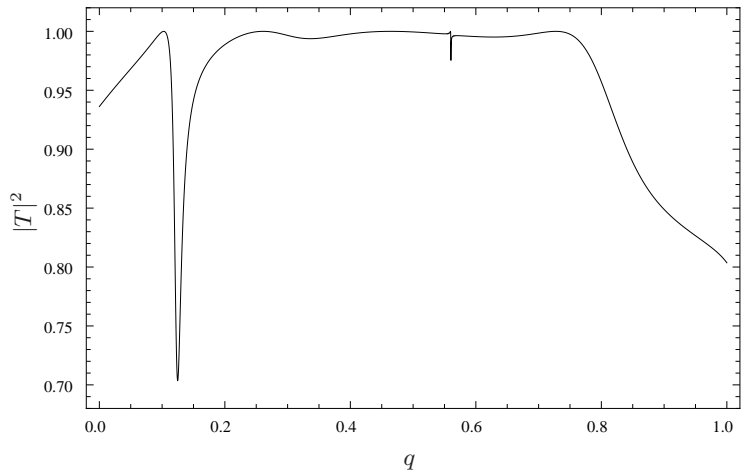}
 \caption{Transmission coefficient $|T_\om|^2$ 
as a function of $q$, at constant $\omega/\kappa=0.100$, and for the other parameters as in the 
bottom plot of  Fig.~\ref{fig:lorentzians}.
The sharp minima at $q=0.1248$ and $q=0.56023$ correspond to complex frequencies at $\om_a/\kappa=0.100$, whereas the broad ones at $q=0.3363$ and $q=0.6285$ correspond to frequencies respectively at $\om_a/\kappa=0.090$ and $\om_a/\kappa=0.103$.}
 \label{fig:Tq}
\end{figure}

The mixing between the $u$-modes, which are right going in the lab in subsonic flows, and the $v$-modes is controlled by the transmission coefficient $T_\om$ of Eq.~\eqref{RT}.
In Fig.~\ref{fig:Tq}, $|T_\om|^2$ is plotted as a function of $q$ for a fixed frequency $\om/\kappa=0.100$. The transmission coefficient is almost 1 between $q=0.25$ and $q=0.75$. Thus, for values of $q$ in this range the $u$--$v$ mixing is negligible, as was noticed in~\cite{MacherBEC} for a single black (or white) hole.
In this regime the approximation discussed in Sec.~\ref{sec:wkb} is valid.
There is, however, an important difference with respect to the single black hole case.
In black-hole--white-hole geometries, the transmission coefficient deviates from zero when approaching a resonance.
When working with a fixed $\omega$, we found two sharp spikes Fig.~\ref{fig:Tq} for $q=0.1248$ and $q=0.56023$ which exactly correspond to two unstable modes.
Moreover, the width of the spikes is almost equal to the corresponding $\Gamma_a$. We also looked for complex eigenfrequency associated with the other two local and broad minima of $|T|^2$, at $q=0.3363$ and $q=0.6285$, and we found two eigenmodes with respectively $\om_a/\kappa=0.090$ and $\om_a/\kappa=0.103$, \ie\/ in the neighborhood of the chosen frequency $\om/\kappa=0.100$.
These considerations show that close to resonances one cannot completely neglect the $u$--$v$ mixing.
Nevertheless, as shown in Fig.~\ref{fig:peakswkb}, the improved treatment of the BS equation produces very good estimates for the values of the real part $\omega_a$ of the complex frequencies.

\subsubsection{The role of the maximal frequency \texorpdfstring{$\ommax$}{}}	%
\label{sec:ld}									%

\begin{figure}
 \centering
 \psfrag{oma}[c][c]{\small $\om/\ommax$}
 \psfrag{b2}[c][c]{\small $|{\cal B}_\om^{(2)}|^2$}
 \psfrag{L}[c][c]{\small $L\kb/\chor$}
 \psfrag{r}[c][c]{}
  \includegraphics[width=\textwidth]{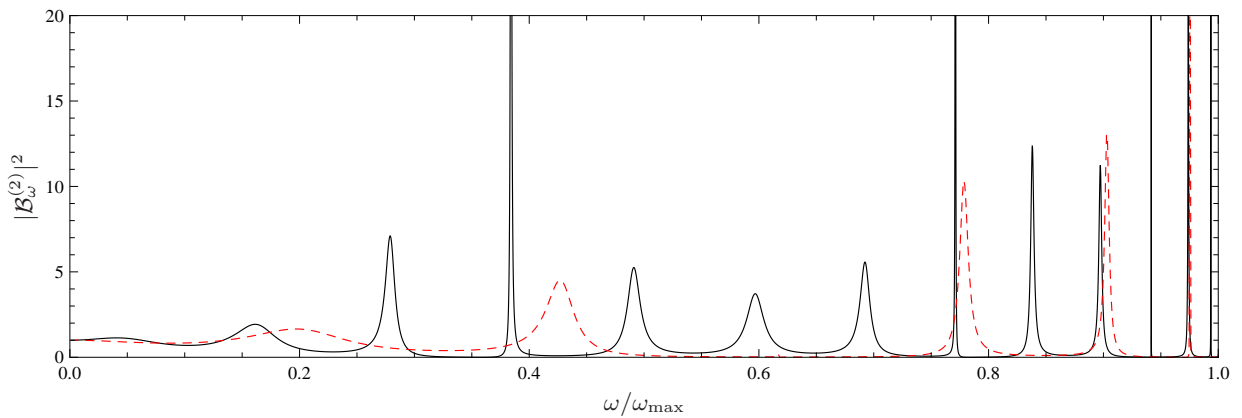}
 \caption{$|{\cal B}_\om^{(2)}|^2$ as a function of $\omega/\ommax$ for different values of $D$ and $\Lambda$ giving rise to the same $\ommax/\kappa=0.19$: black solid line for $D=0.33$, $\Lambda/\kappa=2$, and  red dashed line for $D=0.7$, $\Lambda/\kappa=0.695$.
The other parameters are those of the bottom plot of Fig.~\ref{fig:lorentzians}.}
 \label{fig:ld}
\end{figure}

In \cite{MacherBEC,MacherRP1}, it was shown that the deviations, due to dispersion and with respect to the standard Planckian distribution, of the spectrum  emitted by a single black hole, or white hole, are mainly governed by $\ommax$. However, the latter depends both on the ultraviolet scale $\Lambda$ and the velocity profile (see Sec.~\ref{sec:laser}). The relationship takes the form
\begin{equation}
 \ommax=\Lambda f(D,q).
\end{equation}
Hence the same value of $\ommax$ can be reached from very different cases, and yet it was found that the fluxes are hardly sensitive to this. In the present case however, this insensitivity is lost because the number of resonances directly depends on $\Lambda$. This can be understood by studying Eq.~\eqref{eq:bs}, and it is manifest in Fig.~\ref{fig:ld}, where $|{\cal B}_\om^{(2)}|^2$ is plotted as a function of $\omega/\kappa$ for two different values of $\Lambda$.

%

\subsection{Growth of the asymptotic phonon fluxes}	%
\label{sec:growthofn}					%

\begin{figure}
 \centering
 \psfrag{om}[c][c]{\scriptsize $\om/\ommax$}
 \psfrag{t}[c][c]{\scriptsize $T_{\om_a}^{\rm b}$}
 \psfrag{gamma}[c][c]{\scriptsize $\Gamma_a/\kappa$}
 \psfrag{n}[c][c]{\scriptsize $\dd P/\dd T$}
 \includegraphics[width=.5\textwidth]{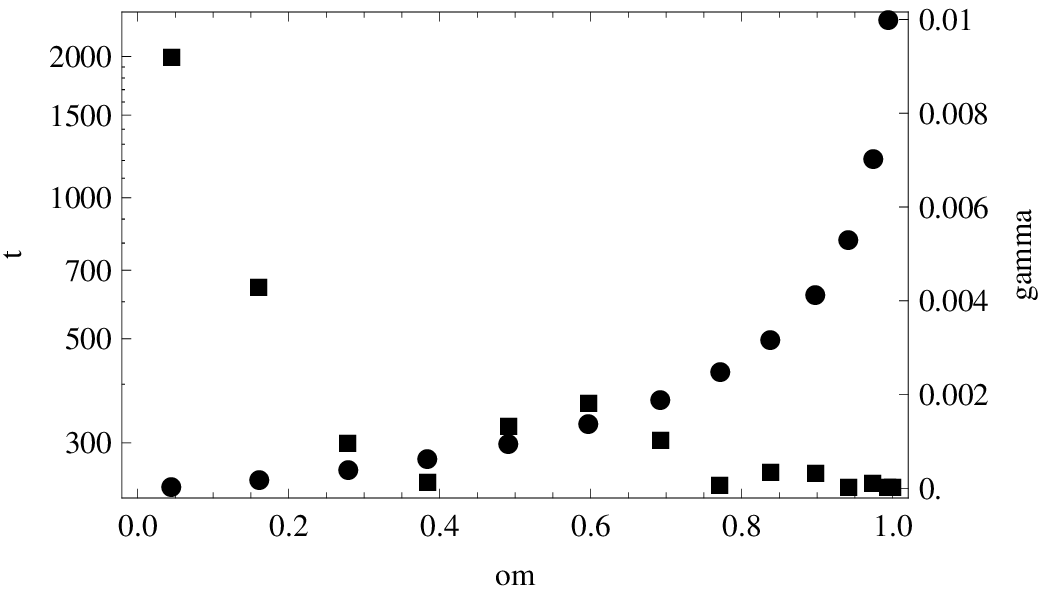}
 \hspace{.01\textwidth}
 \includegraphics[width=.45\textwidth]{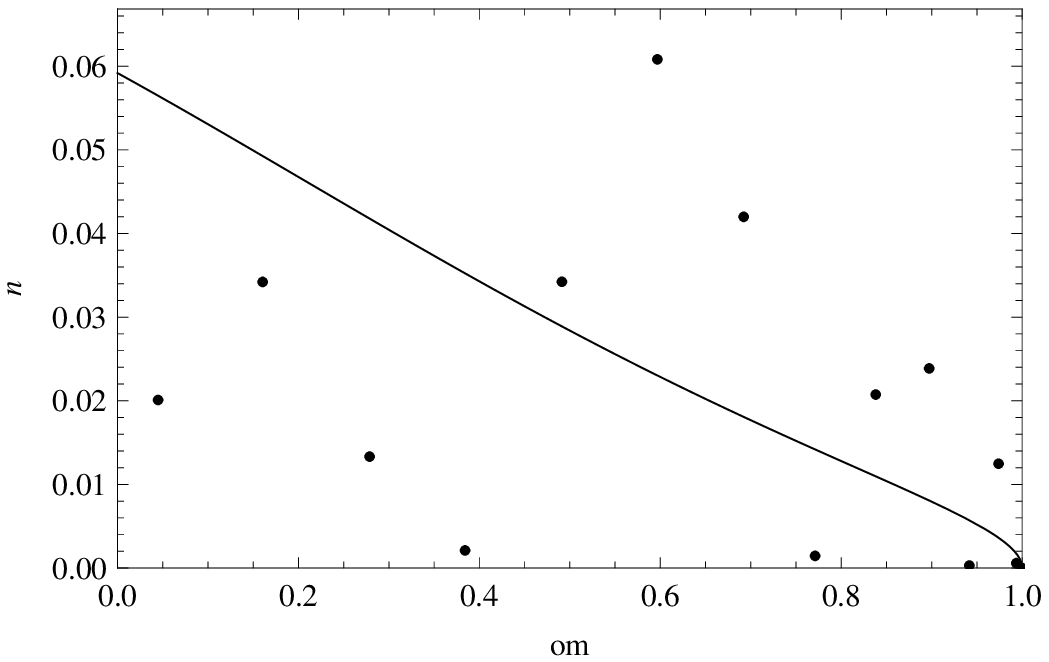}\\
 \psfrag{pr}[c][c]{\scriptsize $P(\om,T=30\kappa^{-1})$}
 \includegraphics[width=.45\textwidth]{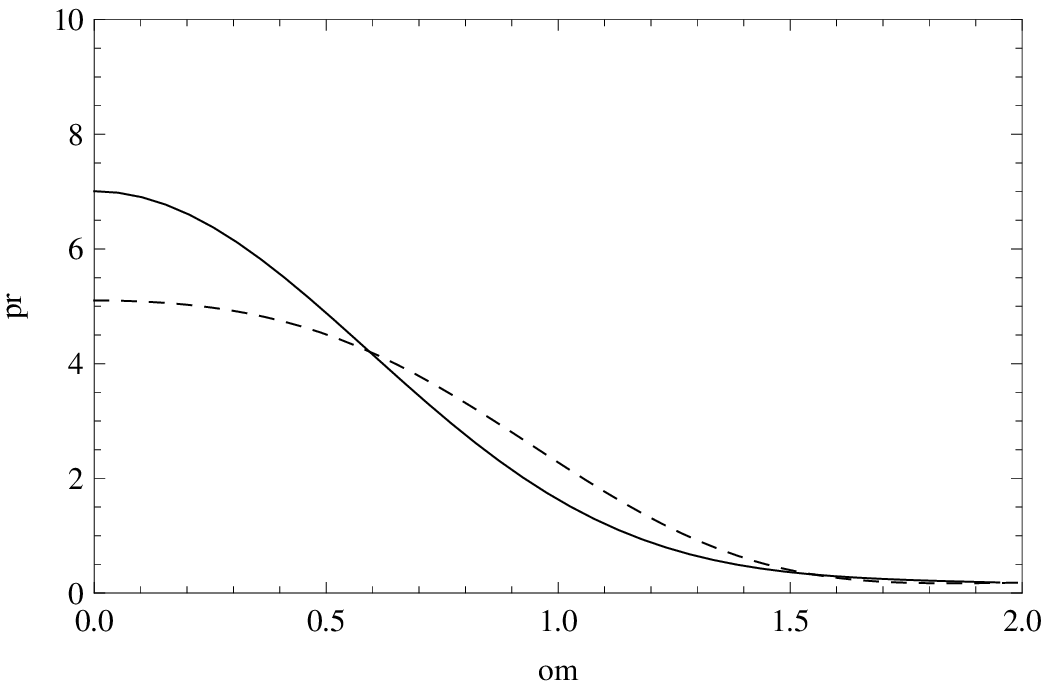}
 \hspace{.06\textwidth}
 \psfrag{pr}[c][c]{\scriptsize $P(\om,T=200\kappa^{-1})$}
 \includegraphics[width=.45\textwidth]{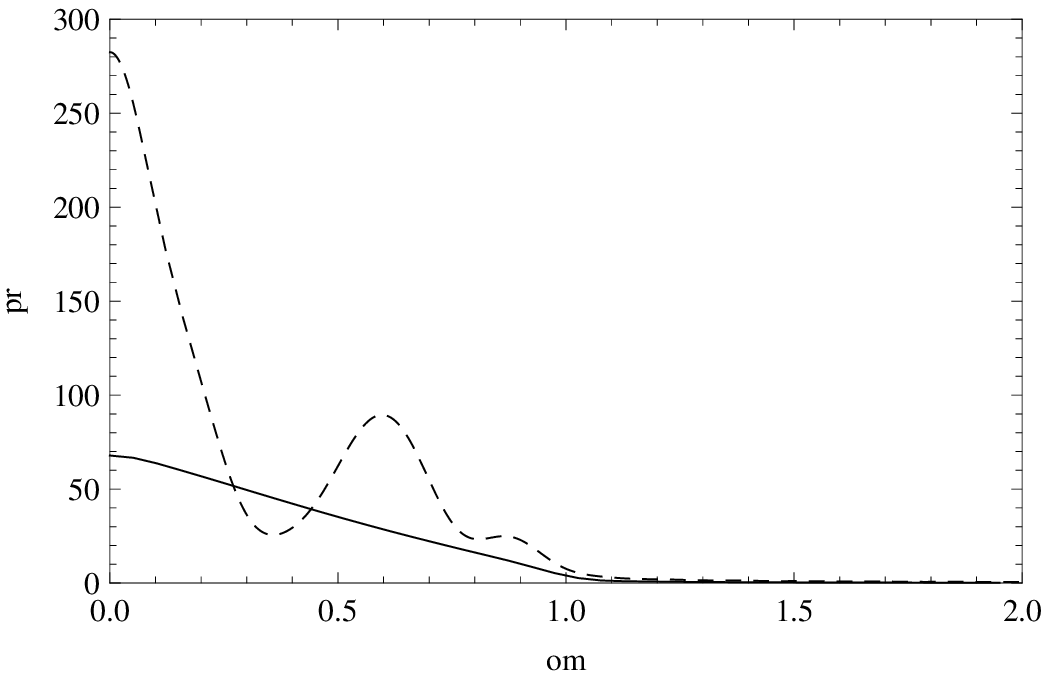}
 \caption{Upper left panel: bouncing time $T_{\om_a}^{\rm b}$ of Eq.~\eqref{eq:tb} as a function of $\om/\ommax$ in unit of $\kappa^{-1}$ (points) and corresponding values of $\Gamma_a$ (boxes), when $L \kappa /\chor = 25$.
The condensate parameters are as in Fig.~\ref{fig:lorentzians}, lower plot.
Upper right panel, solid line:  transition rate associated with~\eqref{rate} due to the radiation emitted by the sole black hole as a function of $\om/\ommax$ (solid line); corresponding quantity due to the discrete set of complex frequency modes in the black-hole--white-hole geometry.
The oscillations are due to the cosine in Eq.~\eqref{eq:gamma}.
Lower panels: $P(\om, T)$ of Eq.~\eqref{rate} as a function of $\om/\ommax$ for an isolated black hole (solid line) and for a black-hole--white-hole pair (dashed line), respectively at time $T=30\kappa^{-1}$ (left plot) and $T=200\kappa^{-1}$ (right plot).}
 \label{fig:fluxes}
\end{figure}

In Fig.~\ref{fig:fluxes}, we represent the relevant quantities that govern the properties of the ``probability'' function introduced in Eq.~\eqref{rate}.
On the left upper plot, we present the growth rates $\Gamma_a$ and the corresponding times $T_{\om_a}^{\rm b}$ of Eq.~\eqref{eq:tb} for the 13 complex frequency modes that exist in the condensate flow of Fig.~\ref{fig:lorentzians}, lower plot.
On the right upper plot, we show the golden-rule transition rate associated with~\eqref{rate} in that black-hole--white-hole geometry (dots), and the rate one would obtain if the white hole were not present. In that case, the quantity plotted is proportional to $\om \times \bar n_\om$, where $\bar n_\om$ is the mean occupation number in the black hole geometry as computed in~\cite{MacherBEC}.
It is given by $\bar n_\om = \vert w_\om \vert^2/(1 - \vert w_\om \vert^2 )$, see $U_3$ of Eq.~\eqref{eq:u3}, where $\vert w_\om \vert^2 $ is well approximated by
\begin{equation}
\vert w_\om \vert^2 = \ee^{- 2\pi \om/\kappa} (1 -\om/\ommax)^{1/2}.
\end{equation}
The prefactor of $\om$ is due to the normalization of the $\chi$ modes of Eq.~\eqref{chimodes}.
In a black-hole--white-hole, the quantity that corresponds to $\om \times \bar n_\om$ is $\om_a \times 2\Gamma_a T_{\om_a}^{\rm b}$. In both cases, these quantities vanish for $\om>\ommax$.

In the lower plots, we show $P(\om,T)$ [see Eq.~\eqref{rate}], at two different times, and for both the black-hole--white-hole and the isolated black hole flows. As expected, the exponential growth of the laser effect does not show up for times $T$ smaller than the inverse of the maximal $\Gamma_a$ (here $T \sim 110/\kappa$).
Moreover, the discreteness of the spectrum is not visible either at early times.
To be resolved it requires times larger than $2\pi \Delta \om_a$, where $\Delta \om_a$ is the frequency gap between neighboring BS frequencies $\om_a$. It is here of the order of $\Delta \om_a \sim \ommax/10 \sim 0.02 \kappa$.

On the contrary, for times of the order of $1/{\max}(\Gamma_a)$ or greater, both the exponential growth of the laser effect and the discreteness of the spectrum show up. These allow one to distinguish the phonon flux emitted by the black-hole--white-hole pair from that emitted by the sole black hole that grows linearly in $T$ for all values of $\om$.
This linear growth can be observed by comparing the continuous lines of the right upper and lower plots, and constitutes a numerical validation of Fermi's golden rule!

\subsection{Spatial properties of complex frequency modes and correlation patterns}	%
\label{sec:modesandcorrelation}								%

The density--density correlation function can be calculated following the procedure described in Sec.~\ref{sec:correlations}. As outlined there, only the largest-$\Gamma_a$ mode significantly contributes to the pattern at late time, because this mode grows fastest than those with lower $\Gamma_a$, having the shortest instability timescale $1/\Gamma_a$. However, at early times, also modes with lower $\Gamma$ contribute to the formation of the total correlation pattern as well as to the flux of phonons (see Fig.~\ref{fig:fluxes}).
For this reason and to appreciate the variety of cases, it is worth studying the modes and the corresponding patterns also for lower $\Gamma$.

\begin{figure}
  \centering
   \psfrag{x}[c][c]{\small $x\kappa/\chor$}
 \psfrag{mode}[c][c]{\small $\xl$}
  \includegraphics[width=.7\textwidth]{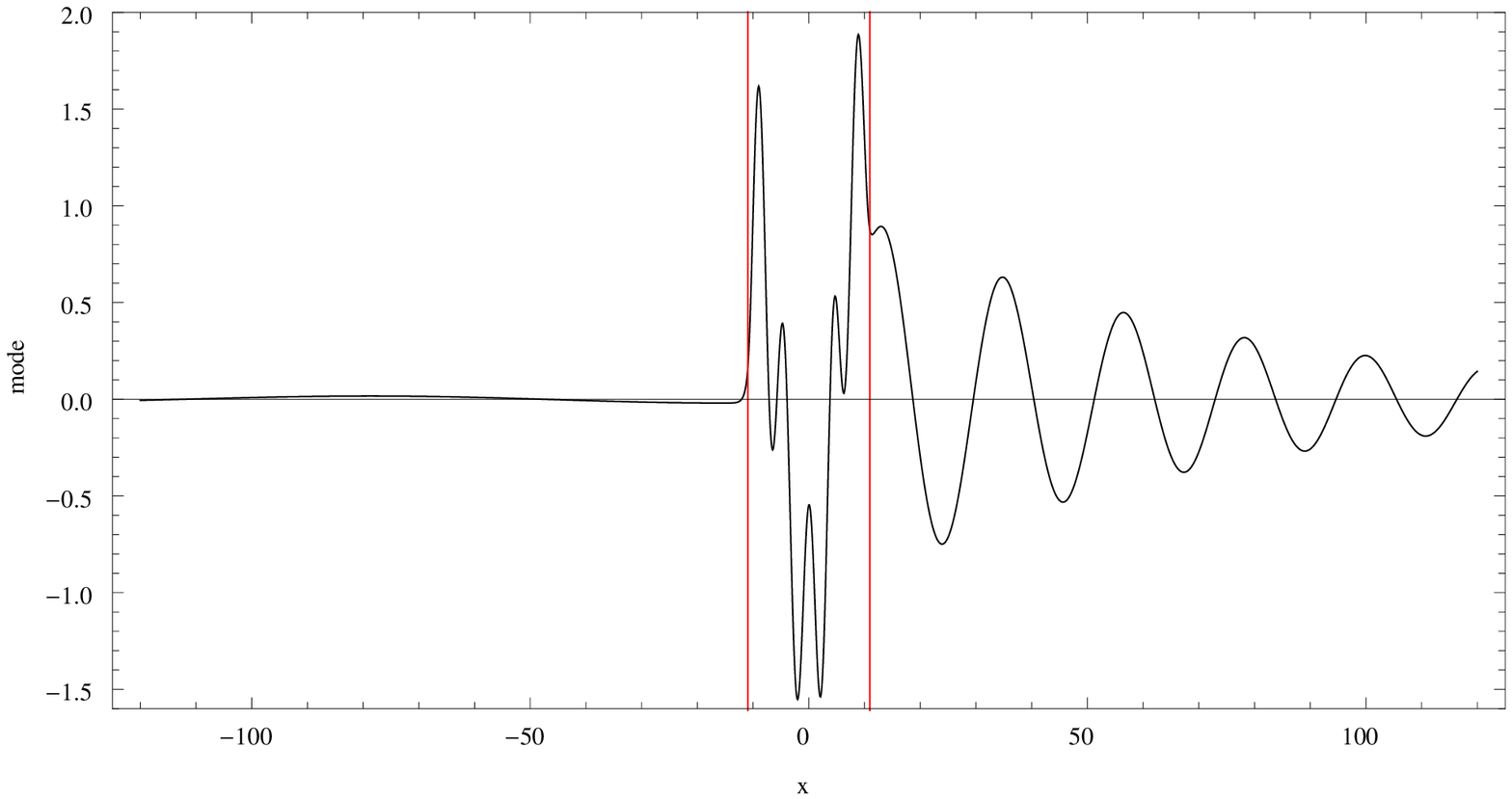}
  \includegraphics[width=.7\textwidth]{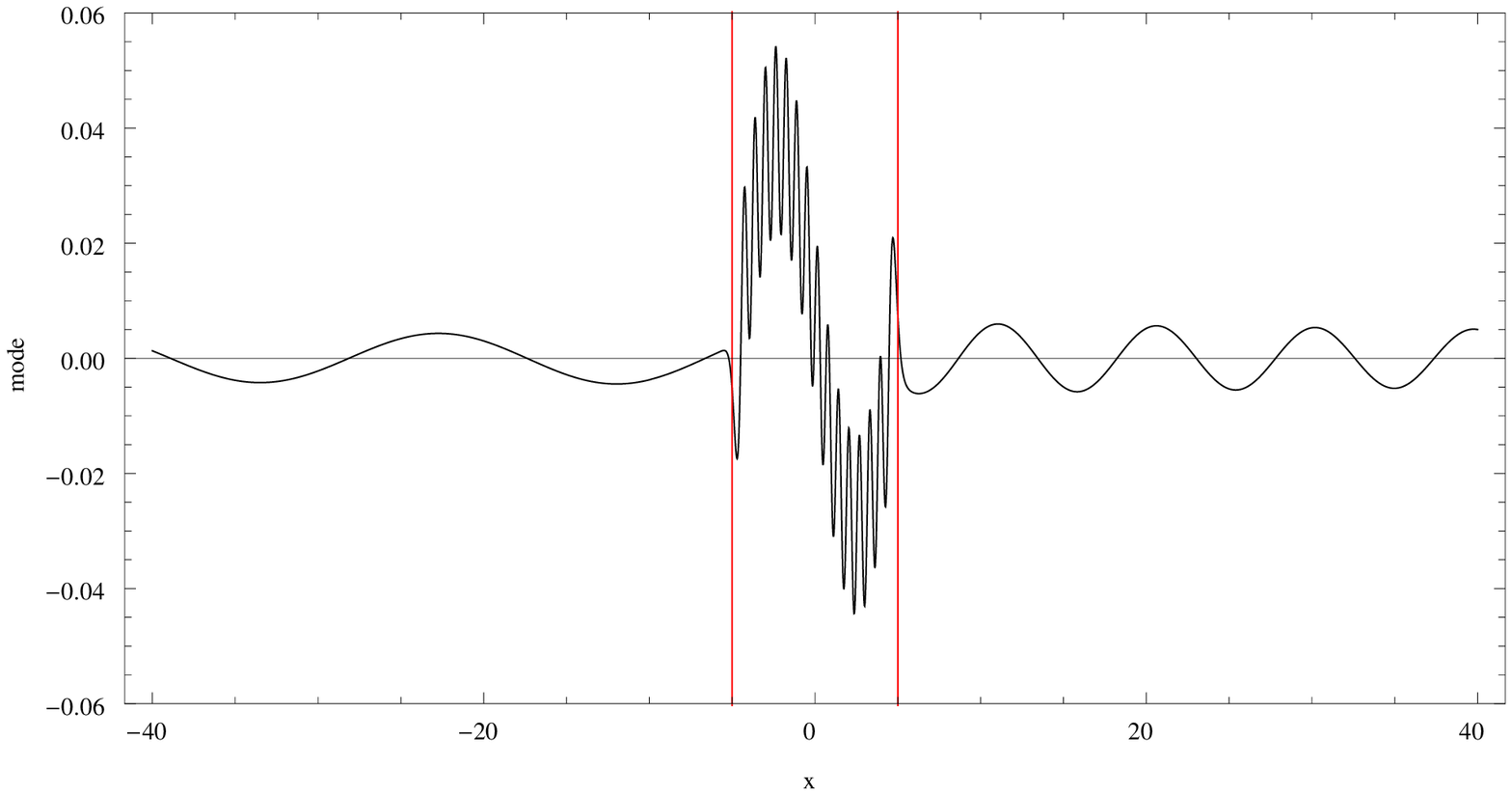}\\
  \hspace{0.01\textwidth}\includegraphics[width=.68\textwidth]{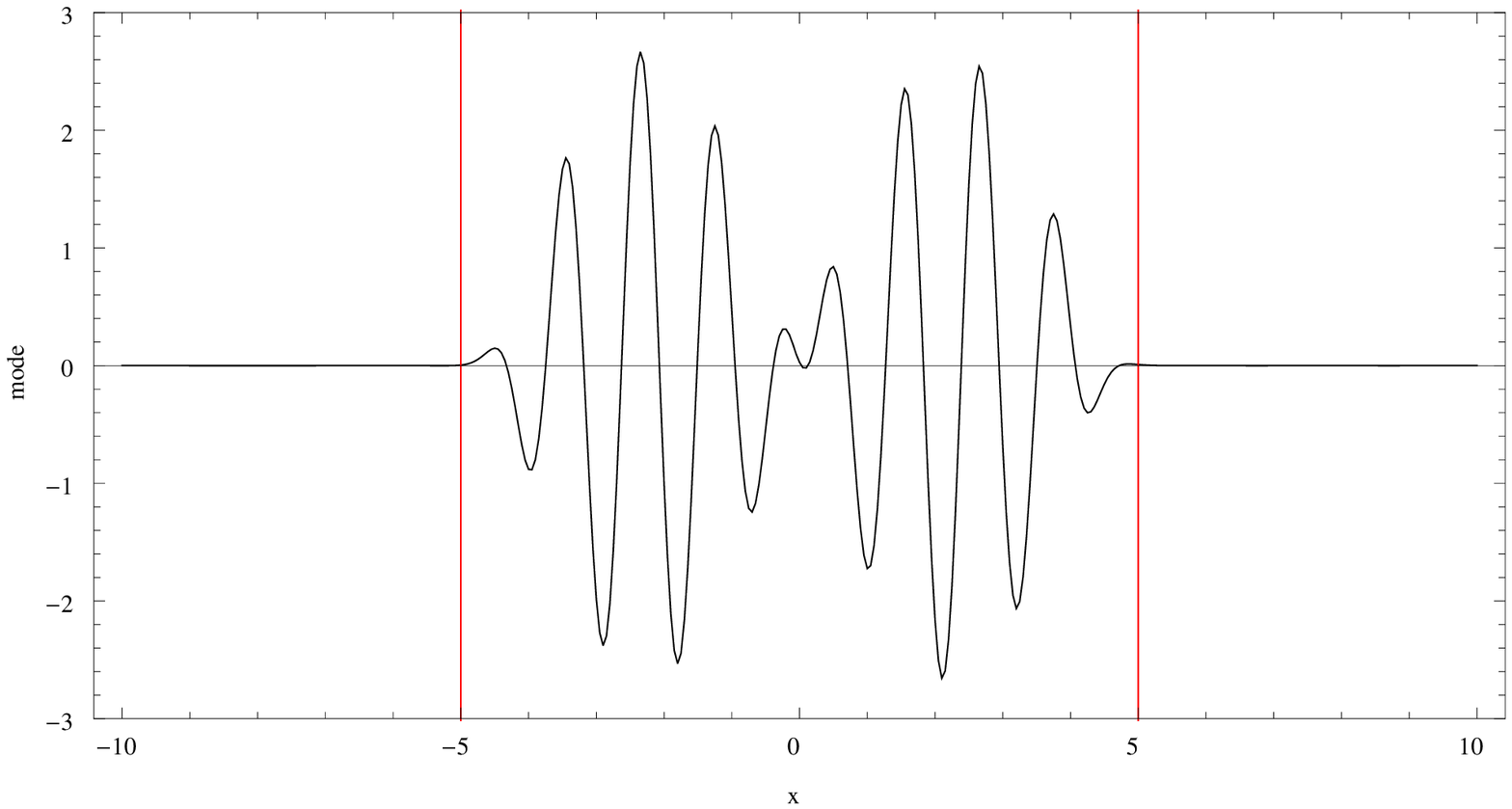}
  \caption{Real part of the growing mode $\xl$ of Eq.~\eqref{eq:phiexpansion} for respectively high, moderate and low values of $\Gamma$.
Red lines: white and black horizons.
Upper panel: fourth eigenmode with $\om_4/\kappa=0.098$, $\Gamma_4/\kappa=0.0056$, with $L\kappa/\chor=10.95$ and the other parameters as those of Fig.~\ref{fig:lorentzians}.
Central panel: eigenmode with $\nbs=14$, $\om_{14}/\kappa=0.46$ and $\Gamma_{14}/\kappa=0.004$.
$\ommax/\kappa=2.53$, $q=0.7$, $D=0.7$, $n=2$, $\kw=\kb=\kappa$, $\Lambda/\kappa=8$, $L\kappa/\chor=5$.
Lower panel: eigenmode in the same background with $\nbs = 2$, $\om_2/\kappa=2.48$, $\Gamma_2/\kappa=2.\times 10^{-6}$.
The importance of $\Gamma$ can be estimated from the relative mode amplitude in the right outside region versus that inside in the trapped region.
Online movies showing the evolution of $\xl$ as a function of $t$, respectively \href{http://people.sissa.it/finazzi/bec_bhlasers/movies/laser_1.gif}{laser\_1.gif} (from $t=-400/\kappa$ to $400/\kappa$), \href{http://people.sissa.it/finazzi/bec_bhlasers/movies/laser_2.gif}{laser\_2.gif} (from $t=-200/\kappa$ to $200/\kappa$), \href{http://people.sissa.it/finazzi/bec_bhlasers/movies/laser_3.gif}{laser\_3.gif} (from $t=-400/\kappa$ to $400/\kappa$)(available from~\urlnjp).
}
 \label{fig:mode}
\end{figure}

As a typical example of a mode with a large $\Gamma$, we consider the fourth BS mode in a configuration with fivecomplex eigenfrequencies. In Fig.~\ref{fig:mode} (upper panel) the real part of this mode is represented, while the supplementary movie \href{http://people.sissa.it/finazzi/bec_bhlasers/movies/laser_1.gif}{laser\_1.gif} (available from~\urlnjp) gives its evolution in time.
As a second example, we choose a very different situation with $q=0.7$ and a dispersion relation with higher $\Lambda$.
In this case, the spectrum of complex frequencies is large. In Fig.~\ref{fig:mode} two modes are plotted: one with $\nbs=14$ and a moderate value of $\Gamma$ (central panel), and one with $\nbs=2$ and a very small value of $\Gamma$ (lower panel).
When comparing the modes and their correlation patterns, some features remain the same, whereas others significantly differ.

\begin{figure}
  \centering
  \includegraphics[width=.9\textwidth]{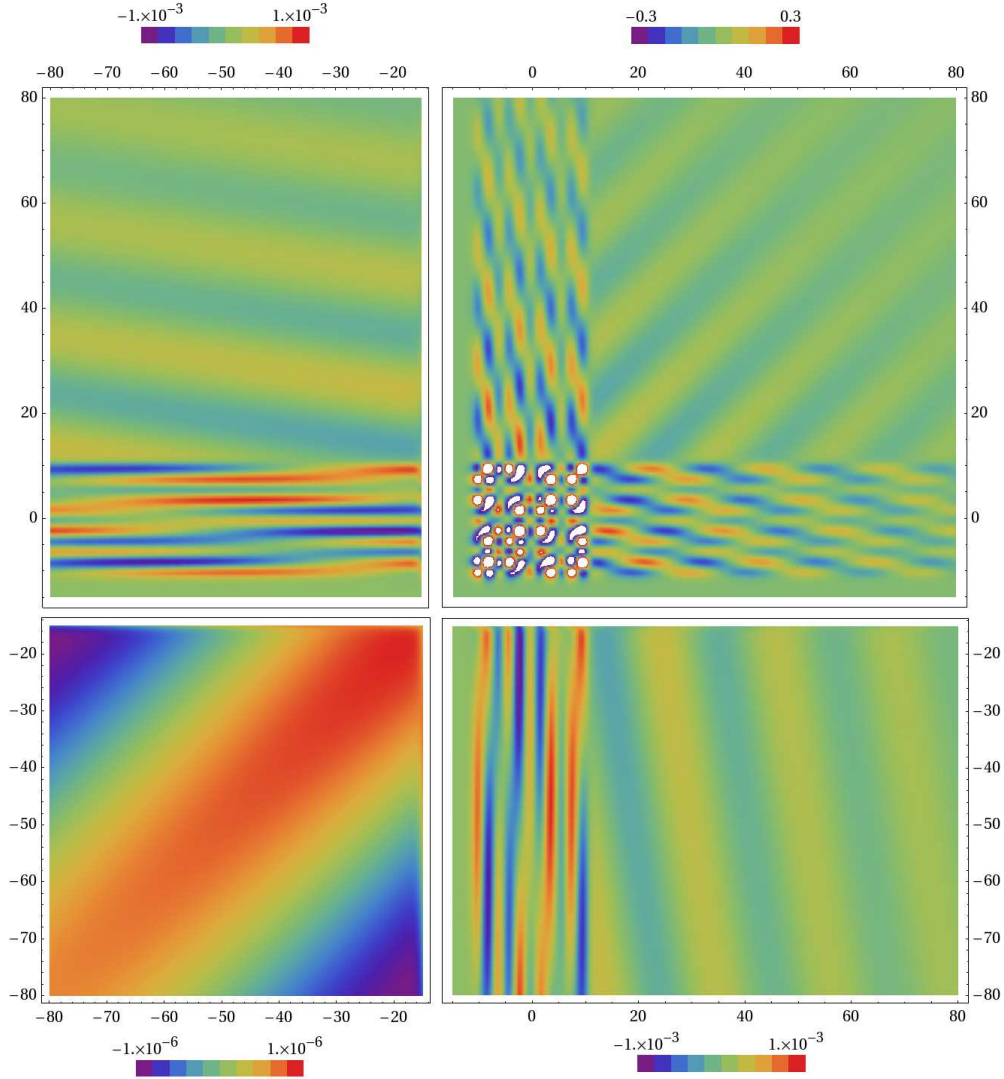}
 \caption{Density--density correlation function~\eqref{eq:densitycorrelation} divided by $e^{2\Gamma_a t}$ in the plane $(x,x')$, for the mode with a high $\Gamma$ represented in the upper plot of Fig.~\ref{fig:mode}, with $x$ and $x'$ in units of $\chor/\kappa$. We have divided the expression by $e^{2\Gamma_a t}$ in order to suppress the dependence on time.
Since the correlations between different regions have very different amplitudes, we were obliged to use different scales to reveal them.
So we split the plot in four quadrants with different ranges of color coding, corresponding to $x<-15\chor/\kappa$, $x>-15\chor/\kappa$, and $x'<-15\chor/\kappa$, $x'>-15\chor/\kappa$.
The central square from $-11$ till $11$ is the trapped region.
It contains the strongest correlations since the mode amplitude is highest in it.
The two (symmetrical) bands of width $\sim 22$ in the right upper plot describe the correlations between the trapped region and the external region on the right of the black hole, while the other two symmetric bands in the left upper panel and in the right lower panel describe those with the left external region. The correlations in the lower left panel are those on the left of the white hole.
They are much weaker and of greater wavelength, in agreement with the mode properties on the left side of the upper plot of Fig.~\ref{fig:mode}.}
 \label{fig:correlations}
\end{figure}
\begin{figure}
  \centering
  \includegraphics[width=.9\textwidth]{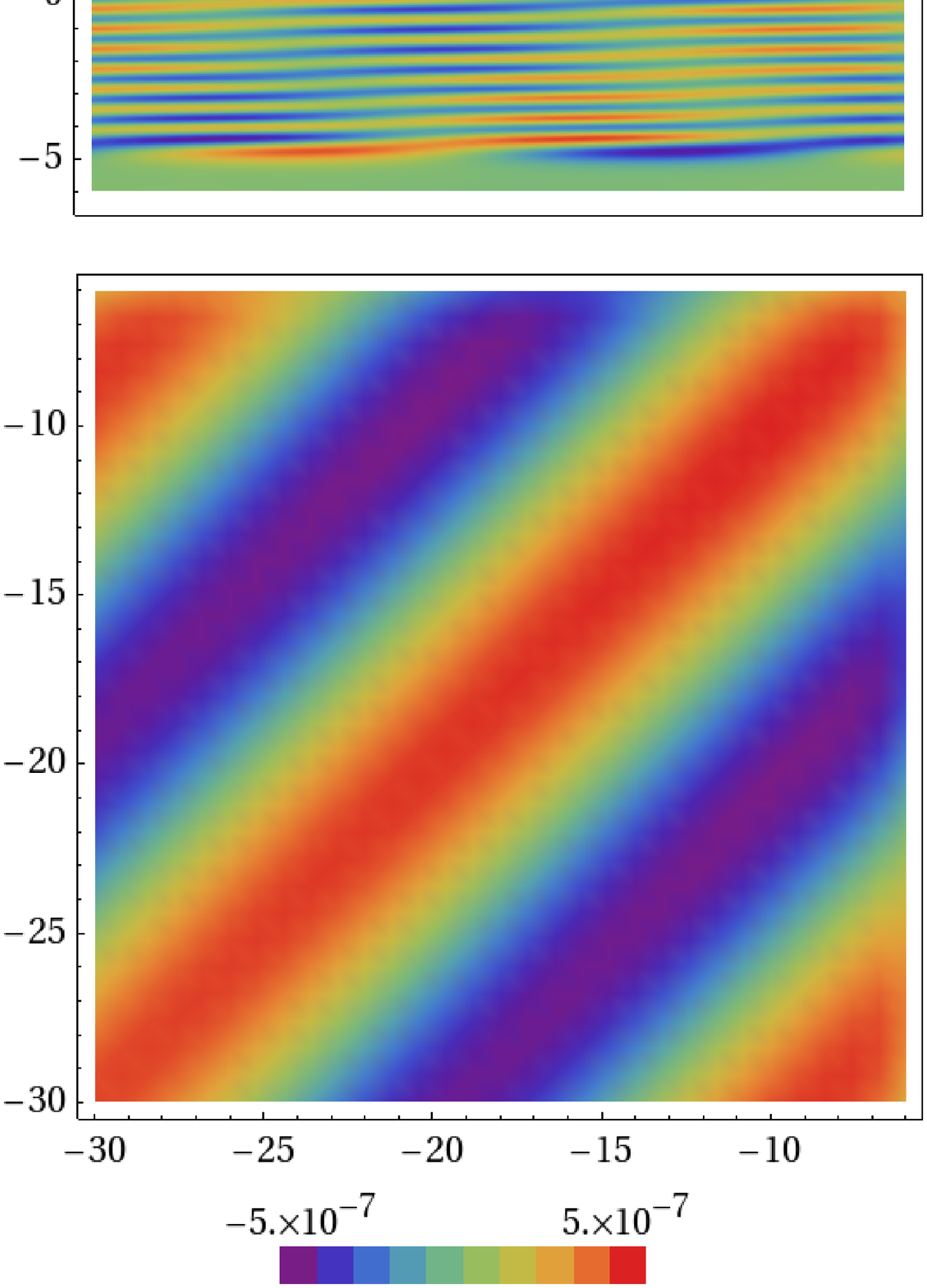}
 \caption{Density--density correlation function~\eqref{eq:densitycorrelation} divided by $e^{2\Gamma_a t}$ in the plane $(x,x')$, for the mode with a moderate $\Gamma$ represented in the central plot of Fig.~\ref{fig:mode}, with $x$ and $x'$ in units of $\chor/\kappa$. The plot region is split in four quadrants with different scales, since the correlations between different regions have very different amplitudes (see the analogous comment in the caption of Fig.~\ref{fig:correlations}).
Similarities with the former figure are manifest, and concern both the spatial properties of the pattern
and the amplitudes of the correlations.
The main difference concerns the high-frequency modulations of the inside--inside and inside--outside correlations
which are due to the fast oscillations of the mode amplitude in the trapped region, which are visible in the central plot of Fig.~\ref{fig:mode}, and which are due to the fact that $\nbs$ is high: $\nbs = 14$.}
 \label{fig:correlations3}
\end{figure}
\begin{figure}
  \centering
  \includegraphics[width=\textwidth]{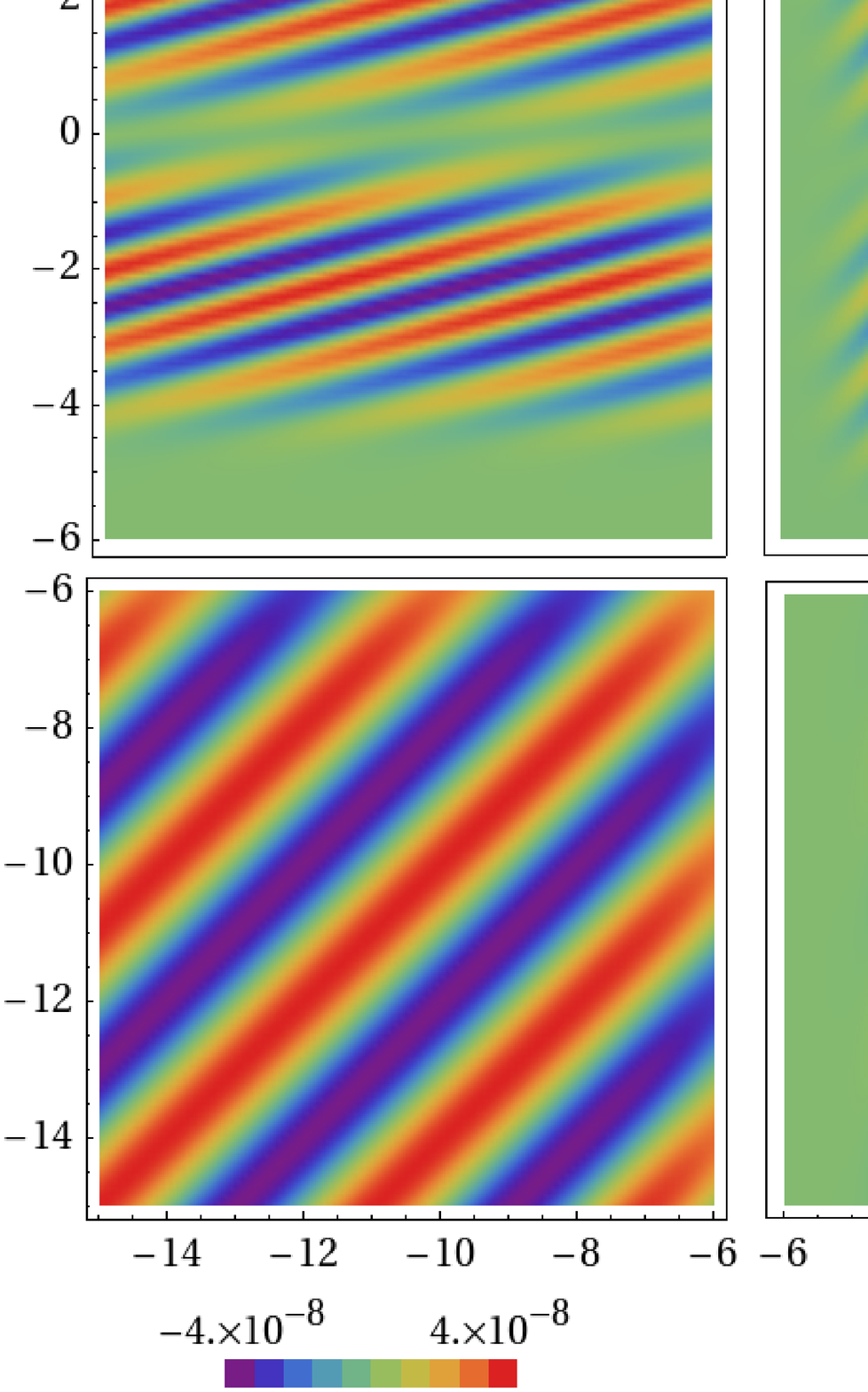}
 \caption{Density--density correlation function~\eqref{eq:densitycorrelation} divided by $e^{2\Gamma_a t}$ in the plane $(x,x')$, for the mode with a low $\Gamma$ represented in the lowest panel of Fig.~\ref{fig:mode}, with $x$ and $x'$ in units of $\chor/\kappa$. The plot region is split in nine quadrants with different scales, since the correlations between different regions have very different amplitudes (see the analogous comment in the caption of Fig.~\ref{fig:correlations}).
The bottom legend corresponds to the central region of the plot. One clearly sees the two nodes in the trapped region.
One also finds them in the inside--outside correlations, both on the left and on the right of the trapped region.
Finally, one notices that in the present case the $u$--$u$ correlations on the right upper square are weaker than the $v$--$v$ correlations on the lower left panel. This follows from the high $u$--$v$ coupling between the
trapped mode, which is a $u$-mode, and the outside $v$-mode.}
 \label{fig:correlations2}
\end{figure}

In the first example, since $\Gamma$ is high, the laser effect is strong as can be seen from the movie: in the right asymptotic region there is an exponentially growing right-going flux.
This growth in time can also be inferred from the spatial decrease of the mode amplitude on the right of the black hole, because these are linearly related, see equation~(54) of \cite{cp}.
Considering the correlation pattern of this mode in Fig.~\ref{fig:correlations}, one sees that the dominant signals come from the inside region (the central square of size $\sim 22$) where the amplitude of the mode is highest, and from the horizontal and vertical bands which describe the correlations between the escaping flux and the trapped mode.

From the mode itself and from its correlation pattern in Fig.~\ref{fig:correlations}, one can see that the $u$--$v$ mixing is very low, as expected from the choice of $q=0.5$. The left-going mode coming out of the white hole is about 2.5 orders of magnitudes smaller than the right-going one that comes out of the black hole, as can be seen from the ratio of amplitudes between the top left part and top right part in Fig.~\ref{fig:correlations}.
To understand how the modes propagate, it is appropriate to consider the top left part.
This region represents the correlations between $u$ (right-going) and $v$ (left-going) modes. The slope of the highest/lowest correlation lines gives the ratio between the velocities of the two modes. 
When the modes form a continuous set, the two velocities are the group velocities. 
However, in the present case, when considering a single mode,
the slope of these lines corresponds instead to the ratio between the phase velocities of $u$- and $v$-modes.
Similarly, the pattern in the top right part are simply due to the nodes of the right moving mode evaluated at a given time,
see equation~(61) in \cite{cp}.

We now consider the correlation pattern of Fig.~\ref{fig:correlations3}, corresponding to a mode with a moderate value of $\Gamma$ and $\nbs$ very large. The overall properties of the pattern are basically the same as those of Fig.~\ref{fig:correlations}. The main differences come from the high wave-vector content of the trapped mode, which produces short distance features in the central square and in the bands. Outside these regions the patterns are very similar.

In the last pattern of Fig.~\ref{fig:correlations2}, corresponding to a mode with a very small value of $\Gamma$ and $\nbs=2$, we get a different picture. The pattern is basically concentrated in the central square since the amplitude of the mode outside is related to its amplitude inside by the square root of $\Gamma$, as can be seen from~\eqref{eq:gamma}. In the present case, the $u$--$v$ mixing is so high that the pattern on the left of the white hole is higher (by a factor of about 5) than on the right of the black hole.
One also notices that in the central panel ($\nbs=2$) there is only one node. Moreover, the beat-like shape shows that this wave is the result of the superposition of two waves with very similar and relatively high wave vectors.

\subsection{The Technion experiment}	%
\label{sec:technion}			%

Quite recently a black-hole--white-hole flow has been experimentally realized~\cite{technion}.
Our numerical code can describe this configuration, up to some limitation.
In fact, the velocity profile [Fig. 3(g) of~\cite{technion}] is rather irregular while our code works accurately only for velocity profiles as in Fig.~\ref{fig:qD}, right panel, with an almost-flat internal region.
Nevertheless, we shall proceed in order to obtain at least an estimation on the number of the unstable modes and their growing rate.
As far as we know, this is the first estimate of this quantity, which is very relevant from an experimental standpoint.

The main parameters of a typical experimental setting are\footnote{\label{foot:private}Private communication of J. Steinhauer to SF and R. Parentani (2010).}
\begin{align}
 &\chor = 7.3\times10^{-4}\;{\rm m/s},\nonumber \\
 &\kb = \kw/2 = \kappa = 213.7\;{\rm s}^{-1},\nonumber \\
 &v(x=0) = -3.6\times10^{-3}\;{\rm m/s} = -4.9\,\chor,\nonumber \\
 &c(x=0) = 3.5\times10^{-4}\;{\rm m/s} = 0.48\,\chor,\label{eq:c0}\\
 &c(x\to\infty) = 8\times10^{-4}\;{\rm m/s} = 1.1\,\chor,\label{eq:cinf} \\
 &2L = 10^{-5}\;{\rm m}=1.5\,\chor/\kappa,\nonumber \\
 &\Lambda/\kappa=6.\nonumber 
\end{align}

\begin{figure}
 \centering
 \psfrag{om}[c][c]{\small $\omega/\ommax$}
 \psfrag{b2}[c][c]{\small $|{\cal B}_\om^{(2)}|^2$}
 \includegraphics[width=0.45\textwidth]{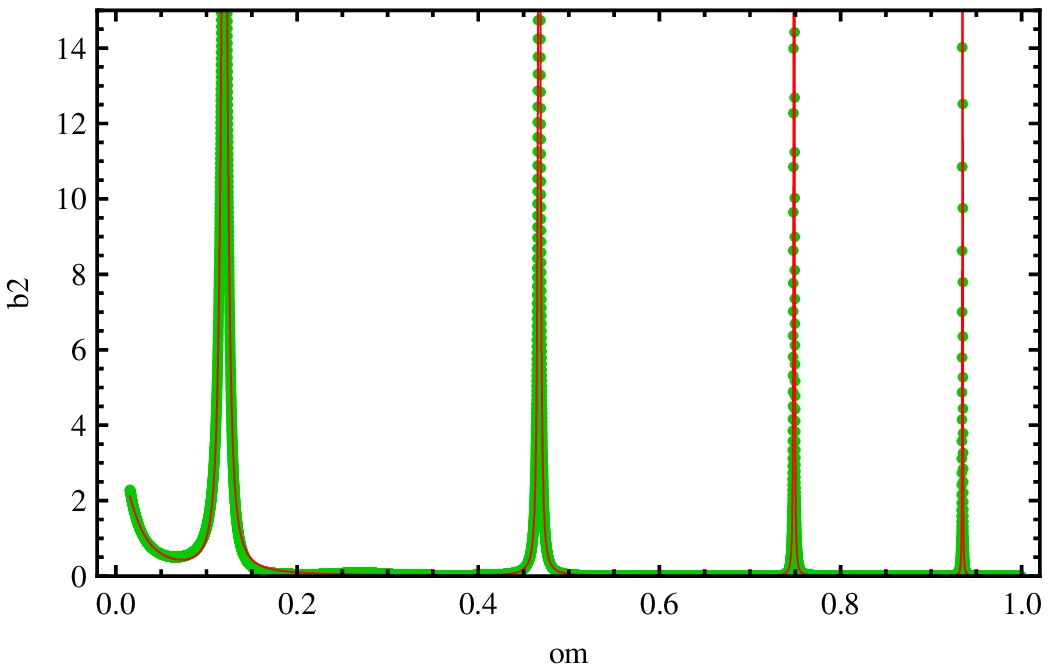}
 \includegraphics[width=0.45\textwidth]{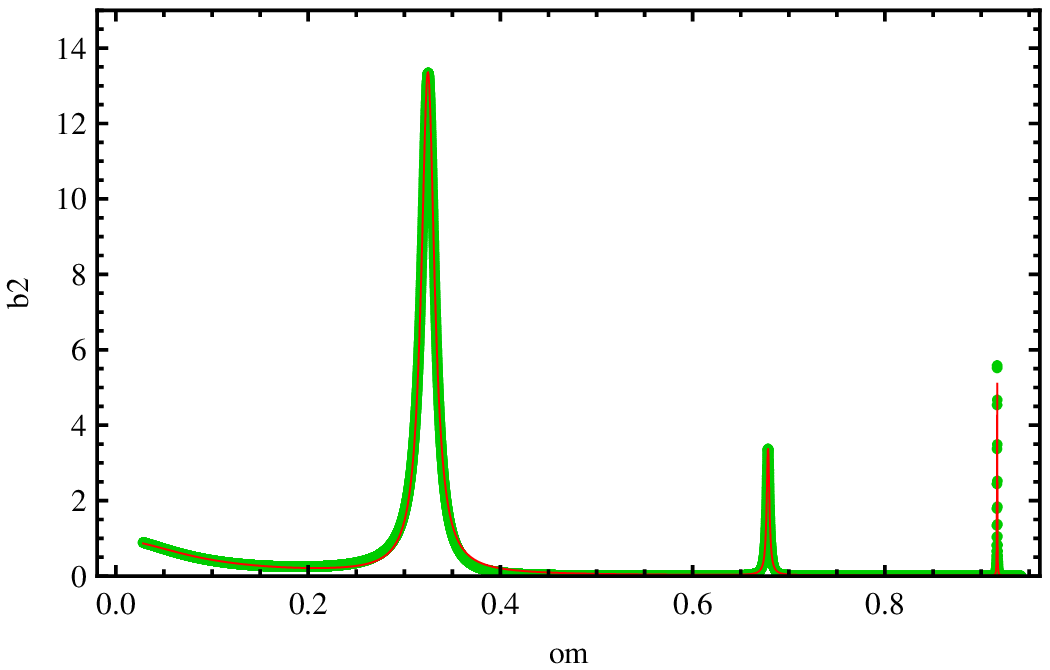}
 \caption{$|{\cal B}_\om^{(2)}|^2$ as a function of $\omega/\ommax$ for  $q=0.9$, $\ommax/\kappa\approx3.4$ (left panel) and $q=0.48$, $\ommax/\kappa\approx2.7$ (right panel).
The values of the other parameters are given in the text. Green points: numerical simulation; red lines: fitted series of Lorentzians.
$\omega_a$ and $\Gamma_a$ from the fit are reported in Tab.~\ref{tab:tech}.}
 \label{fig:tech}
\end{figure}

To reproduce the above setting we perform two simulations with fixed $D=1$.
In Fig.~\ref{fig:tech}, $|{\cal B}_\om^{(2)}|^2$ is plotted as a function of $\omega/\ommax$ for $q=0.9$ (left panel) and $q=0.48$ (right panel). These two values are obtained by using, respectively, $c(x=0)$ and $c(x\to\infty)$ equal to their experimental values in Eqs.~\eqref{eq:c0} and~\eqref{eq:cinf}.%
\footnote{We used these two values because, in our analysis, it is not possible to fix independently $c(x=0)$ and $c(x\to\infty)$.
We expect that the actual value of the complex frequencies would lie in between the two sets we obtained.}
The small distance between the two horizons implies that few trapped modes are present in this situation.
The values of $\Gamma_a$ and $\omega_a$ of the complex eigenfrequencies are reported in both cases in Tab.~\ref{tab:tech}.
The dominant contribution comes from the eigenfrequency with the largest $\Gamma_a$.
In the two cases this corresponds to an instability time scale, respectively $\tau\approx0.06\;{\rm s}$ and $\tau\approx0.02\;{\rm s}$. Even if the two simulations are generated with very different values of $q$, the two results agree in order of magnitude. Note that the estimated instability time scale is more than twice the time scale for which the horizons are maintained in the experimental configuration ($\approx0.08\;{\rm s}$ for the experiment settings considered in the present analysis).
In order to see the laser effect, it would be needed to maintain the system for a longer time, or to increase $\kappa$ by a factor of 10, without changing the dimensionless parameters of the system such as the ratio $L\kappa/\chor$.

\begin{table}
\centering
\begin{tabular}{ccccccc}
\hline\hline\noalign{\smallskip}
     \multicolumn{3}{c}{$q=0.9$}			&&		\multicolumn{3}{c}{$q=0.48$}		\\
 \noalign{\smallskip}\cline{1-3}\cline{5-7}\noalign{\smallskip}
 $\omega_a/\ommax$ & $\omega_a/\kappa$ & $\Gamma_a/\kappa$ && $\omega_a/\ommax$ & $\omega_a/\kappa$ & $\Gamma_a/\kappa$ \\
 \noalign{\smallskip}\cline{1-3}\cline{5-7}\noalign{\smallskip}
 $2\times10^{-9}$ & $7\times10^{-9}$ & 0.076 && $2\times10^{-8}$ & $5\times10^{-8}$ & 0.23 \\
 0.12 & 0.41 & 0.012 && 0.32 & 0.86 & 0.022 \\
 0.47 & 1.58 & 0.005 && 0.68 & 1.80 & 0.004 \\
 0.75 & 2.54 & 0.0007 && 0.92 & 2.43 & 0.0005 \\
 0.93 & 3.17 & 0.0004 && & & \\
 \noalign{\smallskip}\hline\hline
\end{tabular}
\caption{\label{tab:tech}$\omega_a$ and $\Gamma_a$ for the complex eigenfrequencies in the systems described in the caption of Fig.~\ref{fig:tech}.}
\end{table}

\section{Summary and discussion}	%
\label{sec:conclusion_bhlaser}		%

We analyzed the spectrum of phonons in a BEC when the stationary flow of the condensate crosses twice the speed of sound. Our analysis is based on the BdG equation~\eqref{eq:BdG1}. Hence, even though the analogy with light propagation in a pair of black and white horizons is manifest, none of our results rely on this gravitational analogy.

In the limit where the condensate can be considered as infinite, \ie\/ when no periodic boundary condition is introduced, the spectrum of bounded modes contains real frequency modes that are only elastically scattered, see Eq.~\eqref{RT}, plus a discrete and finite set of pairs of complex frequency modes. These modes can be seen as the resonances of the cavity bordered by the two sonic horizons. They lead to dynamical instabilities because the scattering through each sonic horizon is anomalous, in that it mixes positive and negative norm modes, see~Eq.~\eqref{eq:expnov}.
Then, using semi-classical techniques, we compute the real and the imaginary part of the complex frequencies. The real part $\om_a$ obeys a BS condition that introduces the discreteness of the spectrum, whereas the imaginary part $\Gamma_a$ is related to the norm of the scattering coefficients across the horizons, see~Eq.~\eqref{eq:gamma}.
In the case of toroidal geometry, the analysis would instead be complicated by the discreteness of the wave vectors.
The instability appears only when one $\om_a$ approximately matches one of the frequencies associated with that discrete set of wave vectors. This explains the presence of the instability bands found in~\cite{Garay,Gardiner}.

In Sec.~\ref{sec:numerics} we numerically solved the BdG equation for real and complex frequencies. By comparing the numerical properties with those derived using the BS and~Eq.~\eqref{eq:expnov}, we validated the use of semi-classical methods, see Figs.~\ref{fig:lorentzians}--\ref{fig:omega}. In Sec.~\ref{sec:growthofn} we numerically studied the growth of number of phonons emitted by the black-hole--white-hole system. In spite of the discrete character of the trapped modes, at early time, this quantity behaves very similarly to the flux that the black hole horizon would emit in the absence of the white hole horizon. Instead, for larger time both the instability and the discreteness show up. Finally, we studied the equal time correlation pattern of density--density fluctuations associated with three different complex frequency modes, respectively with high, moderate and low value of the imaginary part $\Gamma_a$.
In all cases, the instability shows up most clearly in the supersonic region where the amplitude of the trapped mode is the highest one. The correlations between the trapped mode and the emitted modes display very specific properties, see Figs.~\ref{fig:correlations}--\ref{fig:correlations2}.

Finally, we studied (within some limitations) the experimental situation realized in June 2009 in the Technion~\cite{technion}.
The results are summarized in Fig.~\ref{fig:tech} and show that few unstable modes are found, and that the instability time scale is about ten times larger than the life time of the condensate.
This means that an increase by a factor of ten of the surface gravity (without changing the dimensionless parameters of the system such as the ratio $L\kappa/\chor$) would lead to comparable time scales. The laser effect could then be observable.

\cleardoublepage
\part{Bose--Einstein condensates as Emergent~Gravity~models}
\label{part:BECemergent}

\chapter{The~cosmological~constant: a~lesson~from~Bose--Einstein~condensates}	
\label{chap:cosmoconstant}							
\chaptermark{Cosmological constant in BECs}					

The cosmological constant~\cite{carroll} has been one of the most mysterious and fascinating objects for both cosmologist and theoretical physicists since its introduction almost one century ago~\cite{firstcosm}. Once called by Einstein his greatest blunder, it seems nowadays the driving force behind the current accelerated expansion of the universe.  The explanation of its origin is considered one of the most fundamental issues for our comprehension of general relativity (GR) and quantum field theory (QFT).

Since this constant appears in Einstein's equations as a source term present even in the absence of matter and with the symmetries of the vacuum ($T_{\mu\nu}^\Lambda\propto g_{\mu\nu}$), it is usually interpreted as a ``vacuum energy''. Unfortunately, this has originated the so-called ``worst prediction'' of physics. In fact, the estimated value, which is na\"ively obtained by integrating the zero-point energies of modes of quantum fields below Planck energy, is about 120 orders of magnitude larger than the measured value. 
Despite the large number of attempts (most notably supersymmetry~\cite{susy}, which, however, must be broken at low energy) this problem is still open. {We can summarize the situation by saying that, given the absence of custodial symmetries protecting the cosmological term from large renormalization effects, the only option we have to explain observations is fine tuning~\cite{finetuning, rovelli}.}

This huge discrepancy is plausibly due to the use of effective field theory (EFT) calculations for a quantity which can be computed only within a full quantum gravity (QG) theory. Unfortunately, to date, we do not have any conclusive theory at our disposal. However, the possibility of a failure of our EFT-based intuition is supported by what can be learned from analogue models of gravity~\cite{livrev}, given that, in these models, the way in which the structure of the spacetime emerges from the microscopic theory is fully under control. In~\cite{volovik1,volovik2,volovikbook} it was shown that a na\"ive computation of the ground state energy using the EFT (the analogue that one would do to compute the cosmological constant), would produce a wrong result. The unique way to compute the correct value seems to use the full microscopic theory.

Given the deep difference in the structure of the equations of fluid dynamics and those of GR (and other gravitational theories), it is not possible to have an accurate analogy at the dynamical level: indeed, this is forbidden by the absence of diffeomorphism invariance and of local Lorentz invariance. However, in~\cite{gravdynam} it has been shown for the first time that the evolution of part of the acoustic metric in a Bose--Einstein condensate (BEC) is described by a Poisson equation for a non-relativistic gravitational field, thus realizing a (partial) dynamical analogy with Newtonian gravity. Noticeably, this equation is endowed with a source term which is there even in the absence of real phonons and can be naturally identified as a cosmological constant.

In this chapter we will consider such analogue model for gravity and directly show that the cosmological constant term cannot be computed through the standard EFT approach, confirming the conjecture of~\cite{volovik1,volovik2}. However, we find that also the total ground state energy of the condensate does not give the correct result: indeed, the cosmological constant is comparable with that fraction of the ground state energy corresponding to the quantum depletion of the condensate, \ie\/ to the fraction of atoms inevitably occupying excited states of the single particle Hamiltonian.
In conclusion, the origin and value of such term teach us some interesting lessons about the cosmological constant in emergent gravity scenarios.

\setcounter{section}{1}					%
\newpage						%
\sectionmark{The problem in EFT}			%
\setcounter{section}{0}					%
\section{The cosmological constant problem in EFT}	%
\label{sec:effective field theory}			%
\sectionmark{The problem in EFT}			%

Since the cosmological constant $\Lambda$ enters Einstein's equation
\begin{equation}
 R_{\mu\nu}-\frac{1}{2}g_{\mu\nu}R+\Lambda\,g_{\mu\nu}=\frac{8\pi G}{c^4}T_{\mu\nu}
\end{equation}
as a term multiplying the metric tensor $g_{\mu\nu}$, anything that contributes to the energy density of the vacuum acts just like $\Lambda$. In fact, by Lorentz invariance, vacuum energy must have a stress energy tensor of the form
\begin{equation}
 T_{\mu\nu}=-{\cal E}\,g_{\mu\nu}.
\end{equation}
On the other side this argument implies that the cosmological constant may be interpreted not only as an arbitrary constant appearing in Einstein's equations, but as a vacuum energy
\begin{equation}
 {\cal E}_\Lambda=\frac{c^4\Lambda}{8\pi G}.
\end{equation}

Since every vacuum energy must contribute to the cosmological constant, problems arise when computing this energy using EFT. Summing up all the zero-point energies of all normal modes of some quantum field of mass $m$ up to a cutoff energy $\mu$, one obtains~\cite{weinbergcc}
\begin{equation}
 {\cal E}=\int_0^{\mu/\hbar c}\frac{4\pi\,k^2\dd k}{(2\pi)^3}\,\frac{1}{2}\hbar c\sqrt{k^2+\frac{m^2 c^2}{\hbar^2}}
 \approx\frac{\mu^4}{16\pi^2(\hbar c)^3}.
\end{equation}
If we trust the EFT up to Planck scale, we obtain~\cite{carroll}
\begin{equation}
 {\cal E}_{\rm P}\approx 10^{110}\,\mbox{erg}/\mbox{cm}^3,
\end{equation}
while, if we only consider zero point-energy in quantum chromodynamics (QCD), we expect
\begin{equation}
 {\cal E}_{\rm QCD}\approx 10^{36}\,\mbox{erg}/\mbox{cm}^3,
\end{equation}
but cosmological observations unfortunately give
\begin{equation}
 {\cal E}_{\rm obs}\approx 10^{-10}\,\mbox{erg}/\mbox{cm}^3.
\end{equation}

It is clear that EFTs with a cutoff either at Planck scale or QCD scale cannot anyhow explain the observed value. In a proper semiclassical renormalization procedure of the whole theory (gravity plus quantum fields) a suitable bare cosmological constant $\Lambda_{\rm b}$ might be introduced to tune the renormalized $\Lambda$ to match its observed value~\cite{birreldavies}. However, since the order of magnitudes of $\Lambda_{\rm  EFT}$ and the observed one are so different, such a procedure would require an extremely fine tuning of $\Lambda_{\rm b}$.

Furthermore, another dark issue should be clarified. In EFT, the zero-point energy has no particular meaning, since energy is defined up to an arbitrary constant. Indeed, the zero-point energy of the modes is usually renormalized away from the total energy via normal ordering. On the opposite, in gravity, the zero-point cannot be arbitrary, since energy directly enters Einstein's equations: different zero-point energies yield different stress-energy tensors $T_{\mu\nu}$ and, therefore, different solutions of Einstein's equations.
So one may wonder what is the correct vacuum energy that must be put into Einstein's equations. There is no first principle telling us that this energy must exactly be the EFT zero-point energy.

From such considerations, it seems that there is no way out to this problem. The only possibility to compute $\Lambda$ would be to directly know the underlying QG theory from which both GR and EFT emerge in the low energy limit. Under this hypothesis there is no hope of computing $\Lambda$ from scratch only through GR and EFT. In a certain sense $\Lambda$ is an emergent quantity that parametrizes in one number the whole microscopic structure of the spacetime and influences the macroscopic equations of GR and EFT.

As an example of this emergence mechanism, let us mention the propagation of phonons in BECs. In that case, the equation of motion of phonons~\eqref{eq:BdG1} can be written using only macroscopic quantities (speed of sound, velocity and density of the fluid) plus the healing length~\eqref{eq:healing1} which, through the combination of microscopic quantities, defines the scale at which the dispersion relation~\eqref{eq:dispersion1} is no more in the relativistic regime $\om^2=c^2k^2$. If an observer could only measure phonons, the healing length could be measured, but it could not be computed from first principles, just because only the phonon EFT would be knwon but not the whole theory of BECs.

To shed some light on this problem, it is interesting to have a toy model of gravity where one can compute from first principles both the vacuum energy and the cosmological constant. In this way one can compare them and check if they are the same quantity or they are unrelated instead. In order to do this, one needs a model where it is possible to derive not only the dynamics of a field leaving in an effective geometry described by some metric $g_{\mu\nu}$, as usual in analogue models (see Sec.~\ref{sec:generalmetric}), but also the equation governing the dynamics of the metric itself. That is, analogue Einstein's equations are needed, so that the analogue cosmological constant can be directly read from them. Such a model was studied in~\cite{gravdynam}, using a BEC with $U(1)$-symmetry breaking. The computation of the analogue cosmological constant is performed in~\cite{cosmobec} and reported in Sec.~\ref{sec:cosmobec}, together with a revision of the model of~\cite{gravdynam}.

\section{Volovik's proposal}	%
\label{sec:volovik}		%

The idea of using analogue models to understand the origin of the cosmological constant is not original of~\cite{cosmobec} but was firstly developed by Volovik~\cite{volovik1,volovik2} for a quantum liquid and presented in~\cite{volovikbook} also for a Bose gas. However, his approach is completely different from that of~\cite{cosmobec}. It is worth briefly reviewing it, to make a comparison with ours. Even if Volovik did not have an analogue model with an equation describing the dynamics of the geometry as in~\cite{gravdynam}, he nevertheless used a nice argument to determine the analogue cosmological constant.

The basis of Volovik's argument is the identification of the proper thermodynamical potential for the particular considered problem. In this case, we are interested in the emergence of an analogue QFT in condensed matter. The many-body system of identical atoms constituting the quantum liquid is described by the grand canonical Hamiltonian
\begin{equation}\label{eq:volovikH}
 {\cal \hH}=\hH-\mu\hat N,
\end{equation}
where $\hH$ is the second-quantized Hamiltonian, $\mu$ is the chemical potential, and $\hat N$ the number operator.

Volovik's proposal is that the correct vacuum energy density for the QFT emerging in the many-body system (corresponding to the analogue cosmological constant) is the expectation value of Eq.~\eqref{eq:volovikH} on a state $|0\rangle$ with no phonons, in the thermodynamic limit in which both the volume $V$ and the particle number $N$ goes to infinity
\begin{equation}
 {\cal E}_{\rm vac}=\frac{1}{V}\vev{\hH-\mu\hat N}.
\end{equation}
Using the Gibbs--Duhem relation of thermodynamics~\cite{huang}, stating that at thermodynamic equilibrium
\begin{equation}
 E-TS-\mu N=-pV,
\end{equation}
where $E$ and $N$ are the expectation value of $\hH$ and $\hat N$, respectively, $T$ is the temperature of the system, $S$ its entropy, and $p$ its pressure, one obtains
\begin{equation}\label{eq:statevolovik}
 {\cal E}_{\rm vac}=-p,
\end{equation}
because $T=0$ in the phonon ground state $|0\rangle$.

Equation~\eqref{eq:statevolovik} is the key result of this analysis, since it exactly represents the correct equation of state for the cosmological constant. To summarize, if the vacuum energy is the density of the expectation value of the grand canonical Hamiltonian on the state at zero temperature (no excitations), then $p=-{\cal E}_{\rm vac}$ by thermodynamic relations.
The second interesting feature is that, if $\cal H=\langle{\cal\hH}\rangle$ of Eq.~\eqref{eq:volovikH} were really the quantity that gravitates in place of the energy $E=\langle\hH\rangle$, the freedom in the choice of the zero-point energy would not affect the gravitating quantity $\cal H$. If the energy of each atoms were shifted by a factor of $\alpha$, the Hamiltonian would be shifted by $\alpha\hat N$. However, also the chemical potential would have to be shifted by $\alpha$, such that the grand canonical Hamiltonian, being the difference of $\hH$ and $\mu\hat N$, would not change under this transformation. In so doing, there would be a definite gravitating quantity, allowing at the same time for the freedom in the choice of the zero-point energy.

However, this argument does not use any dynamical equation for the spacetime geometry. This is, in our opinion, the weak point of the above treatment. In fact, the cosmological constant is the quantity that gravitates in the absence of matter, encoding the microscopic properties of the spacetime structure. It may be a vacuum energy of some field, but this interpretation may be also wrong. In that case, as discussed in the previous section, the only way to obtain a reliable result for the cosmological constant would be to derive a dynamical equation for the metric, from which $\Lambda$ might be directly read off.

\section{A lesson from BECs}	%
\label{sec:cosmobec}		%

As broadly discussed above, to compute the cosmological constant from scratch, one should know the microscopic structure of the spacetime and how to derive Einstein's equations from it. We show in this section that this procedure can be applied from first principles to a particular analogue system, a BEC with $U(1)$ breaking~\cite{gravdynam}, for which an equation describing the analogue gravitational dynamics exists. We first review this model in Sec.~\ref{subsec:gravdyn}, then we compute both the ground state energy of the system and its grand canonical ground state energy (see Sec.~\ref{sec:volovik}) in Sec.~\ref{subsec:groundbec}. Finally, in Sec.~\ref{subsec:cosmobec}, we compare those results with the analogue gravitational constant.

\subsection{Analogue gravitational dynamics}	%
\label{subsec:gravdyn}				%

The model used in~\cite{gravdynam} is a modified BEC system including a soft breaking of the $U(1)$ symmetry associated with the conservation of particle number. This unusual choice is a simple trick to give mass to quasiparticles that are otherwise massless by Goldstone's theorem. In second quantization, such a system is described by a canonical field $\hPd$, satisfying
$[\hP(t,\bx),\hPd(t,\bx')]=\delta^3(\bx-\bx'),$
whose dynamics is generated by the grand canonical Hamiltonian
\begin{equation}
{\cal\hH}=\hH-\mu\hat N,
\end{equation}
where
\begin{equation}
 \hH = \int\! \dx \left[\frac{\hbar^2}{2m} \nx \hPd \, \nx\hP + V  \hPd\hP 
 + \frac{g}{2}\hPd\hPd\hP\hP-\frac{\lambda}{2}\left(\hP\hP+\hPd\hPd\right)\right]\label{eq:HU1breaking}
\end{equation}
is the Hamiltonian and
\begin{equation}
 \hat N=\int \!\dx\hPd\hP
\end{equation}
is the standard number operator for $\hP$.
For further details on this model and on possible physical realizations, see \cite{gravdynam}. See also \cite{nbec} for a generalization to condensates with many components.

We describe the formation of a BEC at low temperature through the
complex function $\Pn$ for the condensate and the operator $\hp$ for the perturbations on top of it, as defined in Eq.~\eqref{eq:defphi1}
\begin{equation}\label{eq:defphicc}
 \hP=\Pn(1+\hp).
\end{equation}
The canonical commutation relation is [see Eq.~\eqref{eq:commphi1}]
\begin{equation}
 \comm{\hp(t,x)}{\hpd(t,x')}=\frac{1}{\rn(x)}\delta(x-x').
\end{equation}
Using Eq.~\eqref{eq:defphicc}, it is convenient to expand the grand canonical Hamiltonian $\cal\hH$ of Eq.~\eqref{eq:HU1breaking} up to second order in $\hp$
\begin{equation}\label{eq:Hexpansion}
{\cal\hH}\approx{\cal H}_0+{\cal\hH}_1+{\cal\hH}_2,
\end{equation}
where
\begin{align}
{\cal H}_0&= \int\!\dx \,\left[\Pn^*\left(- \hbm\pdx^2+V-\mu+\frac{g}{2}\rn\right)\Pn-\frac{\lambda}{2}\left(\Pn^2+{\Pn^*}^2\right)\right], \label{eq:h0U1breaking}\\
{\cal\hH}_1	
 	&= \int\!\dx \,\left[\Pn^*\hpd\left(- \hbm\pdx^2+V-\mu+g\rn\right)\Pn-\lambda{\Pn^*}^2\hpd\right]+{\rm h.c.}, \label{eq:h1U1breaking}\\
{\cal\hH}_2	
 	&= \int\!\! \dx\,\rn\!\left\{\hpd\!\left[T_\rho-\ii v\hbar\pdx-\hbm\frac{\pdx^2\Pn}{\Pn}+V-\mu+2g\rn\right]\!\hp + \frac{\rho}{2}\left(\hpds+\hp^2\right)+\frac{\lambda}{2\rn}\left(\Pn^2{\hp}^2+{\Pn^*}^2{\hpds}\right)\right\}.\label{eq:h2U1breaking}
\end{align}

For a stationary condensate, $\pdt\Pn=0$ and the grand canonical Hamiltonian~\eqref{eq:h1U1breaking} generates a modified Gross--Pitaevskii equation
\begin{equation}
  \left[-\hbm\pdx^2+V-\mu+g\rn-\lambda\frac{\Pn^*}{\Pn}\right]\Pn=0. \label{eq:GPcc}
\end{equation}

To compute the analogue cosmological constant, it is enough to consider only homogeneous backgrounds. Thus, one can assume that $V=0$ and the condensate is at rest. $\Pn$ has a constant phase, that one can put to $0$ ($\Pn^*=\Pn=\sqrt{\rn}$), and Eq.~\eqref{eq:GPcc} simplifies to
\begin{equation}
 \mu=g\rn-\lambda.
\end{equation}
Under the same assumptions, the equation of motion of the quasiparticles is generated by the second order Hamiltonian~\eqref{eq:h2U1breaking}
 \begin{equation}
   \ii\hbar\pdt\hp = \left[-\frac{\hbar^2}{2m} \nabla^2 + g\rn+\lambda
\right]
\hp + \left(g\rn-\lambda\right)
\hpd.\label{eq:hp}
\end{equation}
This equation is solved via Bogoliubov transformation involving the Fourier expansion
\begin{equation}
\hp
=\int\!\frac{\dk}{\sqrt{\rn(2\pi)^3}}\left[\uk\ee^{-\ii\omega t+\ii{\bf k}\cdot x}\ak+\vks\ee^{+\ii\omega t-\ii{\bf k}\cdot x}\akd\right],
\end{equation}
where $\ak$ and $\akd$ are quasiparticles' operators and the factor $\sqrt{\rn(2\pi)^3}$ has been inserted such that the Bogoliubov coefficients $\uk$ and $\vk$ obey the standard normalization $|\uk|^2-|\vk|^2=1$.

The dispersion relation is
\begin{equation}\label{eq:disp}
 \hbar^2\omega^2=4\lambda g\rn+\frac{g\rn+\lambda}{m}\hbar^2k^2+\frac{\hbar^4k^4}{4m^2},
\end{equation}
describing massive phonons with ultraviolet corrections, mass $\cal M$, and speed of sound $c_s$~\cite{gravdynam}
\begin{equation}
 {\cal M}=\frac{2\sqrt{\lambda g\rn}}{g\rn+\lambda}m,\qquad c_s^2=\frac{g\rn+\lambda}{m}.\label{eq:csM}
\end{equation}
Finally, standard manipulations give (see Sec.~\ref{subsec:mode3d})
\begin{gather}
 \uk^2=\frac{1}{1-D_\bk^2},
 \qquad
 \vk^2=\frac{D_\bk^2}{1-D_\bk^2},
 \label{eq:bogocoeff}\\
 D_\bk\equiv\frac{\hbar\omega-\left(\hbar^2\bk^2/2m+g\rn+\lambda\right)}{g\rn-\lambda},\label{eq:dk}
\end{gather}
where both $\uk$ and $\vk$ are chosen to be real.

When the homogeneous condensate background is perturbed by small inhomogeneities, the Hamiltonian for the quasi-particles can be written as (see~\cite{gravdynam})
\begin{equation}\label{eq:nonrelhquasip}
\hH_{\rm quasip.} \approx {\cal M} c_s^2- \frac{\hbar^2 \nabla^2}{2{\cal M}} + {\cal M}\Phi_{\rm g}.
\end{equation}
$\hH_{\rm quasip.}$ is the non-relativistic Hamiltonian for particles of mass ${\cal M}$ [see Eq.~\eqref{eq:csM}] in a gravitational potential
\begin{equation}\label{gravitationalpotential}
\Phi_{\rm g}(\bx) = \frac{(g\rn+3\lambda)(g\rn+\lambda)}{2\lambda m} u(\bx)
\end{equation}
and $u(\bx)=[(\rn(\bx)/\rho_\infty)-1]/2$, where $\rho_\infty$ is the asymptotic density of the condensate.
Moreover, the dynamics of the potential $\Phi_{\rm g}$ is described by a Poisson-like equation
\begin{equation}
 \left[\nabla^2-\frac{1}{L^2}\right]\Phi_{\rm g}=4\pi G_{N}\rho_{\rm p}+C_\Lambda,\label{eq:poisson}
\end{equation}
which is the equation for a non-relativistic short-range field with length scale
\begin{equation}\label{eq:L}
 L=\frac{a}{\sqrt{16\pi\rn a^3}}
\end{equation}
and gravitational constant
\begin{equation}
 G_{\rm N}=\frac{g(g\rn+3\lambda)(g\rn+\lambda)^2}{4\pi\hbar^2m\lambda^{3/2}(g\rn)^{1/2}}.
\end{equation}
Despite the obvious difference between $\Phi_{\rm g}$
and the usual Newtonian gravitational potential, we insist in calling it the {\it Newtonian potential}
because it enters the acoustic metric exactly as
the Newtonian potential enters the metric tensor  in the Newtonian limit of GR.

The source term in Eq.~\eqref{eq:poisson} contains both the contribution of real phonons (playing the role of matter)
\begin{equation}
 \rho_{\rm p}  = {\cal M}\rn\left[\left(\sev{\hpd\hp}-\vev{\hpd\hp}\right)+
 \frac{1}{2}\re\left(\sev{\hp\hp}-\vev{\hp\hp}\right)\right],\label{eq:rhophonons}
\end{equation}
where $|\zeta\rangle$ is some state of real phonons and $|0\rangle$ is the Fock vacuum of the quasiparticles ($\ak|0\rangle=0, \: \forall k$), as well as a cosmological constant like term (present even in the absence of phonons/matter)
\begin{equation}\label{eq:Clambda}
 C_\Lambda=\frac{2g\rn (g\rn+3\lambda)(g\rn+\lambda)}{\hbar^2\lambda}
 \re \left[\vev{\hpd\hp}+\frac{1}{2}\vev{\hp\hp}\right].
\end{equation}
Note that the source term in the correct weak field approximation of Einstein's equations is $4\pi G_{N}(\rho+3p/c^2)$. For standard non-relativistic matter, $p/c^2$ is usually negligible with respect to $\rho$. However, it cannot be neglected for the cosmological constant, since $p_\Lambda/c^2=-\rho_\Lambda$. As a consequence the analogue cosmological constant is
\begin{equation}\label{eq:lambdacc}
 \Lambda=-\frac{C_\Lambda}{2c_{s}^{2}}.
\end{equation}

\subsection{BEC ground state energy}	%
\label{subsec:groundbec}		%

\begin{figure}
 \centering
 \includegraphics[width=0.57\textwidth]{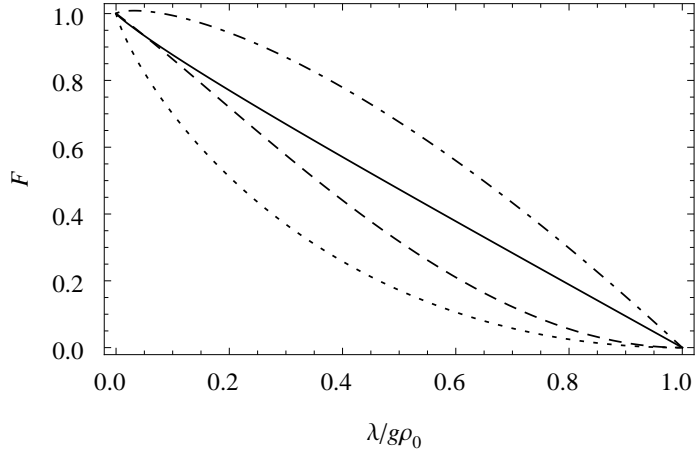}
 \caption{\label{fig:fs}$F_h$ [dashed line, Eq.~\eqref{eq:h2vev2}], $F_\rho$ [dotted line, Eq.~\eqref{eq:depletion}], $F_{\phi\phi}$ [dotdashed line, Eq.~\eqref{eq:phiphi}], and $F_{\Lambda}$ [solid line, Eq.~\eqref{eq:cosmconst}].}
\end{figure}

We can now compute the vacuum expectation value of $\cal\hH$ in the ground state $|0\rangle$. To this aim, it is convenient to use the expansion of $\cal\hH$ in powers of $\hp$ given in Eq.~\eqref{eq:Hexpansion}.
The energy density $h_0$ of the condensate (density of ${\cal H}_0$) and the density $h_2$ of the expectation value of ${\cal\hH}_2$ are given in Appendix~\ref{subsec:homog} [see Eq.~\eqref{eq:h2app}]
\begin{align}
 h_0&=-\frac{g\rn^2}{2},\label{eq:h0vev}\\
 h_2&=-\int\!\frac{\dk}{(2\pi)^3}\hbar\omega|\vk|^2,\label{eq:h2vev}
\end{align}
while the expectation value of ${\cal\hH}_1$ vanishes because it contains only odd powers of $\ak$ and $\akd$.
The integral in Eq.~\eqref{eq:h2vev} is computed by using Eqs.~\eqref{eq:bogocoeff} and~\eqref{eq:dk}. Applying standard regularization techniques (see also~\cite{huang})
\begin{equation}\label{eq:h2vev2}
 h_2=\frac{64}{15\sqrt{\pi}}g\rn^2\sqrt{\rn a^3}\,\,F_h\!\left(\frac{\lambda}{g\rn}\right),
\end{equation}
where $a=4\pi gm/\hbar^2$ is the scattering length, $F_h$ is plotted in Fig.~\ref{fig:fs} (dashed line) and $F_h(0)$=1.
The total grand canonical energy density is therefore
\begin{equation}\label{eq:grandpotential}
 h=h_0+h_2=\frac{g\rn^2}{2}\left[-1+\frac{128}{15\sqrt{\pi}}\sqrt{\rn a^3}\,\,F_h\!\left(\frac{\lambda}{g\rn}\right)\right].
\end{equation}

The number density operator $\hat N$ is analogously expanded in powers of $\hp$ 
\begin{equation}
 \hat N=N_0+\hat N_1+\hat N_2,
\end{equation}
where, as in Eq.~\eqref{eq:Hexpansion}, $N_0$, $\hat N_1$, and $\hat N_2$ contain respectively no power of $\hp$, only first powers, and only second powers.
The density of $N_0$ is simply
\begin{equation}
 \rho_0=|\Pn|^2,
\end{equation}
$\vev{\hat N_1}$ vanishes, and the density $\rho_2$ is [see Eq.~\eqref{eq:dep}]
\begin{equation}\label{eq:depletion}
 \rho_2\equiv\vev{\hat N_2}=\rn\vev{\hpd\hp}=\!\int\!\!\frac{\dk}{(2\pi)^3}|\vk|^2=\frac{8\rn}{3\sqrt{\pi}}\sqrt{\rn a^3}\,\,F_\rho\!\left(\frac{\lambda}{g\rn}\right),
\end{equation}
where $F_\rho$ satisfies $F_\rho(0)=1$ (see Fig.~\ref{fig:fs}, dotted line).
This is the number density of non-condensed atoms ({\it depletion}). Note that the Hamiltonian~\eqref{eq:HU1breaking} holds only if the so called dilution factor $\rn a^3$ is much smaller than 1.

Furthermore, when $\lambda=0$, inverting the expression for total particle density $\rho=\rn+\rho_2$, up to the first order in $\sqrt{\rho a^3}$, one obtains
\begin{equation}
 \rn=\rho\left[1-\frac{8}{3\sqrt{\pi}}\sqrt{\rho a^3}\right],
\end{equation}
which is the density of condensed atoms in terms of the total density~$\rho$ and the scattering length~$a$~\cite{lhy}. In this case, $\mu=g\rn$ and the energy density $\epsilon$ (density of $\vev{\hH}=\vev{{\cal H}+\mu\hat N}$) is
\begin{equation}
 \epsilon=h+\mu\rho= \frac{g\rho^2}{2}\left[1+\frac{128}{15\sqrt{\pi}}\sqrt{\rho a^3}\right].\label{eq:energy}
\end{equation}
This is the well known Lee--Huang--Yang~\cite{lhy} formula for the ground state energy in a condensate at zero temperature.
In general, when the $U(1)$ breaking term is small, this term is expected to be the dominant contribution to the ground state energy of the condensate.

\subsection{What does the cosmological constant correspond to?}	%
\label{subsec:cosmobec}						%

We will now compare the energy density and the grand canonical energy density with the effective cosmological constant $\Lambda$ of Eq.~\eqref{eq:lambdacc}.
$C_\Lambda$ of Eq.~\eqref{eq:Clambda} is computed using Eq.~\eqref{eq:depletion} and
\begin{equation}\label{eq:phiphi}
 \vev{\hp\hp}=\!\int\!\!\frac{\dk}{\rn(2\pi)^3}\uk\vk=\frac{8}{\sqrt{\pi}}\sqrt{\rn a^3}\,F_{\phi\phi}\!\left(\frac{\lambda}{g\rn}\right),
\end{equation}
where $F_{\phi\phi}(0)=1$ (see Fig.~\ref{fig:fs}, dotdashed line). We finally obtain
\begin{equation}\label{eq:cosmconst}
 \Lambda=-\frac{20m\,g\rn\,(g\rn+3\lambda)}{3\sqrt{\pi}\hbar^2\lambda}\sqrt{\rn a^3}\,F_\Lambda\!\left(\frac{\lambda}{g\rn}\right),
\end{equation}
where $F_\Lambda=(2F_\rho+3F_{\phi\phi})/5$ (see Fig.~\ref{fig:fs}, solid line).

Let us now compare the value of $\Lambda$ either with the ground-state grand canonical energy density $h$ [Eq.~\eqref{eq:grandpotential}], which in~\cite{volovik1,volovik2} was suggested as the correct vacuum energy corresponding to the cosmological constant, or to the ground-state energy density $\epsilon$ of Eq.~\eqref{eq:energy}.
Evidently, $\Lambda$ does not correspond to either of them: even when taking into account the correct behavior at small scales, the vacuum energy computed with the phonon EFT does not lead to the correct value of the cosmological constant appearing in Eq.~\eqref{eq:poisson}.
Noticeably, since $\Lambda$ is proportional to $\sqrt{\rn a^3}$, it can even be arbitrarily smaller than both $h$ and $\epsilon$, if the condensate is very dilute. Furthermore, $\Lambda$ is proportional only to the subdominant second order correction of $h$ or $\epsilon$, which is strictly related to the depletion [see Eq.~\eqref{eq:depletion}].

Several scales show up in the emergent system, in addition to the na\"ive Planck scale computed by combining the emergent constants $G_{\rm N}$, $c_s$ and $\hbar$:
\begin{equation}
 L_{\rm P}=\sqrt{\frac{\hbar c_s^5}{G_{\rm N}}}\propto \left(\frac{\lambda}{g\rn}\right)^{-3/4}(\rn a^3)^{-1/4} a.
\end{equation}
For instance, the Lorentz-violation scale (\ie, the healing length of the condensate)
\begin{equation}
 L_{\rm LV}=\xi \propto(\rn a^3)^{-1/2} a
\end{equation}
differs from $L_{\rm P}$, suggesting that the breaking of the Lorentz symmetry might be expected at a much longer scale than the Planck length (energy much smaller than the Planck energy), since the ratio $L_{\rm LV}/L_{\rm P}\propto(\rn a^3)^{-1/4}$ increases with the diluteness of the condensate.

Note that $L_{\rm LV}$ scales with $\rn a^3$ exactly as the range of the gravitational force [see Eq.~\eqref{eq:L}], signaling that this model is too simple to correctly grasp all the desired features. However,
in more complicated systems~\cite{nbec}, this pathology can be cured in the presence of suitable symmetries, leading to long range potentials.

It is instructive to compare the energy density corresponding to $\Lambda$
\begin{equation}
{\cal E}_\Lambda=\frac{\Lambda c_s^4}{4\pi G_{\rm N}}
\end{equation}
to the Planck energy density
\begin{equation}
 {\cal E}_{\rm P}=\frac{c_s^7}{\hbar G_{\rm N}^2}.
\end{equation}
The energy density associated with the analogue cosmological constant is much smaller than the value computed from zero-point-energy calculations with a cut off at the Planck scale. Indeed, the ratio between these two quantities
\begin{equation}
 \frac{{\cal E}_\Lambda}{{\cal E}_{\rm P}}\propto \rn a^3\left(\frac{\lambda}{g\rn}\right)^{-5/2}
\end{equation}
is controlled by the diluteness parameter $\rn a^3$.

\section{Summary and discussion}	%
\label{sec:conclusionscosmobec}		%

Taken at face value, this relatively simple model displays too many crucial differences with respect to any realistic theory of gravity to provide conclusive evidences. However, it displays an alternative path to the cosmological constant,
from the perspective of a microscopic model. The analogue cosmological constant that we have discussed {\it cannot} be computed as the total zero-point energy of the condensed matter system, even when taking into account the natural cut-off coming from the knowledge of the microphysics~\cite{volovik1,volovik2}. In fact the value of $\Lambda$ is related only to the (subleading) part of the zero-point energy proportional to the quantum depletion of the condensate.
Interestingly, this result finds some support from arguments within loop quantum gravity models \cite{Alexander1,Alexander2}, suggesting a BCS energy gap as a (conceptually rather different) origin for the cosmological constant.

The implications for gravity are twofold. First, there could be no {\it a priori} reason why the cosmological costant should be computed as the zero-point energy of the system. More properly, its computation must inevitably pass through the derivation of Einstein's equations emerging from the underlying microscopic system.
Second, the energy scale of $\Lambda$ can be several orders of magnitude smaller than all the other energy scales for the presence of a very small number, non-perturbative in origin, which cannot be computed within the framework of an EFT dealing only with the emergent degrees of freedom (\ie, semiclassical gravity).

The model discussed in this chapter explicitly shows all of this. Furthermore, it strongly supports a picture where gravity is a collective phenomenon in a pregeometric theory.
In fact, the cosmological constant puzzle is elegantly solved in those scenarios.
From an emergent gravity approach, the low energy effective action (and its renormalization group flow) is obviously computed within a framework that has nothing to do with QFT in curved spacetime.
Indeed, if we interpreted the cosmological constant as a coupling constant controlling some self-interaction of the gravitational field, rather than as a vacuum energy, it would straightforwardly follow that the explanation of its value (and of its properties under renormalization) would naturally sit outside the domain of semiclassical gravity.

For instance, in a group field theory scenario (a generalization to higher dimensions of matrix models for two dimensional QG \cite{oriti}), it is transparent that the origin of the gravitational coupling constants has nothing to do with ideas like ``vacuum energy'' or statements like ``energy gravitates'', because energy {\it itself} is an emergent concept. Rather, the value of $\Lambda$ is determined by the microphysics, and, most importantly, by the procedure to approach the continuum semiclassical limit.
In this respect, it is conceivable that the very notion of cosmological constant as a form of energy intrinsic to the vacuum is ultimately misleading.
To date, little is known about the macroscopic regime of models like group field theories, even though some preliminary steps have been recently done~\cite{oritisindoni}.
Nonetheless, analogue models elucidate in simple ways what is expected to happen and can suggest how to further develop investigations in QG models.
In this respect, the reasoning of this chapter sheds a totally different light on the cosmological constant problem, turning it from a failure of EFT to a question about the emergence of the spacetime.

\addtocontents{toc}{\protect\vspace{1.5em}\protect}
\chapter{A~new~system: Relativistic~Bose--Einstein~condensates}	
\label{chap:relbec}						
\chaptermark{Relativistic BECs}					

In Sec.~\ref{subsec:schwarzschild} we saw that it is possible, although rather complicated, to find a fluid metric~\eqref{eq:fluidmetric} reproducing the Schwarzschild geometry. However, it is actually impossible to reproduce the Kerr metric because spatial sections of acoustic geometries are necessarily conformally flat (see Sec.~\ref{subsec:kerr}).
Finally, in Sec.~\ref{subsec:cosmo}, we saw that, for similar reasons, among the cosmological metrics only the de Sitter spacetime and the flat ($k=0$) Friedmann--Robertson--Walker~(FRW) spacetimes can be cast in acoustic form.
In general, the subset of metrics that can be simulated by analogue models is rather small (see discussion in~\ref{sec:generalmetric}), so that it is quite interesting to look for systems allowing to reproduce a wider set of geometries.

Furthermore, except for the recent work on the Abelian Higgs model~\cite{chinese}, analogue models considered so far are characterized by non-relativistic fundamental equations. This would imply, in the emergent gravity perspective, a trans-Planckian world characterized by a preferred system of reference and a Newtonian absolute space and time.
Given the broad interest in these systems as toy models for emergent gravity scenarios~\cite{two-componentBEC,two-componentBEC2,gravdynam,scalargravity,birefringence}, as broadly discussed in the~\nameref{chap:intro} of this Thesis, it would be particularly important to investigate the implications of a system without such a Newtonian preferred time.

Leaded by the above two motivations, in this chapter we consider a new analogue gravity system, namely a relativistic Bose--Einstein condensate (RBEC), for which no absolute time is present. A preferred time foliation is produced through spontaneous symmetry breaking but it is not encoded in the theory {\it ab initio}. From the point of view of emergent gravity analogues, this system is alternative to the standard case with transition from Lorentzian to Galilean relativity in the ultraviolet (UV). This new system shows instead a transition from Lorentz to Lorentz symmetry, encoded in different invariant speeds at different energy regimes.
Moreover, it is richer than its non-relativistic counterpart, since it is characterized by two kinds of propagating modes, a massless and a massive one (which disappears in the non-relativistic limit): we shall show that only the massless one is described by a generalized Klein--Gordon equation in a curved spacetime.
 
Finally, RBECs allow to reproduce novel classes of metrics as, for example, the $k=-1$ FRW metric (see Sec.~\ref{sec:frw}). In general, the spatial slices of acoustic metrics in these systems are not conformally flat, possibly permitting more general mappings.

\section{Relativistic Bose--Einstein condensates}		%
\label{sec:bec}							%

Bose--Einstein condensation, \ie\/ the macroscopic occupation of a single state, may occur both for relativistic and non-relativistic bosons. The main differences between their thermodynamical properties at finite temperature are due both to the different energy spectra and also to the presence, for relativistic bosons, of anti-bosons. These differences result in different conditions for the occurrence of Bose--Einstein condensation, which is possible, \eg, in two spatial dimensions for a homogeneous gas of massless bosons, but not for its massive counterpart---and also, more importantly for our purposes, in the different structure of their excitation spectra.
Note that, in order to see relativistic effects in an experimental realization of a relativistic BEC, a field with a very low mass would be requiered. This might be possible in a condensate of exciton polaritons~\cite{polaritonBEC}, where the polariton mass may be controlled and made sufficiently small (see~\cite{reviacopo} for a recent review).

In this section, under a theoretical perspective, we briefly recall the thermodynamic properties of a relativistic Bose gas~\cite{haber1981,haber1982,kapusta81,singh1,singh2,bernstein91,grether07,witkowska09}, discussing how the non-relativistic limit is obtained and comparing with the results for non-RBECs~\cite{stringari03,pethick08}. The study of the excitation spectrum for a (generally moving and inhomogeneous) condensate is presented in the following section~\ref{sec:perturbations}.

The Lagrangian density for an interacting relativistic scalar Bose field $\hat{\phi}(\bi{x},t)$ may be written as
\begin{equation}
\hat{{\cal L}}=\frac{1}{c^2}\frac{\partial\hat\phi^\dagger}{\partial t}\frac{\partial\hat\phi}{\partial t}-
\bi{\nabla}\hat\phi^\dagger \cdot \bi{\nabla}\hat\phi 
-\left(\frac{m^2c^2}{\hbar^2} + V(t,\bi{x})\right) \hat{\phi}^\dagger \hat{\phi}-U(\hat\phi^\dagger\hat\phi;\lambda_i),
\label{Lagrangian}
\end{equation}
where $V(t,\bi{x})$ is an external potential depending both on time $t$ and position $\bi{x}$, $m$ is the mass of the bosons and $c$ is the velocity of light. $U$ is an interaction term and the coupling constants $\lambda_i(t,\bi x)$ can also depend on both time and position.
$U$ can be expanded as
\begin{equation}
 U(\hat\phi^\dagger\hat\phi;\lambda_i) = \frac{\lambda_2}{2}\hat{\rho}^2 + \frac{\lambda_3}{6}\hat{\rho}^3 + \cdots,
\end{equation}
where $\hat{\rho}=\hat\phi^\dagger\hat\phi$. The usual two-particle $\lambda_2 \hat{\phi}^4$ interaction corresponds to the first term $(\lambda_2/2)\hat{\rho}^2$, while the second term represents the three-particle interaction and so on.

Since the Lagrangian~\eqref{Lagrangian} is invariant under the global $U(1)$ symmetry, it is possible to define the associated Noether current and conserved charge, the latter corresponding to $N-\bar{N}$, where $N$ ($\bar{N}$) is the number of bosons (anti-bosons).

Then, for an ideal homogeneous gas, when both $U$ and $V$ vanish, the relation between the chemical potential $\mu$ and the temperature $T\equiv 1/(k_B\beta)$ is given by~\cite{haber1981,haber1982}
\begin{equation}
N-\bar{N}=\sum_{\bi{k}} \left[ n_{\bi{k}}(\mu,\beta) - \bar{n}_{\bi{k}}(\mu,\beta)\right],
\label{eq:muT}
\end{equation}
where $n_{\bi{k}}(\mu,\beta)=1/\{\exp[\beta (|E_{\bi{k}}|-\mu)]-1\}$ is the average number of bosons in the state of energy  $|E_{\bi{k}}|$, with
\begin{equation}
 E_{\bi{k}}^2= \hbar^2 k^2 c^2+m^2c^4.
\end{equation}
Similarly, $\bar{n}_{\bi{k}}(\mu,\beta)=1/\{\exp[\beta (|E_{\bi{k}}|+\mu)]-1\}$ is the corresponding number of anti-bosons.

Introducing  the number density $n=(N-\bar{N})/\Omega$ (where $\Omega$ is the volume of the system), one obtains in $d$ dimensions for non-interacting bosons the following relation between the critical temperature $T_c$ and the charge density $n$~\cite{haber1981,haber1982,grether07}:
\begin{equation}
n=C \int_{0}^{\infty} \rmd k\, k^{d-1} \frac{\sinh{(\beta_c mc^2)}}
{\cosh{(\beta_c |E_{\bi{k}}|)}-\cosh{(\beta_c mc^2)}},
\label{eq:Tc}
\end{equation}
where $C=1/(2^{d-1} \pi^{d/2}\Gamma(d/2))$ is a numerical coefficient and $d$ is the number of spatial dimensions.

From Eq.~\eqref{eq:Tc} one can readily derive the non-relativistic and the ultra-relativistic limits: the former is obtained when $k_B T_c \ll mc^2$. In this limit the contribution of anti-bosons to Eq.~\eqref{eq:muT} can be neglected (so $n\approx N / \Omega$) and one obtains
\begin{equation}
k_B T_c=\frac{2\pi \hbar^2}{n} \left( \frac{n}{\zeta(d/2)} \right)^{2/d}
\label{T_c_NR}
\end{equation}
($\zeta$ denotes the Riemann zeta function), which is the usual result for the critical temperature of a non-relativistic ideal Bose gas \cite{stringari03,pethick08}.
In the ultra-relativistic limit, $k_B T_c \gg mc^2$ and
\begin{equation}
\left( k_B T_c \right)^{d-1}=\frac{\hbar^d c^{d-2} \Gamma{(d/2)} (2\pi)^d}
{4 m \pi^{d/2} \Gamma(d) \zeta(d-1)}  \,n.
\label{T_c_UR}
\end{equation}
Inspection of~\eqref{T_c_NR} and~\eqref{T_c_UR} makes clear that both non-relativistic and relativistic Bose gases condense for $d \geqslant 3$, since for $d=2$ the critical temperature $T_c$ vanishes in both cases as $\zeta(1)$ diverges. Condensation is possible for $d=2$ only if bosons are massless.
From now on we focus on the case $d=3$, where the homogeneous relativistic ideal Bose gas can undergo Bose--Einstein condensation.

At $T\ll T_c$, when the relativistic bosons condense, it is possible to describe the dynamics of the condensate at the mean-field level by performing the substitution $\hat{\phi} \to \phi$. The order parameter $\phi$ satisfies then the classical equation
\begin{equation}
\frac{1}{c^2} \frac{\partial^2 \phi}{\partial t^2} - \bi{\nabla}^2 \phi
+\left(\frac{m^2 c^2}{\hbar^2} + V(t,\bi{x})\right)\phi
+U'(\rho;\lambda_i(t,\bi{x})) \phi=0,
\label{eq:NLKG}
\end{equation}
where $\rho=\phi^\ast\phi$ and $'$ denotes the derivative with respect to $\rho$. The nonlinear Klein--Gordon equation~\eqref{eq:NLKG} gives the dynamics of the relativistic condensates, and in the non-relativistic limit the Gross--Pitaevskii equation~\eqref{eq:GP1} is retrieved.

Adopting the standard definition for the box operator in flat spacetime,
\begin{equation} 
\Box=\eta^{\mu\nu}\partial_\mu\partial_\nu=-\frac{1}{c^2}\partial_t^2+\nabla^2,
\end{equation}
Eq.~\eqref{eq:NLKG} can be written as
\begin{equation}\label{eq:eqphi_relbec}
\Box\phi-\left(\frac{m^2 c^2}{\hbar^2} + V\right)\phi
-U'(\rho;\lambda_i) \phi=0.
\end{equation}

\section{Analysis of perturbations}	%
\label{sec:perturbations}		%

In this section, we study the excitation spectrum of perturbations on a condensate obeying the classical wave function equation~\eqref{eq:eqphi_relbec}. The field $\hat\phi$ can be written as a classical field (the condensate) plus perturbations:
\begin{equation}\label{eq:exp}
 \hat\phi = \phi(1+\hat\psi).
\end{equation}
It is now worth noting that the expansion in Eq.~\eqref{eq:exp} can be linked straightforwardly to the usual expansion~\cite{livrev,stringari03,pethick08} in phase and density perturbations $\hat\theta_1$, $\hat\rho_1$:
\begin{equation}
 \frac{\hat\rho_1}{\rho}=\frac{\hat\psi+\hat\psi^\dagger}2,\quad \hat\theta_1=\frac{\hat\psi-\hat\psi^\dagger}{2\rmi}.
\end{equation}
The equation for the quantum field $\hat\psi$ describing the perturbations is
\begin{equation}\label{eq:pert}
 \Box\hat\psi+2\eta^{\mu\nu}(\partial_\mu\ln{\phi})\partial_\nu\hat\psi-\rho\, U''(\rho;\lambda_i)(\hat\psi+\hat\psi^\dagger)=0.
\end{equation}
It is now quite convenient to adopt a Madelung representation for the complex mean field $\phi$ and decompose it into two real fields, its modulus $\sqrt{\rho(x,t)}$ and its phase $\theta(x,t)$
\begin{equation}
 \phi=\sqrt{\rho}\,\rme^{\rmi\theta},
\end{equation}
the logarithm in Eq.~\eqref{eq:pert} then becomes
\begin{equation}
 \partial_\mu\ln\phi=\frac{1}{2}\partial_\mu\ln\rho+\rmi\,\partial_\mu\theta.
\end{equation}
For convenience we define the following quantities:
\begin{align}
 u^\mu &\equiv \frac{\hbar}{m}\eta^{\mu\nu}\partial_\nu\theta,\\
 c_{0}^2&\equiv\frac{\hbar^2}{2m^2}\rho\, U''(\rho;\lambda_i),\label{eq:defc0}\\
 T_\rho&\equiv-\frac{\hbar^2}{2m}\left(\Box+\eta^{\mu\nu}\partial_\mu\ln{\rho}\,\partial_\nu\right)=-\frac{\hbar^2}{2m\rho}\eta^{\mu\nu}\partial_\mu\rho\,\partial_\nu,
\end{align}
where the derivatives act on everything on their right, $c_{0}$ encodes the strength of the interactions and has dimensions of a velocity and $T_\rho$ is a generalized kinetic operator that reduces, in the non-relativistic limit $c\to\infty$ and for constant $\rho$, to the standard kinetic energy operator [see Eq.~\eqref{eq:Trho1}]
\begin{equation}
 T_\rho\to-\frac{\hbar^2}{2m\rho}\nabla\rho\nabla=-\frac{\hbar^2}{2m}\nabla^2.
\end{equation}
A straightforward physical interpretation can be given to the four-vector $u^\mu$.
One can introduce the conserved current
\begin{equation}
 j_\mu\equiv \frac{1}{2\rmi}\left(\phi\,\partial_\mu\phi^*-\phi^*\partial_\mu\phi\right),
\end{equation}
and show that
\begin{equation}
j_\mu=\rho\partial_\mu\theta=\rho\,\frac{m}{\hbar}u_\mu,
\end{equation}
hence relating $u^\mu$ to the current associated with the $U(1)$ symmetry.

In terms of these quantities and using the phase-density decomposition, the equation for the condensate classical wave function, Eq.~\eqref{eq:eqphi_relbec}, becomes
\begin{align}
 \partial_\mu (\rho u^\mu)&=0,\label{eq:cont1}\\
 -u_\mu u^\mu &= c^2+\frac{\hbar^2}{m^2}\left[ V(x^\mu) + U'(\rho;\lambda_i(x^\mu))- \frac{\Box \sqrt{\rho}}{\sqrt{\rho}} \right].\label{eq:eul}
\end{align}
The first equation is a continuity equation, which tells that the current $j^\mu$ defined above is conserved. The second one allows one to determine the zero-component of $u^\mu$ and, equivalently, the chemical potential, as a function of the spatial part of the fluid velocity, the strength of the interaction and the condensate density $\rho$.

Multiplying Eq.~\eqref{eq:pert} by $\hbar^2/2m$, one can rewrite the equation for perturbations in a RBEC as
\begin{equation}\label{eq:dirac}
 \left[\rmi\hbar u^\mu\partial_\mu-T_\rho-mc_0^2\right]\hat\psi=mc_0\hat\psi^\dagger.
\end{equation}
Taking the Hermitian conjugate of this equation, we can eliminate $\hat\psi^\dagger$, obtaining a single equation for $\hat\psi$:
\begin{equation}
 \left[-\rmi\hbar u^\mu\partial_\mu-T_\rho-mc_0^2\right]\frac{1}{c_0^2}
 \left[\rmi\hbar u^\mu\partial_\mu -T_\rho-mc_0^2\right]\hat\psi
 =m^2c_0^2\,\hat\psi,
\end{equation}
and, with some simple manipulations, we obtain
\begin{equation}\label{eq:psifluid}
 \left\{\left[\rmi\hbar u^\mu\partial_\mu+T_\rho\right]
 \frac{1}{c_0^2}
 \left[-\rmi\hbar u^\mu\partial_\mu+T_\rho\right]
 -\frac{\hbar^2}{\rho}\eta^{\mu\nu}\partial_\mu\rho\,\partial_\nu\right\}\hat\psi=0.
\end{equation}
This is the generalization to a relativistic condensate of the equation describing the propagation of the linearized perturbations on top of a non-relativistic BEC, in the same form as in~\cite{MacherBEC} [see Eq.~\eqref{eq:eqphi_oper}]. It is worth stressing that Eq.~\eqref{eq:psifluid} is implied by Eq.~\eqref{eq:dirac} but the converse is not true.
Nevertheless Eq.~\eqref{eq:psifluid} is enough for our aim, since we are interested only in the dispersion relation and in the metric describing the field propagation, which can be obtained from the modified Klein--Gordon equation in a curved background.

As an application of Eq.~\eqref{eq:psifluid}, we consider a moving condensate with homogeneous density, $V(t,\bi{x})=0$ and $\lambda_i$ being constant both in space and time.
Let us choose
\begin{equation}
\phi(t,\bi{x})=
\phi_0 \rme^{\rmi \left( \bi{q} \cdot \bi{x} -\mu t/\hbar\right)}.
\end{equation}
One has
\begin{align}
 \mu &\equiv mcu^0,\label{eq:mu}\\
 \bi{q}&\equiv m \bi{u}/\hbar,\\
u^\mu \partial_\mu &=(\mu/mc^2)\partial/\partial t+(\hbar/m)\bi{q} \cdot \bi{\nabla}.
\end{align}
Moreover Eq.~\eqref{eq:eul} reduces to
\begin{equation}\label{eq:homeul}
\mu^2=\hbar^2 c^2 q^2+m^2c^4+\hbar^2 c^2 U'(\rho_0;\lambda_i),
\end{equation}
where $\rho_0=|\phi_0|^2$.

Note that $\mu$ is the relativistic chemical potential of the condensate, which is related to the non-relativistic counterpart $\mu_{\rm NR}$ via the identity $\mu=mc^2+\mu_{\rm NR}$. In particular, from Eq.~\eqref{eq:homeul}, in the non-relativistic limit $\mu_{\rm NR}\ll mc^2$ and $\mu\approx mc^2$. In the general case, when the fluid is not homogeneous, the same argument implies, from Eq.~\eqref{eq:eul}, that $u^0\approx c$.

Setting $\psi \propto {\exp}[\rmi \left( \bi{k} \cdot \bi{x} -\omega t\right)]$ one obtains from Eq.~\eqref{eq:psifluid}
\begin{equation}\label{eq:fourierexpgen}
\left(- \bi{u} \cdot \bi{k}+
\frac{u^0}{c}\omega-\frac{\hbar }{2 mc^2}\omega^2+\frac{\hbar}{2m}k^2\right) 
\left( \bi{u} \cdot \bi{k}
-\frac{u^0}{c}\omega-\frac{\hbar }{2 mc^2}\omega^2+\frac{\hbar}{2m}k^2\right)
 -\left(\frac{c_0}{c}\right)^2\omega^2+c_0^2 k^2=0.
\end{equation}
For a condensate at rest ($\bi{u}=0$), the spectrum~\eqref{eq:fourierexpgen} was derived in \cite{haber1981,haber1982,witkowska09,andersen07}.

The non-relativistic limit of Eq.~\eqref{eq:psifluid} can be recovered by letting $c\to\infty$. This implies $\mu=mc^2+\mu_{\rm NR}\approx mc^2$ and, equivalently $u^0\approx c$. In this limit, the mode equation becomes
\begin{equation}\label{eq:psiNR}
 \left\{\left[\rmi\hbar\left(\partial_t+\ub\cdot\nabla\right)+T_{\rho\,{\rm NR}}\right]
 \frac{1}{c_0^2}
 \left[-\rmi\hbar\left(\partial_t+\ub\cdot\nabla\right)+T_{\rho\,{\rm NR}}\right]
 -\frac{\hbar^2}{\rho}\nabla\rho\,\nabla\right\}\hat\psi=0,
\end{equation}
where
\begin{equation}
 T_{\rho\,{\rm NR}}\equiv-\frac{\hbar^2}{2m\rho}\nabla\rho\nabla.
\end{equation}
Eq.~\eqref{eq:psiNR} reduces to the usual equation describing the propagation of perturbations in a non-relativistic BEC [see Eq.~\eqref{eq:eqphi_oper}].

\section{The dispersion relation}	%
\label{sec:dispersion}			%

Before showing how the propagation of phonons in a RBEC can be described through an effective metric, it is worthwhile to analyze the dispersion relation of such perturbations. To do that we assume in this section that $u$, $\mu$, $\rho$ and $c_0$ are constant both in space and in time. For simplicity we start to analyze the case of background fluid at rest, $\ub=0$. As discussed after Eq.~\eqref{eq:psifluid}, in order to study the dispersion relation, one can directly work on Eq.~\eqref{eq:psifluid}, that becomes in this case
\begin{equation}
 \left[\left(\rmi\frac{u^0}{c}\partial_t-\frac{\hbar}{2m}\Box\right)
 \left(-\rmi\frac{u^0}{c}\partial_t-\frac{\hbar}{2m}\Box\right)
 -c_0^2\,\Box\right]\hat\psi=0,
\end{equation}
which can be rewritten expanding the $\Box$ operator as
\begin{equation}\label{eq:flat}
 \left[\left(\rmi\frac{u^0}{c}\partial_t+\frac{\hbar }{2 mc^2}\partial_t^2-\frac{\hbar\nabla^2}{2m}\right)
 \left(-\rmi\frac{u^0}{c}\partial_t+\frac{\hbar }{2 mc^2}\partial_t^2-\frac{\hbar\nabla^2}{2m}\right)
 +\left(\frac{c_0}{c}\right)^2\partial_t^2-c_0^2\,\nabla^2\right]\hat\psi=0~.
\end{equation}
The previous equation can be solved exactly by Fourier modes $\exp(-\rmi\omega t+\rmi \bi{k}\cdot \bi{x})$
\begin{equation}\label{eq:fourierexp}
\left(\frac{u^0}{c}\omega-\frac{\hbar }{2 mc^2}\omega^2+\frac{\hbar}{2m}k^2\right)
 \left(-\frac{u^0}{c}\omega-\frac{\hbar }{2 mc^2}\omega^2+\frac{\hbar}{2m}k^2\right)
 -\left(\frac{c_0}{c}\right)^2\omega^2+c_0^2 k^2=0,
\end{equation}
whose solution is:
\begin{equation}\label{eq:dispersion_relbec}
 \omega^2_\pm=c^2
 \left\{k^2+2\left(\frac{mu^0}{\hbar}\right)^2 \left[1+\left(\frac{c_0}{u^0}\right)^2\right]
\right.\\ \left. 
 \pm 2\left(\frac{mu^0}{\hbar}\right)
 \sqrt{k^2+\left(\frac{mu^0}{\hbar}\right)^2\left[1+\left(\frac{c_0}{u^0}\right)^2\right]^2}
 \right\}.
\end{equation}
This equation represents the dispersion relation for modes propagating in a RBEC and it is the generalization of the non-relativistic Bogoliubov dispersion relation~\eqref{eq:dispersion1} [see also~Eq.~\eqref{eq:dispersionhom}].

\begin{figure}
\centering
\includegraphics[width=.8\columnwidth]{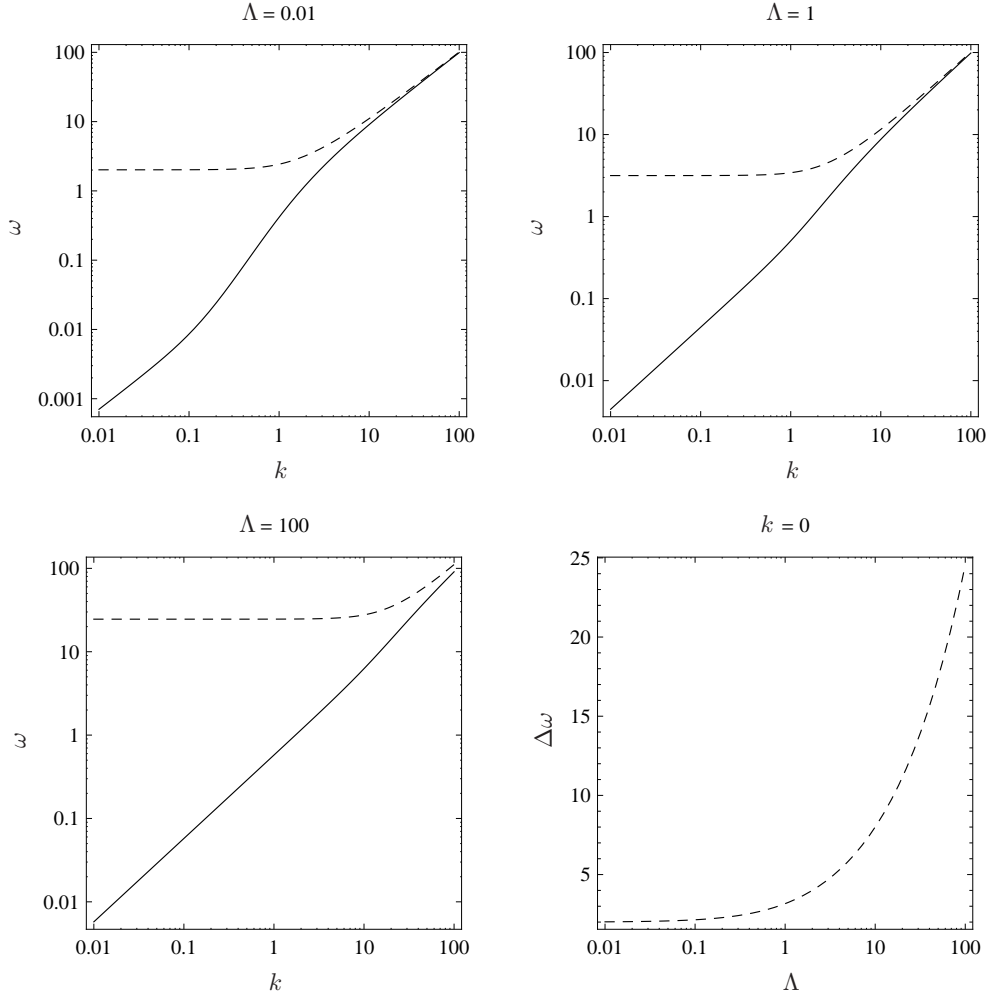}
\caption{Excitation spectrum~\eqref{eq:dispersion_relbec} for a condensate at rest ($\bi{u}=0$) with interaction $U(\rho)=\lambda_2 \rho^2/2$ and $\Lambda = (\hbar/mc)^2\lambda_2\rho_0 =0.01,\,1,\,100$. Solid line: gapless branch $\omega_-$, dashed line: gapped branch $\omega_+$. $\omega$ in units of $mc^2/\hbar$, $k$ in units of $mc/\hbar$. Bottom right: 
plot of the the gap $\Delta \omega=\omega_+-\omega_-$ at $k=0$, which in these units is given by $\Delta \omega(k=0)=\sqrt{4+3\Lambda}$.}
\label{fig:spectrum_at_rest}
\end{figure}

One can see from Eq.~\eqref{eq:dispersion_relbec} that $\omega^2_\pm \geqslant0$, \ie\/ there is no dynamical instability: in figure~\ref{fig:spectrum_at_rest} we plot the dispersion relation~\eqref{eq:dispersion_relbec} for the interaction $U(\rho)=\lambda_2 \rho^2/2$ for different values of the dimensionless parameter $\Lambda \equiv  (\hbar/mc)^2\lambda_2\rho_0$. In figure~\ref{fig:spectrum_at_rest}, we also plot the gap at $k=0$ between $\omega_+(k=0)$ and $\omega_-(k=0)$.

One sees from (\ref{eq:dispersion_relbec}) that the different regimes allowed for the excitations of the systems are determined by the relative strength of the the first two terms on the right-hand side of (\ref{eq:dispersion_relbec}) (note that the same terms enter in the square root).
Hence, in what follows we shall analyze the RBEC excitations dispersion relation~\eqref{eq:dispersion_relbec} in some significant limits. The results are summarized at the end of this section in Tab.~\ref{tab:regimes}.

\subsection{Low momentum limit}	%
\label{sec:low}			%

We begin our study by looking at the low momentum limit of the dispersion relation (\ref{eq:dispersion_relbec}) characterized as
\begin{equation}\label{eq:lowkcond}
|k| \ll \frac{m u^0}{\hbar }\left[1+\left(\frac{c_0}{u^0}\right)^2\right] \equiv  \frac{m u^0}{\hbar }(1+b),
\end{equation}
where, for later convenience, we have introduced the dimensionless parameter
\begin{equation}
b\equiv \left( \frac{c_0}{u^0} \right)^2.
\label{eq:b}
\end{equation}
Under the above assumption (\ref{eq:lowkcond}), the square root in Eq.~\eqref{eq:dispersion_relbec} can be expanded as
\begin{equation}\label{eq:lowk}
 \omega^2_\pm\approx c^2\!
 \left[k^2+2\left(\frac{m u^0}{\hbar }\right)^2\!\!(1+b)
 \pm 2\left(\frac{m u^0}{\hbar }\right)^2\!\!(1+b)
 \pm \frac{k^2}{1+b}
 \mp \frac{k^4}{4(mu^0/\hbar)^2(1+b)^3}
 \right].
\end{equation}
%

\subsubsection{Gapless excitations}	%

Let us focus for the moment on the branch corresponding to the lower sign in the above dispersion relation. One obtains
\begin{equation}\label{eq:lowkm}
 \omega_{-}^2\approx c^2
 \left[\frac{b}{1+b}k^2
 + \frac{k^4}{4(mu^0/\hbar )^2(1+b)^3}
 \right].
\end{equation}
This is the dispersion relation of a massless quasi-particle and has the same form of the Bogoliubov dispersion relation in a non-relativistic BEC [see Eq.~\eqref{eq:dispersion1}].
Let us study now when the $k^2$ term dominates over the $k^4$ one and {\it vice versa}, and check if the conditions obtained for these two regimes are compatible with Eq.~\eqref{eq:lowkcond}.
\paragraph{Phononic (infrared relativistic) regime.} It is easy to see that the quartic term $k^4$ can be neglected whenever
\begin{equation}
 |k|\ll 2\frac{m c_0}{\hbar }(1+b).
\end{equation}
This condition is always compatible with Eq.~\eqref{eq:lowkcond}. Depending on the value of $b$, they can be written in the compact form
\begin{equation}\label{eq:lowlowk}
 |k|\ll \frac{m u^0}{\hbar }(1+b)\min\left\{1,2\frac{c_0}{u^0}\right\}.
\end{equation}
In this limit the dispersion relation reduces to the usual phonon dispersion relation
\begin{equation}\label{eq:lineardisp}
 \omega^2_{-}=c_s^2 k^2,
\end{equation}
where the speed of sound $c_s$ is defined as
\begin{equation}\label{eq:cs}
 c_s^2\equiv\frac{(c\, c_0/u^0)^2}{1+(c_0/u^0)^2}=\frac{c^2 b}{1+b}.
\end{equation}

\paragraph{Newtonian (UV Galilean) regime.} The opposite regime, in which the $k^2$ can be neglected with respect to $k^4$, is defined by the following condition:
\begin{equation}\label{eq:k2negl}
 |k|\gg \left(\frac{mc_0}{\hbar }\right)(1+b).
\end{equation}
This regime is present only in the low-coupling case when $b\ll 1$, since only in this case Eqs.~\eqref{eq:k2negl} and~\eqref{eq:lowkcond} are compatible. They can be summarized as
\begin{equation}
 \frac{mc_0}{\hbar}\ll |k|\ll \frac{m u^0}{\hbar }.
\end{equation}
Under these assumptions the dispersion relation becomes
\begin{equation}\label{eq:nonreldisp}
 \hbar\omega_{-}=\frac{(\hbar k)^2}{2mu^0/c}=\frac{(\hbar k)^2}{2\mu/c^2},
\end{equation}
which represents the Newtonian dispersion relation for massive particles with effective mass $mu^0/c=\mu/c^2$. This is not surprising, since it is known from non-relativistic BEC (see Appendix~\ref{app:BEC}) that decreasing the wavelength of the perturbation, the atomic structure of the condensate emerges.

In this case (a RBEC) we see that, even if the structure of the Lagrangian is fully relativistic, when the energy of the perturbations is not much higher than the chemical potential, the bosons are moving with non-relativistic speed, and the dispersion relation is Newtonian. Noticeably, the mass of the non-relativistic quasi-particles is not the mass of the bosons $m$, but it is an effective mass $\mu/c^2$, the rest mass corresponding to an energy equal to the chemical potential $\mu$. This is not surprising since the chemical potential actually represents the energy needed to add a particle to the condensate.

\paragraph{The non-relativistic limit.} We discuss the non-relativistic limit of the gapless dispersion relation~\eqref{eq:lowk} starting from the behavior of the parameter $b=(c_0/u^0)^2$. As shown in section~\ref{sec:perturbations}, $u^0\to c$ in this limit when the fluid flows with a velocity much smaller than the speed of light. Moreover, in the same limit, the interaction must be weak, namely $c_0\ll c$, such that
\begin{equation}\label{eq:bnonrel}
 b\approx(c_0/c)^2\ll1.
\end{equation}
In this sense we can say that $b$ measures the relativistic nature of the condensate, taking into account both the strength of the interaction and the velocity of the fluid, as it is possible to see from the definition of $b$~\eqref{eq:b}, using Eqs.~\eqref{eq:defc0}, \eqref{eq:mu} and~\eqref{eq:homeul}.
Using Eq.~\eqref{eq:bnonrel} in Eq.~\eqref{eq:lowk} it is easy to obtain the Bogoliubov dispersion relation for a non-relativistic BEC [see Eq.~\eqref{eq:dispersion1}]
\begin{equation}\label{eq:lowkmnonrel}
  \omega_{-}^2\approx c_0^2
 \left[k^2
 + \frac{k^4}{(2mc_0/\hbar )^2}
 \right].
\end{equation}
It is worth noting that $c_s$~\eqref{eq:cs} is, in general, different from the speed of sound $c_0$ in a non-relativistic BEC, but, as it is evident from the above equation, $c_s$ reduces correctly to $c_0$ in the non-relativistic limit. 
The functional form of the dispersion relation~\eqref{eq:lowkm} is unchanged with respect to Eq.~\eqref{eq:lowkmnonrel}. Therefore, at this energy regime, there is no hint of the relativistic nature of the underlying BEC, without knowing the microphysics governing the elementary constituents. In other words, at this energy scale, a phononic observer cannot be aware of the relativistic nature of the condensate.

Finally, let us also note that in this limit, condition~\eqref{eq:lowlowk}, defining the phononic regime, reduces to the usual
\begin{equation}	
 |k|\ll 2\frac{mc_0}{\hbar},
\end{equation}
where $mc_0/\hbar$ is the inverse of the healing length [see Eq.~\eqref{eq:healing1}]. When this condition is satisfied, one obtains the phononic dispersion relation
\begin{equation}
 \omega_-^2 = c_0^2 k^2,
\end{equation}
while, in the opposite regime, the dispersion relation describes Newtonian particles with mass $m$
\begin{equation}
 \hbar\omega_-=\frac{\hbar^2k^2}{2m},
\end{equation}
as expected taking directly the non-relativistic limit of Eq.~\eqref{eq:nonreldisp} and putting $\mu\approx m c^2$.

\subsubsection{Gapped excitations}	%

A very different situation appears for the upper sign solution in Eq.~\eqref{eq:lowk}:
\begin{equation}\label{eq:lowkgap}
 \omega^2_+\approx c^2
 \left[4\left(\frac{m u^0}{\hbar }\right)^2(1+b)
 + \frac{2+b}{1+b}k^2
 - \frac{k^4}{4(m u^0/\hbar )^2(1+b)^3}
 \right].
\end{equation}
The only regime allowed is now the one in which the $k^2$ term dominates over the $k^4$ term.\footnote{In fact, the latter term would dominate if $k^2\gg4(mu^0/\hbar)^2(1+b)^2(2+b)>8(mu^0/\hbar)^2(1+b)^2$, but the above condition is clearly not satisfied when Eq.~\eqref{eq:lowkcond} holds.}
In this limit Eq.~\eqref{eq:lowkgap} becomes
\begin{equation}\label{eq:dispgap}
 \omega^2_+=\frac{m_{\rm eff}^2\,c_{s,{\rm gap}}^4}{\hbar^2}+c_{s,{\rm gap}}^2 k^2,
\end{equation}
where we have defined
\begin{align}
 c_{s,{\rm gap}}^2 &\equiv \frac{2+b}{1+b}\,c^2,\\
 m_{\rm eff} &\equiv 2\frac{(1+b)^{3/2}}{2+b}\frac{u^0}{c}m=2\frac{(1+b)^{3/2}}{2+b}\frac{\mu}{c^2}.
\end{align}
Eq.~\eqref{eq:dispgap} is the dispersion relation for a massive relativistic particle of effective mass $m_{\rm eff}$ and limit speed $c_{s,{\rm gap}}$. Note that, when Eq.~\eqref{eq:lowkcond} holds, and for $b\ll1$, the $k^2$ term is always dominated by the mass term. That is, this mode represents non-relativistic particles with mass $m_{\rm eff}$. When instead $b\gg1$, the $k^2$ term can be of the same order of magnitude as the mass term, and the fully relativistic dispersion relation with mass gap becomes important. The $k^2$ term dominates when
\begin{equation}
 \frac{mu^0}{\hbar }\frac{2(1+b)}{\sqrt{2+b}} \ll |k|\ll \frac{mu^0}{\hbar }(1+b).
\end{equation}

Let us now spend a few words about the physical meaning of this massive mode. Let us assume, for simplicity, that $b$ can be neglected. In this case, from the above formulas, the gap is $\Delta\omega\approx 2 mc^2/\hbar$. This is an excitation of completely different nature with respect to the previously discussed phonons. The mass gap of $2m$ indicates that the lowest possible excitation in this mode requires us to create a boson--anti-boson couple. When $k\neq 0$ the massive excitation propagates: this can be seen as the creation of a couple of particles, then its annihilation, while its energy is used to create another couple close to the former one and so on. Note that, apparently, this mode can make energy travel faster than light. However, one should consider that the dispersion relation~\eqref{eq:dispgap} is valid only for sufficiently low $k$, satisfying Eq.~\eqref{eq:lowkcond}. It is easy to check indeed that under this condition for $k$, the group velocity $\partial\omega/\partial k$ always remains smaller than $c$.

The non-relativistic limit of the gapped dispersion relation~\eqref{eq:lowkgap} is trivial. The zero-order term in $k^2$ becomes now $4m^2c^4/\hbar^2$, which diverges when taking the limit $c\to\infty$. In fact the gapped branch disappears because the energy needed to excite this mode is much larger (indeed infinite in the limit) than the typical energy scales in non-relativistic configurations. In other words, this mode cannot be excited because one cannot create particle--anti-particle couples in the non-relativistic limit.

\subsection{High momentum}	%
\label{sec:high}		%

The situation for high momentum is much simpler. When
\begin{equation}
 |k| \gg \frac{mu^0}{\hbar }\left[1+\left(\frac{c_0}{u^0}\right)^2\right]=\frac{mu^0}{\hbar }(1+b),
\end{equation}
it is easily found that the dispersion relation~\eqref{eq:dispersion_relbec} assumes the very simple form
\begin{equation}
 \omega_\pm^2=c^2k^2.
\end{equation}
This is the standard relativistic dispersion relation of a massless field propagating at the speed of light $c$.
This means that, when the energy of the perturbation is much larger than the chemical potential, the Newtonian particle regime of Eq.~\eqref{eq:nonreldisp} is overcome. The dispersion relation assumes a novel form that was absent in the non-relativistic case. At very high momenta the perturbations therefore probe the relativistic nature of the background condensate.

\begin{table}[t]
\scriptsize
\begin{tabular}{ccccc}
  \hline\hline\noalign{\smallskip}
  						&						&			\multicolumn{2}{c}{Gapless}					& Gapped\\
 \noalign{\smallskip}\cline{3-4}\noalign{\smallskip}
						&						&	$b\ll1$			&	$b\gg1$					&	\\
  \noalign{\smallskip}\hline\noalign{\smallskip}

   \multirow{2}{*}{$|k|\ll mu^0(1+b)/\hbar$}	& 	$|k|\ll2mc^0/\hbar$			&	$\omega^2 = c_s^2 k^2$	&	\multirow{2}{*}{$\omega^2 = c_s^2 k^2$}	&	\multirow{2}{*}{$\omega^2=m_{\rm eff}^2\,c_{s,{\rm gap}}^4/\hbar^2+c_{s,{\rm gap}}^2 k^2 $}	\\ \noalign{\smallskip}
  						&	$2mc^0/\hbar\ll|k|\ll mu^0/\hbar$	&	$\hbar\omega = (\hbar k)^2/2(\mu/c^2)$	&				&	\\
  \noalign{\smallskip}\hline\noalign{\smallskip}
  $|k|\gg mu^0(1+b)/\hbar$ 			&						&	 \multicolumn{3}{c}{$\omega^2 = c^2 k^2$}					\\
 \noalign{\smallskip}\hline\hline
 \end{tabular}
 \caption{\label{tab:regimes}Dispersion relation of gapless and gapped modes in different energy regimes. $c_s^2 = c^2 b/(1+b)$, $c_{s,{\rm gap}}^2 = c^2(2+b)/(1+b)$, $m_{\rm eff} = 2(\mu/c^2)(1+b)^{3/2}/(2+b)$, $b=(c_0/u^0)^2$.}
\end{table}

\section{The acoustic metric}	%
\label{sec:metric_relbec}	%

The propagation of phonons can be described in the formalism of quantum field theory in a curved background only when the relativistic quantum potential $T_\rho$ can be neglected in Eq.~\eqref{eq:psifluid} (see Sec.~\ref{sec:metric} for a comparison with the non-relativistic case).%
\footnote{As discussed after Eq.~\eqref{eq:psifluid}, to read off the dispersion relation and the metric describing the field propagation it is enough to start from Eq.~\eqref{eq:psifluid} instead of the full Eq.~\eqref{eq:dirac}.}
This is the relativistic generalization of neglecting the quantum pressure in the Gross--Pitaevskii equation in a non-relativistic BEC.

This is possible under two conditions. Firstly, the frequency and wave-number on the gapless branch $\omega_-$, Eq.~\eqref{eq:lowkm}, must be sufficiently low [see Eq.~\eqref{eq:lowlowk}]. Secondly, one has to perform an eikonal approximation, \ie\/ one assumes that all the background quantities vary slowly in space and time on scales comparable with the wavelength and the period of the perturbation, respectively. This latter condition is equivalent to requiring that
\begin{equation}
 \left|\frac{\partial_t \rho}{\rho}\right|\ll\omega,\quad \left|\frac{\partial_t c_0}{c_0}\right|\ll\omega,\quad\left|\frac{\partial_t{ u_\mu}}{u_\mu}\right|\ll\omega,
\end{equation}
and the corresponding relations for the variation in space.
Under these assumptions, Eq.~\eqref{eq:psifluid} simplifies to
\begin{equation}
 \left[u^\mu\partial_\mu
 \frac{1}{c_0^2}
 u^\nu\partial_\nu
 -\frac{1}{\rho}\eta^{\mu\nu}\partial_\mu\rho\,\partial_\nu\right]\hat\psi=0,
\end{equation}
where, again, the differential operators act on everything on their right.
In order to recover the acoustic metric, we need to cast the above equation in the form
\begin{equation}\label{eq:metricform}
 \partial_\mu \left(f^{\mu\nu}(x,t)\partial_\nu\hat\psi\right)=0,
\end{equation}
where $f^{\mu\nu}$ is the metric density
\begin{equation}
 f^{\mu\nu}=\sqrt{-g}g^{\mu\nu}.
\end{equation}
To this aim, as in the non-relativistic case (see Sec.~\ref{sec:metric}), one can use the continuity equation~\eqref{eq:cont1}
\begin{equation}\label{eq:contdef}
 \eta^{\mu\nu}\partial_{\mu}j_\nu=\partial_\mu(\rho u^\mu)=0,
\end{equation}
to commute $\rho u^\mu$ with $\partial_\mu$
\begin{equation}\label{eq:metricpsi2}
 \partial_\mu\left[
 \frac{\rho}{c_0^2}
 u^\mu u^\nu
 -\rho\eta^{\mu\nu}\right]\partial_\nu\hat\psi=0,
\end{equation}
from which the metric density is easily read
\begin{equation}
 f^{\mu\nu}=\frac{\rho}{c_0^2}
 \begin{pmatrix}
 -c_0^2-(u^0)^2		&	-u^0\ub^T				\\
 -u^0\ub		&	c_0^2\mathbbm{1}-\ub\otimes\ub^T
 \end{pmatrix}~.
\end{equation}
Finally, the metric describing the propagation of phonons in a RBEC is
\begin{equation}\label{eq:becmetric}
 g_{\mu\nu}=\frac{\rho}{\sqrt{1-u_\sigma u^\sigma/c_o^2}}\left[\eta_{\mu\nu}\left(1-\frac{u_\sigma u^\sigma}{c_0^2}\right)+\frac{u_\mu u_\nu}{c_0^2}\right],
\end{equation}
in coordinates $x^\mu=(x^0,{\bi x})=(ct,{\bi x})$.

The acoustic metric for perturbations in a relativistic, barotropic and irrotational fluid flow has been very recently derived in~\cite{Molina-Visser}. The above metric can be put in the same form as in~\cite{Molina-Visser} by a few variable redefinitions. To achieve this aim, let us define a four-velocity $v^\mu$:
\begin{equation}
 v^\mu \equiv \frac{c}{\|u\|}u^\mu,\quad \|u\| \equiv \sqrt{-u_\sigma u^\sigma}.
\end{equation}
With this definition, it is possible to generalize Eq.~\eqref{eq:cs} when the spatial part of the four-vector is different from zero. To keep all the formulas in a covariant form, the scalar speed of sound $c_s$ must be defined as
\begin{equation}\label{eq:scalarcs}
 c_s^2 = \frac{c^2\, c_0^2/\|u\|^2}{1+c_0^2/\|u\|^2}.
\end{equation}
Using the above definitions, the metric~\eqref{eq:becmetric} reduces to
\begin{equation}\label{eq:mattmetric}
 g_{\mu\nu}=\rho\frac{c}{c_s}\left[\eta_{\mu\nu}+\left(1-\frac{c_s^2}{c^2}\right) \frac{v_\mu v_\nu}{c^2}\right].
\end{equation}
This metric is manifestly conformal to that found in~\cite{Molina-Visser}. However, it is worth checking that they are exactly the same metric, comparing the conformal factors too. We have to show that $\rho c/c_s$ matches perfectly with the $\tilde n_0^2c^2/\tilde c_s(\tilde\rho_0+\tilde p_0)$ in~\cite{Molina-Visser}. All the quantities of~\cite{Molina-Visser} will be tilded to avoid confusion with ours.
From (54) of~\cite{Molina-Visser} we can write
\begin{equation}
 ||u|| = C\frac{\tilde\rho_0+\tilde p_0}{\tilde n_0},
\end{equation}
where $C$ is a constant and, by definition, $||u||=\nabla\tilde\Theta$. Note that from (46) of~\cite{Molina-Visser} $\tilde n_0 = ||u||\rho$, such that, apart from irrelevant constant factors,
\begin{equation}
 \frac{\tilde n_0^2}{\tilde c_s(\tilde\rho_0+\tilde p_0)} \propto \frac{\rho}{c_s},
\end{equation}
and also the conformal factors coincide.

\subsection{An application: \texorpdfstring{$k=-1$}{k = -1} FRW metric}	%
\label{sec:frw}								%

Mapping the metrics that describe expanding universes is one of the most interesting applications of analogue models.
However, only a mapping of the de Sitter spacetime and the $k=0$ FRW has been obtained so far \cite{livrev,expandinguniverse_blv,expandinguniverse_ff,expandinguniverse_silke,expandinguniverse_piyush}.%
\footnote{%
Since every FRW metric is conformally flat, the natural attempt to reproduce such a spacetime within the framework of the standard Eulerian fluid-gravitational analogy is to set $v=0$ and $c_s={\rm constant}$.
However, with this choice only stationary universes can be reproduced, because the acoustic conformal factor $\rho$ cannot depend on time by the continuity equation. Of course, one may try to map a non-flat FRW metric onto an acoustic one by allowing for time-dependent $v$, $c_s$ and $\rho$.
Nevertheless, this task (even though not forbidden by any no-go theorem) would be much more difficult than the natural procedure obtained by using a relativistic fluid.
}
Here we show how the above acoustic metric can be mapped into the FRW metric with $k=-1$. This is just an application, but it is important to stress how the category of metrics that one can map in this acoustic form has been immediately generalized.

We start by rewriting Eq.~\eqref{eq:mattmetric} in spherical coordinates, assuming isotropy and $v^\theta=v^\phi=0$:
\begin{equation}
g_{\mu\nu}=\rho\frac{c}{c_s}
\begin{pmatrix}
 -1+A(v^t)^2/c^2  & - A v^tv^r/c^2  & 0 & 0 \\
 - Av^tv^r/c^2   & 1 + A(v^r)^2/c^2 & 0 & 0 \\
 0 & 0 & r^2 & 0 \\
 0 & 0 & 0 & r^2 \sin ^2(\theta )
 \end{pmatrix}.
\end{equation}
where
\begin{equation}
 A\equiv 1-\frac{c_s^2}{c^2}.
\end{equation}
Defining the new coordinate $\tau$ and $\xi$ as
\begin{equation}
\left\{
\begin{aligned}
 \tau &= \sqrt{c^2 t^2-r^2},\\
 \xi &=  \frac{r}{\sqrt{c^2 t^2-r^2}},
\end{aligned}
\right.
\end{equation}
and choosing the following velocity profile:
\begin{align}\label{eq:vfrw}
 v^t &= c\,\sqrt{1+\xi^2},\\
 v^r &= c\,\xi,
\end{align}
one can write the inverse transformations as
\begin{equation}
\left\{
\begin{aligned}
 t &= \tau \frac{\sqrt{1+\xi^2}}{c} = \tau \frac{v^t}{c}, \\
 r &=  \tau \xi = \tau \frac{v^r}{c}.
\end{aligned}
\right.
\end{equation}

To investigate the physical meaning of these coordinates, it is worth applying the above transformation also to the underlying Minkowski spacetime, seen by the condensate. The fluid velocity $v$ can be transformed to
\begin{align}\label{eq:vfrwtauxi}
 v^\tau &= c,\\
 v^\xi &= v^\theta = v^\phi = 0.
\end{align}
Therefore they represent coordinates comoving with the fluid. The flat Minkowski line element $\rmd s_{\rm M}$ is
\begin{equation}\label{eq:minkowski}
 \rmd s_{\rm M}^2 = -c^2\rmd t^2+\rmd r^2+r^2 \rmd\Omega^2
  = -\rmd\tau^2+\tau^2\left(\frac{\rmd\xi^2}{1+\xi^2}+\xi^2 \rmd\Omega^2 \right),
\end{equation}

Going back to the acoustic metric, with the above choices the acoustic line element $\rmd s^2$ becomes
\begin{equation}\label{eq:lineelementfrw}
 \rmd s^2=\rho\frac{c}{c_s}\left[-\frac{c_s^2}{c^2}\rmd\tau^2+ \tau ^2\left(\frac{\rmd\xi^2}{1+\xi^2}+ \xi^2\rmd\Omega^2\right)\right].
\end{equation}
One may notice that, when both $c_s$ and $\rho$ depend only on $\tau$, the above expression represents the line element of a FRW spacetime with hyperbolic space sections:
\begin{equation}
 \rmd s^2=-\rho(\tau)\frac{c_s(\tau)}{c}\rmd\tau^2
 +\tau^2 \rho(\tau)\frac{c}{c_s(\tau)}\left(\frac{\rmd\xi^2}{1+\xi^2}+ \xi^2\rmd\Omega^2\right).
\end{equation}

However $\rho$ cannot be chosen freely because it is related through the continuity equation~\eqref{eq:cont1} to the velocity $v^\mu = u^\mu / ||u||$, which has already been fixed in Eq.~\eqref{eq:vfrw}. Nonetheless, it is easy to show that assuming $\rho=\rho_0$ constant both in $\tau$ and $\xi$, one can satisfy the continuity equation with an appropriate $||u||$. Using Eqs.~\eqref{eq:vfrwtauxi} and~\eqref{eq:minkowski} the continuity equation simplifies to
\begin{equation}
 \partial_\tau(\tau^3||u||)=0,
\end{equation}
from which
\begin{equation}
 ||u|| = c\frac{\Xi(\xi)}{\tau^3},
\end{equation}
where $\Xi$ is an arbitrary function.

It is therefore possible to define a new time coordinate
\begin{equation}\label{eq:timeT}
\rmd T \equiv \sqrt{\rho_0\frac{c_s(\tau)}{c}}\rmd\tau
\end{equation}
and a function $a(T)$
\begin{equation}\label{eq:Ttau}
 a(T) \equiv \tau \sqrt{\rho_0\frac{c}{c_s(\tau)}},
\end{equation}
where $\tau(T)$ is implicitly defined in Eq.~\eqref{eq:timeT}. In this way, the line element assumes the familiar $k=-1$ FRW form
\begin{equation}\label{eq:frw}
 \rmd s^2=-\rmd T^2+ a(T)^2\left(\frac{\rmd\xi^2}{1+\xi^2}+ \xi^2\rmd\Omega^2\right).
\end{equation}
At least from a mathematical standpoint, one has the freedom to choose an arbitrary form for $a(T)$. This is equivalent to choosing $c_s$ as an arbitrary function of $\tau$. Using Eqs.~\eqref{eq:timeT} and~\eqref{eq:Ttau}
\begin{equation}
 \tau(T)=\sqrt{\frac{2}{\rho_0}\int_0^T{a(T'})dT'},
\end{equation}
from which one can get $T$ as a function of $\tau$ given a generic $a(T)$, and $c_s(\tau)$ is
\begin{equation}
 c_s(\tau)=\frac{\rho_0c\tau^2}{a(T(\tau))^2}.
\end{equation}
Fixing $c_s(\tau)$ means fixing $U(\rho_0;\lambda_i(x^\mu))$ in Eq.~\eqref{eq:defc0}, and hence $\lambda_i$, as given functions of $x^\mu$. Indeed, from Eq.~\eqref{eq:scalarcs}, $c_0(x^\mu)$ depends both on $c_s(\tau)$ and $||u(x^\mu)||$, the latter given by the continuity equation, as discussed above. Finally, going back to the Euler equation~\eqref{eq:eul}, for constant $\rho$, one can determine also $V(x^\mu)$, because all the other functions $||u||$ and $U$ have already been fixed. Using a two-body interaction $U(\rho;\lambda_i)=\lambda_2(x^\mu)\rho^2/2$ one finds
\begin{align}
 \lambda_2(\tau,\xi)	&= \frac{2m^2c^2}{\rho_0\hbar^2}\frac{c_s(\tau)^2}{c^2-c_s^2(\tau)}\left(\frac{\Xi(\xi)}{\tau^3}\right)^2,\\
 V(\tau,\xi)		&= -\frac{m^2c^2}{\hbar^2}\left[1+\frac{c^2+c_s^2(\tau)}{c^2-c_s^2(\tau)}\left(\frac{\Xi(\xi)}{\tau^3}\right)^2\right].
\end{align}
Note that for a given FRW spacetime and hence given $a(T)$ and $c_s(\tau)$, one still has the freedom to choose the most convenient $\Xi(\xi)$ for a possible experimental realization.

\section{Summary and discussion}	%
\label{sec:conclusion_relbec}		%

In this chapter we have applied the analogy between gravity and condensed matter to a RBEC. Starting from the relativistic description of BECs, we have studied the propagation of excitations on top of a RBEC in a very generic framework, allowing for non-homogeneous density and nontrivial flow velocity.
We have analyzed in different regimes (see Tab.~\ref{tab:regimes}) the full dispersion relation, which is made of two branches, a gapless one and one with mass gap. The former one has three different regimes: For low momenta [wavelength larger than the healing length in the non-relativistic limit, see Eq.~\eqref{eq:healing1}] the dispersion relation is linear $\omega^2 = c_s^2 k^2$. Then, for intermediate energies, lower than the chemical potential of the condensate, the dispersion relation becomes that of a Newtonian massive particle, just as for the non-relativistic BEC case, when the wavelength is larger than or comparable with the healing length. Finally, for a RBEC, a third regime appears when the energy of the perturbation is larger than the chemical potential, and the dispersion relation becomes linear again ($\omega^2= c^2 k^2$), but with velocity equal to the speed of light.
The second branch has instead a mass gap, showing two different regimes. In this case, the dispersion relation describes, for low momenta, a relativistic massive particle with some effective mass and an effective limit speed apparently larger than the speed of light. However, this limit speed cannot be reached by such perturbations because at energy exceeding the chemical potential, the dispersion relation enters the second regime, in which the frequency is linearly proportional to the speed of light $c$. In the non-relativistic limit the energy needed to excite this mode diverges such that this branch becomes inaccessible. All these features are evident from the plot of the two branches in figure~\ref{fig:spectrum_at_rest}.

For sufficiently low momenta (a generalization of the non-relativistic condition involving the wavelength of the perturbations and the healing length), it is possible to describe the propagation of the phononic mode (massless branch) through the gravitational analogy, and the corresponding acoustic metric is given. This metric has a wider range of applications than the usual one obtained in non-relativistic fluids. We propose as an example the mapping to the $k=-1$ FRW metric for expanding universes. We also saw how it can be used to map spacetimes whose spatial section is not conformally flat. This could give the possibility to map even more complicated metrics (such as for example the Kerr metric)  or to build novel structures not allowed in the non-relativistic case.

This application has another important feature. Differently from all the previous systems where the analogy with gravity has been applied, the system investigated here is fully relativistic. Therefore, it does not suffer from the troubles related to the breaking of the Lorentz symmetry due to the appearance at sufficiently short wavelengths of the Newtonian structure of the underlying fluid and of its constituents. Hence, the RBEC is the first example of emergent Lorentzian spacetime from a Lorentzian background, showing therefore a Lorentz-to-Lorentz symmetry transition at high frequencies. Transitions from an infrared (IR) Lorentz symmetry to a UV Galilean one are known to be severely constrained (see \eg~\cite{Mattingly,GAC,MacLib}). Furthermore, in effective field theories they do tend to ``percolate'' from the UV to the IR via renormalization group effects, by showing up in non-suppressed corrections to the low energy propagators of elementary particles \cite{Collins,IRS}. It is plausible that situations like the one just exemplified by this novel analogue model of gravity might show a less severe behavior and ameliorate this naturalness problem common to most of UV Lorentz breaking theories.

\cleardoublepage

\pagestyle{headernonumber}

\makeatletter
\renewcommand\Hy@currentbookmarklevel{-2}
\makeatother

\phantomsection							
\addcontentsline{toc}{chapter}{Conclusions and Perspectives}	
\chapter*{Conclusions and Perspectives}				
\label{chap:conclusions}					
\chaptermark{Conclusions and Perspectives}			

In this Thesis, we presented analogue models of gravity and, in particular, we focused on Bose--Einstein condensates (BECs), from three main perspectives:
\begin{enumerate}
 \item as laboratory tests of quantum field theory (QFT) in curved spacetime (CS),
 \item as toy models of quantum gravity (QG),
 \item for the techniques that they provide to address various issues in general relativity (GR).
\end{enumerate}
The first two aspects are the fundamental motivations which have been stimulating the research in this field since the first paper by Unruh in 1981~\cite{unruh}, while, in a certain sense, the third one represents a by-product of all the work done in the last 30 years.
Even if almost only physicists from GR and QFT were originally interested in analogue gravity (AG), a wider community has nowadays been involved, with people from condensed matter or hydrodynamics. Indeed, we may say that the importance of AG does not reside anymore only on the possibility of learning something about GR and QG. As a matter of fact, AG is now providing more insight on phenomena of various fields of physics through their reinterpretation with the tools and the language provided by GR.

A significant example is the description of the propagation of phonons in BEC flows with sonic points, by using the analogy with CS with black/white horizons. Indeed, a series of phenomena once believed peculiar of gravity, such as Hawking radiation, are found also in these systems. In this Thesis, inspired by those results, we assumed a rather different point of view and we investigated the properties of BEC flows with horizons, without using the analogy with gravity but, nevertheless, being guided by that. Directly starting from the fundamental equations describing dilute Bose gases, we numerically solved the Bogoliubov--de Gennes equations for various velocity profiles.

Motivated by the fact that some predictions of QFT in CS are quite unphysical in real systems as BECs, we firstly investigated profiles with a single horizon for velocity profiles that are more asymmetric and irregular with respect to those of~\cite{MacherBEC}. As an example, the temperature of the spectrum of Hawking radiation is generally expected to be proportional to the spatial derivative of the velocity profile calculated at the sonic point. However, if we took a smooth velocity profile and added a very narrow but steep perturbation on top of it, the value of the Hawking temperature would enormously increase with respect to its non-perturbed value. This argument clearly does not make any sense in the limit in which the width of this perturbation goes to zero. We actually show that the quantity that really fixes the Hawking temperature is an average of the spatial derivative of the velocity profile on a region across the horizon whose size is related to the healing length of the condensate. This also 
implies that, regarding the phenomenon of Hawking radiation, the horizon must be considered as a thick layer and not as a two dimensional surface.
In the context of BECs those results are quite well understood and, even if not yet experimentally tested, they are somehow proved by numerical simulations~\cite{carusotto2}. However, the situation is more delicate if one wants to use these systems to understand what would happen for a real black hole originated by a star collapse in a spacetime where quantum fields have modified superluminal dispersion relations. As discussed in Sec.~\ref{subsec:bogo}, it is not completely clear and still controversial~\cite{carlos_stability} whether the quantum state selected by a dynamical collapse would correspond either to an Unruh-like state, showing thermal Hawking radiation or a Boulware-like state, or maybe something in between. We think that further investigations are needed to shed some light on this issue.

Furthermore, we extended our treatment to flow profiles with two horizons. These systems are indeed interesting for both theoretical and experimental reasons. First, they are the acoustic analogues of interesting spacetimes in GR, such as Alcubierre's warp drives~\cite{alcubierre} or Corley and Jacobson's black hole lasers~\cite{cj}. Second, flows with two horizons are more similar to what is expected to be realized in experiments~\cite{technion}. In fact, if a BEC flow was sped up from a subsonic region to a supersonic one, thus creating the analogue of a black horizon, the flow would plausibly slow down again to subsonic velocities, creating also the analogue of a white horizon.

The first two-horizon system considered in this Thesis is an infinite BEC flow with a black and a white horizon, dividing the fluid in three regions: two external supersonic ones and a subsonic one in between the horizons. Considering Hawking-like particle production on outgoing waves, the spontaneous emission of phonons is due to mode conversions of incoming negative norm waves scattering on both the horizons. This mechanism produces an infrared divergent spectrum, originating a flux of phonons linearly growing with time, that triggers a mild instability after the creation of the horizons. Moreover, we found that some radiation is present also when the two horizons collapse into a single point. In that case no horizon is present because the flow is everywhere supersonic. This is a quite important result since it proves that particle production in stationary 
geometries can take place even in the absence of a horizon. The only fundamental requirement at the origin of this phenomenon is the presence of negative norm modes and of a region with non-zero curvature where mode conversion can take place.

Then, we moved to the study of BEC flows where a white and black horizons divide the flow in two subsonic asymptotic regions and a supersonic region in between the horizons. Given that phonons in BECs are described by a supersonic dispersion relation, this flow configuration generates the so-called black hole laser effect, originally discovered by Corley and Jacobson, in the general context of CS with scalar field with modified superluminal dispersion relations. We found that the dynamics of the phonon field is fully described by a continuous set of real frequency eigenmodes plus a discrete set of complex frequency eigenmodes, which give rise to the claimed instability of this system. Interestingly, we found that this instability, if triggered on not too short time scales since the formation of the horizons, does not prevent us from observing Hawking radiation in systems with two horizons. Actually, the standard Hawking radiation and the laser effect are originated exactly by the same mechanism. In fact, we found that the spectrum of emitted particles just after the creation of the two horizons is identical to the continuous Planckian spectrum of Hawking particles. Only after some time the laser effect shows up and the spectrum starts to grow exponentially with time, appearing composed by a set of 
discrete frequencies. Finally, at very late time, the spectrum will be peaked around a unique frequency, corresponding to the complex eigenfrequency with largest imaginary part, \ie, with shortest growing time scale.
The drawback of our analysis is that it has been performed in a stationary background. This approximation is quite accurate for the description of Hawking radiation from a single analogue black horizon. However, it is evident that is not very physical in the case of the presence of dynamical instabilities. In that case, one is no more allowed to describe the system by separating the condensed part from the small amplitude perturbations, because after a short time the perturbations are no more small. It is an interesting open question whether the system is destroyed by the instability or the growing of the modes proceeds only up to some threshold and then saturates. A full analysis taking into account backreaction effect is deserved, possibly using numerical techniques as those adopted in~\cite{carusotto2}. This investigation is of primary relevance, because it would permit to describe in a more accurate way the experimental settings and would provide more solid theoretical predictions to be compared with 
forthcoming experimental data.

Regarding the second main motivation at the basis of the AG program, we showed how to address interesting questions about the fundamental structure of spacetime and the origin of the cosmological constant, by implementing the analogy in BECs. We compared the energy density associated to the analogue cosmological constant appearing in the dynamical equation of the analogue gravitational field (a Poisson-like equation where the source term is given by massive phonons) to other energy scales of the system. In particular we compared it to the Planck energy density computed using the speed of sound in the condensate and the gravitational constant $G_{\rm N}$ emerging from the microphysics of the system and appearing in the analogue Poisson equation. We found that the former is much smaller than the latter, when the condensate is dilute (this hypothesis is by the way required in order to apply the Gross--Pitaevskii analysis). Moreover, we compare the cosmological constant energy density to the ground 
state energy of the condensate (on a state without phonons). Again, the former is much smaller than the latter but, interestingly, it is of the same order of magnitude of the first order correction to the zero-point energy, related to quantum fluctuations of the phonon field. This result does not provide any positive statement on how to compute the actual value of the cosmological constant in gravity. However, it suggests that na\"ive computations assuming that the cosmological constant is somehow related to the cut off scale of QFT or to the zero-point energy of such fields can be wrong or at least misleading. It seems that the only viable procedure to compute the cosmological constant would be starting from fundamental physical principles, which are beyond GR and effective field theory. This statement must not be interpreted as pessimistic, given that we do not have a well-established theory of QG. On the contrary this is very good news, since the comparison of theoretical predictions against the 
observed value of the cosmological constant may represent an important test bench for QG theories and to discriminate among them.

Furthermore, we proposed a new analogue system, namely a system of bosons described by a relativistic Lagrangian, which can condense as well as ordinary non-relativistic BEC. This system is very interesting at least for two theoretical reasons. First, relativistic BECs allow to reproduce a larger set of geometries, because their spatial sections are not flat. As an example, we showed how Friedmann--Roberson--Walker spacetimes can be mapped onto acoustic metrics of relativistic fluids. Second, this system represents a very interesting and peculiar toy model for emergent gravity, since it has no preferred direction of time, because the background is not Galilean invariant but Lorentz invariant. A preferred time foliation appears only from a spontaneous symmetry breaking, given by the choice of the condensate rest frame. This not only avoids a quite unpleasant feature of standard analogue systems, but also suggests a possible solution to 
the naturalness problem of Lorentz violating theories. It is known that every high energy Lorentz violation percolates down to low energy because of renormalization of one-loop self-interaction diagrams. This leads, for different fields, to different values of the speed of light, which is however strongly constrained by observations and experimental evidence. It is plausible that re-establishing Lorentz symmetry at high energies (with a limit speed differing from the low energy limit speed) might tame the naturalness problem. We believe that this system represents a very significant example where the analysis of analogue systems suggests new interesting ideas for the development of a consistent QG theory.

Finally, let us conclude by commenting on the third of the aspects listed at the beginning of this summary. We investigated the causal properties of spacetimes allowing superluminal travel, by using techniques developed for AG models, based on the analysis of the renormalized stress-energy tensor in geometries whose metrics can be put in Painlev\'e--Gullstrand form. Interestingly, we found that warp-drive bubbles are unstable only once they become superluminal. Since opportunely combining two superluminal travels one can build spacetimes with closed timelike curves (\ie, time machines), such a result looks like a low-level implementation of a chronology protection mechanism, by preventing the realization of a spacetime structure that might potentially lead to time travel.
This result might be general, in the sense that chronology protection might be implemented at a low level, by blocking the mechanisms at the basis of the realization of time machines. This issue requires further investigations. Since in analogue models time travel is strictly forbidden by the presence of a background without close timelike curves, it is interesting to check what happens to the effective geometry when dynamically approaching the creation of an analogue time machine. We therefore think that such an analysis may suggest how to address the problem of chronology protection in GR.
The linear divergence in time of the flux of emitted phonons in a BEC flow with two horizons, simulating a warp-drive geometry, is a first result in this direction. Even if this divergence is much milder than the exponential one found for a scalar field with relativistic dispersion relation, it suggests that warp-drive bubbles may be unstable even in the presence of trans-Planckian physics leading to superluminal modifications of the dispersion relation of quantum fields.

As a final remark, we strongly believe that all the experimental data that will be available in the very next future will require a strict collaboration between experimentalists, condensed matter theoreticians, and physicists from GR. Thus, we think that one of the main goals to be pursued in the next years will be the strengthening of the interactions among these communities, in order to interpret experimental results, propose new experiments, investigating new phenomena in various physical systems, and, eventually, understand something more on the microscopic structure of spacetime.

\makeatletter
\renewcommand\Hy@currentbookmarklevel{-2}
\makeatother

\cleardoublepage
\phantomsection								
\addcontentsline{toc}{chapter}{Acknowledgments --- Ringraziamenti}	
\chapter*{Acknowledgments --- Ringraziamenti}				
\label{chap:ack}							
\chaptermark{Acknowledgments --- Ringraziamenti}			

The Italian part is not a translation of the English one. I have decided to use two languages to acknowledge different people, for different motivations.
\\

La parte in italiano non \`e una traduzione di quella in inglese. Ho scelto di usare due lingue per ringraziare persone diverse, con motivazioni diverse.

\subsubsection*{Italiano}	%

Questa tesi \`e dedicata ai miei genitori e a Donata, le persone che pi\`u mi sono state vicine in questi quattro anni, in cui sono stati realizzati i lavori presentati in questa tesi. Con loro ho condiviso gioe e difficolt\`a. Grazie al loro appoggio \`e stato pi\`u facile affrontare dubbi e incertezze. Per varie ragioni \`e stato molto difficile scegliere di continuare a fare questo lavoro: i loro consigli mi hanno aiutato a capire quello che volevo per mio futuro.
In particolare voglio ringraziare Donata per aver compreso e condiviso questa scelta e per il coraggio di aver accettato, per qualche anno, il rischio di spostamenti in luoghi non ancora definiti.

Un aiuto fondamentale per chiarirmi le idee \`e venuto da Stefano, che ho annoiato con lunghe discussioni, e da Guido, che mi ha offerto la possibilit\`a di utilizzare le mie competenze al di fuori della fisica e con cui spero di continuare a collaborare, insieme a tutti i ragazzi del GiViTI. Li ringrazio entrambi per il loro grande entusiasmo nella ricerca, anche se in campi diversi. Ringrazio anche Daniele per la sua capacit\`a di tradurre, interpretare e analizzare dati clinici con strumenti matematici e statistici. Da lui ho imparato a cogliere molteplici aspetti di un problema e a leggerli con linguaggi diversi. E infine ringrazio Iacopo, con cui lavorer\`o per i prossimi due anni, che ha sicuramente contribuito alla mia decisione.

Voglio anche ringraziare tutte le persone con cui ho passato questi anni a Trieste, in particolare i miei coinquilini: Michele, Franco e Pietro, con i quali ho condiviso nove anni della mia vita, Marco, l'ultimo arrivato Tommaso, Aurora e Luigi, che spesso sono stati in casa con noi.
Ringrazio anche Marina, Fulvio e Lorenzo e tutte le persone conosciute a Trieste, amici e amiche di Donata e i colleghi della SISSA.
Grazie anche a tutti i parenti e gli amici di Bergamo, alcuni dei quali sono anche venuti a trovarmi a Trieste per qualche giorno: Manuela, che conosco da pi\`u di vent'anni, Silvia e Paolo, Luis e Alessia, la mia cuginetta Diana e mio cugino Daniele. Un grazie speciale a mio nonno Pierino, per i due giorni che abbiamo passato, anche insieme a mio padre, in giro per il Carso a cercare di capire dove combatt\`e suo padre Giovanni, quasi cent'anni fa. Grazie anche alla nonna Dina, che \`e rimasta per tre giorni a casa senza di lui. Un pensiero ai miei nonni Gianni e Michelina che se ne sono andati ormai da pi\`u di dieci anni e ai miei cuginetti che invece sono nati da meno di un anno, Davide, Elisa e Michelle.

\subsubsection*{English}	%

First, I wish to thank my supervisor Stefano Liberati for his invaluable support and advice.
I have appreciated his enthusiasm and his broad view on physics in many discussions for the last four years.
It has been a true pleasure to have had the opportunity to work with him. I also enjoyed the freedom that he gives to his students, allowing them to find their own way of doing research and to collaborate with other people all around the world. I also thanks him for his help in the organization of the material of this Thesis.

I thank all the people at SISSA, that I have met since I arrived here, all the professors, the researchers, and the students of the Astrophysics Sector and the Astroparticle Curriculum, in particular the former and present students of Stefano Liberati, for very stimulating discussions: Thomas Sotiriou, Christoph Rahmede, Luca Maccione, Lorenzo Sindoni, Goffredo Chirco, Vincenzo Vitagliano, Angus Prain, and Eolo Di Casola.
Among all the SISSA students, I wish to specially thank the students of the Astrophysics Sector of my year, Laura, Lulu and specially Goffredo for all the time spent discussing together in our office.

I also thank all the people I have been collaborating with, in particular Carlos Barcel\'o and Renaud Parentani, who hosted me at their institutions in Granada and Paris for a few weeks during my PhD.
Without them, almost all the work of this Thesis would not have been realized.
I hope to continue those collaborations in the future.
I am grateful to all the other coauthors of my papers: Antonin Coutant, Serena Fagnocchi, Marton Kormos, Lorenzo Sindoni, Andrea Trombettoni. I warmly thank all the people I discussed with about my work and those who read and gave helpful comments about my manuscripts: Roberto Balbinot, Iacopo Carusotto, Ted Jacobson, Pyush Jain, Gil Jannes, Sebastiano Sonego, Ralph Sch\"utzhold, Jeff Steinhauer, William G. Unruh, Matt Visser, and Silke Weinfurtner.

A special acknowledge to Jean Macher, who wrote the first version of the code I used for the computations of this Thesis and some of my papers. I wish to thanks him for the careful explanations and his very helpful advice about how to use his code.

This work has been possible also thanks to the economic support for scientific collaborations and for the participation to conferences from SISSA, INFN, and MICINN. I also acknowledge FQXi and SVCF for a mini-grant (FQXi-MGA-1002).
I thank Lorena, the secretary of the astrophysics sector, for her help with bureaucracy, and the secretaries of the students' secretariat Riccardo and Federica. I want also to thank Alex and Tatiana, for their help with the grant from FQXi, and Sara for the English translation of the SISSA statutes.

Finally, I want to thank Stefano and Donata for having carefully proofread the draft of this Thesis. I also thank the two referees, Iacopo Carusotto and Luis Garay, for a frutiful discussion (Carlos Barcel\'o was also present) on the SISSA terrace the day before my PhD defense, when they suggested me several improvements to this Thesis.

\cleardoublepage
\appendix %

\pdfbookmark[-1]{Appendixes}{appendix}

\pagestyle{header}

\chapter{Theory of Bose--Einstein~condensates}	
\label{app:BEC}					
\chaptermark{Theory of BECs}			

\section{The fundamental equations}	%
\label{sec:BEC}				%

We present the main steps leading to the equations describing Bose--Einstein condensates (BECs)~\cite{Dalfovo} and linear perturbations on top of them, following the approach of~\cite{ulf} and~\cite{MacherBEC}.

At low temperature, the quantum properties of a gas of weakly interacting atoms are efficiently described by a second quantized field satisfying the commutation relation
\begin{equation}\label{eq:comm}
 \comm{\hP(t,\bx)}{\hPd(t,\bx')}=\delta^3(\bx-\bx'),
\end{equation}
and by its Hamiltonian
\begin{equation}\label{eq:Hsc}
 \hH = \int\! \dtx \left[\frac{\hbar^2}{2m} \nx \hPd \, \nx\hP + V  \hPd\hP  + \frac{g}{2} 
 \hPd\hPd\hP\hP    \right],
\end{equation}
where $m$ is the atom mass, $V$ the external potential and $g$ the effective coupling, which is related to the scattering length $a$ through the following relation
\begin{equation}\label{eq:scatteringlength}
 g=\frac{4\pi a\hbar^2}{m}.
\end{equation}

When a significant fraction of the atoms condenses, it is meaningful to expand $\hP$ in a $c$-number function $\Pn$, describing the condensed part, plus a field operator $\hp$, describing (relative) density perturbations over the condensate
\begin{equation}\label{eq:defphi}
 \hP=\Pn(1+\hp).
\end{equation}
Given the above definition, $\hp$ satisfies the following canonical commutation relation
\begin{equation}\label{eq:commphi}
 \comm{\hp(t,\bx)}{\hpd(t,\bx')}=\frac{1}{\rnbx}\delta(\bx-\bx').
\end{equation}

We assume that the condensate is stationary, which means that $\Pn$ is of the form
\begin{equation}\label{eq:phasedensity}
 \Pn(t,\bx)=\ee^{-\ii\mu t/\hbar}\times\sqrt{\rnbx}\ee^{\ii\theta_0(\bx)},
\end{equation}
where $\mu$ is the chemical potential, $\rnbx$ the mean density of condensed atoms and
\begin{equation}
 \bv(\bx)=\frac{\hbar}{m}\nx\theta_0(\bx)
\end{equation}
is their mean velocity.

Defining the number operator
\begin{equation}\label{eq:N}
  \hN\equiv\int\!\dtx\,\hPd\hP,
\end{equation}
the grand canonical Hamiltonian is
\begin{equation}\label{eq:grandHsc}
 \hGH\equiv\hH-\mu\hN=\int\! \dtx \left[\frac{\hbar^2}{2m} \nx \hPd \, \nx\hP + (V-\mu)  \hPd\hP + \frac{g}{2} \hPd\hPd\hP\hP\right].
\end{equation}

We expand $\hH$ of Eq.~\eqref{eq:Hsc} in powers of $\hp$ [see Eq.~\eqref{eq:defphi}] to the second order:
\begin{align}
 H_0	&= \int\!\dtx \left[ \hbm\nx\Pn^*\nx\Pn+V|\Pn|^2+\frac{g}{2}|\Pn|^4\right]=
 		 \int\!\dtx \,\Pn^*\left[- \hbm\nx^2+V+\frac{g}{2}\rn\right]\Pn, \label{eq:h0}\displaybreak[0]\\
 \hH_1	&= \int\!\dtx\left[\hbm\nx(\Pn^*\hpd)\nx\Pn+\hbm\nx\Pn^*\nx(\Pn\hp)+V|\Pn|^2(\hp+\hpd)+g|\Pn|^4(\hp+\hpd)\right]\nonumber\\
 	&= \int\!\dtx \,\Pn^*\hpd\left[- \hbm\nx^2+V+g\rn\right]\Pn+\mbox{h.c.}, \label{eq:h1}\displaybreak[0]\\
 \hH_2	&= \int\!\dtx \left[\hbm\nx(\Pn^*\hpd)\nx(\Pn\hp)+V|\Pn|^2\hpd\hp+\frac{g}{2}\left(\hpds+\hp^2+4\hpd\hp\right)\right]\nonumber\\
 	&= \int\! \dtx\,\rn\left\{\hpd\left[T_\rho-\ii \hbar\bv\!\cdot\!\nx-\hbm\frac{\nx^2\Pn}{\Pn}+V+2g\rn\right]\hp + g\frac{\rn}{2}\left(\hpds+\hp^2\right)\right\},
\end{align}
where
\begin{equation}\label{eq:Trho}
 T_\rho\equiv -\frac{\hbar^2}{2m} \frac {1}{\rn}\nx\!\cdot\!\rn\nx
\end{equation}
reduces to the usual kinetic operator when the background condensate is homogeneous.

In the same fashion we can expand $\hN$:
\begin{align}
 N_0&=\int\!\dtx\,\rn,\displaybreak[0]\\
 \hN_1&=\int\!\dtx\,\Pn\,\hpd+\mbox{h.c.},\displaybreak[0]\\
 \hN_2&=\int\!\dtx\,\rn\,\hpd\hp.\label{eq:N2}
\end{align}
$N_0$ is the number of condensed bosons, while $\hN_2$ counts the number of particles out of the condensates (depletion).

Expanding also the Heisenberg equation of motion for $\hP$,
\begin{equation}
 \ii\hbar\pdt\hP(t,\bx)=\comm{\hP(t,\bx)}{\hH},
\end{equation}
one obtains
\begin{align}
 &\ii\hbar\pdt\Pn = \Pn\comm{\hp}{\hH_1},\\
 &\ii\hbar\pdt\hp = \comm{\hp}{\hH_2}-\ii\hbar\hp\frac{\pdt\Pn}{\Pn}.\label{eq:ptp}
\end{align}
The former equation gives the Gross--Pitaevskii equation
\begin{equation}
 \ii\hbar\pdt\Pn=\left[-\hbm\nx^2+V+g\rn\right]\Pn,
\label{eq:GP}
\end{equation}
which, for a stationary condensate, reduces to
\begin{equation}
 \mu = \frac{1}{2}mv^2-\hbm\frac{\nx^2\sqrt{\rn}}{\rn}+V+g\rn,\qquad
 \nx\!\cdot\!(\rn\bv)=0,\label{eq:cont}
\end{equation}
where the second equation is the continuity equation for a stationary flow.

Defining
\begin{equation}
 \delta\hH\equiv\int\!\dtx\,\rn\left[-\hbm\frac{\nx^2\Pn}{\Pn}+V+g\rn\right]\hpd\hp=\mu\int\!\dtx\,\rn\hpd\hp=\mu\hN_2,
\end{equation}
where we used Eqs.~\eqref{eq:GP}, \eqref{eq:phasedensity} and~\eqref{eq:N2}, we obtain
\begin{equation}
 \ii\hbar\hp\frac{\pdt\Pn}{\Pn}=\hp\left[-\hbm\frac{\nx^2\Pn}{\Pn}+V+g\rn\right]=\comm{\hp}{\delta\hH}.
\end{equation}
Eq.~\eqref{eq:ptp} can now be rewritten as
\begin{equation}\label{eq:phi}
 \ii\hbar\pdt\hp = \comm{\hp}{\hHe},
\end{equation}
where
\begin{equation}
\hHe\equiv\hH_2-\delta\hH=\hH_2-\mu\hN_2=\int\! \dtx\,\rn\left\{\hpd\left[T_\rho-\ii\hbar \bv\!\cdot\!\nx +g\rn\right]\hp 
 + g\frac{\rn}{2}\left(\hpds+\hp^2\right)\right\}.
\end{equation}
Remarkably, $\hHe$ is independent of the external potential $V$ and of ${\nx^2\Pn}/{\Pn}$. Furthermore, from the stationarity assumption, it does not depend explicitly on time. When symmetrizing it with respect to $\hp$ and $\hpd$, one obtains
\begin{equation}\label{eq:heffsym}
 \hHe=\frac{1}{2}\int\! \dtx\,\rn\left\{\hpd\left[T_\rho-\ii \hbar\bv\!\cdot\!\nx +g\rn\right]\hp +[(T_\rho+\ii \hbar\bv\!\cdot\!\nx +g\rn)\hpd]\hp 
+g{\rn}\left(\hpds+\hp^2\right)\right\},
\end{equation}
which is manifestly Hermitian.

Finally, from Eq.~\ref{eq:phi}, one derives the Bogoliubov--de Gennes (BdG) equation, which is obeyed by $\hp$ at the linear level:
\begin{equation}
 \ii\hbar\pdt\hp = \left[T_\rho-\ii \hbar\bv\!\cdot\!\nx + m c^2
\right]
\hp + m c^2
\hpd,\label{eq:BdG}
\end{equation}
where we introduced the $x$-dependent speed of sound
\begin{equation}\label{eq:sound}
 c^2(\bx)\equiv\frac{g(\bx)\rn(\bx)}{m}.
\end{equation}
Eq.~\eqref{eq:BdG} tells us that the perturbation $\hp$ is coupled to the condensate only through the functions $v$ and $c$.
The disappearance from Eq.~\eqref{eq:BdG} of the potential $V$, the quantum potential $-(\hbar^2/2m)(\nx^2\sqrt{\rn}/\rn)$, and the coupling $g$ constitutes an essential step towards the notion of the acoustic metric (see Sec.~\ref{sec:metric}).

For further convenience in computing $\hHe$, we rewrite Eq.~\eqref{eq:heffsym} in a more handling form by using the BdG equation~\eqref{eq:BdG}
\begin{equation}\label{eq:ht}
 \hHe = \frac{\ii\hbar}{2}\int\!\dtx\,\rnbx\left[\hpd(\pdt\hp)-(\pdt\hpd)\hp\right].
\end{equation}

We conclude this section with a remark. We generated the equations of motion using the Hamiltonian $\hH$, Eq.~\eqref{eq:Hsc}. Since the condensate is stationary one might also have used the grand canonical Hamiltonian $\hGH$, Eq.~\eqref{eq:grandHsc}. However one should have coherently suppressed the explicit phase-dependence on time of $\Pn$
\begin{equation}
 \Pn(t,\bx)=\sqrt{\rnbx}\ee^{\ii\theta_0(\bx)}.
\end{equation}
The two approaches are completely equivalent and, not surprisingly, the equation of motion for the perturbation field $\hp$ are generated by the same Hamiltonian $\hGH_2$.

\section{Phonons in BECs}	%
\label{sec:BECphonons}		%

\subsection{The quantization procedure}	%
\label{subsec:quantization}		%

In stationary condensates, the spectrum of Eq.~\eqref{eq:heffsym} can be characterized using very general properties.
These are, first, the fact that the eigenmodes are asymptotically bounded (to guarantee that they have a well-defined norm
since the spatial domain of $\hp$ is infinite), second, the Hermiticity of~\eqref{eq:heffsym}, \ie, its self-adjointness with respect to the scalar product defining the eigenmodes norm, and third, that this scalar product is {\it not} positive definite, as it is the case here, see~Eq.~\eqref{eq:scalar} (see also \eg~\cite{Greinerbook}).

To understand how to expand the complex field operator $\hp$ in modes and creation/destruction operators, 
it is convenient to define a two-component field~\cite{ulf}
\begin{equation}
 \hW\equiv\spin{\hp}{\hpd}.
\end{equation}
With this notation Eq.~\eqref{eq:BdG} becomes
\begin{align}
 & \ii\hbar\pdt\hW = B\hW,\label{eq:spineq}\\
 & B = (T_\rho+g\rn)\sigma_3 -\ii \bv\!\cdot\!\nx +\ii g\rn\sigma_2,
\end{align}
where $\sigma_i$ are the the Pauli matrices
\begin{equation}
 \sigma_1=
  \begin{pmatrix}
  0 & 1 \\
  1 & 0
 \end{pmatrix},
 \quad
 \sigma_2=
 \begin{pmatrix}
  0 & -\ii\\
  \ii & 0
 \end{pmatrix},
 \quad
 \sigma_3=
 \begin{pmatrix}
  1 & 0\\
  0 & -1
 \end{pmatrix}.
\end{equation}
Since the field $\hW$ is invariant under the conjugation operation defined by
\begin{equation}\label{eq:hWsymm}
 \bar{S}\equiv\sigma_1 S^\dagger,
\end{equation}
the structure of $\hW$ must be
\begin{equation}
 \hW=\sum_n (W_n \ha_n + \bar W_n \had_n),
\label{eq:hW}
\end{equation}
where $W_n$ are doublets of $\mathbb{C}$-functions, and $\sum_n$ denotes the summation over a (possibly continuous) complete sets of modes. As evident from the above expansion, the symmetry of the field $\hW$ described by Eq.~\eqref{eq:hWsymm} has a direct physical interpretation as the coincidence of phonons and anti-phonons.

Using $\nx^\tra=-\nx$, $T_\rho^\tra=T_\rho$ and the properties of Pauli matrices, one verifies that the scalar product
\begin{equation}\label{eq:scalar}
 \scal{W_1}{W_2}\equiv\int\!\dtx\,\rnbx \, W_1^{*}(t,\bx)\sigma_3 W_2(t,\bx)
\end{equation}
is conserved under time evolution when $W_i$ are solutions of Eq.~\eqref{eq:spineq}, since
\begin{equation}
 B^{*\tra}\sigma_3=\sigma_3 B.
\end{equation}
In terms of the two components of the doublet $W$
\begin{equation}
 W=\spin{\phi}{\varphi},
\end{equation}
the scalar product becomes
\begin{equation}\label{eq:scalarphi}
 \left\langle{\spin{\phi_1}{\varphi_1}}\right|\left.{\spin{\phi_2}{\varphi_2}}\right\rangle
 =\int\!\dtx\,\rnbx \left[\phi_1^{*}(t,\bx)\phi_2(t,\bx)-\varphi_1^{*}(t,\bx)\varphi_2(t,\bx)\right].
\end{equation}

For later convenience, we state here some properties of the scalar product that can be verified using the anticommutation relation of Pauli matrices and Eq.~\eqref{eq:commphi}:
\begin{align}
 &\scal{W_1}{W_2}=\scal{W_2}{W_1}^*,\\
 &\scal{\bar W_1}{\bar W_2}=-\scal{W_2}{W_1}=-\scal{W_1}{W_2}^*,\\
 &\hW\hW^\tra-(\hW\hW^\tra)^\tra=\frac{1}{\rnbx}\delta(\bx-\bx')\ii\sigma_2,\\
 &\comm{\scal{W_1}{\hW}}{\scal{W_2}{\hW}}=-\scal{W_1}{\bar W_2}=\scal{W_2}{\bar W_1}.\label{eq:commfor}
\end{align}

Assuming that the doublets $W_n$ have a non-zero norm, we define an orthonormal basis:
\begin{align}
 \scal{W_n}{W_m} &= -\scal{\bar W_n}{\bar W_m}=\delta_{nm},\label{eq:wnorm}\\
 \scal{W_n}{\bar W_m} &= 0,
\end{align}
where the Kronecker $\delta$ is replaced by a $\delta$-distribution in the case of a continuous set of modes.
Using Eq.~\eqref{eq:hW} and the above identities, one obtains
\begin{equation}
 \comm{\ha_n}{\had_{m}}=\comm{\scal{W_n}{\hW}}{-\scal{\bar W_m}{\hW}}=\scal{W_n}{W_m}=\delta_{nm},
\end{equation}
which shows that $\ha_n$ and $\had_n$ are in fact destruction and creation operators.

When the condensate is stationary, one can work with eigenmodes $W_\lambda^\alpha$ of frequency $\lambda$, where the index $\alpha$ describes the set of modes with the same frequency.
By definition, one has
\begin{equation}
 B\, W_\lambda^\alpha=\hbar\lambda\, W_\lambda^\alpha.
\end{equation}
The conservation of the scalar product gives
\begin{equation}\label{eq:nocomplex}
 0=\pdt\scal{W_\lambda^\alpha}{W_{\lambda'}^{\alpha'}}=-\ii(\lambda^*-\lambda')\scal{W_\lambda^\alpha}{W_{\lambda'}^{\alpha'}}.
\end{equation}
From this it is clear that only real frequency modes can be normalized as in Eq.~\eqref{eq:wnorm}.
However, in the presence of dynamical instabilities, also complex frequency eigenmodes are present.
These modes form in general a discrete and finite set of pairs of modes with conjugated frequencies that we call
$\{\lambda_a=\omega_a+\ii\Gamma_a, a=1, 2, ... N\}$.
In place of Eq.~\eqref{eq:wnorm}, we choose the following pseudo-normalization for each pair,
\begin{equation}
 \scal{W_{\lambda_a}}{W_{\lambda_{a'}^*}}=\ii\delta_{aa'},
\end{equation}
and the other scalar products must vanish because of Eq.~\eqref{eq:nocomplex}.
In fact, the Hermiticity of $\hH$ implies that if $\lambda_a$ is an eigenfrequency, then $\lambda_a^*$ is an eigenfrequency too~\cite{ulf}.
We call $V_{\lambda_a}$ and $Z_{\lambda_a}$ the doublets corresponding to $\lambda_a$ and $\lambda_a^*$, respectively.

Below we analyze two specific situations in details: the infinite homogeneous 3-dimensional condensate in Sec.~\ref{subsec:mode3d} and the inhomogeneous 1-dimensional case in Sec.~\ref{subsec:mode3d}.

\subsection{Infinite homogeneous condensates}	%
\label{subsec:homog}				%

When the condensate is homogeneous and infinite, the spectrum of~\eqref{eq:heffsym} is completely characterized by a 3-dimensional set of momentum eigenmodes labelled by $\bk$. In such a condensate the spectrum cannot contain complex frequency modes which are not asymptotically bounded, given that, in an homogeneous background, a complex frequency mode is also an eigenmode of the momentum operator with complex eigenvalue.

In this situation, the field $\hW$ can be expanded as
\begin{equation}\label{eq:fieldexpansionhom}
 \hW=\int\!\dtk\left[\wk\ak+\wkb\akd\right].
\end{equation}
The modes satisfy the following normalization rules,
\begin{equation}
 \scal{\wk}{\wkp}=-\scal{\wkb}{\wkbp}=\delta^{(3)}(\bk-\bk')\label{eq:normWk}
\end{equation}
and the operators $\ha$ satisfy
\begin{equation}
 \comm{\ak}{\akdp}=\delta^{(3)}(\bk-\bkp).
\end{equation}
All the other scalar products and commutators vanish.

By homogeneity and stationarity, the doublets must have the following form
\begin{equation}
 \wk=\frac{\ph{-\ii\om t+\ii\bk\cdot\bx}}{\knorm}\spin{\uk}{\vk},
\end{equation}
where $\knorm$ is a convenient normalization factor and $\uk$ and $\vk$ are constant.
The field $\hp$ can be expanded as
\begin{equation}\label{eq:phiexpansionk}
\hp(t,\bx) =\int\frac{\dtk}{\knorm}\left[\ph{-\ii\om t+\ii\bk\cdot\bx}\uk\ak+\ph{+\ii\om t-\ii\bk\cdot\bx}\vks\akd\right].
\end{equation}
Eqs.~\eqref{eq:phiexpansionk} and~\eqref{eq:normWk} give the normalization condition
\begin{equation}\label{eq:normuv}
 |\uk|^2-|\vk|^2=1.
\end{equation}

Putting the field expansion~\eqref{eq:phiexpansionk} into Eq.~\eqref{eq:ht} and using the above given commutation relations and the scalar products, one obtains
\begin{align}
 \hHe &= \frac{\hbar}{2} \int\!\frac{\dtk\,\dtkp}{(2\pi)^3}\int\!\dtx
 \nonumber\\
 &\qquad\times\left[(\omp+\om)\ph{\ii(\om-\omp)t-\ii(\bk-\bkp)\cdot\bx}\uks\ukp\akd\akp\right.
 -(\omp+\om)\ph{-\ii(\om-\omp)t+\ii(\bk-\bkp)\cdot\bx}\vk\vksp\ak\akdp
 \nonumber \\
 &\qquad+\left.(\om-\omp)\ph{\ii(\om+\omp) t-\ii(\bk+\bkp)\cdot\bx}\uks\uksp\akd\akdp
  +(\omp-\om)\ph{-\ii(\om+\omp) t+\ii(\bk+\bkp)\cdot\bx}\vk\ukp\ak\akp
 \right]\\
 &=\int\!\dtk\,\hbar\om\left[\akd\ak-\int\frac{\dtx}{(2\pi)^3}\,|\vk|^2\right].
\end{align}
The first term is the usual sum over harmonic oscillators of real frequency, while the second term is a $c$-number and it diverges when the condensates is infinite. This term, present even when there are no phonons, represents a homogeneous density equal to
\begin{equation}\label{eq:h2app}
 h_2=-\int\frac{\dtk}{(2\pi)^3}\,\hbar\om|\vk|^2,
\end{equation}
which is a correction to the ground state energy $H_0$ due to interactions.
Indeed, this quantity is strictly related to the depletion of the condensate, that is the number of particles which are out of the condensate when there are no phonons,
\begin{equation}
 \vev{\hN_2},
\end{equation}
where $|0\rangle$ is the state annihilated by the destruction operators $\ak$. The ratio between the first order (in powers of $\hp$) number of non-condensed particles and the total number $\vev{\hN}$ is called depletion factor (see~\cite{Dalfovo})
\begin{equation}\label{eq:dep}
 d_f\equiv\frac{\vev{\hN_2}}{\vev{\hN}}\approx\vev{\frac{\hN_2}{N_0}}=\vev{\hpd\hp}=\int\frac{\dtk}{(2\pi)^3}\,|\vk|^2,
\end{equation}
where we used the field expansion~\eqref{eq:phiexpansionk} in Eq.~\eqref{eq:N2}.

\subsubsection{Mode analysis}	%
\label{subsec:mode3d}		%

When the condensate is homogeneous, the quantum potential vanishes, the Gross--Pitaevskii equation~\eqref{eq:GP} simplifies to
\begin{equation}
 \mu = \frac{1}{2}mv^2+V+g\rn,
\end{equation}
and the BdG equation~\eqref{eq:BdG} is solved by Fourier analysis, using the field expansion~\eqref{eq:phiexpansionk}:
\begin{equation}\label{eq:uvequations}
\begin{aligned}
 &\left[\hbar(\om-\bv\cdot\bk)-\left(\frac{\hbar^2 k^2}{2m}+g\rn\right)\right]\uk=g\rn\,\vk,\\
 &\left[\hbar(\om-\bv\cdot\bk)+\left(\frac{\hbar^2 k^2}{2m}+g\rn\right)\right]\vk=-g\rn\,\uk.
 \end{aligned}
\end{equation}
By imposing that this system admits nontrivial solutions, one obtains the dispersion relation
\begin{equation}\label{eq:dispersionhom}
 (\om-\bv\cdot\bk)^2=c^2k^2+\frac{c^4 k^4}{\Lambda^2}\equiv\Omega^2(\bk) ,
\end{equation}
where $c$ is the speed of sound [see Eq.~\eqref{eq:sound}], $\Omega$ is the frequency in the comoving frame and
\begin{equation}\label{eq:lambda}
 \Lambda\equiv\frac{2m c^2}{\hbar}
\end{equation}
is the ``healing frequency'', which gives the characteristic dispersive scale and is related to the healing length~\cite{Dalfovo}
\begin{equation}\label{eq:healing}
 \xi=\frac{\hbar}{\sqrt{2}mc}=\frac{\sqrt{2}c}{\Lambda}.
\end{equation}
When sending $\Lambda \to \infty$, Eq.~\eqref{eq:dispersionhom} becomes the relativistic equation $\Omega^2 = c^2 k^2$ since the quartic term drops out.

Finally, the values of $\uk$ and $\vk$ are fixed by Eq.~\eqref{eq:uvequations} and by the normalization condition~\eqref{eq:normuv}:
\begin{align}
 \uk&=\frac{1}{\sqrt{1-D_\bk^2}},\\
 \vk&=\frac{D_\bk}{\sqrt{1-D_\bk^2}},
\end{align}
where their phase has been arbitrarily chosen to make them real and
\begin{equation}\label{eq:dbk}
 D_\bk\equiv\frac{\vk}{\uk}=\frac{\hbar(\om-\bv\cdot\bk)-\left(\hbar^2 k^2/2m+g\rn\right)}{g\rn}=\frac{1}{mc^2}\left[\hbar\sqrt{c^2k^2+\frac{\hbar^2k^4}{4m^2}} - \frac{\hbar^2k^2}{2m} - mc^2\right].
\end{equation}

\subsection{1D inhomogeneous condensates}	%
\label{subsec:1D}				%

When the flow is not homogeneous, the spectrum may contain complex eigenfrequencies. This happens, for instance, when there is a compact region where the flow is superluminal, as shown in Chap.~\ref{chap:bhlaser}.
For the sake of simplicity, we study only 1-dimensional flows, since, in that case, the spectrum of~\eqref{eq:heffsym} is completely characterized by a continuous set of real frequency eigenmodes labelled by $\om$ and a discrete index $\alpha$, plus, {\it possibly}, a discrete and finite set of complex frequency modes that appear in pairs with complex conjugated frequencies. [If the scalar product were positive definite, as for fermions, the spectrum of~\eqref{eq:heffsym} would be purely real.]
We shall write complex frequencies as $\la =\om_a + i \Gamma_a $, where $a$ is a positive integer which labels the discrete set of pairs, and where $\om_a$ and $\Gamma_a $ are both real and positive.
The other cases are reached by complex conjugation and/or multiplication by $-1$.
The discrete index $\alpha$ describes the subset of modes with the same real frequency, but different wavenumbers.

The field $\hW$ can thus be expanded as
\begin{equation}\label{eq:fieldexpansion}
 \hW=\int\!\dom\sum_{\alpha}\!\left[\wom\aom+\womb\aomd\right]+\sum_a\!\left[\vl\bl+\zl\cl+\vlb\bld+\zlb\cld\right],
\end{equation}
where we have defined $\vl\equiv V_{\la}$, $\zl\equiv Z_{\la}$, $\vlb\equiv\bar V_{\la}$, and $\zlb\equiv\bar Z_{\la}$, in order to make the notation more compact.
We also define $\lambda_a$ to have a positive imaginary part $\Gamma_a$. The modes satisfy the following normalization rules,
\begin{align}
 &\scal{\wom}{\womp}=-\scal{\womb}{\wombp}=\delta(\omega-\omega')\delta_{\alpha\alpha'},\label{eq:normWom}\\
 &\scal{\vl}{\zlp}=\scal{\vlb}{\zlbp}=-\scal{\zlp}{\vl}=-\scal{\zlbp}{\vlb}=\ii\delta_{aa'}.\label{eq:normVZ}
\end{align}
Using
\begin{align}
 \bl&=\ii\scal{\zl}{\hW},\\
 \bld&=\ii\scal{\zlb}{\hW},\\
 \cl&=-\ii\scal{\vl}{\hW},\\
 \cld&=-\ii\scal{\vlb}{\hW}
\end{align}
and Eq.~\eqref{eq:commfor}, one obtains
\begin{equation}
 \comm{\bl}{\cldp}=\comm{\ii\scal{\zl}{\hW}}{-\ii\scal{\vlbp}{\hW}}=-\scal{\zl}{\vlp}=\ii\delta_{a a'},\label{eq:combcd}
\end{equation}
whereas the operators $\ha$ satisfy
\begin{equation}\label{eq:commaaapp}
 \comm{\aom}{\aomdp}=\delta_{\alpha\alpha'}\delta(\om-\omp).
\end{equation}
All the other scalar products and commutators vanish. In particular, $\comm{\bl}{\bldp}= \comm{\cl}{\cldp}= 0$.
Hence $\hb_a$ and $\hc_a$ are not destruction operators. In fact, the operators $\hb_a$ and $\hc_a$ form pairs~\cite{cp} that characterize complex, \ie, with two degrees of freedom, harmonic oscillators that are unstable, and for which therefore there is no notion of quanta.

Decomposing the doublets as
\begin{equation}\label{eq:doublets}
 \wom=\phm{\om t}\spin{\pom}{\vpom},\quad \vl=\phm{\la t}\spin{\xl}{\el},\quad \zl=\phm{\las t}\spin{\psl}{\zel},
\end{equation}
the field $\hp$ is expanded as
\begin{multline}\label{eq:phiexpansion}
\hp(t,x) =\int\!\dom\sum_{\alpha}\left[\phm{\om t}\pom\aom+\php{\om t}\vpoms\aomd\right]
\\ 
 +\sum_a\left[\phm{\la t}\xl\bl+\phm{\las t}\psl\cl+\php{\las t}\els\bld+\php{\la t}\zels\cld\right].
\end{multline}
Using the scalar product of Eq.~\eqref{eq:scalarphi}, Eq.~\eqref{eq:phiexpansion}, and Eqs.~\eqref{eq:normWom} and~\eqref{eq:normVZ} give
\begin{align}
 &\int\!\dx\,\rn\left[\poms\pomp-\vpoms\vpomp\right]=\delta_{\alpha\alpha'}\delta(\om-\omp),\label{eq:normphi}\\
 &\int\!\dx\,\rn\left[\xls\pslp-\els\zelp\right]=\ii\delta_{\lambda_a\lambda_{a'}},\label{eq:normpsi}\\
 &\int\!\dx\,\rn\left[\vert \xlsx \vert^2- \vert \elsx \vert^2\right]=0.
\end{align}

$\hHe$ is now easily computed by putting the field expansion~\eqref{eq:phiexpansion} in Eq.~\eqref{eq:ht}:
\begin{align}
 \hHe &=\, \frac{\hbar}{2}\int\!\dx\,\rnx
 \left[\int\!\dom\sum_{\alpha}\left(\php{\om t}\poms\aomd+\phm{\om t}\vpom\aom\right)
 \right.\nonumber\\ 
  &\qquad\left.
 +\sum_{a}\left(\php{\las t}\xls\bld+\phm{\la t}\el\bl
 +\php{\la t}\psls\cld+\phm{\las t}\zel\cl\right)\right]
 \nonumber\\
  &\qquad\times\left[\int\!\dom'\sum_{\alpha'}\omp\left(\phm{\omp t}\pomp\aomp-\php{\omp t}\vpomsp\aomdp\right)
 \right.\nonumber\\ 
  &\qquad\left.+\sum_{a'}\left(\lap\phm{\lap t}\xlp\blp-\lasp\php{\lasp t}\elsp\bldp
  +\lasp\phm{\lasp t}\pslp\clp-\lap\php{\lap t}\zelsp\cldp\right)\right]
 \nonumber\displaybreak[0]\\
 &+
 \frac{\hbar}{2}\int\!\dx\,\rnx
 \left[\int\!\dom\sum_{\alpha}\om\left(\php{\om t}\poms\aomd-\phm{\om t}\vpom\aom\right)
 \right.\nonumber\\ 
 &\qquad\left.+\sum_a\left(\las\php{\las t}\xls\bld
 -\la\phm{\la t}\el\bl+\la\php{\la t}\psls\cld-\las\phm{\las t}\zel\cl\right)\right]
 \nonumber\\
 &\qquad\times\left[\int\!\dom'\sum_{\alpha'}\left(\phm{\omp t}\pomp\aomp+\php{\omp t}\vpomsp\aomdp\right)
 \right.\nonumber\\ 
 &\qquad\left.+\sum_{a'}\left(\phm{\lap t}\xlp\blp+\php{\lasp t}\elsp\bldp
  +\phm{\lasp t}\pslp\clp+\php{\lap t}\zelsp\cldp\right)\right],
\end{align}
which is a sum of three terms, containing respectively only real frequency modes, only complex frequency modes and mixing real and complex frequencies:
\begin{equation}
\hHe =\hHr+\hHc+\hHm.
\end{equation}
Using the commutation relations and the scalar products of Sec.~\ref{subsec:quantization}, we obtain
\begin{align}
 \hHr &= \frac{\hbar}{2} \int\!\dom\dom'\sum_{\alpha\alpha'}\int\!\dx\,\rnx
 \left[
  (\omp+\om)\php{\om t}\poms\phm{\omp t}\pomp\aomd\aomp
  \right. \nonumber \\  &\qquad\left.
 +(\om-\omp)\php{\om t}\poms\php{\omp t}\vpomsp\aomd\aomdp
  +(\om'-\om)\phm{\om t}\vpom\phm{\omp t}\pomp\aom\aomp
  \right. \nonumber \\ & \qquad\left.
 -(\omp+\om)\phm{\om t}\vpom\php{\omp t}\vpomsp\aom\aomdp
 \right]
 \nonumber \displaybreak[0] \\
 &=\frac{\hbar}{2}\int\!\dom\dom'\sum_{\alpha\alpha'} (\om+\omp) \scal{\wom}{\womp} \aomd\aomp
 \nonumber \\
 &\qquad-\frac{\hbar}{2}\int\!\dom\dom'\sum_{\alpha\alpha'} (\om+\omp) \delta(\om-\omp)\delta_{\alpha\alpha'}
 \int\!\dx\,\rnx \phm{(\omp-\om)t}(\vpoms\vpomp)
  \nonumber \\
 &\qquad+\frac{\hbar}{2}\int\!\dom\dom'\sum_{\alpha\alpha'} \om \scal{\wom}{\wombp} \aomd\aomdp
 +\frac{\hbar}{2}\int\!\dom\dom'\sum_{\alpha\alpha'} \om \scal{\womb}{\womp} \aom\aomp
 \nonumber \\
 &=\int\!\dom\sum_\alpha\hbar\om\left[\aomd\aom-\int\!\dx\,\rnx|\vpom|^2  \right].
\end{align}
Using the same techniques, it is easy to show that $\hHm=0$ and
\begin{equation}
 \hHc=\ii\hbar\sum_a \left\{\las\!\left[\bld\cl\!+\!\!\int\!\dx\rnx\zel\els\right]- \mbox{h.c.} \right\}\!.
\end{equation}
Putting everything together, one obtains
\begin{multline}\label{eq:hamiltonianabc}
 \hHe = \int\!\dom\sum_\alpha\hbar\om\left(\aomd\aom-\int\!\dx\,\rnx|\vpom|^2  \right)
 \\
 \quad  \quad \quad +\sum_a  \left[\ii\hbar\las\left(\bld\cl+\int\!\dx\,\rnx\zel\els\right)+\mbox{h.c.} \right].
\end{multline}
The first line contains again the usual sum over harmonic oscillators of real frequency and the $c$-number term accounting for the depletion, while the second line is due to the complex frequency modes.
It should be noticed that the unusual form of the second $c$-number term follows from the unusual scalar product of Eq.~\eqref{eq:normpsi}.
When complex eigenfrequencies $\la$ are present, this Hamiltonian is not bounded from below due to the $b_a^\dagger c_a$ terms.
Hence the vacuum can no longer be defined as the ground state of $\hH$,
as one might have expected since one is dealing with unstable oscillators.
This gives rise to some ambiguity when choosing the initial ``vacuum'' state (see Sec.~\ref{sec:correlations}).

Eqs.~\eqref{eq:phiexpansion} and~\eqref{eq:hamiltonianabc} can be rewritten in a more familiar form by decomposing each couple ($\hb_a, \hc_a$) into two couples of destruction/creation operators $\dlpl$, $\dlpld$ and $\dlmi$, $\dlmid$,
\begin{equation}\label{eq:defd}
 \dlpl\equiv\ppl\frac{\bl+\ii\cl}{\sqrt{2}},\qquad \dlmi\equiv\pmi\frac{\bld+\ii\cld}{\sqrt{2}},
\end{equation}
which in fact satisfy the commutation relations
\begin{equation}
 \comm{\dlpl}{\dlpldp}=\comm{\dlmi}{\dlmidp}=\delta_{aa'},
\end{equation}
and all the other commutators vanish. Eq.~\eqref{eq:fieldexpansion} becomes
\begin{equation}
  \hW=\int\!\dom\sum_{\alpha}\left[\wom\aom+\womb\aomd\right]+\sum_a\left[\wlpl\dlpl+\wlmi\dlmi+\wlplb\dlpld+\wlmib\dlmid\right],
\end{equation}
where
\begin{equation}
 \wlpl\equiv\pplm\frac{\vl-\ii\zl}{\sqrt{2}},\qquad\wlmi\equiv\pmim\frac{\vlb-\ii\zlb}{\sqrt{2}},
\end{equation}
whose normalization is
\begin{equation}
\scal{\wlpl}{\wlplp}=\!\scal{\wlmi}{\wlmip}=\!-\scal{\wlplb}{\wlplbp}=\!-\scal{\wlmib}{\wlmibp}=\!\delta_{aa'},
\end{equation}
and the other scalar products vanish. Notice that $W_{a\pm}$ are no longer frequency eigenmodes.
Decomposing $\wlpl$ and $\wlmi$ as
\begin{equation}
 \wlpl=\spin{\plpl}{\vplpl},\quad \wlmi=\spin{\plmi}{\vplmi},
\end{equation}
the field expansion~\eqref{eq:phiexpansion} becomes
\begin{multline}
  \hp(t,x)=\int\!\dom\sum_{\alpha}\left[\phm{\om t}\pom\aom+\php{\om t}\vpoms\aomd\right]
 \\ 
 +\sum_a\left[\plpl\,  \dlpl+\plmi\, \dlmi+\vplpls\, \dlpld+\vplmis\, \dlmid\right]
\label{eq:phiadd}
\end{multline}
and the Hamiltonian~\eqref{eq:hamiltonianabc} is
\begin{align}
 \hHe &= \int\!\dom\sum_\alpha\hbar\om\left(\aomd\aom-\int\!\dx\,\rnx|\vpom|^2  \right)
  \nonumber\\
 &\qquad+\sum_a\hbar\om_a\left[\dlpld\dlpl-\dlmid\dlmi-\int\!\dx\,\rnx\left(|\vplpl|^2-|\vplmi|^2\right)\right]
 \nonumber\\
 &\qquad+
\sum_a\ii\hbar\Gamma_a\left[\dlpld\dlmid
  +\int\!\dx\,\rnx\left(\plpl\vplmi \right) - \mbox{h.c.} \right].
\label{eq:hamiltonianadd}
\end{align}
when the (arbitrary) phases $\theta_\pm$ are chosen to be $\theta_\pm=0$.

We conclude with two remarks.
Firstly, if one can describe the discrete set using the $\hat d_a$ operators, there is a price to pay as the associated modes $\plplsx$, $\vplplsx$, $\plmisx$, $\vplmisx$ are not frequency eigenmodes. Hence their time dependence cannot be factorized.
Secondly, in usual circumstances, the discrete set of complex frequency modes is empty, as in homogeneous and in near-homogeneous condensates. Nevertheless, when the condensate crosses twice the speed of sound, the discrete set is (generally) nonempty.
To explain why, we first need to study the spectrum in homogeneous subsonic and supersonic flows (see below), and then understand how to paste the corresponding modes when the flow crosses the speed of sound (see Sec.~\ref{sec:laser}).
When solving this, the dispersive properties of $\hp$ become essential, as we now recall, and as first understood in~\cite{cj}.

\subsubsection{Mode analysis}	%
\label{subsec:mode1d}		%

Inserting the field expansion~\eqref{eq:phiexpansion} in Eq.~\eqref{eq:BdG}, one obtains a system of two $c$-number equations
\begin{equation}\label{eq:eqphi}
\begin{aligned}
 \left[\hbar(\lambda +\ii v\pdx) - T_\rho - mc^2\right]\phi_\lambda &= mc^2\varphi_\lambda,\\
 \left[-\hbar(\lambda +\ii v\pdx) - T_\rho - mc^2\right]\varphi_\lambda &= mc^2\phi_\lambda.
\end{aligned}
\end{equation}
The couple $(\phi_\lambda,\varphi_\lambda)$ can be formed by either the real frequency modes $(\pom,\vpom)$, or the complex frequency modes $(\xl,\el)$ or $(\psl,\zel)$ and, accordingly, $\lambda$ can be real or complex. Eliminating $\varphi_\lambda$ from the above system, one obtains~\cite{MacherBEC}
\begin{equation}\label{eq:modes}
 \left\{\left[\hbar(\lambda +\ii v\pdx) + T_\rho\right] \frac{1}{c^2}\left[-\hbar(\lambda +\ii v\pdx) + T_\rho \right]-\hbar^2 v \pdx\frac{1}{v}\pdx\right\}\phi_\lambda=0,
\end{equation}
where we use the following expression
\begin{equation}
 T_\rho=-\frac{\hbar^2 }{2m} \, v\pdx \frac{1}{v}\pdx,
\end{equation}
that can be derived from Eq.~\eqref{eq:Trho} in the 1-dimensional case by using the stationary continuity equation~\eqref{eq:cont}.

When the background quantities are also independent of $x$, that is the condensate is homogeneous (see Sec.~\ref{subsec:mode3d}), Fourier modes $\phi_\lambda\propto\exp(\ii k_\lambda x)$ are solutions of Eq.~\eqref{eq:modes}, provided $k_\lambda$ is a (possibly complex) root of the dispersion relation
\begin{equation}\label{eq:dispersion}
 (\lambda-vk)^2=\Omega^2(k)=c^2k^2+\frac{c^4 k^4}{\Lambda^2},
\end{equation}
where, again, $\Omega$ is the frequency in the comoving frame [see eq.~\eqref{eq:dispersionhom}] and the dispersive scale $\Lambda$ is defined in Eq.~\eqref{eq:lambda}.
We notice that, when keeping this term, the dispersion relation~\eqref{eq:dispersion} possesses four roots, some of which can be complex. As we saw in Chap.~\ref{chap:bhlaser}, the two extra roots are at the origin of a laser-like effect.

In subsonic flows, $|v|< c$, for real $\lambda=\omega$, two roots are real, and describe the right and left moving solutions.
The other two are complex, conjugated to each other, and correspond to modes that are asymptotically growing or decaying, say to left.
Only the first two should be used in Eq.~\eqref{eq:phiexpansion}, as the last two are not asymptotically bounded.

In supersonic flows, the situation is quite different.
For real $\lambda=\omega$, there exists a critical frequency $\om_{\max}$ such that the four roots of Eq.~\eqref{eq:dispersion} are~\cite{MacherBEC}
\begin{itemize}
\item $\om<\ommax$: all real;
 \item $\om>\ommax$: two real and two complex, as in subsonic flows.
\end{itemize}
See Fig.~\ref{fig:dispersion} for a graphical solution of the dispersion relation~\eqref{eq:dispersion}.
At $\om = \omm$ the two extra roots merge and the curve $\Omega(k)$ is tangent to the line $\om - vk$.
Hence the group velocity in the lab
\begin{equation}\label{eq:gr_vel}
v_{\rm gr} = 
\frac{\partial\omega}{\partial k} =
\frac{\partial}{\partial k} (\Omega + vk), 
\end{equation}
vanishes at that frequency.
Solving this equation one finds that $\omm$ is
proportional to $\Lambda$, but depends also on the supersonic velocity excess $|v| - c$.

When $\om$ is real, the normalization of the (bounded) modes can be easily worked out (see also Sec.~\ref{subsec:mode3d} for the 3-dimensional case), as it is, up to a trivial factor, independent of the constant velocity $v$ (because of Galilean invariance). Let us write
\begin{align}
 \pom\ee^{-\ii\om t}&=\frac{\ee^{-\ii\om t+\ii k_\om^\alpha x}}{\sqrt{2\pi\rn}}u_\om^\alpha,\\
 \vpom\ee^{-\ii\om t}&=\frac{\ee^{-\ii\om t+\ii k_\om^\alpha x}}{\sqrt{2\pi\rn}}v_\om^\alpha,
\end{align}
where $\alpha$ spans over four values if $|v|>c$ and $\om<\ommax$ and over two values elsewhere.
Using the mode normalization~\eqref{eq:normphi}, we obtain
\begin{equation}\label{eq:uvcond}
 (|u_\om^\alpha|^2-|v_\om^\alpha|^2)\frac{\partial\om}{\partial k}=1.
\end{equation}
This equation slightly differs from the normalization condition~\eqref{eq:normuv} found in the 3-dimensional case. The discrepancy is due to the choice in Eq.~\eqref{eq:normWom} of normalizing the modes with $\delta(\om-\omp)$, instead of the more standard $\delta(k-k')$ [see Eq.~\eqref{eq:normWk}]. This choice generates the appearance of the factor $\partial\om/\partial k$ in Eq.~\eqref{eq:uvcond}, that is the wave group velocity [see Eq.~\eqref{eq:gr_vel}].

From the mode equation~\eqref{eq:eqphi} and the dispersion relation~\eqref{eq:dispersion}, when the comoving frequency $\Omega=\om-vk$ is positive, one has
\begin{equation}\label{eq:dku}
 D_{k_\om^\alpha} u_\om^\alpha = v_\om^\alpha,
\end{equation}
where
\begin{equation}\label{eq:dk_1D}
 D_k=\frac{1}{mc^2}\left[\hbar\sqrt{c^2k^2+\frac{\hbar^2k^4}{4m^2}} - \frac{\hbar^2k^2}{2m} - mc^2\right]
\end{equation}
is independent of $v$ and coincedes with $D_\bk$ defined in Eq.~\eqref{eq:dbk} for the 3-dimensional flow.
Eqs.~\eqref{eq:uvcond} and~\eqref{eq:dku} fix again $u_\om^\alpha$ and $v_\om^\alpha$ except for a common phase.
Taking both $u_\om^\alpha$ and $v_\om^\alpha$ real, one obtains
\begin{equation}\label{eq:normmode}
\begin{aligned}
 \pom&=\sqrt{\frac{\partial k_\om^\alpha}{\partial\om}}\frac{1}{\sqrt{1-D_{k_\om^\alpha}^2}}\frac{\ee^{\ii k_\om^\alpha x}}{\sqrt{2\pi\rn}},\\
 \vpom&=\sqrt{\frac{\partial k_\om^\alpha}{\partial\om}}\frac{D_{k_\om^\alpha}}{\sqrt{1-D_{k_\om^\alpha}^2}}\frac{\ee^{\ii k_\om^\alpha x}}{\sqrt{2\pi\rn}}.
\end{aligned}
\end{equation}

Summarizing, in subsonic flows there are only two real roots and the field $\hp$ is accordingly expanded as
\begin{equation}\label{eq:expsub}
 \hp_{\rm sub}=\int_0^{\infty}\!\dom\left[ \ee^{-\ii\om t}  \left(\phi_\om^u \ha_\om^u+\phi_\om^v\ha_\om^v\right)
+ \ee^{+ \ii\om t} \left((\varphi_\om^u)^*\ha_\om^{u\, \dagger}+(\varphi_\om^v)^*\ha_\om^{v\, \dagger}\right)
 \right]
\end{equation}
or, in terms of the doublets $\hW$ defined in Sec.~\ref{subsec:quantization},
\begin{equation}
\hW_{\rm sub}=\int_{0}^{\infty}\!\dom\left[W_\om^u\ha_\om^u+W_\om^v\ha_\om^v+\bar W_\om^u\ha_\om^{u\dagger}+\bar W_\om^v\ha_\om^{v\dagger}\right].
\end{equation}

In supersonic flows, instead, when $0< \om < \ommax$, there are two extra real roots of Eq.~\eqref{eq:dispersion}.
Unfortunately, the doublets $W^{(i)}_\om$ associated with those two extra roots have negative norm, as one can directly check by using the expression for $\phi$ and $\varphi$, given in the above equations~\eqref{eq:normmode}. As a consequence, the associated destruction and creation operators $\hat a^{(i)}_\om$ and $\hat a^{(i)\dagger}_\om$ do not satisfy the standard commutation relations of Eq.~\eqref{eq:commaaapp}, so that they cannot enter the field expansion of Eq.~\eqref{eq:fieldexpansion}.
As discussed in Sec.~\ref{subsec:bogo}, the two solutions $k_\om^{(1)}$ and $k_\om^{(2)}$ of the dispersion relation have negative comoving frequency $\Omega(k_\om^{(i)})$ (in Fig.~\ref{fig:dispersion}, right panel, they are in fact located in the third quadrant).
Because of that, the norm obtained by putting~\eqref{eq:normmode} in the expression of the scalar product~\eqref{eq:scalar} is negative.
To obtain positive norm modes $W$, corresponding to the two new solutions $k_\om^{(i)}$ of the dispersion relation, it is enough to pick up solutions with positive $\Omega$. 
This aim is achieved by solving the mode equation for negative values $-\om$ of the conserved frequency (we define $\om$ to be always positive), such that the two solutions $k_{-\om}^{(i)}=-k_{\om}^{(i)}$ are now located in the first quadrant.
The corresponding modes $W^{(i)}_{-\om}$ have now positive norm and the destruction and creation operators $\hat a^{(i)}_{-\om}$ and $\hat a^{(i)\,\dagger}_{-\om}$ satisfy the proper commutation relations~\eqref{eq:commaa}.

Accordingly, the field $\hW$ is expanded in supersonic flows as
\begin{equation}
 \hW_{\rm sup}=\int_{0}^{\infty}\!\dom\left[W_\om^u\ha_\om^u+W_\om^v\ha_\om^v+\bar W_\om^u\ha_\om^{u\dagger}+\bar W_\om^v\ha_\om^{v\dagger}\right]
 +\int_{0}^{\ommax}\!\dom\sum_{i=1,2}\left[W_{-\om}^{(i)}\ha_{-\om}^{(i)}+
\bar W_{-\om}^{(i)}\ha_{-\om}^{{(i)}\dagger}\right].
\end{equation}
Because the modes $W^{(i)}_{-\om}$ are negative frequency eigenmodes, while $\bar W^{(i)}_{-\om}$ are positive frequency eigenmodes, this yields the following expansion for the phonon field $\hp$:
\begin{multline}\label{eq:expsup}
 \hp_{\rm sup}=\int_0^{\infty}\!\dom\left\{ \ee^{-\ii\om t} \left[ \phi_\om^u \ha_\om^u+\phi_\om^v\ha_\om^v
+ \theta(\ommax - \om)
\sum_{i=1,2}(\varphi_{-\om}^{(i)})^*\,\ha_{-\om}^{{(i)}\dagger}\right] \right.
 \\
\left. + \ee^{+ \ii\om t} \left[ (\varphi_\om^u)^*\,\ha_\om^{u\, \dagger}+(\varphi_\om^v)^*\,\ha_\om^{v\, \dagger}
+ \theta(\ommax - \om)
\sum_{i=1,2}\phi_{-\om}^{(i)}\ha_{-\om}^{{(i)}}\right] \right\}, 
\end{multline}
where we inserted the Heaviside function $\theta(\omm-\om)$, because the two extra roots exist only for $\om<\omm$.

To conclude, we go back to the inhomogeneous case. When $v$ varies slowly with respect to the wavelength of the perturbations, the exact solutions of Eq.~\eqref{eq:eqphi} are well approximated by their WKB approximation, which can be directly inferred from Eq.~\eqref{eq:normmode}
\begin{align}
 \pom&=\sqrt{\frac{\partial k_\om^\alpha(x)}{\partial\om}}\frac{1}{\sqrt{1-D_{k_\om^\alpha}^2(x)}}\frac{\exp\left[\ii\int^x \dx' k_\om^\alpha(x')\right]}{\sqrt{2\pi\rn(x)}},\label{eq:phiWKB}\\
 \vpom&=\sqrt{\frac{\partial k_\om^\alpha(x)}{\partial\om}}\frac{D_{k_\om^\alpha}(x)}{\sqrt{1-D_{k_\om^\alpha}^2(x)}}\frac{\exp\left[\ii\int^x \dx' k_\om^\alpha(x')\right]}{\sqrt{2\pi\rn(x)}},\label{eq:vphiWKB}
\end{align}
and the $x$-dependent wave numbers $k_\om^\alpha(x)$ are real roots of the dispersion relation~\eqref{eq:dispersion} in a stationary inhomogeneous flow characterized by $v(x)$ and $c(x)$:
\begin{equation}\label{eq:dispersionWKB}
 (\om-v(x)k(x))^2=c^2(x) k^2(x)+\frac{\hbar^2 k^4(x)}{4 m^2}. 
\end{equation}
Ignoring the quartic term, one obtains the dispersion relation of a massless field propagating in a curved geometry and, similarly, Eq.~\eqref{eq:modes} becomes the Euler equation governing sound waves.
We show in the next section how the metric describing such a geometry can be directly obtained from the BdG equation~\eqref{eq:BdG}.

\section{The acoustic metric}	%
\label{sec:metric}		%

Under certain assumptions, the propagation of phonons on top of a BEC can be described with the formalism of quantum field theory on a curved background, that is the BdG equation~\eqref{eq:BdG} can be written in the form of a modified Klein--Gordon equation
\begin{equation}\label{eq:KGhp}
 \frac{1}{\sqrt{|g|}}\partial_\mu \left(\sqrt{|g|}g^{\mu\nu}(t,\bx)\partial_\nu\hp\right)=0,
\end{equation}
or, equivalently [see Eqs.~\eqref{eq:KGfluid}--\eqref{eq:fieldeq} in Sec.~\ref{sec:generalmetric}]
\begin{equation}\label{eq:metricformhp}
 \partial_\mu \left(f^{\mu\nu}(t,\bx)\partial_\nu\hp\right)=0.
\end{equation}

Equation~\eqref{eq:BdG} and its Hermitian conjugate can be written as
\begin{align}
 \left[\ii\hbar\left(\pdt+\bv\!\cdot\!\nx\right) -  T_\rho - m c^2 \right]\hp &= m c^2 \hpd,\\
 \left[\ii\hbar\left(\pdt+\bv\!\cdot\!\nx\right) +  T_\rho + m c^2 \right]\hpd &= -m c^2 \hp,\\
\end{align}
and, eliminating $\hpd$, one obtains a single equation
\begin{equation}\label{eq:eqphi_oper}
 \left[\ii\hbar\left(\pdt+\bv\!\cdot\!\nx\right) +  T_\rho + m c^2 \right]\frac{1}{c^2}\left[\ii\hbar\left(\pdt+\bv\!\cdot\!\nx\right) -  T_\rho - m c^2 \right]\hp = -m^2 c^2 \hp,
\end{equation}
which, using the definition of $T_\rho$ of Eq.~\eqref{eq:Trho} becomes
\begin{equation}\label{eq:eqhp}
 \left[\ii\hbar\left(\pdt+\bv\!\cdot\!\nx\right) +  T_\rho\right]\frac{1}{c^2}\left[\ii\hbar\left(\pdt+\bv\!\cdot\!\nx\right) -  T_\rho \right]\hp  +\frac{\hbar^2}{\rn}\nx\!\cdot\!\rn\nx \hp=0.
\end{equation}
It is worth stressing that Eq.~\eqref{eq:eqhp} is implied by Eq.~\eqref{eq:BdG} but the converse is not true.\footnote{This is similar to what happens in standard quantum field theory where the Dirac equation implies the Klein--Gordon equation but not the other way around.}
Nevertheless Eq.~\eqref{eq:eqhp} is enough for our aim, since we are interested only in the dispersion relation and in the metric describing the field propagation, which can be obtained from the modified Klein--Gordon equation in a curved background.

To obtain a second order equation from Eq.~\eqref{eq:eqhp}, one must neglect all the terms with higher order derivatives. This is possible only under two conditions.
First, a WKB approximation must be performed, \ie\/ all the background quantities must vary slowly on scales comparable with the characteristic wavelength of the perturbation. A local frequency $\om(\bx)$ and a local wavenumber $\bk(\bx)$ can then be defined. In term of $\bk(\bx)$ this condition reads
\begin{equation}
 \left|\frac{\partial_{x_i} \rho}{\rho}\right|\ll |k_i|,\quad \left|\frac{\partial_{x_i} v_j}{v_j}\right|\ll |k_i|. 
\end{equation}
Then, a local dispersion relation is directly inferred from Eq.~\eqref{eq:dispersionhom}, generalizing the 1-dimensional example of Sec.~\ref{subsec:mode1d}:
\begin{equation}\label{eq:dispersionWKBgen}
 \Omega^2(\bx)=\left[\om-\bv(\bx)\cdot\bk(\bx)\right]^2=c^2(\bx)k^2(\bx)+\frac{\hbar^2 k^4(\bx)}{4m^2}.
\end{equation}
Note that $\om$ is constant because the condensate is stationary.

Second, the quartic order in Eq.~\eqref{eq:dispersionWKBgen} must be negligible with respect to the second order one, namely
\begin{equation}
 k^2(\bx)\ll\frac{4m^2 c^2(x)}{\hbar^2}=\frac{\Lambda^2(\bx)}{c^2(\bx)}=\frac{2}{\xi^2(\bx)},
\end{equation}
\ie, the wavelength of the perturbation must be much larger than the healing length of the condensate [see Eq.~\ref{eq:healing}].

When these two conditions hold the quantum potential $T_\rho$ can be neglected in Eq.~\eqref{eq:eqhp}, which thus becomes
\begin{equation}
 \rn\left(\pdt+\bv\!\cdot\!\nx\right)\frac{1}{c^2}\left(\pdt+\bv\!\cdot\!\nx\right)\hp-\nx\!\cdot\!\rn\nx \hp=0.
\end{equation}
Using the continuity equation
\begin{equation}
 \pdt\rn+\nx\!\cdot\!(\rn\bv)=0,
\end{equation}
we obtain
\begin{equation}\label{eq:eqhp2}
 \left(\pdt+\nx\!\cdot\!\bv\right)\frac{\rn}{c^2}\left(\pdt+\bv\!\cdot\!\nx\right)\hp-\nx\!\cdot\!\rn\nx \hp=0.
\end{equation}
Equation~\eqref{eq:eqhp2} is now exactly in the soughtafter form of~\eqref{eq:metricformhp} with
\begin{equation}\label{eq:fmunu}
 f^{\mu\nu}=\frac{\rn}{c^2}
 \begin{pmatrix}
 -1	&	-\bv^T	\\
 -\bv	&	c^2{\mathbbm 1}_{d\times d}-\bv\otimes\bv^T
 \end{pmatrix},
\end{equation}
where ${\mathbbm 1}_{d\times d}$ is the identity matrix in $d$ dimensions. From Eq.~\eqref{eq:deff}
\begin{equation}
 \sqrt{|g|}=\left[\mbox{det}(f^{\mu\nu})\right]^{1/(d-1)}=\left(\frac{\rn^{d+1}}{c^2}\right)^{1/(d-1)}
\end{equation}
and
\begin{equation}
 g_{\mu\nu}=(g^{\mu\nu})^{-1}=\sqrt{|g|}(f^{\mu\nu})^{-1}.
\end{equation}
From eq.~\eqref{eq:fmunu}
\begin{equation}
 (f^{\mu\nu})^{-1}=\frac{1}{\rn}
  \begin{pmatrix}
 -(c^2-v^2)	&	-\bv^T	\\
 -\bv	&	{\mathbbm 1}_{d\times d},
 \end{pmatrix},
\end{equation}
from which the effective acoustic metric is
\begin{equation}\label{eq:metricBEC}
 g_{\mu\nu}=\left(\frac{\rn}{c}\right)^{2/(d-1)}
  \begin{pmatrix}
 -(c^2-v^2)	&	-\bv^T	\\
 -\bv	&	{\mathbbm 1}_{d\times d}
 \end{pmatrix}.
\end{equation}
%

\chapter{Phonon spectral analysis: the code}		
\label{app:code}					
\chaptermark{The code}					

The code used for the numerical computation of this Thesis is based on the original code by Jean Macher~\copyright~2008~\cite{MacherBEC}, distributed under the {\it GNU General Public License (version 2)}. The code has been broadly adapted and extended in various versions for the computations of~\cite{bhlasers,smearhor,robustness,cfp,4x4}. We present the basic structure of the code in Sec.~\ref{appsec:structure} and the main features of the three different versions used to study
\begin{itemize}
 \item modifications of Hawking spectrum for asymmetrical velocity profiles~\cite{robustness} and profiles with small scale perturbations~\cite{smearhor,cfp} (see Chap.~\ref{chap:hawking}) in Sec.~\ref{appsec:3x3},
 \item the spectrum of emitted particles from a warp-drive bubble in a Bose--Einstein condensate (BEC)~\cite{4x4} (see Chap.~\ref{chap:warpdriveBEC}) in Sec.~\ref{appsec:4x4},
 \item the set of complex frequency modes at the basis of the black-hole-laser effect in BECs~\cite{bhlasers,cfp}) in Sec.~\ref{appsec:complex}.
\end{itemize}

\section{The general structure}	%
\label{appsec:structure}	%

\begin{figure}
 \centering
 \includegraphics[width=0.85\textwidth]{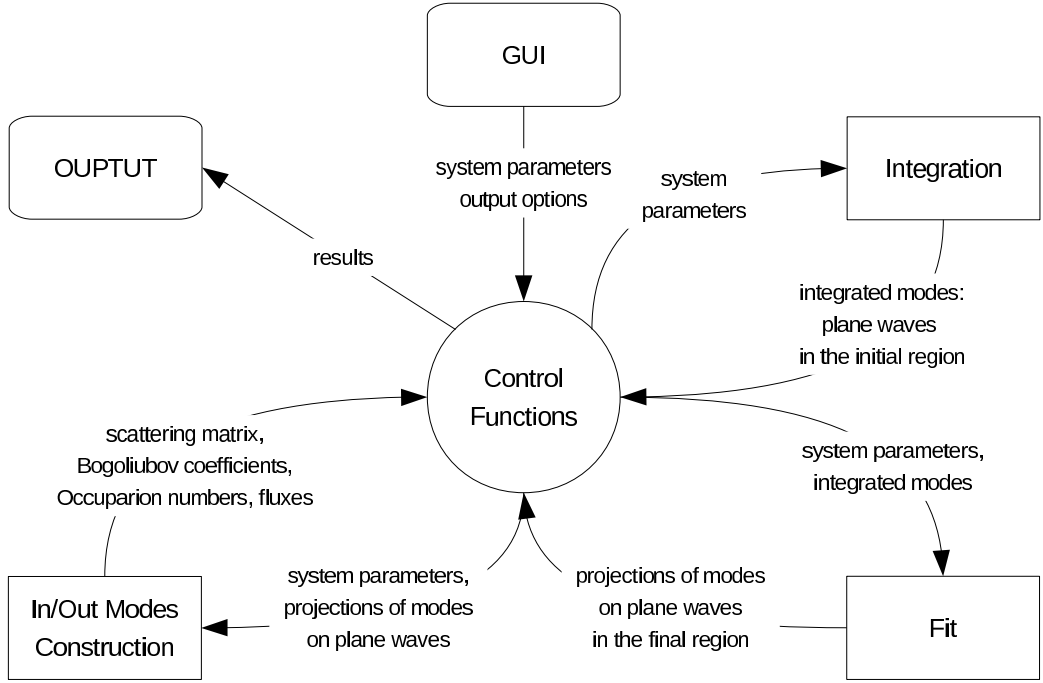}
 \caption{Graphical representation (read clockwise) of the structure of the code used for the computation of this Thesis and of~\cite{bhlasers,smearhor,robustness,cfp,4x4}.}
 \label{fig:code}
\end{figure}

The core of the code is made of six main parts, that are briefly listed below, together with their respective functions. See Fig.~\ref{fig:code} for a graphical representation
\begin{itemize}
 \item {\it Graphical user interface (GUI):} sets up the values of the initial parameters (physical parameters, precisions in numerical integration and fit, \dots) and the output options (saving or not all the modes, computing the scattering matrix and the Bogoliubov coefficients, \dots).
 \item {\it Control functions:} increment the values of the variables and call the functions that execute specific tasks (integration, fit, output saving, \dots).
 \item {\it Integration functions:} determine the initial conditions of modes that are plane waves in one of the asymptotic regions and integrate them from this region to the region/s on the other side of the horizon/s.
 \item {\it Fit functions:} extract the projections of the initial modes on plane waves defined in the final region/s.
 \item {\it Functions for in and out modes construction:} compute the scattering in and out modes and the Bogoliubov coefficients (when required) from the coefficients obtained from the fit procedure.
 \item {\it Output functions:} save the integrated modes, the fitted parameters, the Bogoliubov coefficients, and the values of the complex frequency poles (only for the black hole laser version, Sec~\ref{appsec:complex}).
\end{itemize}

\section{Three versions of the code}	%
\label{appsec:versions}			%

\subsection{Single horizon profile: 3\texorpdfstring{$\times$}{x}3 scattering}	%
\label{appsec:3x3}								%

We first consider profiles with a single horizon. In this case the computation of observables as the final occupations numbers on an intially vacuum quantum state [see Eqs.~\eqref{eq:occvacuum}, \eqref{eq:occvacuum1} and~\eqref{eq:occvacuum2}] is quite simple.

For each value of the system parameters and for each frequency, the code integrates the mode equation~\eqref{eq:eqphi} starting from the subsonic region $x>0$ to the supersonic region for three modes:
\begin{itemize}
 \item the two modes $u$ and $v$ that reduce to planewaves at $x\to+\infty$ corresponding to the two real roots of the dispersion relation~\eqref{eq:dispersion1} (see Fig.~\ref{fig:dispersion}, left panel)
 \item the mode with imaginary $k$ which decays at $x\to+\infty$.
\end{itemize}

The three modes are integrated well inside the supersonic region, where the velocity profile of Eq.~\eqref{eq:velocity_simp} is flat. In this region, the modes are decomposed in four plane waves corresponding to the four real-$k$ solutions of Eq.~\eqref{eq:dispersion1} (see Fig.~\ref{fig:dispersion}, right panel).

Using this decomposition, one can combine the three modes to obtain pure in or out modes, namely modes whose asymptotic expansion (both in the supersonic and supersonic region) contains only a single ingoing or outgoing branch (see Sec.~\ref{subsec:bogo}), respectively.

Ingoing modes are finally projected on outgoing modes as in Eq.~\eqref{eq:bog_transf} and the corresponding transformation matrix is read off. The computation of the occupation numbers~\eqref{eq:occvacuum}, \eqref{eq:occvacuum1}, and~\eqref{eq:occvacuum2}, and of the observables described in Sec.~\ref{subsec:observables} is straightforward.

\subsection{Two horizon profile: 4\texorpdfstring{$\times$}{x}4 scattering}	%
\label{appsec:4x4}								%

The construction of a $4\times4$ scattering matrix for a flow profile with a supersonic region between two horizons and two supersonic asymptotic regions exactly follows the procedure described in the previous section for $3\times3$ scattering.

The main difference is that all the four real-$k$ solutions of the dispersion relation~\eqref{eq:dispersion1} are integrated from one of the two supersonic regions, then propagated through the subsonic region till the other supersonic region. Here, they are projected on the four planes waves corresponding to the four real-$k$ solutions in the other supersonic region.

In and out modes are constructed, as above, out of these four modes by selecting modes with a single ingoing or out outgoing branch, respectively. The $4\times4$ matrix is obtained, again, by projecting in modes on out modes.

\subsection{Two horizon profile: complex eigenfrequencies}	%
\label{appsec:complex}						%

\begin{figure}
 \centering
 \includegraphics[width=0.5\textwidth]{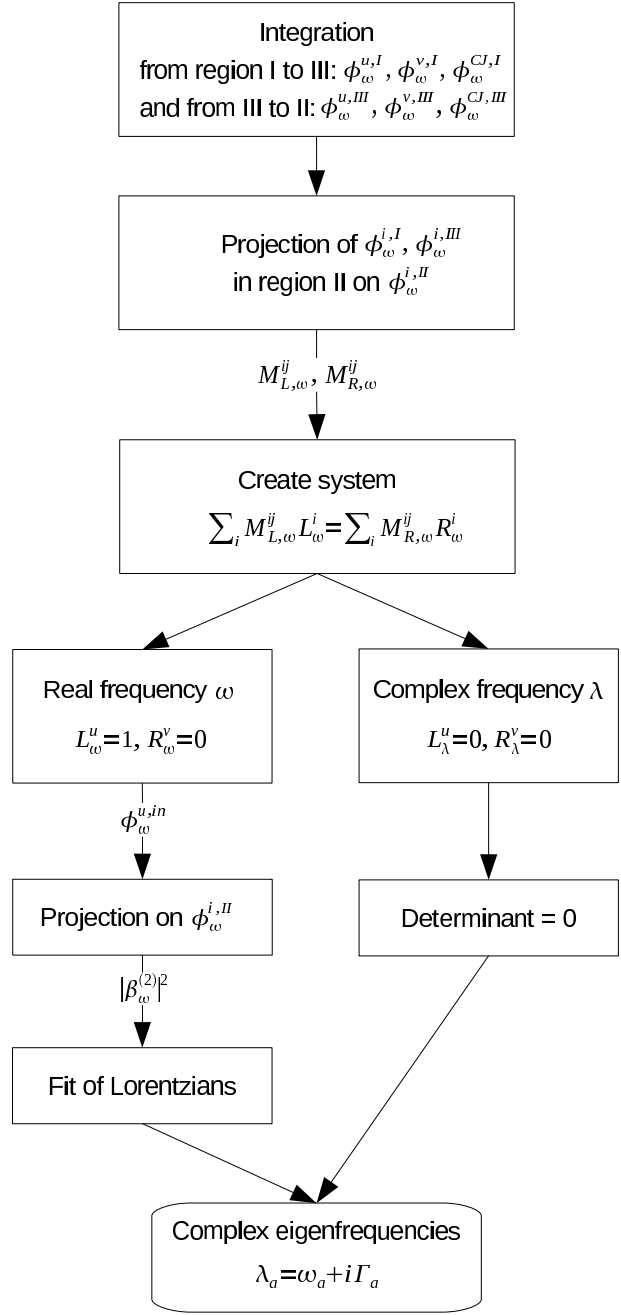}
 \caption{Graphical representation of the strategy for the computation of the set of complex eigenfrequencies of a black hole laser.}
 \label{fig:bhlasercode}
\end{figure}

The computations of the complex eigenfrequencies is a bit more involved (see Fig.~\ref{fig:bhlasercode} for a graphical representation). Since the velocity profile is made of two subsonic asymptotic regions plus a supersonic compact region between the horizons, it is not possible to integrate the modes starting from one of the two subsonic asymptotic regions to the other one passing through the internal supersonic region, as done for the supersonic-subsonic-supersonic profiles of Sec.~\ref{appsec:4x4}. In fact, in the subsonic region there are imaginary solutions of $k$ even for real values of $\omega$. If we integrated some mode from the internal supersonic region to one of the two external subsonic regions, this mode would generally appear as a superposition of all the four solutions of the dispersion relations, included the two solutions with imaginary $k$. Thus, it would be impossible to project such a mode on all the four modes in the second subsonic region, because one of the modes with imaginary-$k$ 
diverges exponentially in space, then masking the other three modes.

For this reason, for real frequency ($\Gamma=0$) one can integrate the two real-$k$ modes and the asymptotically decaying complex-$k$ mode starting from one of the two subsonic region only toward the supersonic region as in Sec.~\ref{appsec:3x3}, but without arriving to the second subsonic region, \ie\/ the integration must stop in the region between the horizons. In so doing, one obtains six modes $\phi_\om^{i,{\rm I}}$ and $\phi_\om^{i,{\rm III}}$, starting from the left and the right asymptotic regions, respectively [see Eqs.~\eqref{eq:phileft} and~\eqref{eq:phiright}]. Then, we project those six modes on the four plane waves solutions $\phi_\om^{i,{\rm II}}$ in the internal region%
\footnote{This is possible only if the distance between the horizon $2L$ is sufficiently large, such that there is a region where the velocity profile~\eqref{eq:velocity} is effectively flat, that is when $D\chor/\kw$ and $D\chor/\kb$ are sufficiently smaller then $L$.}
and extract the coefficients $M_{L,\om}^{ij}$ and $M_{R,\om}^{ij}$. This computation is stable also when $\Gamma$ is non-vanishing but small.

From this point on, the code does different things depending on whether the value of the frequency is real ($\Gamma=0$) or complex ($\Gamma\neq0$).

\begin{itemize}

\item For real frequencies ($\Gamma=0$), it is possible to construct in and out modes as $\phi_\om^{u,{\rm in}}$ [as in Eq.~\eqref{eq:expnov}, but without neglecting the $v$-mode in the internal region] by solving the system~\eqref{eq:system} with $L_\om^u=1$, $R_\om^v=0$, and computing the remaining four values of $L_\om^i$ and $R_\om^i$. From $L_\om^i$, $R_\om^i$, $M_{L,\om}^{ij}$, and $M_{R,\om}^{ij}$, the coefficients ${\cal A}_\om^{u}$, ${\cal A}_\om^{v}$, ${\cal B}_\om^{(1)}$, and ${\cal B}_\om^{(2)}$ are easily computed.
Following the evolution of $|{\cal B}_\om^{(2)}|^2$ with $\om$, it is possible to determine the complex eigenfrequencies $\om_a+\ii\Gamma_a$ by fitting a series of Lorentzians~\eqref{eq:lorentzian} as in Fig.~\ref{fig:lorentzians}.

\item If the mode has complex frequency, one may still, in principle, proceed as above. However, there is no point in computing the above coefficients when $\Gamma\neq0$, since, in general, $\lambda=\om+\ii\Gamma$ is not an eigenfrequency of the system. Instead, it is more interesting to solve the system~\eqref{eq:system} with initial conditions $L_{\lambda}^u=0$, $R_{\lambda}^v=0$, which correspond to physical solutions (vanishing at $x\to\pm\infty$ and $t\to-\infty$) with $\Gamma>0$, representing the unstable exponentially growing (in time) modes. With such initial condition, the system admits solutions if and only if its determinant vanishes. Given an initial guess for $\lambda_a=\om_a+\ii\Gamma_a$, the code computes the determinant of the system and varies $\om$ and $\Gamma$ to find the closest zero of the determinant. The resulting  $\om_a+\ii\Gamma_a$ is a precise estimate of the complex eigenfrequency.

\end{itemize}

The first method (first computing $|{\cal B}_\om^{(2)}|^2$ for real frequencies and then calculating $\om_a+\ii\Gamma_a$ by a fit) is less precise but allows to easily determine all the complex eigenfrequencies. The second method is much more precise and can be used to refine the estimates obtained with the first method and to follow the values of $\om_a$ and $\Gamma_a$, when one of the parameters of the system is changed in a continuous way. The plots of Figs.~\ref{fig:omega} and~\ref{fig:kw} have been realized by using the latter technique.

We conclude by stressing that both methods fail when $\om\to0$. While this limit would be interesting to be explored, our code cannot go to very small values of $\om$. In fact, at low frequency, the wavelength of some modes becomes very large and, when it is much larger than $2L$, $M_{L,\om}^{ij}$ and $M_{R,\om}^{ij}$ cannot be reliably determined.

\cleardoublepage

\makeatletter
\renewcommand\Hy@currentbookmarklevel{-2}
\makeatother

\phantomsection
\addcontentsline{toc}{chapter}{Bibliography}
\bibliographystyle{stefano}
\pagestyle{headernonumber}
\bibliography{thesis}

\end{document}